%% file: Blowup-main.tex
\documentclass[12pt,a4paper]{article}
\pdfoutput=1

\usepackage[T1]{fontenc}
\usepackage[nomath]{lmodern}
\usepackage{microtype}
\usepackage{amsmath}
\usepackage{amsfonts}
\usepackage{amssymb}
\usepackage{mathrsfs}
\usepackage{physics}
\usepackage{graphicx}
\usepackage{caption}
\usepackage{subcaption}
\usepackage{jheppub}
\usepackage{enumitem}
\usepackage{framed,float}
\usepackage{booktabs}
\usepackage{array}
\usepackage{soul} 
\usepackage{tikz}
\usetikzlibrary{calc,arrows,decorations.markings}
\usepackage{multirow}

\newcommand{\PE}{\operatorname{PE}}
\newcommand{\PLog}{\operatorname{PLog}}
\newcommand{\vol}{\operatorname{vol}}
\newcommand{\inner}[2]{\left\langle {#1}, {#2} \right\rangle}

\newcolumntype{L}[1]{>{\raggedright\let\newline\\\arraybackslash\hspace{0pt}}m{#1}}
\newcolumntype{C}[1]{>{\centering\let\newline\\\arraybackslash\hspace{0pt}}m{#1}}
\newcolumntype{R}[1]{>{\raggedleft\let\newline\\\arraybackslash\hspace{0pt}}m{#1}}

\title{Bootstrapping BPS spectra of 5d/6d field theories}

\author[a,b]{Hee-Cheol Kim,}
\author[a]{Minsung Kim,}
\author[c]{Sung-Soo Kim,}
\author[b]{Ki-Hong Lee}

\affiliation[a]{Department of Physics, POSTECH, Pohang 790-784, Korea}
\affiliation[b]{Asia Pacific Center for Theoretical Physics, Postech, Pohang 37673, Korea}
\affiliation[c]{School of Physics, University of Electronic Science and Technology of China,\\
Chengdu, Sichuan 611731, China}

\abstract{We propose a systematic approach to computing the BPS spectra of any 5d/6d supersymmetric quantum field theory in Coulomb phases, which admits either gauge theory descriptions or geometric descriptions, based on the Nakajima-Yoshioka's blowup equations. We provide a significant generalization of the blowup equation approach in terms of both properly quantized magnetic fluxes on the blowup $\hat{\mathbb{C}}^2$ and the effective prepotential for 5d/6d field theories on the Omega background which is uniquely determined by the Chern-Simons couplings on their Coulomb branches. We employ our method to compute BPS spectra of all rank-1 and rank-2 5d Kaluza-Klein (KK) theories descending from 6d $\mathcal{N}=(1,0)$ superconformal field theories (SCFTs) compactified on a circle with/without twist. We also discuss various 5d SCFTs and KK theories of higher ranks, which include a few exotic cases such as new rank-1 and rank-2 5d SCFTs engineered with frozen singularity as well as the 5d $SU(3)_8$ gauge theory currently having neither a brane web nor a smooth shrinkable geometric description. The results serve as non-trivial checks for a large class of non-trivial dualities among 5d theories and also as independent evidences for the existence of certain exotic theories.}

\begin{document}

\maketitle
\input{sec-intro.tex}

\input{sec-5dSCFTs.tex}

\input{sec-blowupEq.tex}

\input{sec-examples.tex}

\input{sec-conclusion.tex}

\input{sec-appendix.tex}

\input{sec-appendix2.tex}

\bibliographystyle{JHEP}
\bibliography{ref}
\end{document}

%% file: sec-intro.tex
\section{Introduction}\label{sec:intro}

Supersymmetric theories with eight supercharges in five and six dimensions are a very rich subject that has been investigated over the past decades. Lots of recent progress in this subject have been made in the classification of 5d and 6d superconformal theories (SCFTs) \cite{Heckman:2013pva, Heckman:2015bfa, Bhardwaj:2015xxa} and 6d little string theories (LSTs) \cite{Bhardwaj:2015oru}. Such classifications have been carried out based on geometric properties of F-theory compactified on local elliptic Calabi-Yau (CY) three-folds. Also, a large class of 5d SCFTs has been classified using gauge theoretic constructions \cite{Seiberg:1996bd, Intriligator:1997pq, Jefferson:2017ahm} and M-theory compactified on local CY 3-folds \cite{Douglas:1996xp, Intriligator:1997pq, DelZotto:2017pti, Xie:2017pfl, Jefferson:2018irk, Bhardwaj:2018yhy, Bhardwaj:2018vuu, Apruzzi:2018nre,Apruzzi:2019vpe, Apruzzi:2019opn, Bhardwaj:2019fzv, Apruzzi:2019kgb, Apruzzi:2019enx}. These higher dimensional theories turn out to exhibit several fascinating features of quantum field theories (QFTs) such as the existence of tensionless strings, dualities, and symmetry enhancements. Moreover, they have played a pivotal role in constructing and studying lower dimensional QFTs via compactifications.

Supersymmetric partition functions have provided unexpectedly powerful methods for exploring such features of higher dimensional field theories. Of particular interest are the partition functions on $\mathbb{R}^4\times S^1$ in 5d and $\mathbb{R}^4\times T^2$ (or the elliptic genera) in 6d, which are the Witten indices on the $\Omega$-deformed $\mathbb{R}^4$ computing the number of the BPS states weighted by their electric charges and angular momenta. Throughout the paper, we refer to them as the BPS partition functions, though it may be an abuse of notation. Protected by the supersymmetry, the BPS spectra encoded in the partition function can be used to confirm non-trivial dualities and symmetry enhancements at various points on the moduli space \cite{Bao:2011rc,Hayashi:2013qwa,Taki:2014pba,Mitev:2014jza,Hayashi:2014wfa,Gaiotto:2015una,Hayashi:2015xla,Hayashi:2016abm,Yun:2016yzw,Hayashi:2017jze,Kim:2018gjo,Chen:2020jla}. Also, the Seiberg-Witten prepotential in the Coulomb phase can be reproduced from the BPS partition function by taking the limit $\epsilon_1,\epsilon_2\rightarrow 0$ of two $\Omega$-deformation parameters \cite{Nekrasov:2002qd,Nekrasov:2003rj}. In particular, other supersymmetric observables such as the superconformal index \cite{Bhattacharya:2008zy,Kim:2012gu,Kim:2013nva,Iqbal:2012xm} and the $S^5$ partition function \cite{Kallen:2012va,Kim:2012ava,Kallen:2012zn,Lockhart:2012vp,Imamura:2013xna,Kim:2012qf} can be factorized into a product of several BPS partition functions under localization, which signifies importance of the partition function on the $\Omega$-background as a building block for other observables in 5d and 6d.

Various computational tools for the 5d and 6d BPS partition functions have been developed. First, the partition function on the $\Omega$-background for 5d $\mathcal{N}=1$ gauge theories is closely related to Nekrasov instanton partition function counting the BPS bound states of instantons with other charged particles on the Coulomb branch of the moduli space. The Nekrasov instanton partition functions for gauge theories with the classical gauge groups are computed through localization based on the ADHM constructions of the instanton moduli space \cite{Atiyah:1978ri,Nekrasov:2002qd,Nekrasov:2003rj}. (See also \cite{Marino:2004cn,Nekrasov:2004vw,Fucito:2004gi,Hwang:2014uwa} for various generalizations). The ADHM constructions have also been used to compute the elliptic genera of the self-dual strings in 6d SCFTs (in the tensor branch) \cite{Haghighat:2013tka,Kim:2014dza, Haghighat:2014vxa,Gadde:2015tra}. Even though the ADHM method is systematic and applicable for many 5d and 6d gauge theories with classical gauge groups, ADHM construction for the exceptional gauge groups is however still missing.%
\footnote{The ADHM-like construction for the moduli space of instantons in the $G_2$ gauge theory with fundamental matters was proposed in \cite{Kim:2016foj,Kim:2018gjo}.}
Moreover, the ADHM method is not applicable when a gauge theory of classical group is coupled to a large number of matter fields in generic representations. Hence, there are still challenges and difficulties in the ADHM method when computing the BPS partition functions for other more generic gauge theories in 5d and in 6d.

Topological vertex method \cite{Aganagic:2003db,Iqbal:2007ii,Awata:2008ed}\footnote{See also the Ding-Iohara-Miki (DIM) algebra \cite{Ding:1996mq, doi:10.1063/1.2823979} and its relation to topological vertex \cite{Aganagic:2012hs, Awata_2012,Bourgine:2017jsi}.} provides yet another systematic way of computing the BPS partition functions when the 5d/6d theories are realized by Type IIB 5-brane webs that are toric or toric-like \cite{Aharony:1997bh, Benini:2009gi}, by which we mean those 5-brane webs which can be constructed from toric 5-brane webs through Higgsing procedures \cite{Hayashi:2013qwa}. Topological vertex has been further developed so that it is also applicable for 5-brane webs with $O5$-planes \cite{Kim:2017jqn, Hayashi:2020hhb} or $ON$-planes \cite{Bourgine:2017rik, Kim-Wei2020}. Though fairly many 5-brane webs are known or discovered for 5d theories and for 6d theories on a circle \cite{Kim:2015jba, Hayashi:2015fsa, Bergman:2015dpa, Zafrir:2015rga, Hayashi:2015zka, Hayashi:2015vhy, Zafrir:2015ftn, Zafrir:2016jpu, Hayashi:2018lyv, Hayashi:2019yxj, Kim:2019dqn}, there still remain challenges when the number of hypermultiplets is large or the Chern-Simons (CS) level is high. For instance, 5-brane webs for the 5d $SU(3)_8$ and $SU(4)_8$ gauge theories are unknown, and 5-brane web realizations for gauge theories of exceptional groups are still far from clear except for $G_2$ gauge theories \cite{Hayashi:2018bkd,Kim:2019dqn}. 

In \cite{Nakajima:2003pg}, Nakajima-Yoshioka formulated the so-called blowup equations to compute the Nekrasov instanton partition functions for four-dimensional $\mathcal{N} = 2\ SU(N)$ gauge theories. The K-theoretic blowup equations were established soon in \cite{Nakajima:2005fg, Gottsche:2006bm} for five-dimensional $\mathcal{N} = 1\ SU(N)$ gauge theories. This blowup equation approach has been further generalized to compute the instanton partition functions for other simple gauge groups including exceptional gauge groups in 5d in \cite{Keller:2012da,Kim:2019uqw}. Moreover, a geometric generalization of the blowup equation approach was formulated for certain local Calabi-Yau 3-folds \cite{Gu:2017ccq, Huang:2017mis}. This geometric formulation has been extensively studied recently for computing refined BPS invariants of various 5d SCFTs and elliptic genera of 6d SCFTs admitting geometric constructions in M-/F-theory on local (toric or elliptic) CY 3-folds \cite{Gu:2017ccq, Huang:2017mis, Gu:2018gmy, Gu:2019dan, Gu:2019pqj, Gu:2020fem}.

Despite the fact that the blowup method is a very powerful tool for computing BPS spectra of a broad class of 5d/6d SCFTs even beyond the scope of other computational methods, there is as yet no complete formulation of the blowup equations that can be applicable for all supersymmetric theories in five and six dimensions. For example, though the blowup equations obtained in \cite{Kim:2019uqw} cover a large set of 5d gauge theories with simple gauge group including some theories whose ADHM descriptions are not known, it still requires further studies for plenty of other interesting theories such as 5d quiver gauge theories, 5d theories with half-hypermultiplets, and 5d Kaluza-Klein (KK) theories arising from 6d SCFTs compactified on a circle with outer-automorphism twists. In particular, there exist some 5d/6d gauge theories, for instance, the 5d $SU(3)_8$ gauge theory, that currently have neither ADHM constructions nor shrinkable geometric descriptions nor associated 5-brane webs, so that all the computational methods introduced above including the blowup method cannot be used to compute their BPS partition functions.

The main aim of this paper is to devise a complete blowup formalism that enables one to compute BPS spectra of {\it all} supersymmetric field theories having UV completions in five or six dimensions. In this paper, we will generalize the Nakajima-Yoshioka's blowup equations in \cite{Nakajima:2003pg} to arbitrary 5d/6d gauge theories (including quiver gauge theories and twisted circle compactifications of 6d theories) with matter fields in arbitrary representations, and also extend the work of \cite{Gu:2017ccq,Huang:2017mis} for a geometric application of the blowup equations to any local Calabi-Yau 3-folds (including elliptic and non-toric ones) based on novel geometric constructions of 5d SCFTs/KK theories introduced in \cite{Jefferson:2018irk,Bhardwaj:2019fzv}.

The blowup equation is a functional equation identifying two partition functions on different backgrounds, one is the $\Omega$-deformed $\mathbb{C}^2$ and another one is the one point blow-up $\hat{\mathbb{C}}^2$ of the $\mathbb{C}^2$, that are related to each other by a smooth blow-up or blow-down transition. Each 5d/6d field theory can have a number of blowup equations depending on background magnetic fluxes residing on the blown-up $\mathbb{P}^1$ in $\hat{\mathbb{C}}^2$. As we will discuss in section~\ref{sec:BlowupEq}, such blowup equations can be solved recursively by expanding them in terms of K\"ahler parameters of the theory.

The main input in our recursion process is the {\it effective prepotential $\mathcal{E}$} of a given 5d/6d SQFT evaluated on the $\Omega$-background. The effective prepotential, as we will illustrate more precisely in section~\ref{sec:5dthyOmega}, is fully determined by effective cubic and mixed Chern-Simons terms in the low energy theory on the Coulomb branch which can be systematically calculated by computing the induced Chern-Simons terms after integrating out charged fermions \cite{Witten:1996qb} (and also by collecting classical Green-Schwarz contributions for a 6d theory on a circle with/without a twist \cite{Bhardwaj:2019fzv}). See \cite{Maldacena:1997de,Katz:2020ewz} for the geometric counterparts of the effective Chern-Simons terms. Therefore, the effective prepotential, which is one of the main ingredients for our blowup formula, can be systematically computed for every 5d/6d field theory having a UV completion. Here we assume that every UV finite 5d/6d theory either has a gauge theory description in 5d or in 6d on a circle with/without twist, or has a geometric description as a local (elliptic) Calabi-Yau 3-fold, or can be obtained by RG-flows thereof.

By seeding the effective prepotential as well as a consistent choice of background magnetic fluxes on the blown-up $\mathbb{P}^1$ into the blowup equations, we can {\it bootstrap} the spectrum of charged BPS states in a given 5d/6d theory. Since the effective prepotential can be easily prepared for any arbitrary 5d/6d field theory, we now make a bold conjecture that we can compute BPS spectra of all 5d/6d field theories by employing our bootstrap method to formulate and solve blowup equations in those theories.

As concrete examples for our conjecture, we will apply our method to all rank-1 and rank-2 5d supersymmetric field theories including also KK theories to obtain their BPS spectra explicitly. This will involve the $SU(3)_{15/2}+1{\bf F}$ theory (dual to the $\mathcal{N}=2$ $G_2$ gauge theory), the $SU(3)_8$ theory, and new rank-1 and rank-2 5d SCFTs, which we call the local $\mathbb{P}^2+1{\bf Adj}$ and the local $\mathbb{P}^2\cup \mathbb{F}_3+1{\bf Sym}$, obtained by mass deformations of the 5d $\mathcal{N}=2$ $Sp(N)_\pi$ gauge theories with $N=1,2$ respectively first introduced in \cite{Bhardwaj:2019jtr}. We emphasize that the partition functions of these theories cannot be obtained by other means since they have none of ADHM constructions, conventional geometric constructions, and also brane webs (but we will introduce brane webs for the new rank-1 and rank-2 theories in this paper).

We will also compute BPS spectra of some higher rank theories. In particular, we will show that the blowup equations for the $SU(5)_8$ theory can be solved. The result shows that this theory may have no physical Coulomb branch and thus be inconsistent in UV limit. This theory was `undetermined' to exist in \cite{Bhardwaj:2020gyu} because its existence was neither confirmed nor ruled out with currently known techniques. Our computation provides a supporting evidence that the $SU(5)_8$ theory has no UV completion. In this sense, our bootstrap approach can be used to confirm or disprove the existence of certain 5d and 6d QFTs. \bigskip

The organization of the paper is as follows. In section~\ref{sec:5dthyOmega}, we review salient features of 5d/6d supersymmetric gauge theories and their geometry engineering. In section~\ref{sec:BlowupEq}, we explain the blowup equation as a tool for bootstrapping BPS spectra of 5d/6d supersymmetric theories, and discuss our main conjecture with instructive examples. Section~\ref{sec:rank1theories} and section~\ref{sec:rank2 theories} are devoted to cover all rank-1 and rank-2 5d theories including KK theories. We also discuss some interesting higher rank theories, in section~\ref{sec:higher rank theories}, including $SU(4)_8$ and $SU(5)_8$. We then conclude with some subtle issues. In Appendix~\ref{appendix:1-loop}, we further discuss 1-loop partition functions of 6d SCFTs on a circle with twists. In Appendix~\ref{sec:appendix2}, some new 5-brane webs associated with frozen singularity are presented. \bigskip

\paragraph{Notation:} To avoid the cluttering of theories, we denote by $G+N_{\mathbf{r}}\,\mathbf{r}$, the theory of gauge group $G$ with $N_\mathbf{r}$ number of hypermultiplets in the representation $\mathbf{r}$. Gauge group $G$ can be any classical groups, $SU(N)$, $SO(N)$, $Sp(N)$, and exceptional groups $G_2$, $F_4$, $E_6$, $E_7$, $E_8$ as well as a quiver gauge group. For hypermultiplets in the $\mathbf{r}$ representation of $G$, we use the following shorthand notation: $\mathbf{F}$ for fundamental, $\mathbf{bi\text{-}F}$ for bi-fundamental, $\mathbf{\Lambda}^n$ for rank-$n$ antisymmetric, $\mathbf{Sym}$ for symmetric, $\mathbf{Adj}$ for adjoint, $\mathbf{S}$ for spinor, and $\mathbf{C}$ for conjugate spinor. For example, $SU(2)+8\mathbf{F}$ means $SU(2)$ gauge theory with 8 hypermultiplets in the fundamental representation. For the $\Omega$-deformation parameters $\epsilon_{1}, \epsilon_{2}$, we frequently use $\epsilon_+=\frac{\epsilon_1+\epsilon_2}{2}$ and the fugacities associated with them are denoted by $p_1 =e^{-\epsilon_1}$ and $p_2= e^{-\epsilon_2}$. We also denote the set of complex, real, rational, integer numbers by $\mathbb{C}$, $\mathbb{R}$, $\mathbb{Q}$, $\mathbb{Z}$, respectively.

%% file: sec-5dSCFTs.tex
\section{\texorpdfstring{5d theories on $\Omega$-background}{5d theories on Omega-background}}\label{sec:5dthyOmega}

In this section, we review some basic properties of 5d $\mathcal{N}=1$ QFTs that have UV completions.

\subsection{Gauge theories and effective prepotential}

A large class of 5d SQFTs admits mass deformations that lead to non-Abelian gauge theory descriptions at low energy. Consider a 5d $\mathcal{N}=1$ gauge theory with a non-Abelian gauge group $G$. The theory consists of the vector multiplet $\Phi$ for the gauge group $G$ and charged matter hypermultiplets. The vector multiplet contains a real scalar field $\phi$ as well as the vector field $A_\mu$. On the Coulomb branch of the moduli space, the scalar field $\phi$ gets the expectation value in the Cartan subalgebra of the gauge group $G$. This breaks the gauge group to its Abelian subgroup $U(1)^r$, with $r={\rm rank}(G)$. Then the low-energy theory is described by an effective theory with the Abelian groups. The scalar expectation values $\phi^i, i=1,\cdots, r$ for the Abelian gauge groups parameterize the Coulomb branch of the moduli space. 

The effective Abelian theory is characterized by a prepotential $\mathcal{F}(\Phi)$ which is a cubic polynomial in the Abelian vector multiplets $\Phi^I$ for both the dynamical gauge symmetry and the non-dynamical flavor symmetry where the index $I$ labels both the dynamical and the background vector multiplets. The exact prepotential can be computed by 1-loop calculations. For a general gauge group $G$ and matter hypermultiplets in generic representations, the cubic prepotential in terms of the scalar components is given by \cite{Witten:1996qb, Intriligator:1997pq}
\begin{equation}\label{eq:preF}
	\mathcal{F} \!=\! \sum_a\!\Big(\frac{m_a}{2}K^a_{ij}\phi^a_i\phi^a_j + \frac{\kappa_a}{6}d^a_{ijk}\phi^a_i\phi^a_j\phi^a_k\Big) +\frac{1}{12}\bigg(\sum_{e\in{\bf R}}|e\cdot \phi|^3-\!\sum_f\!\sum_{w\in{\bf w}_f}|w\cdot\phi+m_f|^3\!\bigg),
\end{equation}
where $a$ runs over all non-Abelian subgroups $G_a\subset G$. Here, $m_a= 1/g_a^2$ is the inverse gauge couplings squared and $\kappa_a$ is the classical Chern-Simons level, which is non-zero only for $G_a=SU(N)$ with $N\ge3$, for the group $G_a$. $K^a_{ij}={\rm Tr} (T^a_iT^a_j)$ is the Killing form of $G_a$ and $d^a_{ijk}=\frac{1}{2}{\rm Tr}T^a_i\{T^a_j,T^a_k\}$ with the generator $T^a_i$ in the fundamental representation of $G_a$. ${\bf R}$ and ${\bf w}_f$ are the roots and the weights for the $f$-th hypermultiplet with mass $m_f$ of $G$, respectively. The mass parameters $m_a$ and $m_f$ can be regarded as the scalar components in the background vector multiplets for the topological symmetries and the flavor symmetries rotating hypermultiplets respectively. We note from the prepotential that the Coulomb branch is divided into distinct sub-chambers (or phases) distinguished by the signs of masses inside absolute values in \eqref{eq:preF}, and accordingly the prepotential takes different values in the different sub-chambers.

The prepotential $\mathcal{F}$ determines the gauge kinetic terms in the effective action with the gauge coupling 
\begin{equation}
	\tau_{ij}^{\rm eff} = (g^{-2}_{\rm eff})_{ij} = \partial_i \partial_j \mathcal{F} \ , 
\end{equation}
which sets the metric on the Coulomb branch and the cubic Chern-Simons terms of the form
\begin{equation}\label{eq:cubic-CS}
	S_{CS}=\frac{C_{IJK}}{24\pi^2}\int A^I\wedge F^J\wedge F^K \ , \quad C_{IJK} = \partial_I\partial_J\partial_K \mathcal{F} \ ,
\end{equation}
with the level $C_{IJK}$ quantized as $C_{IJK}\in \mathbb{Z}$ due to gauge invariance of the Abelian symmetries \cite{Witten:1996qb}.

Other topological terms in the effective action are also important in our discussion later on the blowup equations. First, the effective action contains the mixed gauge/gravitational Chern-Simons terms of the form, \cite{Bonetti:2011mw,Bonetti:2013ela,Grimm:2015zea}
\begin{equation}\label{eq:grav-CS}
	S_{\rm grav} = -\frac{1}{48}\int C^G_i A^i\wedge p_1(T) \ ,
\end{equation}
where $p_1(T)$ is the first Pontryagin class of the tangent bundle on the 5d spacetime. Here $C_i^G$ is the level for the mixed Chern-Simons term and it is quantized as $C_i^G\in \mathbb{Z}$ \cite{Chang:2019uag,Katz:2020ewz}. The mixed Chern-Simons terms are induced at low energy by integrating out the charged fermions. The induced level from the fermion 1-loop calculations is \cite{Bonetti:2013ela,Grimm:2015zea}
\begin{equation}
	C_i^G = -\partial_i\bigg(\sum_{e\in {\bf R}}|e\cdot \phi| - \sum_f\sum_{w\in{\bf w}_f}|w\cdot\phi+m_f|\bigg) \ .
\end{equation}

There also exists the mixed gauge/$SU(2)_R$ Chern-Simons terms of the form,
\begin{equation}\label{eq:R-CS}
	S_{R} = \frac{1}{2}\int C_i^R A^i\wedge c_2(R)\ ,
\end{equation}
where $c_2(R)$ is the second Chern class of the $SU(2)_R$ R-symmetry bundle. Due to the gauge invariance, the level $C^R_i$ is quantized as $C^R_i\in 2\mathbb{Z}$. Note that the gauginos in the vector multiplets are doublets, while the matter fermions are singlets under the $SU(2)_R$ R-symmetry. Thus this term receives 1-loop contributions only from the charged gauginos and therefore it is independent of the number of hypermultiplets. The mixed gauge/$SU(2)_R$ Chern-Simons level induced from the gaugino 1-loop calculation is \cite{BenettiGenolini:2019zth}
\begin{align}\label{eq:CS-R}
	C^R_i &= \frac{1}{2} \partial_i \sum_{e\in {\bf R}}|e\cdot \phi| \ .
\end{align}
We remark here that, in the {\it Dynkin basis} where the rows of the Cartan matrix $A_{ij}$ for a gauge group $G$ are given by the simple roots, this level in the low-energy effective theory is fixed to be $C^R_i=2$ for all $i$'s.

In this paper, we are interested in the partition functions of 5d $\mathcal{N}=1$ theories on $\Omega$-deformed $\mathbb{R}^4\times S^1$. This partition function is a Witten index counting BPS states in the 5d theory, which is defined as\footnote{We can also define another Witten index as $\hat{Z}(\phi,m;\epsilon_1,\epsilon_2) \equiv Z(\phi,m;\epsilon_1,\epsilon_2)|_{(-1)^F\rightarrow (-1)^{2J_R}}$ by replacing $(-1)^F$ in $Z$ by $(-1)^{2J_R}$. This index will be used later when a 5d theory is put on the blowup $\hat{\mathbb{C}}^2$.} \cite{Nekrasov:2002qd}
\begin{equation}\label{eq:Z}
	Z(\phi,m;\epsilon_1,\epsilon_2) = {\rm Tr}\left[(-1)^Fe^{-\beta\{Q,Q^\dagger\}}e^{-\epsilon_1(J_1+J_R)}e^{-\epsilon_2(J_2+J_R)}e^{-\phi\cdot \Pi}e^{-m\cdot H}\right] \ ,
\end{equation} 
where $J_1,J_2$ are the Cartan generators of the $SO(4)$ Lorentz rotation and $J_R$ is the Cartan of the $SU(2)_R$ R-symmetry, and $\Pi$ and $H$ are the gauge and the flavor charges respectively. $Q$ is a supercharge commuting with $J_1+J_R$ and $J_2+J_R$, and $Q^\dagger$ is its conjugate. $\beta$ is the radius of $S^1$ and $\epsilon_1,\epsilon_2$ are the $\Omega$-deformation parameters. We denote by $\phi$ and $m$ the chemical potentials for the gauge and the flavor symmetries, respectively. The index computes the BPS spectrum annihilated by the supercharges $Q$ and $Q^\dagger$. So the index is independent of $\beta$.

The index can be represented by a path integral of the 5d theory on the $\Omega$-background, which can be evaluated using localization \cite{Kim:2011mv,Kim:2012gu,Hwang:2014uwa}. We will call this path integral representation of the Witten index the partition function on the $\Omega$-background or just {\it BPS partition function}.

We compute the path integral on a vacuum on the Coulomb branch specified by the expectation values $\phi$, which are now complexified by combining the scalar vevs with the gauge holonomies around $S^1$ at infinity of $\mathbb{R}^4$. The chemical potential $\phi$ in the above index is identified with the complexified expectation value $\phi$ in the path integral. Similarly, we identify the chemical potential $m$ with the complexified background gauge field for a flavor symmetry. In the following discussions, however, we shall take the chemical potentials $\phi$ and $m$ to be pure real values.

In the localization, the BPS partition function receives perturbative and non-perturbative instanton contributions which factorizes as
\begin{equation}
	Z=Z_{\rm pert}\cdot Z_{\rm inst} \ .
\end{equation}
The perturbative partition function $Z_{\rm pert}$ consists of the classical action contribution and the 1-loop contribution. Actually, it depends on the boundary condition at infinity of $\mathbb{R}^4$. We need to consider boundary conditions preserving two supercharges $Q,Q^\dagger$ and being compatible with the vacuum on the Coulomb branch. We shall choose the following boundary condition at infinity: the vector multiplets associated with the positive roots of gauge group $G$ survives, and the chiral halves of hypermultiplets with positive masses, {i.e.,} $w\cdot \phi +m_f>0$ survive. With this boundary condition, the perturbative partition function can be written as
\begin{align}\label{eq:Zpert=cx1-loop}
	Z_{\rm pert} &= Z_{\rm class}\cdot Z_{\text{1-loop}} \\
	&= e^{\mathcal{E}} \cdot {\rm PE}\bigg[-\frac{1+p_1p_2}{(1-p_1)(1-p_2)}\!\sum_{e\in{\bf R}^+}\!e^{-e\cdot\phi}+\frac{(p_1p_2)^{1/2}}{(1-p_1)(1-p_2)}\sum_f\!\sum_{w\in{\bf w}_f}e^{-|w\cdot\phi+m_f|}\bigg], \nonumber
\end{align}
where $p_{1,2}\equiv e^{-\epsilon_{1,2}}$ and ${\bf R}^+$ denotes the positive roots for the gauge group, and PE means the Plethystic exponential of a letter index $f(\mu)$ with a chemical potential $\mu$ defined as
\begin{equation}
	\PE\left[f(\mu)\right] \equiv \exp\bigg(\sum_{n=1}^\infty \frac{1}{n}f(n\mu)\bigg) \ .\label{eq:PE}
\end{equation}

In \eqref{eq:Zpert=cx1-loop}, $\mathcal{E}=\mathcal{E}(\phi,m;\epsilon_1,\epsilon_2)$ in the prefactor is a combination of the classical contribution and the Casimir energy contribution coming from regularization of infinite products in the 1-loop part. In fact, $\mathcal{E}$ is {\it effective prepotential}, which includes the cubic and mixed Chern-Simons terms \eqref{eq:cubic-CS}, \eqref{eq:grav-CS}, \eqref{eq:R-CS}, and their SUSY completions, evaluated on the $\Omega$-background. We find, for the boundary condition we chose above,
\begin{align}\label{eq:E-func}
	\mathcal{E}(\phi,m;\epsilon_1,\epsilon_2)&=i\left(S_{CS}+S_{\rm grav}+S_R + \cdots \right)|_{\phi, m,\epsilon_1,\epsilon_2} \nonumber\\
	&= \frac{1}{\epsilon_1\epsilon_2}\left[\mathcal{F}+\frac{1}{48}C_i^G \phi^i (\epsilon_1^2+\epsilon_2^2)+\frac{1}{2}C^R_i \phi^i \epsilon_+^2\right] \ ,
\end{align}
where $\cdots$ denotes the SUSY completions of the Chern-Simons terms and $\epsilon_+\equiv \frac{\epsilon_1+\epsilon_2}{2}$. The first term $\mathcal{F}$ in the bracket is the cubic prepotential \eqref{eq:preF} on the Coulomb branch, and the other two terms are the contributions from the mixed gauge/gravitational CS terms and the mixed gauge/$SU(2)_R$ CS terms, respectively. 	This factor $\mathcal{E}$ can also be considered as an equivariant integral of effective Chern-Simons terms by making the replacements
\begin{equation}
	p_1(T)\rightarrow -(\epsilon_1^2+\epsilon_2^2) \ , \qquad c_2(R) \rightarrow \epsilon_+^2 \ ,
\end{equation}
with the equivariant parameters $\epsilon_{1,2}$ and $\phi,m$. A similar interpretation for the Casimir energy of superconformal indices has been proposed in \cite{Bobev:2015kza}.

The instanton contribution $Z_{\rm inst}$ is in general given by a power series expansion by
the instanton numbers $k_a$ for each non-Abelian gauge group factor. It can thus be written as
\begin{equation}
	Z_{\rm inst} = \sum_{k_a=0}^\infty \prod_aq_a^{k_a} Z_{\{k_a\}} \ ,
\end{equation}
where $q_a=e^{-m{}_a}$ is the instanton fugacity for the $a$-th gauge group, and $Z_{\{k_a\}}$ denotes the path integral over instanton moduli space with instanton numbers $\{k_a\}$. When a UV completion for the instanton moduli space is known, for example by using ADHM construction (see \cite{Nekrasov:2002qd,Nekrasov:2003rj,Nekrasov:2004vw,Shadchin:2005mx} for some early works), we can use it to compute the instanton partition function $Z_{\{k_a\}}$ by localization. However, unfortunately, such ADHM constructions for general gauge group $G$ and matter representations are yet unknown. 

Recently, there has been some progress on computation of the instanton partition functions for more general gauge groups by using Nakajima-Yoshioka's blowup equations \cite{Nakajima:2003pg}. See also \cite{Huang:2017mis,Gu:2018gmy,Kim:2019uqw} and the references therein. Still, this method is applicable only for very limited cases. In this paper, we will propose a new and simple strategy to compute the instanton partition functions for arbitrary gauge groups and matter representations based on the blowup formula. We expect that our strategy can be applied to {\it all} the 5d gauge theories that have UV completions as 5d SCFTs or 5d KK theories coming from 6d $\mathcal{N}=(1,0)$ theories on a circle with/without twists.

\subsection{Geometric engineering}

Many examples of 5d $\mathcal{N}=1$ theories have been engineered in M-theory on local Calabi-Yau threefolds \cite{Morrison:1996xf,Douglas:1996xp,Intriligator:1997pq}. In this subsection we review some basic features of M-theory compactification on a smooth non-compact 3-fold $X$ which gives rise to a 5d SCFT or a 6d SCFT on $S^1$ (possibly with twists) in a singular limit. See \cite{Jefferson:2018irk,Bhardwaj:2019fzv} for more details.

A smooth 3-fold $X$ can be locally described as a neighborhood of a collection of K\"ahler surfaces $S_i$. A K\"ahler surface $S_i$ inside $X$ is represented by either a local $\mathbb{P}^2$ or a Hirzebruch surface with blowups $\mathbb{F}_{n}^b$ where $n$ is the degree of the Hirzebruch surface and $b$ is the number of blowups. More explanation on the local $\mathbb{P}^2$ and Hirzebruch surfaces can be found in Appendix A of \cite{Bhardwaj:2019fzv}.

The volumes of complex $p$-cycles in $X$ are controlled by K\"ahler deformations. There are normalizable K\"ahler deformations parameterized by dynamical K\"ahler parameters $\phi_i$ assigned for each compact surface $S_i$. The number $r$ of independent compact surfaces is the rank of the CY 3-fold $X$. Upon M-theory compactification, the K\"ahler moduli space of the dynamical parameters $\phi_{i=1,\cdots,r}$ is identified with the Coulomb branch moduli in the low-energy 5d theory, where $r$ is the rank of the gauge group in the field theory.

There are also non-compact K\"ahler deformations parameterized by non-dynamical K\"ahler parameters $m_{j=1,\cdots,r_F}$ where $r_F=h^{1,1}(X)-r$. These non-dynamical parameters are identified with the mass parameters in the 5d theory. For a given basis $S_i,N_j \in H^{1,1}(X)$, one can then express the K\"ahler form $J$ of $X$ as a linear sum of the K\"ahler parameters
\begin{equation}
	J = \sum_{I=1}^{h^{1,1}(X)}\phi_I D_I= \sum_{i=1}^r\phi_i S_i + \sum_{j=1}^{r_F} m_j N_j \ ,
\end{equation}
where $D_{I=1,\cdots,r}=S_{i=1,\cdots,r}$ and $D_{I=r+1,\cdots, h^{1,1}(X) }=N_{j=1,\cdots,r_F}$ are the divisors for the compact and the non-compact 4-cycles inside $X$, respectively. In particular, the K\"ahler parameters $\phi_i$ for elementary surfaces $S_i$ in this geometric basis are directly mapped to the Coulomb branch parameters $\phi_i$ in the Dynkin basis of the associated gauge group $G$. We will only use the Dynkin basis for gauge groups in the following discussions.

The volumes of $p$-cycles in $X$ are measured with respect to the K\"ahler form $J$. The total volume of the 3-fold $X$ is 
\begin{equation}
	{\rm vol}(X) = \frac{1}{3!}\int_X J^3  \ .
\end{equation}
This is identified with the 5d cubic prepotential given in \eqref{eq:preF}, i.e. $\mathcal{F}={\rm vol}(X)$. Therefore the cubic Chern-Simons coefficients in the 5d theory are geometrically determined by the triple intersections of divisors in $X$ \cite{Cadavid:1995bk,Ferrara:1996hh,Witten:1996qb,Ferrara:1996wv} (See also \cite{Grimm:2011fx,Bonetti:2011mw,Bonetti:2013ela})
\begin{equation}
	C_{IJK} = D_I\cdot D_J\cdot D_K \equiv \int_X D_I\wedge D_J\wedge D_K \ .
\end{equation}

The low-energy effective action in the 5d theory is also characterized by the coefficients of the mixed Chern-Simons terms. The mixed gauge/gravitational Chern-Simons level for a divisor $S_i$ is determined by its intersection with $c_2(X)\in H^4(X,\mathbb{Z})$, the 2nd Chern-class of the 3-fold $X$, as \cite{Cadavid:1995bk,Ferrara:1996hh,Bonetti:2011mw,Bonetti:2013ela,Katz:2020ewz}
\begin{equation}
	C_i^G = c_2(X)\cdot S_i \ .\label{eq:C_i}
\end{equation}
For a local $\mathbb{P}^2$ and a Hirzebruch surface $\mathbb{F}_n^{b}$ with $b$ blowups,
\begin{align}
	c_2(X)\cdot \mathbb{P}^2 = -6, \qquad c_2(X)\cdot \mathbb{F}_n^{b} = -4+2b \, .
\end{align}

In addition, we {\it propose} that the level $C^R_i$ of the mixed gauge/$SU(2)_R$ Chern-Simons term is always
\begin{equation}
	C_i^R = 2 \ ,
\end{equation}
for {\it all} the basis surfaces $S_i$ represented by either a local $\mathbb{P}^2$ or a Hirzebruch surface with blowups, in a Calabi-Yau 3-fold. This is motivated by the field theory result in \eqref{eq:CS-R}; the level $C^R_i$ is always $2$ in the Dynkin basis of the gauge groups.

Therefore the cubic and the mixed Chern-Simons terms in the low-energy effective field theory are fully expressed in terms of the topological data in the CY $3$-fold $X$. The effective prepotential $\mathcal{E}$ in \eqref{eq:E-func} for a 5d field theory can then be readily computed from the associated 3-fold $X$. As we will see below, this effective prepotential $\mathcal{E}$ together with few more information about primitive 2-cycles in a CY $3$-fold allows us to compute the BPS partition function of the corresponding 5d theory by solving the blowup equations.

The BPS spectrum in the 5d SQFT involves electric particles and (dual) magnetic strings charged under the gauge groups. In the M-theory compactification on $X$, these states arise from M2-branes and M5-branes wrapping holomorphic curves and holomorphic surfaces respectively. Their masses and tensions are determined by the volumes of the corresponding $p$-cycles. The volume of a 2-cycle (or a curve) $C$ is
\begin{equation}
	{\rm vol}(C) = -J\cdot C \ , 
\end{equation}
and the volume of a 4-cycle $S_i$ is given by
\begin{equation}
	{\rm vol}(S_i) = J\cdot J\cdot S_i \equiv \partial_i\mathcal{F}= \frac{1}{2}\int_X J^2\wedge S_i \ .
\end{equation}

The K\"ahler surfaces are glued to each other by identifying a pair (or multiple pairs) of holomorphic curves at the intersections as
\begin{equation}
	C_{ij}^\alpha \sim C_{ji}^\alpha \ ,
\end{equation}
where $C_{ij}^\alpha$ is a curve in $S_i$ and $C_{ji}^\alpha$ is a curve in $S_j$ at the intersection of two adjacent surfaces $S_i\cap S_j$, and $\alpha$ labels a pair of gluing curves. In order to be consistent with the Calabi-Yau structure of $X$, a pair of gluing curves should satisfy the condition
\begin{equation}\label{eq:CY-cond}
	(C^\alpha_{ij})^2 + (C^\alpha_{ji})^2 = 2g-2 \ ,
\end{equation} 
where $g$ is the genus of curve $C_{ij}^\alpha$ and $C_{ji}^\alpha$. 

It is also possible that two curves in a single surface are glued together while satisfying the Calabi-Yau condition \eqref{eq:CY-cond}. Such gluing is often called {\it self-gluing} \cite{Jefferson:2018irk}. A surface can have multiple self-glued curves. With $s$ self-gluings, the canonical class $K_S$ of a surface $S$ changes to
\begin{equation}
	K_S' = K_S + \sum_{i=1}^s (x_i+y_i) \ ,
\end{equation}
where $(x_i,y_i)$ is $i$-th pair of self-glued curves.

The volume of a 2-cycle can in fact be written in terms of its intersection with the canonical classes $K'_S$ as
\begin{equation}
	{\rm vol}(C) = -J\cdot C = -\sum_I\phi_I D_I\cdot C = -\sum_I\phi_I K_{D_I}' \cdot C \ .
\end{equation}
Also, the genus $g$ of a curve $C$ in a surface $S$ can be determined by the modified adjunction formula \cite{Bhardwaj:2019fzv,Bhardwaj:2020gyu}:
\begin{equation}
	C\cdot (K_S +C)+\sum_{i=1}^s{\rm min}(C\cdot x_i,C\cdot y_i)=2g-2\ ,
\end{equation}
where $K_S$ is the original canonical divisor class of $S$.

We can now easily compute the triple intersection product $S_i\cdot S_j \cdot S_k$ among three surfaces in a 3-fold $X$. For three distinct surfaces, their intersection product is given by
\begin{equation}
	S_i\cdot S_j\cdot S_k = C_{ij} \cdot C_{ik} = C_{ji} \cdot C_{jk} = C_{ki} \cdot C_{kj} \ \quad \ {\rm for}\ i\neq j\neq k \ .
\end{equation}
The triple intersection product of two distinct surfaces is 
\begin{equation}
	S_i\cdot S_i \cdot S_j = K'_i\cdot C_{ij} \ \quad \ {\rm for}\ i\neq j \ .
\end{equation}
Lastly, the triple intersection of a single surface $S_i$ is given by
\begin{equation}
	S_i^3 = K_i'^2\ .\label{eq:triple intersection}
\end{equation}
It is now straightforward to obtain the full effective prepotential $\mathcal{E}$ on the $\Omega$-background for the low-energy theory from a geometric construction.

\subsection{\texorpdfstring{6d SCFTs on $S^1$ with/without twists}{6d SCFTs on S1 with/without twists}}

Compactification of 6d SCFTs on a circle with/without outer automorphism twists provides concrete UV completions of a large class of 5d field theories. These 5d theories are often called 5d Kaluza-Klein (KK) theories. The effective prepotential of such a 5d theory can be easily obtained from the 6d classical action on the tensor branch and the action of outer automorphism as well as matter content. The detailed procedure has been introduced in \cite{Bhardwaj:2019fzv}. We will now generalize this and propose full effective prepotentials for 5d KK theories on the $\Omega$-background including the contributions from mixed gauge/gravitational and gauge/$SU(2)_R$ Chern-Simons terms.

\paragraph{5d reductions without twist}
Let us first consider the compactification of a 6d SCFT without a twist on its tensor branch. We shall also consider a generic point on the Coulomb branch of the resulting 5d theory where both tensor scalar fields and gauge holonomies are turned on. The effective prepotential of the 5d theory can be written in terms of the Chern-Simons coefficients which can be exactly calculable from the classical action of the original 6d SCFT and 1-loop computations for charged fermions.

The 6d tree-level action on the tensor branch is given by
\begin{equation}
	S_{\rm tree}=\int -\frac{1}{4}\Omega^{\alpha\beta} G_\alpha \wedge * G_\beta - \Omega^{\alpha\beta}B_{\alpha} \wedge X_{4\beta} \ , \label{eq:S_tree-GS}
\end{equation}
and supersymmetric completions, where $\Omega^{\alpha\beta}$ is the negative-definite, symmetric bilinear form of $T$ tensor fields and $G_\alpha$ is the gauge-invariant field strength for the 2-form tensor field $B_\alpha$. The second term in \eqref{eq:S_tree-GS} is the Green-Schwarz term with the 4-form $X_{4\beta}$ defined as
\begin{equation}
	dG_\alpha = X_{4\alpha} \ , \qquad X_{4\alpha} = -\frac{1}{4}a_\alpha p_1(T) +\frac{1}{4}\sum_a b_{a,\alpha}{\rm Tr}F_{a}^2 +c_\alpha c_2(R) \ ,
\end{equation}
where $a_\alpha,b_{a,\alpha}, c_\alpha$ are fixed to cancel 1-loop gauge anomalies via the Green-Schwarz-Sagnotti mechanism \cite{Sagnotti:1992qw}, the summation index $a$ runs over all the gauge and the flavor groups, and $F_a$ is the field strength for the $a$-th symmetry group. The circle reduction of the classical action then gives rise to the following terms
\begin{equation}
	S_{\rm tree} = \int -\frac{\tau}{4} \Omega^{\alpha\beta} F_{\alpha}\wedge *F_\beta - \Omega^{\alpha\beta}A_{0,\alpha} \wedge X_{4\beta} + \cdots \ ,
\end{equation}
where $\tau\equiv 1/R$ is the inverse radius of the 6d circle and $F_{\alpha}$ denotes the $U(1)_\alpha$ field strength for the gauge field $A_{0,\alpha}$ obtained by reducing the tensor field $B_\alpha$ on the circle.

In order to compute the matter contribution to the effective Chern-Simons terms, we need to perform fermion 1-loop computations including Kaluza-Klein momentum states \cite{Bonetti:2011mw,Bonetti:2013ela,Grimm:2015zea}. The cubic Chern-Simons terms are captured by the 1-loop prepotential which is given by
\begin{equation}
	\mathcal{F}_{\text{1-loop}} = \frac{1}{12}\sum_{n \in \mathbb{Z}}\left(\sum_{e \in{\bf R}} |n\tau + e\cdot \phi|^3 - \sum_f \sum_{w\in {\bf w}_f}|n\tau + w\cdot \phi + m_f|^3\right) \ ,
\end{equation}
where the sum for an integer $n$ is performed over all KK charges $n$, ${\bf R}$ means collectively the roots of the 6d gauge groups, ${\bf w}_f$ is the weight and $m_f$ is the mass parameter for the $f$-th hypermultiplet in 6d. $\phi$'s are the gauge holonomies that become the scalar fields in the 5d vector multiplets. The infinite sums in the prepotential can be regularized using the zeta function regularization. In the 5d limit where $\tau\gg \phi_i,m_f$, the regularized cubic prepotential is given by \cite{Bonetti:2013ela,Grimm:2015zea}
\begin{align}
	\mathcal{F}_{\text{1-loop}} &= \frac{1}{12}\left(\sum_{e \in {\bf R}}|e\cdot \phi|^3 - \sum_f \sum_{w\in{\bf w}_f} |w\cdot \phi + m_f|^3\right) \nonumber \\
	& \ \ \ \ - \frac{\tau}{24}\left(\sum_{e \in {\bf R}}(e\cdot \phi)^2 - \sum_f \sum_{w\in{\bf w}_f} (w \cdot \phi + m_f)^2\right) \ .
\end{align}
The first line comes from the zero KK momentum modes and the second line is the contribution from the KK momentum states after the regularization. We omitted the terms independent of the dynamical parameters $\phi$. 

The mixed gauge/gravitational and gauge/$SU(2)_R$ Chern-Simons terms can be computed in a similar manner. Note here that the contributions from the positive and the negative KK momentum states cancel each other. So these Chern-Simons terms receive the contribution only from the zero modes on a circle.
Therefore, the coefficients for the mixed Chern-Simons terms are, respectively, \cite{Bonetti:2013ela,Grimm:2015zea}
\begin{align}
	C^G_i =& -\partial_i\left(\sum_{e\in {\bf R}}|e\cdot \phi| - \sum_f\sum_{w\in{\bf w}_f}|w\cdot\phi+m_f|\right) \ , \crcr
	C^R_i =&~ \frac{1}{2} \partial_i \sum_{e\in {\bf R}}|e\cdot \phi| \ .
\end{align}
Here, $e$ and $w$ run over all the zero modes of the vectors and the hypers, respectively, from the 6d theory.

The effective action of a 5d KK theory on the Coulomb branch can be obtained by collecting the above tree-level and the 1-loop Chern-Simons terms. On the $\Omega$-background, the full effective action is then given by
\begin{align}\label{eq:6dE}
	\mathcal{E}_{6d} &= i(S_{\rm tree} + S_{\text{1-loop}} + S_{\rm grav} + S_R) |_{\epsilon_{1,2},\phi,m,\tau} \nonumber \\
	&= \frac{1}{\epsilon_1\epsilon_2}\left(\mathcal{E}_{\rm tree}+\mathcal{F}_{\text{1-loop}}+\frac{1}{48} C^G_i\phi^i (\epsilon_1^2+\epsilon_2^2) + \frac{1}{2}C_i^R \phi^i\epsilon_+^2\right) \ ,  \\
	\mathcal{E}_{\rm tree} &\equiv iS_{\rm tree}|_{\epsilon_{1,2},\phi,m,\tau}\cr
	&= -\frac{\tau}{2}\Omega^{\alpha\beta}\phi_{\alpha,0}\phi_{\beta,0} - \Omega^{\alpha\beta}\phi_{\alpha,0}\left(\frac{a_\beta}{4}(\epsilon_1^2+\epsilon_2^2)+\frac{b_{a,\beta}}{2}K_{a,ij}\phi_{a,i}\phi_{a,j}+c_\beta \epsilon_+^2\right) \nonumber \ ,
\end{align}
where $K_{a,ij}$ is the Killing form for the $a$-th symmetry group $G_a$. If $G_a$ is a gauge group, then $\phi_{a,i}$ is the Coulomb branch parameter for it, otherwise, it is the mass parameter for the flavor symmetry. Consequently, we can compute the effective prepotentials on the Coulomb branch evaluated on the $\Omega$-background directly from the knowledge of the 6d SCFTs.

As discussed in \cite{Bhardwaj:2019fzv}, when we compare this effective prepotential of the 5d KK theory with geometry, in which $b_{a,\beta}=\delta_{a,\beta}$ for gauge groups $G_a$, we need to shift the Coulomb branch parameters $\phi_{\alpha,i}$ as
\begin{equation}\label{eq:shift}
	\phi_{\alpha,i} \rightarrow \phi_{\alpha,i} - d_i^{\vee} \phi_{\alpha,0} \ ,
\end{equation}
for all $1\le i \le r_\alpha$ as well as for all $\alpha$, where $d_i^{\vee}$ is the dual Coxeter label for the gauge group $G_\alpha$. After this shift, the tensor parameter $\phi_{\alpha,0}$ becomes the K\"ahler parameter for the affine node of the associated affine gauge algebra $\hat{\mathfrak{g}}_\alpha$.

We claim that the function $\mathcal{E}_{6d}$ in \eqref{eq:6dE} with the shift \eqref{eq:shift} is the full effective prepotential on the Coulomb branch in the 5d reduction of the 6d SCFT without a twist. This contains all the terms of the dynamical Coulomb branch parameters that do not vanish on the $\Omega$-background.

As an example, consider the 6d minimal $SU(3)$ gauge theory consists of a tensor multiplet (so $\alpha=1$) coupled to an $SU(3)$ vector multiplet. The 1-loop anomalies are cancelled by the Green-Schwarz-Sagnotti mechanism with the data
\begin{equation}\label{eq:GS-SU3}
	\Omega^{11} = -3 \ , \quad a_1 = -\frac{1}{3} \ , \quad b_1 = 1 \ , \quad c_1 = 1 \ .
\end{equation}
Thus, the effective prepotential on the $\Omega$-background before the shift \eqref{eq:shift} is given by
\begin{align}
\epsilon_1\epsilon_2\,\mathcal{E} 
=&~ \frac{3}{2} \,\tau\phi_0^2 +3\phi_0\Big(-\frac{1}{12}(\epsilon_1^2+\epsilon_2^2)+\frac{1}{2}K_{ij}\phi_i\phi_j+\epsilon_+^2\Big)\cr
&+\frac{1}{6}\Big(8\phi_1^3+8\phi_2^3-3\phi_1\phi_2(\phi_1+\phi_2)\Big) -\frac{\tau}{4}K_{ij}\phi_i\phi_j\cr	
&-\frac{1}{12}(\phi_1+\phi_2)(\epsilon_1^2+\epsilon_2^2) + (\phi_1+\phi_2)\epsilon_+^2 \ ,
\end{align}
where $K_{ij}$ is the Killing form of the $SU(3)$ group. To compare with its geometric construction, we first shift $\phi_0\rightarrow \phi_0+\tau/6$ and then perform the shift \eqref{eq:shift}, $\phi_{1,2}\rightarrow \phi_{1,2}-\phi_0$. The result is then
\begin{align}
	\epsilon_1\epsilon_2 \,\mathcal{E}' =& ~\frac{1}{6}\bigg(9\tau\phi_0^2+3\tau^2\phi_0 + 8(\phi_0^3+\phi_1^3+\phi_2^3)\cr
	&\quad -3\phi_0^2(\phi_1+\phi_2)-3\phi_1^2(\phi_0+\phi_2)-3\phi_2^2(\phi_0+\phi_1)-6\phi_0\phi_1\phi_2\bigg) \nonumber \\
	& -\frac{1}{12}(\phi_0+\phi_1+\phi_2)(\epsilon_1^2+\epsilon_2^2) + (\phi_0+\phi_1+\phi_2)\epsilon_+^2 \ .
\end{align}
This is in perfect agreement with the effective prepotential for the geometry for the minimal $SU(3)$ gauge theory that consists of three $\mathbb{F}_1$ surfaces glued along their $-1$ curves and one non-compact surface with K\"ahler parameter `$-\tau$' glued to the $\mathbb{F}_1$ surface for $\phi_0$ along the base $+1$ curve as constructed in \cite{DelZotto:2017pti}.

\paragraph{5d reductions with twist}
When a 6d SCFT has a discrete global symmetry $\Gamma$, we can compactify the theory on a circle with a discrete holonomy $\gamma$ for the background gauge field of the symmetry $\Gamma$. This is often called automorphism twist on $\Gamma$ around the circle. Such twists for 6d SCFTs are well described in section 3 of \cite{Bhardwaj:2019fzv}. We will employ this prescription to compute the effective prepotentials and the partition functions of 6d SCFTs compactified on a circle with twists.

There are two kinds of discrete symmetries in 6d SCFTs. The first kind is the symmetry arising from outer automorphism of gauge algebra $\mathfrak{g}$ which permutes matter representations of the gauge algebra. The second one is the symmetry from permutations of tensor fields. A general discrete symmetry is generated by a combination of these two kinds of discrete symmetries.

\begin{table}[t]
\centering
\begin{tabular}{|c|c|c|c|c|c|}
	\hline
	$\mathfrak{\hat{g}}$ & $A_{2\ell}^{(2)}$ & $A_{2\ell-1}^{(2)}$ & $D^{(2)}_{\ell+1}$ & $E_6^{(2)}$ & $D_4^{(3)}$  \\
	\hline 
	$\mathfrak{h}$ & $C_\ell$ & $C_\ell$ & $B_\ell$ & $F_4$ & $G_2$  \\ 
	\hline
\end{tabular}
\caption{Twisted affine Lie algebra $\mathfrak{\hat{g}}$ and invariant subalgebra $\mathfrak{h}$ under outer automorphism. The superscript ${}^{(p)}$ denotes order $p=2,3$ of twist.}\label{tb:g-h}
\end{table}

The twist of the first kind on a gauge algebra $\mathfrak{g}$ splits representations of  $\mathfrak{g}$ into representations of the invariant subalgebra $\mathfrak{h}$, listed in Table \ref{tb:g-h}, under the outer automorphism. (If the twist does not act on a gauge algebra, then the invariant subalgebra is the same as the original gauge algebra, i.e. $\mathfrak{h}=\mathfrak{g}$.) A 6d state in a representation of $\mathfrak{g}$ is then decomposed into several representations of the subalgebra $\mathfrak{h}$. The decomposed states will carry shifted KK momentum as $n\rightarrow n+r$ depending on their charge under the discrete symmetry, where $n\in \mathbb{Z}$ and $r$ is a fractional KK charge. See Appendix \ref{appendix:1-loop} for more details.

For example, a 5d KK state from twisting a 6d state in the adjoint representation of the original gauge algebra will carry the following charges \cite{Kac:1990gs}
\begin{align}
	A_{2\ell}^{(2)}	\ \ &: \ \ {\bf Adj} \ {\rm of} \ A_{2\ell} \ \rightarrow \ {\bf Adj}_0\oplus {\bf F}_{1/4}\oplus {\bf F}_{3/4} \oplus {\bf \Lambda}^2_{1/2} \oplus {\bf 1}_{1/2} \ \ {\rm of} \  C_{\ell} \ ,\nonumber \\
	A_{2\ell-1}^{(2)} \ \ &:  \ \ {\bf Adj} \ {\rm of} \  A_{2\ell-1} \ \rightarrow \ {\bf Adj}_0 \oplus {\bf \Lambda}^2_{1/2}  \ \ {\rm of} \  C_{\ell} \ , \nonumber \\
	D_{\ell+1}^{(2)} \ \ &:  \ \ {\bf Adj} \ {\rm of} \  D_{\ell+1} \ \rightarrow \ {\bf Adj}_0 \oplus {\bf F}_{1/2} \ \ {\rm of} \ B_\ell \ , \nonumber \\
	E_6^{(2)} \ \ &:  \ \ {\bf Adj} \ {\rm of} \  E_6 \ \rightarrow \ {\bf Adj}_0\oplus{\bf F}_{1/2}  \ \ {\rm of} \ F_4 \ , \nonumber \\
	D_{4}^{(3)} \ \ &:  \ \ {\bf Adj} \ {\rm of} \  D_4 \ \rightarrow \  {\bf Adj}_0 \oplus{\bf F}_{1/3} \oplus {\bf F}_{2/3} \ \ {\rm of} \ G_2 \ ,
\end{align}
where $\mathbf{Adj}_r$, $\mathbf{F}_r$, and $\mathbf{\Lambda}^2_r$ refer to the adjoint, fundamental, and 2nd rank antisymmetric representation of subalgebra $\mathfrak{h}$, carrying shifted KK charge $r$ for the KK state.

The second kind of discrete symmetries is generated by a permutation $S$ acting on tensor nodes as $\alpha \rightarrow S(\alpha)$. The twist with $S$ then identifies the tensor nodes $\alpha$ and $S(\alpha)$. This brings the intersection form $\Omega^{\alpha\beta}$ into another matrix $\Omega_S^{\alpha'\beta'}$ where $\alpha',\beta'$ parametrize orbits of tensor nodes permuted by the action of $S$. The intersection matrix after the twist is determined by 
\begin{equation}
	\Omega_S^{\alpha'\beta'} = \sum_{\beta\in \beta'}\Omega^{\alpha\beta} \ ,
\end{equation}
where $\alpha$ is any node in a given orbit $\alpha'$. See \cite{Bhardwaj:2019fzv} for more details.

Now consider a 5d KK theory obtained by a twisted compactification of a 6d SCFT. The 5d gauge group and matter content as well as their KK charges are now fully fixed by the action of the twists discussed above. Knowing this, we can compute the effective prepotential for any twisted KK theory. The tree-level part $\tilde{\mathcal{E}}_{\rm tree}$ after a certain twist is given by
\begin{align}
	\tilde{\mathcal{E}}_{\rm tree} &\equiv i\tilde{S}_{\rm tree}|_{\epsilon_{1,2},\phi,m,\tau} \nonumber \\
	&= -\frac{\tau}{2}K^{\alpha'\beta'}_S\phi_{\alpha',0}\phi_{\beta',0} - \Omega^{\alpha'\beta'}_S\phi_{\alpha',0}\left(\frac{a_{\beta'}}{4}(\epsilon_1^2+\epsilon_2^2)+\frac{b_{a,\beta'}}{2}\tilde{K}_{a,ij}\phi_{a,i}\phi_{a,j}+c_{\beta'} \epsilon_+^2\right) \nonumber \ .
\end{align}
Here $K_S^{\alpha'\beta'}\equiv\sum_{\alpha\in\alpha',\beta\in\beta'}\Omega^{\alpha\beta}$ and $\tilde{K}_{a,ij}$ is the Killing form for an invariant subalgebra $\mathfrak{h}_a$ and $a$ runs over the gauge and the flavor symmetry groups.

The cubic prepotential receives contributions from the KK momentum states. We compute, in the parameter regime $\tau\gg \phi,m_f$,
\begin{align}
\tilde{\mathcal{F}}_{\text{1-loop}} 
&= \frac{1}{12}\sum_{n\in \mathbb{Z}} \bigg(\sum_{e\in \oplus{\bf R}_r} \big|(n+r)\tau + e\cdot \phi\big|^3 - \sum_f \sum_{w\in \oplus{\bf w}_{r,f}} \big|(n+r)\tau +w\cdot \phi+m_f\big|^3\bigg) \nonumber \\
&= \frac{1}{12}\bigg(\sum_{e\in {\bf R}_0} | e\cdot \phi|^3 - \sum_f \sum_{w\in {\bf w}_{0,f} } |w\cdot \phi+m_f|^3\bigg) \\
&\quad -\frac{\tau}{24}\bigg(\sum_{e \in \oplus{\bf R}_r} \!\big(6r(r\!-\!1)\!+\!1\big)(e\cdot \phi)^2 - \sum_f\!\!\sum_{w \in \oplus {\bf w}_{r,f}}\!\!\big(6r(r\!-\!1)\!+\!1\big)(w\cdot \phi+m_f)^2\bigg) \ .\nonumber
\end{align}
Here in the first line, the first term comes from the 6d vector multiplets decomposed into $\oplus {\bf R}_r$ representations of invariant subalgebras $\otimes_a \mathfrak{h}_a$, and the second term corresponds to the 6d hypermultiplets decomposed into $\oplus {\bf w}_{r,f}$ representations of $\otimes_a \mathfrak{h}_a$. We used the zeta function regularization in the second and third lines. Note that the cubic terms in the Coulomb branch parameters $\phi$ receive contributions only from the zero KK momentum states because the contributions from other higher KK towers are all cancelled each other. Similarly, the mixed gauge/gravitational and the mixed gauge/$SU(2)_R$ Chern-Simons coefficients receive contributions only from zero KK momentum states. We compute
\begin{align}
	\tilde{C}^G_i = ~&-\partial_i\bigg(\sum_{e\in {\bf R_0}}|e\cdot \phi| - \sum_f\sum_{w\in{\bf w}_{0,f}}|w\cdot\phi+m_f|\bigg) \ , \nonumber \\
	\tilde{C}^R_i =~& \frac{1}{2}\, \partial_i \sum_{e\in {\bf R}_0}|e\cdot \phi| \ .
\end{align}

Then the effective prepotential on the $\Omega$-background for a 6d SCFT on a circle with general outer automorphism twists is given by
\begin{align}\label{eq:E-twist}
\mathcal{E}_{{\rm twist}} &= i(\tilde{S}_{\rm tree} + \tilde{S}_{\text{1-loop}} + \tilde{S}_{\rm grav} + \tilde{S}_R) |_{\epsilon_{1,2},\phi,m,\tau} \nonumber \\
	&= \frac{1}{\epsilon_1\epsilon_2}\left(\tilde{\mathcal{E}}_{\rm tree}+\tilde{\mathcal{F}}_{\text{1-loop}}+\frac{1}{48} \tilde{C}^G_i\phi^i (\epsilon_1^2+\epsilon_2^2) + \frac{1}{2}\tilde{C}_i^R \phi^i\epsilon_+^2\right) \ .
\end{align}
Again, we need to perform the shift in \eqref{eq:shift} for the Coulomb branch parameters when we compare this 6d result with the effective prepotential of a geometric description or with the effective prepotentials of other 5d dual gauge theories. 

As an example, consider the 6d minimal $SU(3)$ theory and its twisted compactification. This theory has discrete $\mathbb{Z}_2$ global symmetry acting on $SU(3)$ representations. We can therefore compactify this theory on a circle with twist of the $\mathbb{Z}_2$ symmetry. The invariant subalgebra under the twist is $\mathfrak{h}=\mathfrak{su}(2)$ and the KK states in the 5d theory take representations of $\mathfrak{su}(2)$. The vector multiplet of the $\mathfrak{su}(3)$ algebra in 6d reduces to the following combination of KK momentum states.
\begin{equation}
	{\bf 8} \ {\rm of} \ \mathfrak{su}(3) \ \rightarrow \ {\bf 3}_0 \oplus {\bf 2}_{1/4} \oplus {\bf 2}_{3/4} \oplus {\bf 1}_{1/2} \ {\rm of} \ \mathfrak{su}(2) \ .
\end{equation}
From this together with the data \eqref{eq:GS-SU3}, we can compute the tree-level and the loop contributions to the effective prepotential as
\begin{align}
	\tilde{\mathcal{E}}_{\rm tree} =~& \frac{3}{2}\tau \phi_0^2 +\phi_0\left(3\phi_1^2 -\frac{1}{4}(\epsilon_1^2+\epsilon_2^2) + 3\epsilon_+^2\right) \ , \nonumber \\
	\tilde{\mathcal{F}}_{\text{1-loop}} = ~&\frac{4}{3}\phi_1^3 -\frac{5}{16}\tau \phi_1^2 \ ,
\end{align}
and the mixed Chern-Simons coefficients as
\begin{equation}
	\tilde{C}^G_1 = -4 \ , \qquad \tilde{C}^R_1 = 2 \ .
\end{equation}
Plugging all these terms into the formula \eqref{eq:E-twist}, one obtains the effective prepotential for the twisted minimal $SU(3)$ SCFT as
\begin{align}\label{eq:E-SU3-Z2}
	\epsilon_1\epsilon_2\, \mathcal{E} = &~\frac{3}{2}\tau \phi_0^2 +\phi_0\left(3\phi_1^2 -\frac{1}{4}(\epsilon_1^2+\epsilon_2^2) + 3\epsilon_+^2\right)+\frac{4}{3}\phi_1^3 -\frac{5}{16}\tau \phi_1^2 \nonumber \\
	&-\frac{1}{12}\phi_1(\epsilon_1^2+\epsilon_2^2)+\phi_1\epsilon_+^2 \ .
\end{align}
As this theory is dual to the 5d pure $SU(3)$ gauge theory at the Chern-Simons level $\kappa=9$, we can compare the effective prepotential $\mathcal{E}$ in \eqref{eq:E-SU3-Z2} with that for the dual 5d theory. For the comparison, we should shift the Coulomb branch parameters as $\phi_0 \rightarrow \phi_0 -\frac{1}{16}\tau$ and consecutively as $\phi_1 \rightarrow \phi_1 - 2\phi_0+\frac{\tau}{4}$ as guided by \eqref{eq:shift}. Then the shifted effective prepotential $\mathcal{E}'$ becomes
\begin{align}
\epsilon_1\epsilon_2\, \mathcal{E}' =&~ \frac{\tau}{2}(\phi_0^2-\phi_0\phi_1+\phi_1^2)
+\frac{1}{6}\left(8\phi_0^3+8\phi_1^3 +24\phi_0^2\phi_1 - 30\phi_0\phi_1^2\right) \nonumber \\
& -\Big(\frac{1}{12}(\epsilon_1^2+\epsilon_2^2)-\epsilon_+^2 \Big) (\phi_0+\phi_1) \ ,
\end{align}
up to terms independent of $\phi_i$. The result is precisely the effective prepotential for the pure $SU(3)_9$ theory with gauge coupling $\frac{1}{g^2} = \tfrac{\tau}{2}$.

%% file: sec-blowupEq.tex
\section{Blowup equations}\label{sec:BlowupEq}

In this section, we review and generalize the Nakajima-Yoshioka's blowup equations which are the main tool for computing the BPS partition functions of 5d/6d SQFTs on $\mathbb{C}^2\times S^1 ({\rm or~} \mathbb{C}^2\times T^2)$, and propose our main claim which enables one to compute the BPS spectrum of any SQFT. We also present some instructive examples for computing the partition functions.

\input{sec-BE-review}

\subsection{Instructive examples}\label{sec:instructive}

In this section, we present instructive examples which illustrate our proposal in section~\ref{sec:BlowupEqReview} for bootstrapping the BPS partition functions.

\subsubsection{5d pure \texorpdfstring{$SU(2)_\theta$}{SU2theta} with \texorpdfstring{$\theta=0,\pi$}{theta0pi}}

As the simplest example, consider the 5d $\mathcal{N}=1$ pure $SU(2)_\theta$ gauge theory with a discrete theta angle $ \theta = 0 $ or $ \pi $. As the perturbative descriptions of two theories are the same, they have the same prepotential. It follows from \eqref{eq:preF} that the cubic prepotential is given by
\begin{align}
6\mathcal{F}
= 6 m \phi^2+8\phi^3 \ ,
\end{align}
where $ m= 1/g^2 $ is the gauge coupling and $\phi$ is the Coulomb branch parameter for the $SU(2)$ gauge symmetry. The effective prepotential defined in \eqref{eq:E-func} evaluated on the $\Omega$-background for both theories takes the form of
\begin{align}
\mathcal{E}
= \frac{1}{\epsilon_1 \epsilon_2} \qty(\mathcal{F} - \frac{\epsilon_1^2  + \epsilon_2^2}{12}\phi+\epsilon_+^2 \phi) \ ,
\end{align}
where the second term comes from the mixed gauge/gravitational CS term with $C^G_\phi=-4$ in \eqref{eq:C_i}, and the third term comes from the mixed gauge/$SU(2)_R$ CS term with $C^R_\phi=2$. The mixed CS terms are also insensitive to the discrete theta angle.

We now use the effective prepotential $\mathcal{E}$ as an initial input for the blowup equation to obtain BPS spectrum of this theory. Recall that though $\mathcal{E}$ for both $SU(2)_0$ and $SU(2)_\pi$ theories are the same, they have different UV fixed points and thus they have different BPS spectra. This is reflected in the blowup equation by two distinct choices of background magnetic flux $B_m$ for the topological $U(1)$ symmetry: 
\begin{equation}
B_m \in \left\{ \begin{array}{ll} \mathbb{Z} &\ {\rm for} \ SU(2)_0\  \\ \mathbb{Z}+1/2 & \ {\rm for} \ SU(2)_\pi \end{array}\right.\ .
\end{equation}

To choose the flux $B_m$ properly, it is convenient to consider geometric construction of two theories. The $SU(2)_0$ theory is geometrically engineered by a CY 3-fold containing a single Hirzebruch surface $ \mathbb{F}_0 $, while the $SU(2)_\pi$ theory is engineered by another 3-fold containing a Hirzebruch surface $ \mathbb{F}_1 $.
\begin{figure}
	\centering
	\begin{subfigure}{0.45\textwidth}
		\centering
		\begin{tikzpicture}
		\draw[thick] (0, 0) -- (1.5, 0) -- (1.5, 1.5) -- (0, 1.5) -- (0, 0);
		\draw[thick] (0, 0) -- (-0.5, -0.5);
		\draw[thick] (1.5, 0) -- (2, -0.5);
		\draw[thick] (1.5, 1.5) -- (2, 2);
		\draw[thick] (0, 1.5) -- (-0.5, 2);
		\draw (-0.2, 0.75) node {$ _f $} (0.75, 1.7) node {$ _e $};
		\end{tikzpicture}
		\caption{$ \mathbb{F}_0 $}
	\end{subfigure}
	\begin{subfigure}{0.45\textwidth}
		\centering
		\begin{tikzpicture}
		\draw[thick] (0, 0) -- (0, 1.5) -- (1, 1.5) -- (2, 0) -- (0, 0);
		\draw[thick] (0, 0) -- (-0.5, -0.5);
		\draw[thick] (0, 1.5) -- (-0.5, 2);
		\draw[thick] (1, 1.5) -- (1, 2);
		\draw[thick] (2, 0) -- (2.6, -0.3);
		\draw (-0.2, 0.75) node {$ _f $} (0.5, 1.7) node {$ _e $} (1, -.3) node {$ _h $};
		\end{tikzpicture}
		\caption{$ \mathbb{F}_1 $}
	\end{subfigure}
	\caption{Geometric constructions of (a) $ SU(2)_0 $ and (b) $ SU(2)_\pi $. The fiber $\mathbb{P}^1$ in a Hirzebruch surface $\mathbb{F}_n$ is denoted by $f$, the base $\mathbb{P}^1$ is denoted by $e$, and $h=e+f$ for $\mathbb{F}_1$. } \label{fig:SU2}
\end{figure}
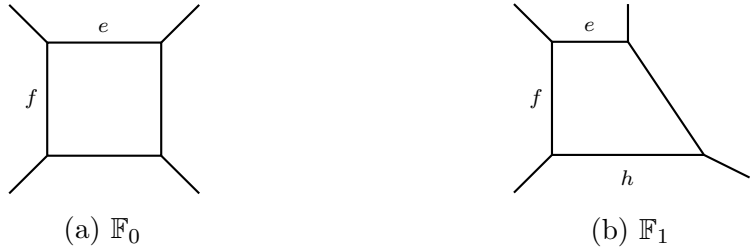
The primitive 2-cycles (or Mori cone generators) in a Hirzebruch surface are the fiber $f$ and the base $e$. The volumes of these 2-cycles are given by
\begin{align}
\left\{
\begin{array}{ll}
\mathbb{F}_0 &: \quad \vol(f) = 2\phi, \quad \vol(e) = 2\phi + m\,, \\
\mathbb{F}_1 &: \quad \vol(f) = 2\phi, \quad \vol(e) = \phi + m\,.
\end{array} \right. 
\end{align}
Here $m$ is a K\"ahler parameter for a non-compact 4-cycle that is identified with the gauge coupling in the gauge theory. This implies that the curve $e$ in each geometry carries charge $+1$ for the $U(1)$ topological symmetry. 

Let us consider a canonically quantized magnetic flux $n\in\mathbb{Z}$ for the $SU(2)$ gauge symmetry in both $\mathbb{F}_0$ and $\mathbb{F}_1$ theories. This magnetic flux can suitably couple to the state coming from a wrapped M2-brane on the fiber curve $f$. This state is a W-boson with gauge charge $+2$ and it feels an integer magnetic flux $2n$, which is compatible with the quantization condition in \eqref{eq:quantization}.

Two theories have different quantization conditions on the background flux $B_m$ for the $U(1)$ topological symmetry. Consider first the theory of $\mathbb{F}_0$. The $e$ curve in this geometry gives rise to a vector multiplet with charge $+2$ for the gauge symmetry and charge $+1$ for the topological symmetry. From the quantization condition \eqref{eq:quantization} with $n\in\mathbb{Z}$, the background flux should be quantized as $B_m\in\mathbb{Z}$. On the other hand, the state associated to the $e$ curve in the theory of $\mathbb{F}_1$ is a hypermultiplet with gauge charge $+1$ and topological charge $+1$. Then the quantization condition \eqref{eq:quantization} requires the background flux to be half-integrally quantized, $B_m\in\mathbb{Z}+1/2$. Hence we find the following quantizations for the magnetic fluxes:
\begin{align}\label{eq:su2_e_shift}
\left\{
\begin{array}{ll}
\mathbb{F}_0 &: \quad n\in\mathbb{Z} \ , \quad B_m \in \mathbb{Z} \\
\mathbb{F}_1 &: \quad n\in\mathbb{Z} \ , \quad B_m \in \mathbb{Z}+1/2\,.
\end{array}\right.
\end{align}

We note however that not all integer/half-integer fluxes $B_m$ are allowed in the blowup equations; the blowup equations only with the consistent magnetic fluxes can be solved to produce the correct BPS spectrum, as discussed in the previous subsection. If one uses other sets of magnetic fluxes, then the solution to the blowup equation will contain some BPS states with negative volumes ${\rm vol}(C)<0$ on the Coulomb branch, which is surely inconsistent. So we first need to identify the consistent magnetic fluxes $B_m$.

Let us try to solve the blowup equation with magnetic fluxes $n\in\mathbb{Z}$ and some $B_m$. The magnetic fluxes result in the shifts of parameters in the North and the South poles as $\phi \to \phi + n \,\epsilon_{1,2}$, $m \to m + B_m\,\epsilon_{1,2}$. The GV-invariant (or the Witten index) can be decomposed into two parts, the perturbative and the instanton parts, as
\begin{align}
	Z_{GV}(\phi,m;\epsilon_1,\epsilon_2) &= \mathcal{Z}_{\rm pert}(\phi;\epsilon_1,\epsilon_2)\cdot \mathcal{Z}_{\rm inst}(\phi,m;\epsilon_1,\epsilon_2) \ , \nonumber \\
	\mathcal{Z}_{\rm pert}(\phi;\epsilon_1,\epsilon_2) & ={\rm PE}\left[-\frac{1+p_1p_2}{(1-p_1)(1-p_2)}e^{-2\phi}\right] \ , \nonumber \\
	\mathcal{Z}_{\rm inst}(\phi,m;\epsilon_1,\epsilon_2) &= \sum_{k=0}^\infty q^k Z_k(\phi; \epsilon_1, \epsilon_2) \ ,
\end{align}
with the instanton fugacity $q\equiv e^{-m}$ and $Z_0=1$. The perturbative part comes from the W-boson (or the M2-brane state on $f$) on the Coulomb branch. Plugging the GV-invariant into \eqref{eq:bleq-GV}, one finds a unity blowup equation that can be expanded in terms of the instanton fugacity as
\begin{align}
\sum_{k,k'=0}^\infty q^{k+k'}\Lambda_{k'} \hat{Z}_k(\phi; \epsilon_1, \epsilon_2)=& \
\sum_{n \in \mathbb{Z}}  \sum_{k_1, k_2=0}^\infty (-1)^n e^{-V}\frac{\hat{\mathcal{Z}}_{\text{pert}}^{(N)}\hat{\mathcal{Z}}_{\text{pert}}^{(S)}}{\hat{\mathcal{Z}}_{\text{pert}}} (q \,p_1^{B_m})^{k_1} (q\, p_2^{B_m})^{k_2}  \\
&\  \times \hat{Z}_{k_1}^{(N)}(\phi + n\epsilon_1 ; \epsilon_1, \epsilon_2 - \epsilon_1) \cdot \hat{Z}^{(S)}_{k_2}(\phi + n\epsilon_2; \epsilon_1 - \epsilon_2, \epsilon_2)\,, \nonumber
\end{align}
with $V$ defined in \eqref{eq:GV-V}, where $\Lambda_0=1$ which is fixed by the zeroth order equation in the expansion. Here, the hatted functions are defined with the shift in the parameter $\epsilon_1$ as
\begin{equation} 
	\hat{f}(\phi,m;\epsilon_1,\epsilon_2)\equiv f(\phi,m;\epsilon_1+2\pi i,\epsilon_2) \ .
\end{equation}

One can extract the $k$-th order of the instanton expansion of this equation, which can be schematically written as
\begin{align}
\Lambda_k(\epsilon_1,\epsilon_2) + \hat{Z}_k(\phi; \epsilon_1, \epsilon_2)
=& ~p_1^{kB_m}\, \hat{Z}_k(\phi; \epsilon_1, \epsilon_2 - \epsilon_1) + p_2^{kB_m}\, \hat{Z}_k(\phi; \epsilon_1 - \epsilon_2, \epsilon_2) \cr
&+ (\text{terms with } \hat{Z}_{r<k} \ \text{and } \Lambda_{r<k})\ .
\end{align}
Here the $k$-instanton partition function $\hat{Z}_k$ at the $k$-th order in the expansion is independent of the background magnetic flux $B_m$ and appears only with a trivial gauge flux $n=0$.
All $\phi$ independent terms in the second line (first by expanding it in $e^{-\phi}$) will be absorbed into $\Lambda_k$. This equation can be solved once we know the solutions $Z_{r}$ at the lower orders $r<k$. In particular, since $\hat{Z}_k$ is independent of $B_m$, when there exist 3 (or more) distinct consistent fluxes $B_m$, we can exactly solve three linearly independent equations for 3 unknown functions, $\hat{Z}_k,\hat{Z}_k^{(N)},$ and $\hat{Z}^{(S)}_k$, and hence find a closed expression for the $k$-instanton partition function $\hat{Z}_k$ at each $k$.

The first order of the blowup equation can be written explicitly as
\begin{align}\label{eq:1-inst-su2}
	\Lambda_1+\hat{Z}_1 =& \  p_1^{B_m}\hat{Z}_1^{(N)}+p_2^{B_m}\hat{Z}_1^{(S)} - \frac{(p_1p_2)^{B_m+1}e^{-2(2+B_m)\phi_1}}{(1-e^{-2\phi})(1-p_1e^{-2\phi})(1-p_2e^{-2\phi})(1-p_1p_2e^{-2\phi})} \nonumber \\
	&- \frac{(p_1p_2)^{B_m-1}e^{-2(2-B_m)\phi_1}}{(1-e^{-2\phi_1})(1-p_1^{-1}e^{-2\phi_1})(1-p_2^{-1}e^{-2\phi_1})(1-(p_1p_2)^{-1}e^{-2\phi_1})} \ .
\end{align}
The last two terms on the right-hand side come from the perturbative parts with the gauge fluxes $n=1$ and $n=-1$ respectively. One can then easily see that if $B_m>2$ or $B_m<-2$, then the last two terms will start with a negative power of $e^{-\phi}$ in the expansion. This means that the solution $\hat{Z}_1$ to the above equation will involve a BPS state of mass $|M|=m+ a \phi$ with a negative coefficient $a$. This is problematic because this state violates the unitarity condition on its mass, i.e. $a\phi\ge0$, on the Coulomb branch parametrized by $\phi>0$ in the UV limit where $m\rightarrow0$. In addition, one can check that this state is not a hypermultiplet and thus there is no associated flop transition. Thus the solution is inconsistent due to the states with  negative masses (or volumes) in the UV limit. This tells us that solving the blowup equations with flux $|B_m|>2$ cannot give the correct BPS partition function for the pure $SU(2)_\theta$ gauge theory.

On the other hand, if $-2\le B_m\le 2$, then the solution will not contain any states with ${\sf{e}}\cdot\phi<0$. Therefore we conclude that the consistent magnetic fluxes for the $SU(2)_0$ and $SU(2)_\pi$ theories are
\begin{align}
\left\{
\begin{array}{ll}
\mathbb{F}_0:~~SU(2)_0 & \quad n\in\mathbb{Z} \ , \ \ B_m = -2,\ -1,\ 0,\ 1, 2 \\
\mathbb{F}_1:~~SU(2)_\pi & \quad n\in\mathbb{Z} \ , \ \  B_m =-\frac{3}{2},\ -\frac{1}{2},\ \frac{1}{2},\ \frac{3}{2}\ .
\end{array}\right. \label{eq:su2 three choices of B}
\end{align}

Since we have more than three sets of consistent fluxes $(n,B_m)$ for both cases, we can compute  a closed expression of the partition function at each instanton order. For example, the 1-instanton partition functions are given as follows:
\begin{align}
Z_1^{SU(2)_0}(\phi; \epsilon_1, \epsilon_2)
&= \frac{p_1 p_2 (1+p_1 p_2)e^{-2\phi}}{(1-p_1)(1-p_2)(1-p_1 p_2 e^{-2\phi})(e^{-2\phi}-p_1p_2)}\ , \label{eq:SU2_0Z-1instanton}\\
Z_1^{SU(2)_\pi}(\phi; \epsilon_1, \epsilon_2)
&= -\frac{p_1^{3/2} p_2^{3/2}(1+e^{-2\phi})e^{-\phi}}{ (1-p_1)(1-p_2) (1-p_1 p_2 e^{-2\phi})(e^{-2\phi}-p_1 p_2)}\ . \label{eq:SU2_piZ-1instanton}
\end{align}
By repeating this procedure for the blowup equations to higher orders in $q$, it is straightforward to obtain the $k$-instanton partition functions $Z_k$. 

We note that though there are many theories which possess three different sets of consistent magnetic fluxes, many of which are discussed in \cite{Kim:2019uqw}, a much larger class of theories do not allow such three distinct sets. However, as discussed in the previous subsection and also in \cite{Huang:2017mis,Gu:2019pqj}, a single unity blowup equation with the consistent magnetic fluxes $(\vec{n},\vec{B})$ is enough to compute the BPS partition function. We will now illustrate this by computing the BPS spectra of the pure $SU(2)_\theta$ theories, using only a single background flux $B_m$ giving a unity blowup equation.

\paragraph{Solving a unity blowup equation}

Let us start with the following ansatz for the GV-invariant.
\begin{align}
	Z_{GV}(\phi,m;\epsilon_{1,2})&={\rm PE}\left[\sum_{j_l,j_r}\sum_{d_1,d_2=0}^\infty(-1)^{2(j_l+j_r)}N_{j_l,j_r}^{(d_1,d_2)} A_{j_l,j_r}(\epsilon_1,\epsilon_2) e^{-d_1{\rm vol}(e)-d_2{\rm vol}(f)}\right] \nonumber \\
	A_{j_l,j_r}(\epsilon_1,\epsilon_2) &\equiv\frac{\chi_{j_l}^{SU(2)}(p_1/p_2)\chi_{j_r}^{SU(2)}(p_1p_2)}{(p_1^{1/2}-p_1^{-1/2})(p_2^{1/2}-p_2^{-1/2})} \ .
\end{align}
Here ${\bf d}=(d_1,d_2)$ denotes the degrees $d_1$ and $d_2$ of 2-cycles $e$ and $f$ respectively. The BPS states with $d_1=0$ at 0-instanton sector are all captured by the perturbative spectrum of the $SU(2)$ theory. The perturbative spectrum then fixes $N_{0,\frac12}^{(0,1)}=1$ and $N_{j_l,j_r}^{(0,d_2)}=0$ for $d_2>1$.

We shall now put the ansatz to the blowup equation and solve it order by order in $d_1$ expansion. Let us first consider the $SU(2)_0$ theory with the background flux $B_m=0$. The equation at $d_1=1$ order is given in \eqref{eq:1-inst-su2}. To solve this equation, we further expand the equation in $d_2$ as
\begin{align}
	\Lambda_1 = &~\hat{Z}^{(N)}_1 + \hat{Z}^{(S)}_1 -\hat{Z}_1 -\big(p_1p_2+(p_1p_2)^{-1}\big)e^{-4\phi}\cr
	&-\frac{(1+p_1)(1+p_2)(1+p_1^3p_2^3)}{p_1^2p_2^2}e^{-6\phi}+\mathcal{O}(e^{-8\phi}) \ .
\end{align}
This equation can be easily solved order by order in $d_2$ or in the powers of $e^{-\phi}$. 

It turns out that this equation uniquely determines all the degeneracies $N_{j_l,j_r}^{(1,d_2)}$, except for the degeneracies $N_{0,\frac12}^{(1,d_2)}$ for all $d_2\ge0$. The result is listed in Table~\ref{table:SU2_0_1inst}. 
\begin{table}[H]
	\centering
	\begin{tabular}{|c|C{25ex}||c|C{25ex}|} \hline
		$\mathbf{d}$ & $\oplus N_{j_l, j_r}^{\mathbf{d}} (j_l, j_r)$ & $\mathbf{d}$ & $\oplus N_{j_l, j_r}^{\mathbf{d}} (j_l, j_r)$ \\ \hline
	 	$ (1, 0) $ & $ N_{0, \frac{1}{2}}^{(1, 0)}(0, \frac{1}{2}) $ & $ (1, 1) $ & $ N_{0, \frac{1}{2}}^{(1, 1)}(0, \frac{1}{2}) \oplus (0, \frac{3}{2}) $ \\ \hline
		$ (1, 2) $ & $ N_{0, \frac{1}{2}}^{(1, 2)}(0, \frac{1}{2}) \oplus (0, \frac{5}{2}) $ & $ (1, 3) $ & $ N_{0, \frac{1}{2}}^{(1, 3)}(0, \frac{1}{2}) \oplus (0, \frac{7}{2}) $ \\ \hline
	\end{tabular}
	\caption{The result of solving the $ SU(2)_0 $ blowup equation at $d_1=1$ order for the background flux $ B_m = 0 $.} \label{table:SU2_0_1inst}
\end{table}
The degeneracies $N_{0,\frac12}^{(1,d_2)}$ are not determined in the 1st order computation because they vanish in the combination $\hat{Z}^{(N)}_1 + \hat{Z}^{(S)}_1 -\hat{Z}_1$ and thus do not appear in the above equation. These undetermined degeneracies at $d_1=1$ order are all fixed by solving the blowup equation in the next order or higher orders: $N_{0,\frac12}^{(1,0)}=1$ and $N_{0,\frac12}^{(1,d_2)}=0$ for $d_2>0$.

We can solve the blowup equation iteratively and compute the degeneracies of higher degree curves.  
The resulting BPS spectra for $ d_1 \leq 2 $, $ d_2 \leq 3 $ are summarized in  Table~\ref{table:SU2_0}. One can readily see that this result agrees with the above result obtained by solving three blowup equations in \eqref{eq:SU2_0Z-1instanton}.
\begin{table}[H]
	\centering
	\begin{tabular}{|c|C{25ex}||c|C{25ex}|} \hline
		$\mathbf{d}$ & $\oplus N_{j_l, j_r}^{\mathbf{d}} (j_l, j_r)$ & $\mathbf{d}$ & $\oplus N_{j_l, j_r}^{\mathbf{d}} (j_l, j_r)$ \\ \hline
		$ (1, 0) $ & $ (0, \frac{1}{2}) $ & $ (1, 1) $ & $ (0, \frac{3}{2}) $ \\ \hline
		$ (1, 2) $ & $ (0, \frac{5}{2}) $ & $ (1, 3) $ & $ (0, \frac{7}{2}) $ \\ \hline
		$ (2, 1) $ & $ (0, \frac{5}{2}) $ & $ (2, 2) $ & $ (0, \frac{5}{2}) \oplus (0, \frac{7}{2}) \oplus (\frac{1}{2}, 4) $ \\ \hline
		$ (2, 3) $ & \multicolumn{3}{c|}{$ (0, \frac{5}{2}) \oplus (0, \frac{7}{2}) \oplus 2(0, \frac{9}{2}) \oplus (\frac{1}{2}, 4) \oplus (\frac{1}{2}, 5) \oplus (1, \frac{11}{2}) $} \\ \hline
	\end{tabular}
	\caption{BPS spectrum of the $ SU(2)_0 $ theory for $ d_1 \leq 2 $, $ d_2 \leq 3 $, where $ \mathbf{d} = (d_1, d_2) $ labels the states from an M2-brane wrapping $ d_1 e + d_2 f $ curve in $ \mathbb{F}_0 $.} \label{table:SU2_0}
\end{table}
\noindent 
Next, consider the $ SU(2)_\pi $ theory with a background magnetic flux $B_m = \frac{1}{2}$. The computation is basically the same as the previous case with $\theta=0$. The blowup equation at $d_1=1$ order is expanded as
\begin{align}
	\Lambda_1 = &~ p_1^{1/2}\hat{Z}^{(N)}_1 + p_2^{1/2}\hat{Z}^{(S)}_1 -\hat{Z}_1 -(p_1p_2)^{1/2}e^{-3\phi}\cr
	&-\frac{1+(1+p_1)(1+p_2)p_1^2p_2^2}{(p_1p_2)^{3/2}}e^{-5\phi}+\mathcal{O}(e^{-7\phi}) \ .
\end{align}
All the degeneracies $N_{j_l,j_r}^{(1,d_2)}$ are uniquely determined by solving this equation, except for the degeneracies $N_{0,0}^{(1,d_2)}$ for all $d_2\ge0$. The undetermined degeneracies $N_{0,0}^{(1,d_2)}$ are also fixed uniquely by solving the equations in higher orders. We list the result for $d_1\le2,d_2\le3$ in Table~\ref{table:SU2_pi}. Like the $SU(2)_0$ case, this result agrees with the expansion of \eqref{eq:SU2_piZ-1instanton}.

\begin{table}[H]
	\centering
	\begin{tabular}{|c|C{25ex}||c|C{25ex}|} \hline
		$\mathbf{d}$ & $\oplus N_{j_l, j_r}^{\mathbf{d}} (j_l, j_r)$ & $\mathbf{d}$ & $\oplus N_{j_l, j_r}^{\mathbf{d}} (j_l, j_r)$ \\ \hline
		$ (1, 0) $ & $ (0, 0) $ & $ (1, 1) $ & $ (0, 1) $ \\ \hline
		$ (1, 2) $ & $ (0, 2) $ & $ (1, 3) $ & $ (0, 3) $ \\ \hline
		$ (2, 2) $ & $ (0, \frac{5}{2}) $ & $ (2, 3) $ & $ (0, \frac{5}{2}) \oplus (0, \frac{7}{2}) \oplus (\frac{1}{2}, 4) $ \\ \hline
	\end{tabular}
	\caption{ BPS spectrum of the $ SU(2)_\pi $  theory for $ d_1 \leq 2 $, $ d_2 \leq 3 $, where $ \mathbf{d} = (d_1, d_2) $ labels the states from an M2-brane wrapping $ d_1 e + d_2 f $ curve in $ \mathbb{F}_1 $.} \label{table:SU2_pi}
\end{table}

\paragraph{Solving a vanishing blowup equation}
In the $ SU(2) $ gauge theories, it is possible to turn on non-canonically quantized magnetic flux as $ n \in \mathbb{Z}+1/2 $ because the W-boson carries charge +2 for the gauge symmetry. Then we can choose $ B_m \in \mathbb{Z} $ which is a consistent quantization for the instanton state corresponding to the curves $ e $  in both $ \mathbb{F}_0 $ and $ \mathbb{F}_1 $. So the magnetic fluxes
\begin{align}\label{eq:su2_vanishing_flux}
n \in \mathbb{Z} + 1/2 \ , \quad
B_m = 0 \ ,
\end{align}
satisfy the quantization conditions for both $ \theta = 0 $ and $ \pi $. These fluxes with the effective prepotential for the $SU(2)$ theories lead to a vanishing blowup equation. We will now show that this vanishing blowup equation alone can be solved with additional geometric data. In fact, we find two independent solutions to the blowup equation and they correspond to respectively the BPS spectrum of the $SU(2)_0$ theory and that of the $SU(2)_\pi$ theory.

Solving the vanishing blowup equation is much harder than solving unity equations because of huge cancellations in the expansion and also non-uniqueness of solutions. Let us start with an ansatz for the GV-invariant as
\begin{align}
Z_{\mathrm{GV}} = \PE\qty[ \,\sum_{j_l, j_r} \sum_{d_1, d_2 = 0}^\infty (-1)^{2(j_l + j_r)} N_{j_l, j_r}^{(d_1, d_2)} A_{j_l, j_r}(\epsilon_1, \epsilon_2) e^{-(d_1 m + d_2 \phi)} ] \, .
\end{align}
As discussed in \cite{Gu:2017ccq}, the spins of a degree $(d_1,d_2)$ state are bounded as
\begin{align}
j_l &\leq \frac{C^2\! +\! K_S \cdot C}{2} \!+\! 1 = \frac{d_1 d_2 \!-\! 2d_1^2 \!-\! d_2\!+\!2}{2} \, , \cr
 j_r &\leq \frac{C^2 \!-\! K_S \cdot C}{2} = \frac{d_1 d_2 \!-\! 2d_1^2 \!+\! d_2}{2}\, ,
\end{align}
where $ C $ is a curve with volume $ d_1 m + d_2 \phi $ and $ K_S $ is the canonical class in $ \mathbb{F}_0 $ or $ \mathbb{F}_1 $.

The zeroth-order in the expansion of the blowup equation is
\begin{align}
\sum_{n \in \{ \pm 1/2 \}} (-1)^n e^{-V} \frac{\hat{\mathcal{Z}}_{\mathrm{pert}}^{(N)} \hat{\mathcal{Z}}_{\mathrm{pert}}^{(S)}}{\hat{\mathcal{Z}}_{\mathrm{pert}}} = 0 \, ,
\end{align}
which is automatically satisfied by the perturbative spectrum $ N_{0, 1/2}^{(0, 2)} = 1 $. The 1st order blowup equation with $d_1=1$ is then given by
\begin{align}
\sum_{n \in \{ \pm 1/2 \}} (-1)^n e^{-V} \frac{\hat{\mathcal{Z}}_{\mathrm{pert}}^{(N)} \hat{\mathcal{Z}}_{\mathrm{pert}}^{(S)}}{\hat{\mathcal{Z}}_{\mathrm{pert}}} \qty(\hat{Z}_1^{(N)} + \hat{Z}_1^{(S)}) = 0 \, .
\end{align}
This equation for each $ d_2 $ can be rephrased in terms of the multiplicities $ N_{j_l,j_r}^{(1, d_2)} $ as
\begin{align}\label{eq:su2_vanishing_1st}
& \sum_{j_l, j_r} (-1)^{2(j_l + j_r)} N_{j_l, j_r}^{(1, d_2)} \qty( A_{j_l, j_r}(\epsilon_1, \epsilon_2/\epsilon_1) p_1^{-d_2/2} + A_{j_l, j_r}(\epsilon_1/\epsilon_2, \epsilon_2) p_2^{-d_2/2} -  A_{j_l, j_r}(\epsilon_1, \epsilon_2)) \nonumber \\
& = \sum_{j_l, j_r} (-1)^{2(j_l + j_r)} N_{j_l, j_r}^{(1, d_2)} \qty( A_{j_l, j_r}(\epsilon_1, \epsilon_2/\epsilon_1) p_1^{d_2/2} + A_{j_l, j_r}(\epsilon_1/\epsilon_2, \epsilon_2) p_2^{d_2/2} -  A_{j_l, j_r}(\epsilon_1, \epsilon_2)) \, .
\end{align}
It turns out that this equation is not enough to fix the multiplicities $ N_{j_l, j_r}^{(1, d_2)} $, and in fact it has infinitely many solutions. For example, if $ N_{j_l,j_r}^{(1, d_2)}=c^{(d_2)} $ with an integer number $c^{(d_2)}$ is a solution, then $ N_{j_l,j_r}^{(1, d_2)} =2c^{(d_2)}$ is also another solution. 
We thus need to solve higher order equations to determine them.

The 2nd order of the blowup equation can be written as
\begin{align}
\sum_{n \in \{ \pm 1/2 \}} (-1)^n e^{-V} \frac{\hat{\mathcal{Z}}_{\mathrm{pert}}^{(N)} \hat{\mathcal{Z}}_{\mathrm{pert}}^{(S)}}{\hat{\mathcal{Z}}_{\mathrm{pert}}} & \qty( \hat{Z}_2^{(N)} +  \hat{Z}_2^{(S)} +  \hat{Z}_1^{(N)} \hat{Z}_1^{(S)} ) \nonumber \\
& + \sum_{n \in \{\pm 3/2\}} (-1)^n e^{-V} \frac{\hat{\mathcal{Z}}_{\mathrm{pert}}^{(N)} \hat{\mathcal{Z}}_{\mathrm{pert}}^{(S)}}{\hat{\mathcal{Z}}_{\mathrm{pert}}}
= 0 \ .
\end{align}
Unfortunately, this equation does not help us to fix the degeneracies $ N_{j_l, j_r}^{(1, d_2)} $ because they do not appear in this order. We need the blowup equation at higher orders to fully fix $ N_{j_l, j_r}^{(1, d_2)}$. We find that at the order of the blowup equation involving $ N_{j_l, j_r}^{(4, 10)} $, the degeneracy $ N_{j_l, j_r}^{(1, 1)} $ is first fixed to be
\begin{align}
N_{j_l, j_r}^{(1, 1)} =0\ ,\quad\text{except for} ~~
N_{0, 0}^{(1, 1)} = \Bigg\{
\begin{array}{ll}
1 & \quad \text{when~} \theta=\pi \\
0 & \quad \text{when~}\theta=0
\end{array} \ .
\end{align}
Here we fixed the discrete theta angles from the geometric information: $ N_{j_l, j_r}^{(1, 1)} = 0 $ for the $ SU(2)_0 $ theory, while $ N_{0, 0}^{(1, 1)} =1 $ for the $ SU(2)_\pi $ theory. 

Now consider the $ SU(2)_0 $ theory. Solving the blowup equation at the order having terms of $ N_{j_l, j_r}^{(4, 12)} $ again gives two sets of solutions: one with $ N_{j_l, j_r}^{(1, 2)} = 0 $ for all $ (j_l, j_r) $ and the other with $ N_{j_l, j_r}^{(1, 2)} = 0 $ but $ N_{0, \frac{1}{2}}^{(1, 2)} = 1 $. However, the geometric realization suggests that there exists a BPS state with charge $ m + 2\phi $, so only the second solution is acceptable. The next order blowup equation which contains $ N_{j_l, j_r}^{(4, 13)} $ gives $ N_{j_l, j_r}^{(1, 3)} = 0 $ for all $ (j_l, j_r) $. Now to fix $ N_{j_l, j_r}^{(1, 4)} $, instead of calculating higher order equations, one can use other consistency conditions. First, we know from geometry that there are effective curves with volume $ m + 4\phi $. In addition, the blowup equation yields that $ N_{j_l, j_r}^{(1, 4)} = 0 $ for $ (j_l, j_r) \neq (0, \frac{3}{2}) $ and  $N_{1,\frac{15}{2}}^{(3,12)} = 1 - N_{0,\frac{3}{2}}^{(1, 4)} $. Notably, $ N_{1,\frac{15}{2}}^{(3,12)} $ must be a non-negative integer and $ N_{0,\frac{3}{2}}^{(1, 4)} \neq 0 $ from geometry. Therefore the only possible solution is $ N_{0,\frac{3}{2}}^{(1, 4)} = 1 $. This result reproduces all the BPS states of the $SU(2)_0$ theory up to $(d_1,d_2)=(1,4)$. In this way, a vanishing blowup equation together with other consistency conditions can determine the BPS spectrum in this theory. We expect other higher degree states can also be captured by solving higher order blowup equations in this manner. 

Next, consider the $ SU(2)_\pi $ theory. The blowup equation at the order containing $ N_{j_l, j_r}^{(4, 11)} $ fixes $ N_{j_l, j_r}^{(1, 2)} = 0 $ for all $ (j_l, j_r) $. The higher order equations containing $ N_{j_l, j_r}^{(4, 14)} $ has two sets of solutions: one with $ N_{j_l, j_r}^{(1, 3)} = 0 $ for all $ (j_l, j_r) $, and the other one with $ N_{j_l, j_r}^{(1, 3)} = 0 $ but $ N_{0, 1}^{(1, 3)} = 1 $. Also, the geometry of $\mathbb{F}_1$ says that there is a state with charge $ m + 3\phi $, so only the second one is acceptable. The next order equation containing $ N_{j_l, j_r}^{(4, 15)} $ fixes $ N_{j_l, j_r}^{(1, 4)} = 0 $. It is also possible to fix $ N_{j_l, j_r}^{(1, 4)} $ by consistency conditions instead of calculating higher order equations. The blowup equation fixes $ N_{j_l, j_r}^{(1, 4)} = 0 $  for $ (j_l, j_r) \neq (0, \frac{3}{2}) $, and $N_{1,\frac{15}{2}}^{(3,12)} = - N_{0,\frac{3}{2}}^{(1, 4)}$, from which the only possible solution is $ N_{0,\frac{3}{2}}^{(1, 4)} = 0 $. This result agrees with the spectrum of the $ SU(2)_\pi $ theory that we computed above using unity bloup equations. 

These computations show that, although the BPS spectrum is not completely determined only by solving vanishing blowup equations, we may be able to uniquely determine the BPS degeneracies by requiring the solution of the vanishing blowup equations to be physically or geometrically consistent.

\subsubsection{\texorpdfstring{5d pure $SU(3)_{\kappa \leq 7}$}{5d pure SU(3)k}}\label{sec:su3_k}

Let us now discuss rank-2 theories. A non-trivial example is the pure $SU(3)$ gauge theory at the Chern-Simons level $ \kappa $. When $ \kappa $ is even, it is geometrically engineered as
\begin{align}\label{eq:su3_even_geo}
\begin{tikzpicture}
\draw[thick](-3,0)--(0,0);	
\node at(-3.6,0) {$\mathbb{F}_{\kappa+1}{}\big|_1$};
\node at(-2.8,0.3) {${}_e$};
\node at(0.5,0) {$\mathbb{F}_{1}{}\big|_2$\ .};
\node at(-0.8,0.3) {${}_{h+(\frac{\kappa}{2}-1)f}$};
\end{tikzpicture}
\end{align}
Whereas, if $\kappa$ is odd, it is engineered by
\begin{align}\label{eq:su3_odd_geo}
\begin{tikzpicture}
\draw[thick](-3,0)--(0,0);	
\node at(-3.6,0) {$\mathbb{F}_{\kappa+1}{}\big|_1$};
\node at(-2.8,0.3) {${}_e$};
\node at(0.5,0) {$\mathbb{F}_{0}{}\big|_2$\ .};
\node at(-0.8,0.3) {${}_{h+\frac{\kappa-1}{2}f}$};
\end{tikzpicture}
\end{align}
Each geometry contains three primitive curves: the fiber $f_1$ in $\mathbb{F}_{n_1}|_1$, the fiber $f_2$ and the base $e_2$ in $\mathbb{F}_{n_2}|_2$.
The volumes of the primitive 2-cycles are
\begin{align}
&\vol(f_1) = 2\phi_1 - \phi_2, \quad
\vol(f_2) = -\phi_1 + 2\phi_2, \nonumber\\
&\vol(e_2) = \left\{ \begin{array}{ll} \left(1-\tfrac{\kappa}{2}\right)\phi_1+\phi_2 + m& \ \ {\rm for \ even}\ \kappa \\
	\frac{1-\kappa}{2}\phi_1+2\phi_2 + m & \ \ {\rm for \ odd}\ \kappa 
	\end{array}\right. \ .
\end{align}
These theories with CS level $|\kappa|\le9$ have UV completions \cite{Jefferson:2018irk,Bhardwaj:2019jtr}. The 5-brane webs for $|\kappa|\le 7$ and $|\kappa|=9$ have been constructed in \cite{Hayashi:2018lyv}. 

Surprisingly the BPS spectra of the  $SU(3)_\kappa$ theories for all $|\kappa|\le 9$ can be computed by employing our bootstrapping approach explained above. The $SU(3)_8$ theory is of particular interest. Unlike other $\kappa$, this theory at $\kappa=8$ has no consistent geometric realization \cite{Jefferson:2018irk,Bhardwaj:2019jtr,Bhardwaj:2020gyu} and also its 5-brane web is unknown yet. Moreover the usual ADHM construction for instantons in this theory does not work. So it was a quite challenging task to compute its BPS partition function. The $SU(3)_8$ theory and its blowup equations will be discussed separately in subsection~\ref{sec:SU3_8}. The $SU(3)_9$ theory, which is a KK theory, will be discussed in subsection~\ref{subsubsec:SU3/Z2}.

In this subsection, we will illustrate how to bootstrap the BPS spectra of  $|\kappa| \le 7$ theories and their dual theories using the blowup equations as well as their geometric descriptions.

Consider the $SU(3)_\kappa$ theory put on the blowup $\hat{\mathbb{C}}_2$. We turn on the magnetic flux $\textsf{F}=(n_1,n_2,B_m)$ for two Hirzebruch surfaces and a non-compact divisor of the non-normalizable K\"ahler parameter $m$. The quantization conditions for the fluxes are given by their intersections with the primitive 2-cycles as
\begin{align}
	&\textsf{F}\cdot f_1 = 2n_1-n_2\ \in \ \mathbb{Z} \ , \qquad \textsf{F}\cdot f_2 = 2n_2-n_1 \ \in \ \mathbb{Z} \nonumber \ ,\\
	&  \textsf{F}\cdot e_2 = \left\{ \begin{array}{ll} \left(1-\tfrac{\kappa}{2}\right)n_1+n_2 + B_m \ \in \ \mathbb{Z}+\frac{1}{2} & \ \ {\rm for \ even}\ \kappa \\
	\frac{1-\kappa}{2}n_1+2n_2 + B_m \ \in \ \mathbb{Z} & \ \ {\rm for \ odd}\ \kappa 
	\end{array}\right. \ ,
\end{align}
which follow from the geometric quantization condition \eqref{eq:quantization} together with the self-intersections of the curves, $f_1^2 = f_2 ^2 = 0$ and $e_2^2 = -1$ for even $\kappa$ and $e_2^2=0$ for odd $\kappa$. Each theory has 3 or more sets of magnetic fluxes satisfying the quantization conditions. We can use, for example, the following fluxes for solving the blowup equations:
\begin{equation}\label{eq:su3_r}
	n_1,n_2 \in \mathbb{Z} \quad {\rm and} \quad B_m = \left\{
\begin{array}{ll}
-\frac{1}{2}, \frac{1}{2}, \frac{3}{2} & \quad \kappa = 0, 2, 4, 6 \\
-1, 0, 1 & \quad \kappa = 1, 3, 5, 7\ .
\end{array}\right.
\end{equation}

Let us now consider the Coulomb branch of the theory where $2\phi_1-\phi_2 \ge0$ and $2\phi_2-\phi_1\ge0$ with $\phi_1,\phi_2>0$. The effective prepotential $\mathcal{E}$ on the Coulomb branch takes the following form
\begin{align}\label{eq:SU3_k_prepotential}
\mathcal{E} &= \frac{1}{\epsilon_1 \epsilon_2} 
\left( \mathcal{F} - \frac{\epsilon_1^2  + \epsilon_2^2}{12}(\phi_1 + \phi_2) +\epsilon_+^2 (\phi_1 + \phi_2) \right), \\ 
6\,\mathcal{F} &= 6 m(\phi_1^2 - \phi_1 \phi_2 + \phi_2^2)+ 3\kappa \phi_1 \phi_2(\phi_1 - \phi_2) + 8 \phi_1^3 - 3\phi_1^2 \phi_2 - 3\phi_1 \phi_2^2 + 8\phi_2^3\, ,\nonumber
\end{align}
where $m$ is the gauge coupling. The GV-invariant of this theory can be factorized as
\begin{align}
Z_{GV}(\phi_i, m; \epsilon_1, \epsilon_2) &= \mathcal{Z}_{\mathrm{pert}}(\phi_i; \epsilon_1, \epsilon_2) \cdot \mathcal{Z}_{\mathrm{inst}}(\phi_i, m; \epsilon_1, \epsilon_2) \ , \nonumber  \\
\mathcal{Z}_{\mathrm{pert}}(\phi_i; \epsilon_1, \epsilon_2) &= \PE \qty[-\frac{1+p_1 p_2}{(1-p_1)(1-p_2)} \qty(e^{-(2\phi_1 - \phi_2)} + e^{-(2\phi_2-\phi_1)} + e^{-(\phi_1 + \phi_2)})] \ , \nonumber \\
\mathcal{Z}_{\mathrm{inst}}(\phi_i, m; \epsilon_1, \epsilon_2) &= \sum_{k=0}^{\infty} q^k Z_k(\phi; \epsilon_1, \epsilon_2) \ ,
\label{eq:SU3_k_GV}
\end{align}
with the instanton fugacity $ q = e^{-m} $ and $ Z_0 = 1 $. 

We can then write the blowup equation for the $SU(3)_\kappa$ theory as
\begin{align}
\sum_{k, k'=0}^{\infty} q^{k+k'} \Lambda_{k'} \hat{Z}_k &= \sum_{\vec{n}\in \mathbb{Z}^2}\sum_{k_1,k_2=0}^\infty (-1)^{\abs{\vec{n}}} e^{-V} \frac{\mathcal{Z}_{\mathrm{pert}}^{(N)} \mathcal{Z}_{\mathrm{pert}}^{(S)}}{\mathcal{Z}_{\mathrm{pert}}} (q \,p_1^{B_m})^{k_1} \hat{Z}^{(N)}_{k_1} \cdot (q \,p_2^{B_m})^{k_2} \hat{Z}^{(S)}_{k_2} \ .
\end{align}
Again, the blowup equation at the $ k $-th order can be written as
\begin{align}\label{eq:su3_k_kinst}
\Lambda_k(m; \epsilon_1, \epsilon_2) + \hat{Z}_k(\phi_i, \epsilon_1, \epsilon_2)
=&~ p_1^{kB_m} \hat{Z}_k(\phi_i, \epsilon_1, \epsilon_2 - \epsilon_1) + p_2^{kB_m}\hat{Z}_k(\phi_i, \epsilon_1 - \epsilon_2, \epsilon_2) \nonumber \\
& + (\text{terms with } Z_{r<k} \text{ and } \Lambda_{r<k})\ .
\end{align}
The $ \Lambda_0 $ factor is then fixed to be 1 from the zeroth order expansion. In $ q^1 $ order, the blowup equation is given by
\begin{align}\label{eq:su3_k_1inst}
\Lambda_1 + \hat{Z}_1(\phi_i, \epsilon_1, \epsilon_2)
= ~& p_1^{B_m} \hat{Z}_1(\phi_i, \epsilon_1, \epsilon_2 - \epsilon_1) + p_2^{B_m} \hat{Z}_1(\phi_i, \epsilon_1 - \epsilon_2, \epsilon_2) \nonumber \\
& +  \sum_{\vec{n} \in S_1} q^{-1} e^{-V} \frac{\mathcal{Z}_{\mathrm{pert}}^{(N)} \mathcal{Z}_{\mathrm{pert}}^{(S)}}{\mathcal{Z}_{\mathrm{pert}}} \ ,
\end{align}
where the last term comes from the contribution with $k_1=k_2=0$ and $n_1^2 - n_1 n_2 +n_2^2 =1$ which is solved for $(n_1,n_2)\in S_1 = \{\pm(1, 1), \pm(1, 0), \pm(0, 1) \}$.

The three choices of $ B_m $ given in \eqref{eq:su3_r} provide three linearly independent  algebraic equations for $\hat{Z}_k(\phi_i, \epsilon_1, \epsilon_2)$ at each $k$-th order.
They allow us to find a closed form of $\hat{Z}_k(\phi_i, \epsilon_1, \epsilon_2)$. Thus, the $k$-instanton partition function $Z_k(\phi_i, \epsilon_1, \epsilon_2)$ for the $SU(3)_\kappa$ theory at $|\kappa|\le 7$ can be computed by recursively solving three blowup equations with $B_m$'s in  \eqref{eq:su3_r}. In \cite{Kim:2019uqw}, the partition functions for $|\kappa|\le 3$ were computed in this manner and checked against the results from the ADHM calculations.

The instanton partition function for the $SU(3)_4$ theory can be computed by solving the blowup equations similarly to the $ \abs{\kappa} \leq 3 $ cases. Using the effective prepotential \eqref{eq:SU3_k_prepotential} and consistent magnetic fluxes \eqref{eq:su3_r}, we solve the blowup equations in terms of volumes of primitive 2-cycles
\begin{align}\label{eq:vol of 2-cycles for SU3_4}
\vol(f_1) = 2\phi_1 - \phi_2 \, , \quad
\vol(f_2) = -\phi_1 + 2\phi_2 \, , \quad
\vol(e_2) = -\phi_1 + \phi_2 + m \, .
\end{align}
Instead of giving an explicit form of the partition function, we list some BPS states in Table~\ref{table:SU(3)_4}.\footnote{As the BPS spectrum in Table~\ref{table:SU(3)_4} is expressed as the expansion of 2-cycles, one can reconstruct the partition function given in terms of dynamical variables, by substituting the 2-cycles with \eqref{eq:vol of 2-cycles for SU3_4}. Likewise, one can also rewrite the partition function given in terms of dynamical variables as the BPS spectrum, by first converting them into the 2-cycles and then expanding the 2-cycles.}
\begin{table}
	\centering
	\begin{tabular}{|c|C{28ex}||c|C{28ex}|} \hline
		$ \mathbf{d} $ & $\oplus N_{j_l, j_r}^{\mathbf{d}} (j_l, j_r)$ & $ \mathbf{d} $ & $\oplus N_{j_l, j_r}^{\mathbf{d}} (j_l, j_r)$ \\ \hline
		$ (1, 0, 0) $ & $ (0, 0) $ & $ (1, 0, 1) $ & $ (0, 1) $ \\ \hline
		$ (1, 0, 2) $ & $ (0, 2) $ & $ (1, 0, 3) $ & $ (0, 3) $ \\ \hline
		$ (1, 1, 0) $ & $ (0, 0) $ & $ (1, 1, 1) $ & $ (0, 0) \oplus (0, 1) $ \\ \hline
		$ (1, 1, 2) $ & $ (0, 1) \oplus (0, 2) $ & $ (1, 1, 3) $ & $ (0, 2) \oplus (0, 3) $ \\ \hline
		$ (1, 2, 1) $ & $ (0, 1) $ & $ (1, 2, 2) $ & $ (0, 0) \oplus (0, 1) \oplus (0, 2) $ \\ \hline
		$ (1, 2, 3) $ & $ (0,1) \oplus (0,2) \oplus (0,3) $ & $ (1, 3, 2) $ & $ (0, 1) \oplus (0, 2) $ \\ \hline
		$ (1, 3, 3) $ & $ (0,0) \oplus (0,1) \oplus (0,2) \oplus (0,3) $ & $ (2, 0, 2) $ & $ (0, \frac{5}{2}) $ \\ \hline
		$ (2, 0, 3) $ & $ (0,\frac{5}{2}) \oplus (0,\frac{7}{2}) \oplus (\frac{1}{2},4) $ & $ (2, 1, 2) $ & $ (0, \frac{3}{2}) \oplus (0, \frac{5}{2}) $ \\ \hline
		$ (2, 1, 3) $ & $ (0,\frac{3}{2}) \oplus 3(0,\frac{5}{2}) \oplus 2(0,\frac{7}{2}) \oplus (\frac{1}{2},3) \oplus (\frac{1}{2},4) $ & $ (2, 2, 2) $ & $ (0,\frac{1}{2}) \oplus (0,\frac{3}{2}) \oplus (0,\frac{5}{2}) $ \\ \hline
		$ (2, 2, 3) $ & $ (0,\frac{1}{2}) \oplus 3(0,\frac{3}{2}) \oplus 4(0,\frac{5}{2}) \oplus 2(0,\frac{7}{2}) \oplus (\frac{1}{2},2) \oplus (\frac{1}{2},3) \oplus (\frac{1}{2},4) $ & $ (2, 3, 2) $ & $ (0,\frac{1}{2}) \oplus (0,\frac{3}{2}) \oplus (0,\frac{5}{2}) $ \\ \hline
		$ (2, 3, 3) $ & $ 3(0,\frac{1}{2}) \oplus 4(0,\frac{3}{2}) \oplus 4(0,\frac{5}{2}) \oplus 2(0,\frac{7}{2}) \oplus (\frac{1}{2},1) \oplus (\frac{1}{2},2) \oplus (\frac{1}{2},3) \oplus (\frac{1}{2},4) $ & $ (3, 0, 3) $ & $ (0, 3) \oplus (\frac{1}{2}, \frac{9}{2}) $ \\ \hline
		$ (3, 1, 3) $ & $ (0,2) \oplus 2(0,3) \oplus (0,4) \oplus (\frac{1}{2},\frac{7}{2}) \oplus (\frac{1}{2},\frac{9}{2}) $ & $ (3, 2, 3) $ & $ (0,1) \oplus 2(0,2) \oplus 3(0,3) \oplus (0,4) \oplus (\frac{1}{2},\frac{5}{2}) \oplus (\frac{1}{2},\frac{7}{2}) \oplus (\frac{1}{2},\frac{9}{2}) $ \\ \hline
		$ (3, 3, 3) $ & \multicolumn{3}{C{68.5ex}|}{$ \! (0,0) \oplus 2(0,1) \oplus 3(0,2) \oplus 3(0,3) \oplus (0,4) \oplus (\frac{1}{2},\frac{3}{2}) \oplus (\frac{1}{2},\frac{5}{2}) \oplus (\frac{1}{2},\frac{7}{2}) \oplus (\frac{1}{2},\frac{9}{2}) \! $} \\ \hline
	\end{tabular}
	\caption{BPS spectrum of $SU(3)_4$ for $ d_i \leq 3 $. Here, $\mathbf{d} = (d_1, d_2, d_3)$ labels the BPS states with charge $d_1 e_2 + d_2 f_1 + d_3 f_2$.} \label{table:SU(3)_4}
\end{table}

The $ SU(3)_5 $ theory is dual to the $ Sp(2)_\pi $ gauge theory, so they can provide a non-trivial check for the validity of the blowup equations. In the geometry \eqref{eq:su3_odd_geo} with $\kappa=5$, this duality is realized as the base-fiber duality exchanging two curve classes $e_2$ and $f_2$ in $\mathbb{F}_0$. The map between the parameters in the $ SU(3)_5 $ theory and those in the $ Sp(2)_\pi $ theory can be easily found from the geometric realization. The volumes of 2-cycles in the $ SU(3) $ frame are
\begin{align}\label{eq:su3_5_vol}
\vol(f_1)= 2 \phi_1 - \phi_2, \quad
\vol(f_2) = -\phi_1 + 2\phi_2, \quad
\vol(e_2) = -2\phi_1 + 2\phi_2 + m\ ,
\end{align}
while those in the $ Sp(2) $ frame after exchanging $e_2$ and $f_2$ are
\begin{align}
\vol(f_1) = 2\phi_1 - \phi_2, \quad
\vol(f_2) = -2\phi_1 + 2\phi_2, \quad
\vol(e_2) = -\phi_1 + 2\phi_2 + m\ .
\end{align}
From this, we find a natural map between the parameters in two dual frames as \cite{Gaiotto:2015una, Hayashi:2016abm, Hayashi:2018lyv}
\begin{align}
\phi_1^{SU} &= \phi_1^{Sp} + \frac{1}{3} m^{Sp} \ , \quad \phi_2^{SU} = \phi_2^{Sp} + \frac{2}{3} m^{Sp} \ ,\quad
m^{SU} = -\frac{2}{3} m^{Sp}\ . \label{eq:SU3 and Sp2 duality map}
\end{align}

The effective prepotential of the $Sp(2)_\pi$ theory on the Coulomb branch is written as
\begin{align}
\mathcal{E} &= \frac{1}{\epsilon_1\epsilon_2}\left(\mathcal{F} -\frac{\epsilon_1^2+\epsilon_2^2}{12}(\phi_1+\phi_2) + \epsilon_+^2(\phi_1+\phi_2)\right) \ , \nonumber \\
6\mathcal{F}
 &= 6m (2\phi_1^2 - 2\phi_1 \phi_2 + \phi_2^2)+8\phi_1^3 + 12\phi_1^2 \phi_2 - 18\phi_1 \phi_2^2 + 8\phi_2^3 \ ,
\end{align}
where $ m $ is the $ Sp(2) $ gauge coupling. One can easily check that this $\mathcal{E}$ of the $ Sp(2)_\pi $ theory coincides with that of the $ SU(3)_5 $ theory in \eqref{eq:SU3_k_prepotential} under the above parameter map up to terms independent of the dynamical K\"ahler parameters. The perturbative GV-invariants of the $Sp(2)_\pi$ gauge theory is
\begin{align}
	\mathcal{Z}_{\rm pert} = {\rm PE} \left[ -\frac{1+p_1p_2}{(1-p_1)(1-p_2)}\left(e^{-2\phi_1}+e^{-\phi_2}+e^{-(2\phi_2-2\phi_1)}+e^{-(2\phi_1-\phi_2)}\right)\right] \ .
\end{align} 

In the $ Sp(2) $ frame, there are three (and more) sets of consistent fluxes respecting the $ Sp(2) $ structure:
\begin{align}
n_1,n_2 \in \mathbb{Z} \ , \quad B_m = -1, 0, 1 \, .
\end{align}
The three sets of unity blowup equations from these magnetic fluxes can be easily solved and the solution provides a closed expression of the instanton partition function of the $ Sp(2)_\pi $ theory at each instanton order. We checked that the result perfectly matches the BPS states captured by the $ SU(3)_5 $ calculation under the parameter map \eqref{eq:SU3 and Sp2 duality map} in the K\"ahler parameter expansion. Instead of giving explicit forms of  the instanton partition functions, we list BPS spectrum up to 3-instanton order in Table~\ref{table:SU(3)_5}.
\begin{table}
	\centering
	\begin{tabular}{|c|C{28ex}||c|C{28ex}|} \hline
		$ \mathbf{d} $ & $\oplus N_{j_l, j_r}^{\mathbf{d}} (j_l, j_r)$ & $ \mathbf{d} $ & $\oplus N_{j_l, j_r}^{\mathbf{d}} (j_l, j_r)$ \\ \hline
		$ (1, 0, 0) $ & $ (0, \frac{1}{2}) $ & $ (1, 0, 1) $ & $ (0, \frac{3}{2}) $ \\ \hline
		$ (1, 0, 2) $ & $ (0, \frac{5}{2}) $ & $ (1, 1, 0) $ & $ (0, \frac{1}{2}) $ \\ \hline
		$ (1, 1, 1) $ & $ (0, \frac{1}{2}) \oplus (0, \frac{3}{2}) $ & $ (1, 1, 2) $ & $ (0, \frac{3}{2}) \oplus (0, \frac{5}{2}) $ \\ \hline
		$ (1, 2, 0) $ & $ (0, \frac{1}{2}) $ & $ (1, 2, 1) $ & $ (0, \frac{1}{2}) \oplus (0, \frac{3}{2}) $ \\ \hline
		$ (1, 2, 2) $ & $ (0, \frac{1}{2}) \oplus (0, \frac{3}{2}) \oplus (0, \frac{5}{2}) $ & $ (2, 0, 1) $ & $ (0, \frac{5}{2}) $ \\ \hline
		$ (2, 0, 2) $ & $ (0, \frac{5}{2}) \oplus (0, \frac{7}{2}) \oplus (\frac{1}{2}, 4) $ & $ (2, 1, 1) $ & $ (0, \frac{3}{2}) \oplus (0, \frac{5}{2}) $ \\ \hline
		$ (2, 1, 2) $ & $ (0, \frac{3}{2}) \oplus 3(0, \frac{5}{2}) \oplus 2(0, \frac{7}{2}) \oplus (\frac{1}{2}, 3) \oplus (\frac{1}{2}, 4) $ & $ (2, 2, 1) $ & $ (0, \frac{1}{2}) \oplus (0, \frac{3}{2}) \oplus (0, \frac{5}{2}) $ \\ \hline
		$ (2, 2, 2) $ & $ (0, \frac{1}{2}) \oplus 3(0, \frac{3}{2}) \oplus 4(0, \frac{5}{2}) \oplus 2(0, \frac{7}{2}) \oplus (\frac{1}{2}, 2) \oplus (\frac{1}{2}, 3) \oplus (\frac{1}{2}, 4) $ & $ (3, 0, 1) $ & $ (0, \frac{7}{2}) $ \\ \hline
		$ (3, 0, 2) $ & $ (0,\frac{5}{2}) \oplus (0,\frac{7}{2}) \oplus 2(0,\frac{9}{2}) \oplus (\frac{1}{2},4) \oplus (\frac{1}{2},5) \oplus (1,\frac{11}{2}) $ & $ (3, 1, 1) $ & $ (0,\frac{5}{2}) \oplus (0,\frac{7}{2}) $ \\ \hline
		$ (3, 1, 2) $ & $ (0,\frac{3}{2}) \oplus 3(0,\frac{5}{2}) \oplus 5(0,\frac{7}{2}) \oplus 3(0,\frac{9}{2}) \oplus	(\frac{1}{2},3) \oplus 3(\frac{1}{2},4) \oplus 2(\frac{1}{2},5) \oplus (1,\frac{9}{2}) \oplus (1,\frac{11}{2}) $ & $ (3, 2, 1) $ & $ (0,\frac{3}{2}) \oplus (0,\frac{5}{2}) \oplus (0,\frac{7}{2}) $ \\ \hline
		$ (3, 2, 2) $ & \multicolumn{3}{C{68ex}|}{$ (0,\frac{1}{2}) \oplus 3(0,\frac{3}{2}) \oplus 8(0,\frac{5}{2}) \oplus 7(0,\frac{7}{2}) \oplus 4(0,\frac{9}{2}) \oplus (\frac{1}{2},2) \oplus 3(\frac{1}{2},3) \oplus 4(\frac{1}{2},4) \oplus 2(\frac{1}{2},5) \oplus (1,\frac{7}{2}) \oplus (1,\frac{9}{2}) \oplus (1,\frac{11}{2}) $} \\ \hline
	\end{tabular}
	\caption{BPS spectrum of $SU(3)_5$ for $d_1 \leq 3$ and $ d_2, d_3 \leq 2 $. Here, $\mathbf{d} = (d_1, d_2, d_3)$ labels the BPS states with charge $d_1 e_2 + d_2 f_1 + d_3 f_2$.} \label{table:SU(3)_5}
\end{table}

In the case of $ \kappa = 6 $, one should be careful about the $ \Lambda $ factor. When background magnetic flux is $ B_m = 3/2 $, the last term of \eqref{eq:su3_k_1inst} in the K\"ahler parameter expansion  contains a $ \phi_i $ independent term :
\begin{align}
-p_1^{1/2} p_2^{1/2}\ \in\ \sum_{\vec{n} \in S_1} q^{-1} e^{-V} \frac{\mathcal{Z}_{\mathrm{pert}}^{(N)} \mathcal{Z}_{\mathrm{pert}}^{(S)}}{\mathcal{Z}_{\mathrm{pert}}} \ .
\end{align}
This term should be absorbed into $ \Lambda_1 $ on the left side of \eqref{eq:su3_k_1inst}. Thus we have
\begin{align}
&\Lambda_1 = 0 \quad \text{for} \quad B_m = \pm \tfrac{1}{2} \ , \qquad \Lambda_1 = -p_1^{1/2} p_2^{1/2} \quad \text{for} \quad B_m = \tfrac{3}{2} \ .
\end{align}
Then the solution $ Z_1(\phi_i; \epsilon_1, \epsilon_2) $ does not contain $ \phi_i $ independent terms in the expansion. Similarly, the last term of the blowup equation at $ q^2 $ order
\begin{align}\label{eq:su3_6_2inst_blowup}
\Lambda_2(m; \epsilon_1, \epsilon_2) + \hat{Z}_2(\phi_i, \epsilon_1, \epsilon_2)
=&~ p_1^{2B_m} \hat{Z}_2(\phi_i, \epsilon_1, \epsilon_2 - \epsilon_1) + p_2^{2B_m}\hat{Z}_2(\phi_i, \epsilon_1 - \epsilon_2, \epsilon_2) \nonumber \\
& + (\text{terms from } \hat{Z}_{1} \text{ and } \Lambda_1)
\end{align}
contains $ \phi_i $ independent terms for three magnetic fluxes $ B_m = -\tfrac12, \tfrac12, \tfrac32 $ as
\begin{align}
	\frac{-2 p_1 p_2}{(1-p_1) (1-p_2) (p_1-p_2)^2} \, , \ \ \frac{-2p_1^2 p_2^2}{(1-p_1)(1-p_2)(p_1-p_2)^2}  \, , \ \ \frac{-F(p_1, p_2)}{(1-p_1)(1-p_2)(p_1 - p_2)^2}
\end{align}
respectively, where
\begin{align}
F(p_1, p_2) = p_1 p_2 &(-p_1^2 -p_1^4 + 2 p_1 p_2 + p_1^3 p_2 + p_1^4 p_2 - p_2^2 \nonumber \\
& + 2 p_1^2 p_2^2 - p_1^3 p_2^2 + p_1 p_2^3 - p_1^2 p_2^3 - p_2^4 + p_1 p_2^4) \ .
\end{align}
After absorbing these terms into $ \Lambda_2 $, the blowup equation at $q^2$ order is solved consistently. However, when we take the plethystic logarithm and extract single particle states from the resulting partition function, it turns out that the partition function contain unexpected $ \phi_i $ independent term given by
\begin{align}
-\frac{p_1 p_2 q^2}{(1-p_1)^2 (1-p_2)^2}\ \in\ \PLog\qty[1 + q Z_1(\phi_i; \epsilon_1, \epsilon_2) + q^2 Z_2(\phi_i; \epsilon_1, \epsilon_2)] \ ,
\end{align}
where $ \PLog $ denotes the plethystic logarithm. The PE of this term provides some $\phi_i$ independent terms and again we put them into the $ \Lambda_2 $ factor. After all, we find that the $ \Lambda_2 $ factor is given by
\begin{align}
&\Lambda_2 = 0 \quad \text{for } B_m = \pm \tfrac{1}{2}, \nonumber \\
&\Lambda_2 = -p_1 p_2 (p_1 + p_2) \quad \text{for} \quad B_m = \tfrac{3}{2} \ .
\end{align}
In this way, we can solve the blowup equation iteratively while absorbing all $\phi_i$ independent terms into the $ \Lambda $ factor. The resulting BPS spectrum of the $ SU(3)_6 $ theory is given in Table~\ref{table:SU3_6}.
\begin{table}
	\centering
	\begin{tabular}{|c|C{28ex}||c|C{28ex}|} \hline
		$ \mathbf{d} $ & $\oplus N_{j_l, j_r}^{\mathbf{d}} (j_l, j_r)$ & $ \mathbf{d} $ & $\oplus N_{j_l, j_r}^{\mathbf{d}} (j_l, j_r)$ \\ \hline
		$ (1, 0, 0) $ & $ (0, 0) $ & $ (1, 0, 1) $ & $ (0, 1) $ \\ \hline
		$ (1, 0, 2) $ & $ (0, 2) $ & $ (1, 0, 3) $ & $ (0, 3) $ \\ \hline
		$ (1, 1, 1) $ & $ (0, 0) \oplus (0, 1) $ & $ (1, 1, 2) $ & $ (0, 1) \oplus (0, 2) $ \\ \hline
		$ (1, 1, 3) $ & $ (0, 2) \oplus (0, 3) $ & $ (1, 2, 0) $ & $ (0, 0) $ \\ \hline
		$ (1, 2, 1) $ & $ (0, 0) \oplus (0, 1) $ & $ (1, 2, 2) $ & $ (0, 0) \oplus (0, 1) \oplus (0, 2) $ \\ \hline
		$ (1, 2, 3) $ & $ (0, 1) \oplus (0, 2) \oplus (0, 3) $ & $ (1, 3, 1) $ & $ (0, 1) $ \\ \hline
		$ (1, 3, 2) $ & $ (0,1) \oplus (0,2) $ & $ (1, 3, 3) $ & $ (0,0) \oplus (0,1) \oplus (0,2) \oplus (0,3) $ \\ \hline
		$ (2, 0, 2) $ & $ (0, \frac{5}{2}) $ & $ (2, 0, 3) $ & $ (0,\frac{5}{2}) \oplus (0,\frac{7}{2}) \oplus (\frac{1}{2},4) $ \\ \hline
		$ (2, 1, 2) $ & $ (0, \frac{3}{2}) \oplus (0, \frac{5}{2}) $ & $ (2, 1, 3) $ & $ (0,\frac{3}{2}) \oplus 3(0,\frac{5}{2}) \oplus 2(0,\frac{7}{2}) \oplus (\frac{1}{2},3) \oplus (\frac{1}{2},4) $ \\ \hline
		$ (2, 2, 1) $ & $ (0, \frac{1}{2}) $ & $ (2, 2, 2) $ & $ (0, \frac{1}{2}) \oplus 2(0, \frac{3}{2}) \oplus (0, \frac{5}{2}) $ \\ \hline
		$ (2, 2, 3) $ & $ (0,\frac{1}{2}) \oplus 3(0,\frac{3}{2}) \oplus 5(0,\frac{5}{2}) \oplus 2(0,\frac{7}{2}) \oplus (\frac{1}{2},2) \oplus (\frac{1}{2},3) \oplus (\frac{1}{2},4) $ & $ (2, 3, 1) $ & $ (0, \frac{1}{2}) $ \\ \hline
		$ (2, 3, 2) $ & $ 2(0,\frac{1}{2}) \oplus 2(0,\frac{3}{2}) \oplus (0,\frac{5}{2}) $ & $ (2, 3, 3) $ & $ 3(0,\frac{1}{2}) \oplus 5(0,\frac{3}{2}) \oplus 5(0,\frac{5}{2}) \oplus 2(0,\frac{7}{2}) \oplus (\frac{1}{2},1) \oplus (\frac{1}{2},2) \oplus (\frac{1}{2},3) \oplus (\frac{1}{2},4) $ \\ \hline
		$ (3, 0, 3) $ & $ (0,3) \oplus (\frac{1}{2},\frac{9}{2}) $ & $ (3, 1, 3) $ & $ (0,2) \oplus 2(0,3) \oplus (0,4) \oplus (\frac{1}{2},\frac{7}{2}) \oplus (\frac{1}{2},\frac{9}{2}) $ \\ \hline
		$ (3, 2, 2) $ & $ (0, 2) $ & $ (3, 2, 3) $ & $ (0,1) \oplus 3(0,2) \oplus 5(0,3) \oplus (0,4) \oplus (\frac{1}{2},\frac{5}{2}) \oplus 2(\frac{1}{2},\frac{7}{2}) \oplus (\frac{1}{2},\frac{9}{2}) $ \\ \hline
		$ (3, 3, 2) $ & $ (0, 1) \oplus (0, 2) $ & $ (3, 3, 3) $ & $ (0,0) \oplus 3(0,1) \oplus 7(0,2) \oplus 6(0,3) \oplus (0,4) \oplus (\frac{1}{2},\frac{3}{2}) \oplus 2(\frac{1}{2},\frac{5}{2}) \oplus 2(\frac{1}{2},\frac{7}{2}) \oplus (\frac{1}{2},\frac{9}{2}) $ \\ \hline
	\end{tabular}
	\caption{BPS spectrum of the $SU(3)_6$ theory for $d_i \leq 3$. Here, $ \mathbf{d} = (d_1, d_2, d_3) $ labels the BPS states with charge $d_1 e_2 + d_2 f_1 + d_3 f_2$.} \label{table:SU3_6}
\end{table}

When $ \kappa = 7 $, the fiber-base duality of $ \mathbb{F}_0 $ in the geometry \eqref{eq:su3_odd_geo}  gives the $ G_2 $ gauge theory description. The $ G_2 $ instanton partition function can be calculated using the ADHM-like method in \cite{Kim:2018gjo} or using the topological vertex in \cite{Hayashi:2018bkd}. Here, we will present the partition function computation of this theory using blowup equations in both the $SU(3)$ and $G_2$ descriptions. In the $ SU(3) $ frame, the volumes of 2-cycles in the geometry are
\begin{align}
\vol(f_1) &= 2\phi_1 - \phi_2\,, &
\vol(f_2) &= -\phi_1 + 2\phi_2\,, &
\vol(e_2) &= -3\phi_1 + 2\phi_2 + m \ .
\end{align}
On the other hand, in the $ G_2 $ frame after exchanging the fiber and the base of $ \mathbb{F}_0 $, the volumes are 
\begin{align}
\vol (f_1) &= 2\phi_1 - \phi_2\,,&
\vol (f_2) &= -3\phi_1 + 2\phi_2 \,,&
\vol (e_2) &= -\phi_1 + 2\phi_2 + m \ .
\end{align}
The parameters in these two descriptions are thus mapped each other as \cite{Hayashi:2018lyv}
\begin{align}
\phi_1^{SU} &= \phi_1^{G_2} + \frac{1}{3} m_0^{G_2}\,, \quad
\phi_2^{SU} = \phi_2 + \frac{2}{3} m_0^{G_2}\,, \quad
m^{SU} = -\frac{1}{3} m^{G_2}\, .
\end{align}

In the $ G_2 $ description, the effective prepotential on the Coulomb branch where $2\phi_1-\phi_2>0$ and $2\phi_2-3\phi_1>0$ with $\phi_1,\phi_2>0$ is given by
\begin{align}
\mathcal{E} &= \frac{1}{\epsilon_1 \epsilon_2} 
\left( \mathcal{F} - \frac{\epsilon_1^2  + \epsilon_2^2}{12}(\phi_1 + \phi_2) +\epsilon_+^2 (\phi_1 + \phi_2) \right) \, , \nonumber \\ 
6\mathcal{F} &= 6 m \qty(3\phi_1^2 - 3\phi_1 \phi_2 + \phi_2^2 )+ 8\phi_1^3 + 18\phi_1^2 \phi_2 - 24\phi_1 \phi_2^2 + 8\phi_2^2  \, .
\end{align}
One can check that this agrees with the $ SU(3)_7 $ effective prepotential under the above parameter map up to terms independent of $\phi_i$. The perturbative GV-invariants of the $G_2$ gauge theory is
\begin{align}
	\mathcal{Z}_{\rm pert} &= {\rm PE} \bigg[ -\frac{1+p_1p_2}{(1-p_1)(1-p_2)}\left(e^{-\phi_1}+e^{-\phi_2} +e^{-(3\phi_1-\phi_2)}+e^{-(2\phi_1-\phi_2)}\right. \nonumber \\
	&\qquad \qquad \qquad \qquad \qquad \qquad \left.+e^{-(\phi_2-\phi_1)}+e^{-(2\phi_2-3\phi_1)}\right)\bigg] \ .
\end{align} 

There are again three sets of consistent magnetic fluxes respecting the $ G_2 $ structure:
\begin{align}
n_1,n_2 \in \mathbb{Z} \ , \quad B_m  = -1, 0, 1 \ .
\end{align}
We computed the unity blowup equations built from these fluxes and checked that two results from the $SU(3)_7$ theory and the $G_2$ theory perfectly agree with each other in the K\"ahler parameter expansion. The BPS spectrum of this theory can be found in Table~\ref{table:SU(3)_7}.
\begin{table}
	\centering
	\begin{tabular}{|c|C{28ex}||c|C{28ex}|} \hline
		$ \mathbf{d} $ & $\oplus N_{j_l, j_r}^{\mathbf{d}} (j_l, j_r)$ & $ \mathbf{d} $ & $\oplus N_{j_l, j_r}^{\mathbf{d}} (j_l, j_r)$ \\ \hline
		$(1, 0, 0)$ & $(0, \frac{1}{2})$ & $(1, 0, 1)$ & $(0, \frac{3}{2})$ \\ \hline
		$(1, 0, 2)$ & $(0, \frac{5}{2})$ & $(1, 1, 0)$ & $(0, \frac{1}{2})$ \\ \hline
		$(1, 1, 1)$ & $(0, \frac{1}{2}) \oplus (0, \frac{3}{2})$ & $(1, 1, 2)$ & $(0, \frac{3}{2}) \oplus (0, \frac{5}{2})$  \\ \hline
		$(1, 2, 0)$ & $(0, \frac{1}{2})$ & $(1, 2, 1)$ & $(0, \frac{1}{2}) \oplus (0, \frac{3}{2})$ \\ \hline
		$(1, 2, 2)$ & $(0, \frac{1}{2}) \oplus (0, \frac{3}{2}) \oplus (0, \frac{5}{2})$ & $(2, 0, 1)$ & $(0, \frac{5}{2})$ \\ \hline
		$(2, 0, 2)$ & $(0, \frac{5}{2}) \oplus (0, \frac{7}{2}) \oplus (\frac{1}{2}, 4)$ & $(2, 1, 1)$ & $(0, \frac{3}{2}) \oplus (0, \frac{5}{2})$ \\ \hline
		$(2, 1, 2)$ & $(0, \frac{3}{2}) \oplus 3(0, \frac{5}{2}) \oplus 2(0, \frac{7}{2}) \oplus (\frac{1}{2}, 3) \oplus (\frac{1}{2}, 4)$ & $(2, 2, 1)$ & $(0, \frac{1}{2}) \oplus (0, \frac{3}{2}) \oplus (0, \frac{5}{2})$ \\ \hline
		$(2, 2, 2)$ & $(0, \frac{1}{2}) \oplus 3(0, \frac{3}{2}) \oplus 4(0, \frac{5}{2}) \oplus 2(0, \frac{7}{2}) \oplus (\frac{1}{2}, 2) \oplus (\frac{1}{2}, 3) \oplus (\frac{1}{2}, 4)$ & $ (3, 0, 1) $ & $ (0, \frac{7}{2}) $ \\ \hline
		$ (3, 0, 2) $ & $ (0,\frac{5}{2}) \oplus (0,\frac{7}{2}) \oplus 2(0,\frac{9}{2}) \oplus (\frac{1}{2},4) \oplus (\frac{1}{2},5) \oplus (1,\frac{11}{2}) $ & $ (3, 1, 1) $ & $ (0, \frac{5}{2}) \oplus (0, \frac{7}{2}) $ \\ \hline
		$ (3, 1, 2) $ & $ (0,\frac{3}{2}) \oplus 3(0,\frac{5}{2}) \oplus 5(0,\frac{7}{2}) \oplus 3(0,\frac{9}{2}) \oplus (\frac{1}{2},3) \oplus 3(\frac{1}{2},4) \oplus 2(\frac{1}{2},5) \oplus (1,\frac{9}{2}) \oplus (1,\frac{11}{2}) $ & $ (3, 2, 1) $ & $ (0, \frac{3}{2}) \oplus (0, \frac{5}{2}) \oplus (0, \frac{7}{2}) $ \\ \hline
		$ (3, 2, 2) $ & \multicolumn{3}{C{68ex}|}{$ (0,\frac{1}{2}) \oplus 3(0,\frac{3}{2}) \oplus 8(0,\frac{5}{2}) \oplus 7(0,\frac{7}{2}) \oplus 4(0,\frac{9}{2}) \oplus (\frac{1}{2},2) \oplus 3(\frac{1}{2},3) \oplus 4(\frac{1}{2},4) \oplus 2(\frac{1}{2},5) \oplus (1,\frac{7}{2}) \oplus (1,\frac{9}{2}) \oplus (1,\frac{11}{2}) $} \\ \hline
	\end{tabular}
	\caption{BPS spectrum of the $SU(3)_7$ theory for $d_1 \leq 3$ and $ d_2, d_3 \leq 2 $. Here, $\mathbf{d} = (d_1, d_2, d_3)$ labels the BPS states with charge $d_1 e_2 + d_2 f_1 + d_3 f_2$.} \label{table:SU(3)_7}
\end{table}

\subsubsection{\texorpdfstring{6d minimal $ SU(3) $ SCFT on a circle with $ \mathbb{Z}_2 $ twist: 5d $SU(3)_9$}{6d SU(3) gauge theory with Z2 twist}} \label{subsubsec:SU3/Z2}

Elliptic genera of many 6d theories, for instance,  E-string, M-string, or 6d SCFTs compactified on a circle without twist, have been evaluated using the blowup equations \cite{Gu:2018gmy, Gu:2019dan, Gu:2019pqj, Gu:2020fem}. In this subsection, we will demonstrate with a simple example  of how to bootstrap BPS partition functions (or elliptic genera) for 6d SCFTs on a circle with outer automorphism twists by solving the blowup equations. 

We will consider a circle compactification of the 6d minimal $SU(3)$ SCFT with $\mathbb{Z}_2$ automorphism twist. This theory is dual to the 5d $SU(3)$ gauge theory at CS-level $\kappa=9$ \cite{Jefferson:2017ahm},
\begin{align}\label{eq:su3_9-6d}
\begin{tikzpicture}
\draw (0, 0) node {$ SU(3)_9 $}
(2, 0) node {$ = $}
(4, 0.3) node {$ \mathfrak{su}(3)^{(2)} $}
(4, -0.3) node {$ 3 $};
\end{tikzpicture}
\end{align}
We follow the notations in \cite{Bhardwaj:2019fzv} to denote 6d theories. It is geometrically engineered by gluing $ \mathbb{F}_{10} $ and $ \mathbb{F}_0 $ \cite{Jefferson:2018irk}:
\begin{align}\label{eq:su3_9_geo}
\begin{tikzpicture}
\draw[thick](-3,0)--(0,0);	
\node at(-3.6,0) {$\mathbb{F}_{10}{}\big|_1$};
\node at(-2.8,0.3) {${}_e$};
\node at(0.5,0) {$\mathbb{F}_{0}{}\big|_2$};
\node at(-0.45,0.3) {${}_{h+4f}$};
\end{tikzpicture} 
\end{align}
In geometry, the duality is simply the exchange of the base and the fiber classes in $\mathbb{F}_0$.
We will solve the blowup equations for this theory in both 5d and 6d frames and compare the results.

In the 5d $SU(3)_9$ theory, the effective prepotential $\mathcal{E}$ is given by
\begin{align}\label{eq:su3_9_prepotential}
\mathcal{E} &= \frac{1}{\epsilon_1 \epsilon_2} 
\left( \mathcal{F} - \frac{\epsilon_1^2  + \epsilon_2^2}{12}(\phi_1 + \phi_2) +\epsilon_+^2 (\phi_1 + \phi_2) \right), \\ 
6 \mathcal{F} &=  6m\big(\phi_1^2 - \phi_1 \phi_2 + \phi_2^2\big) +8\phi_1^3 + 24\phi_1^2 \phi_2 - 30\phi_1 \phi_2^2 + 8\phi_2^2  \ , \nonumber 
\end{align}
where $m$ is the gauge coupling. The GV-invariant of this theory takes the same form of \eqref{eq:SU3_k_GV}. It is convenient to use volumes of three independent 2-cycles in the geometry \eqref{eq:su3_9_geo} as the basis of states in the GV-invariant,
\begin{align}\label{eq:su3_9_vol}
\vol (f_1) =  2\phi_1 - \phi_2, \quad
\vol (f_2) = -\phi_1 + 2\phi_2, \quad
\vol (e_2) = -4\phi_1 + 2\phi_2 + m\ .
\end{align}
Unlike the other $SU(3)_\kappa$ theories with $|\kappa|\le 7$, this theory has only one set of consistent magnetic fluxes respecting the $ SU(3) $ structure:
\begin{align}\label{eq:MFforSU3_9}
	n_1,n_2\in \mathbb{Z} \ , \quad B_m = 0 \ .
\end{align}
This set of fluxes gives rise to a unity blowup equation. As we discussed already, a single unity blowup equation is enough to compute the BPS invariants. By solving the unity blowup equation, we find the BPS spectrum of the $SU(3)_9$ theory listed in Table~\ref{table:SU3_9}.
\begin{table}
	\centering
	\begin{tabular}{|c|C{28ex}||c|C{28ex}|} \hline
		$\mathbf{d}$ & $\oplus N_{j_l, j_r}^{\mathbf{d}} (j_l, j_r)$ & $\mathbf{d}$ & $\oplus N_{j_l, j_r}^{\mathbf{d}} (j_l, j_r)$ \\ \hline
		$(1, 0, 0)$ & $(0, \frac{1}{2})$ &$ (1, 0, 1)$ & $(0, \frac{3}{2})$ \\ \hline
		$(1, 0, 2)$ & $(0, \frac{5}{2})$ & $(1, 1, 0)$ & $(0, \frac{1}{2})$ \\ \hline
		$(1, 1, 1)$ & $(0, \frac{1}{2}) \oplus (0, \frac{3}{2})$ & $(1, 1, 2)$ & $(0, \frac{3}{2}) \oplus (0, \frac{5}{2})$ \\ \hline
		$(1, 2, 1)$ & $(0, \frac{1}{2}) \oplus (0, \frac{3}{2})$ & $(1, 2, 2)$ & $(0, \frac{1}{2}) \oplus (0, \frac{3}{2}) \oplus (0, \frac{5}{2})$ \\ \hline
		$(2, 0, 1)$ & $(0, \frac{5}{2})$ & $(2, 0, 2)$ & $(0, \frac{5}{2}) \oplus (0, \frac{7}{2}) \oplus (\frac{1}{2}, 4)$ \\ \hline
		$(2, 1, 1)$ & $(0, \frac{3}{2}) \oplus (0, \frac{5}{2})$ & $(2, 1, 2)$ & $(0, \frac{3}{2}) \oplus 3(0, \frac{5}{2}) \oplus 2(0, \frac{7}{2}) \oplus (\frac{1}{2}, 3) \oplus (\frac{1}{2}, 4)$ \\ \hline
		$(2, 2, 1)$ & $(0, \frac{1}{2}) \oplus (0, \frac{3}{2}) \oplus (0, \frac{5}{2})$ & $(2, 2, 2)$ & $(0, \frac{1}{2}) \oplus 3(0, \frac{3}{2}) \oplus 4(0, \frac{5}{2}) \oplus 2(0, \frac{7}{2}) \oplus (\frac{1}{2}, 2) \oplus (\frac{1}{2}, 3) \oplus (\frac{1}{2}, 4)$ \\ \hline
		$ (3, 0, 1) $ & $ (0, \frac{7}{2}) $ & $ (3, 0, 2) $ & $ (0,\frac{5}{2}) \oplus (0,\frac{7}{2}) \oplus 2(0,\frac{9}{2}) \oplus (\frac{1}{2},4) \oplus (\frac{1}{2},5) \oplus (1,\frac{11}{2}) $ \\ \hline
		$ (3, 1, 1) $ & $ (0,\frac{5}{2}) \oplus (0,\frac{7}{2}) $ & $ (3, 1, 2) $ & $ (0,\frac{3}{2}) \oplus 3(0,\frac{5}{2}) \oplus	5(0,\frac{7}{2}) \oplus 3(0,\frac{9}{2}) \oplus (\frac{1}{2},3) \oplus 3(\frac{1}{2},4) \oplus 2(\frac{1}{2},5) \oplus (1,\frac{9}{2}) \oplus (1,\frac{11}{2}) $ \\ \hline
		$ (3, 2, 1) $ & $ (0, \frac{3}{2}) \oplus (0, \frac{5}{2}) \oplus (0, \frac{7}{2}) $ & $ (3, 2, 2) $ & $ (0,\frac{1}{2}) \oplus 3(0,\frac{3}{2}) \oplus 8(0,\frac{5}{2}) \oplus 7(0,\frac{7}{2}) \oplus 4(0,\frac{9}{2}) \oplus (\frac{1}{2},2) \oplus 3(\frac{1}{2},3) \oplus 4(\frac{1}{2},4) \oplus 2(\frac{1}{2},5) \oplus (1,\frac{7}{2}) \oplus (1,\frac{9}{2}) \oplus (1,\frac{11}{2}) $ \\ \hline
	\end{tabular}
	\caption{BPS spectrum of $SU(3)_9$ theory for $d_1 \leq 3$ and $ d_2, d_3 \leq 2 $. Here, $\mathbf{d} = (d_1, d_2, d_3)$ labels BPS states with charge $d_1 e_2 + d_2 f_1 + d_3 f_2$ cycle.} \label{table:SU3_9}
\end{table}

We now perform a similar computation in the perspective of the 6d $SU(3)$ gauge theory with $\mathbb{Z}_2$ twist. In the 6d frame, the volumes of 2-cycles in the geometry are 
\begin{equation}
	{\rm vol}(f_1) = \frac{\tau}{4} - \phi_1 \ , \quad {\rm vol}(f_2) = 2\phi_1 \ , \quad {\rm vol}(e_2) = 3\phi_0+2\phi_1-\frac{\tau}{2}\ .
\end{equation}
The effective prepotential in this frame is given in \eqref{eq:E-SU3-Z2} with a shift $\phi_0 \rightarrow \phi_0-\frac{1}{16}\tau$. 

The GV-invariant for this 6d theory can be written as
\begin{align}
Z_{GV}(\Phi, \phi_1, \tau; \epsilon_{1,2})
= \mathcal{Z}_{\mathrm{pert}}(\phi_1, \tau; \epsilon_{1,2})  \qty(1 + \sum_{k=1}^{\infty} e^{-k \Phi} Z_k(\phi_1, \tau, \epsilon_{1,2}))\ ,
\end{align}
where $ Z_k(\phi_1, \tau, \epsilon_{1,2}) $ is the $ k $-string elliptic genus and $\Phi\equiv 3\phi_0-\tau/2$ which is the string number fugacity. Here, the perturbative part can be read off from the 6d $SU(3)$ vector multiplets under the $\mathbb{Z}_2$ automorphism twist, which is given by
\begin{align}
\mathcal{Z}_{\mathrm{pert}}
= \PE \qty[ -\frac{1+p_1 p_2}{(1-p_1)(1-p_2)} \frac{1}{1-q} \qty(e^{-2\phi_1} + (q^{1/4}+q^{3/4})(e^{-\phi_1}+e^{\phi_1}) + q e^{2\phi_1}) ] , 
\end{align}
where $ q = e^{-\tau} $. See Appendix~\ref{appendix:1-loop} for more details.

There are three sets of consistent magnetic fluxes preserving the affine $A^{(2)}_2$ algebra structure:
\begin{align}
	n_1 \in \mathbb{Z}\ , \quad B_\tau = 0  \ , \quad n_0 \in \mathbb{Z}+B_0 \ \ {\rm with } \ \  B_0=0,1/3,2/3 \ .
\end{align}
Then the elliptic genus can be found by solving three unity blowup equations from these three flux sets. For example, the blowup equations at 1-string order are
\begin{align}
&\Lambda(\tau; \epsilon_1, \epsilon_2) \hat{Z}_1(\phi_1, \tau; \epsilon_1, \epsilon_2) \nonumber \\
&= \sum_{\vec{n} = (n, 2n)} e^{-V} \frac{\hat{\mathcal{Z}}_{\mathrm{pert}}^{(N)} \hat{\mathcal{Z}}_{\mathrm{pert}}^{(S)}}{\hat{\mathcal{Z}}_{\mathrm{pert}}} \qty(p_1^{3n + B_0} \hat{Z}_1(\phi_1, \epsilon_1, \epsilon_2 - \epsilon_1) + p_2^{3n + B_0} \hat{Z}_1(\phi_1, \epsilon_1 - \epsilon_2, \epsilon_2) ) \nonumber \\
& \qquad + \sum_{\vec{n} = (n, 2n \pm 1)} e^\Phi e^{-V} \frac{\hat{\mathcal{Z}}_{\mathrm{pert}}^{(N)} \hat{\mathcal{Z}}_{\mathrm{pert}}^{(S)}}{\hat{\mathcal{Z}}_{\mathrm{pert}}} \ .
\end{align}
Here the prefactor can be computed by collecting all $\phi_i$ independent terms in the blowup equation as follows:
\begin{align}
\Lambda(\tau; \epsilon_1, \epsilon_2) = \sum_{\vec{n} = (n, 2n)} e^{-V} \ .
\end{align}
The solution at 1-instanton string is then given by
\begin{align}
Z_1(\phi_1, \tau; \epsilon_1, \epsilon_2)
&= \frac{e^{-2\phi_1} p_1 p_2 (1+p_1 p_2)}{(1-p_1)(1-p_2) (e^{-2\phi_1} - p_1 p_2) (1-e^{-2\phi_1} p_1 p_2)} \nonumber \\
& \quad + \frac{e^{-\phi_1}(1+e^{-2\phi_1}) p_1 p_2 (1+p_1 p_2)}{(1-p_1) (1-p_2) (e^{-2\phi_1} - p_1 p_2) (1-e^{-2\phi_1} p_1 p_2)} q^{1/4} \nonumber \\
& \quad + \frac{e^{-2\phi_1} (1 + 2p_1 p_2 + 2p_1^2 p_2^2 + p_1^3 p_2^3)}{(1-p_1)(1-p_2) (e^{-2\phi_1} - p_1p_2)(1-e^{-2\phi_1} p_1 p_2)} q^{1/2} + \cdots \ .
\end{align}
This is in perfect agreement with the BPS spectrum of the $ SU(3)_9 $ theory given in Table \ref{table:SU3_9} and also topological vertex as well as the ADHM calculations \cite{Kim:2021cua}.

%% file: sec-BE-review.tex
\subsection{Blowup equation review}\label{sec:BlowupEqReview}

To obtain the partition function $Z$ defined in \eqref{eq:Z}, we first consider the partition function $\hat{Z}$ on blowup $\hat{\mathbb{C}}^2$, where the origin of the $\mathbb{C}^2$ is replaced by a compact 2-cycle $\mathbb{P}^1$. The $\hat{\mathbb{C}}^2$ can be parametrized by the projective coordinates $(z_0,z_1,z_2)\sim(\lambda^{-1}z_0,\lambda z_1,\lambda z_2)$ for $\lambda\in\mathbb{C}\setminus\{0\}$. The Lorentz generators $J_{1,2}$ act on the $\hat{\mathbb{C}}^2$ by 
\begin{align}
(z_0,z_1,z_2)\mapsto(z_0,e^{\epsilon_1}z_1,e^{\epsilon_2}z_2)\ ,
\end{align}
with parameters $\epsilon_{1,2}$ for the Cartans of the $SO(4)$ rotations. There are now two fixed points of the Lorentz rotations, the North pole and the South pole of the coordinates $(0,1,0)$ and $(0,0,1)$ respectively on the $\mathbb{P}^1$. Around these fixed points, the local coordinates are given as $(z_0z_1,z_2/z_1)$ and $(z_1/z_2, z_0z_2)$, and thus their weights under $J_{1,2}$ actions can be represented as $(\epsilon_1,\epsilon_2-\epsilon_1)$ and $(\epsilon_1-\epsilon_2,\epsilon_2)$ at the North and South poles respectively. 

By performing the localization the partition function will be given by a sum over magnetic fluxes $\vec{n}$ on the $\mathbb{P}^1$, which is an $r$-dimensional vector $\vec{n}=(n_1,n_2,\cdots,n_r)\in \mathbb{Q}^r$ (running over the coweight lattices $\Gamma^\vee$ of gauge algebras), for the maximal torus $U(1)^r$ of the gauge symmetry group $G$. In geometry, such magnetic flux sum is performed for each compact 4-cycle. Also, background magnetic fluxes $\vec{B}$ for the Abelian subgroup $U(1)^{r_F}$ of global symmetries can be turned on, but we note that they are fixed and not summed over. Quantization conditions for each set of magnetic fluxes $(\vec{n},\vec{B})$ will be discussed shortly. In this paper, a flux set $(\vec{n},\vec{B})$ represents a set of all allowed dynamical magnetic flux vectors $\vec{n}_i$ and a fixed background flux vector $\vec{B}$.

At each flux background labelled by $\vec{n}$ and $\vec{B}$, the partition function is localized at two fixed points on the $\mathbb{P}^1$ and the path integral near each fixed point reduces to that of local $\Omega$-deformed $\hat{\mathbb{C}}^2$ with shifted chemical potentials due to the magnetic fluxes. As a consequence, the partition function $\hat{Z}$ can be written as \cite{Nakajima:2003pg,Nakajima:2005fg,Gottsche:2006bm}
\begin{align}\label{eq:bleq}
&\Lambda(m_j;\epsilon_1,\epsilon_2)\hat{Z}(\phi_i, m_j,B_j; \epsilon_1,\epsilon_2)  \\
&=\sum_{\vec{n}}(-1)^{|\vec{n}|}\hat{Z}^{(N)}(\phi_i\!+\!n_i\epsilon_1,m_j\!+\!B_j\epsilon_1;\epsilon_1,\epsilon_2\!-\!\epsilon_1) \cdot \hat{Z}^{(S)}(\phi_i\!+\!n_i\epsilon_2,m_j\!+\!B_j\epsilon_2;\epsilon_1\!-\!\epsilon_2,\epsilon_2)\ , \nonumber
\end{align}
where $|\vec{n}|=\sum_i n_i$. Here $\hat{Z}^{(N)}$ and $\hat{Z}^{(S)}$ are the partition function $\hat{Z}$ with shifted chemical potentials evaluated near the North and South poles respectively. The shifts in the chemical potentials $\phi_i$ and $m_j$ reflect the fact that magnetically charged states experience angular momentum shifts under the flux background. The prefactor $\Lambda(m_j;\epsilon_1,\epsilon_2)$ does not depend on dynamical parameters $\phi_i$, but depends only on mass parameters $m_j$ as well as $\epsilon_{1,2}$.

Now we will smoothly blow down the $\mathbb{P}^1$ at the origin. This is a smooth transition bringing the blowup geometry $\hat{\mathbb{C}}^2$ back to the flat $\mathbb{C}^2$ (with $\Omega$-deformation) without the $\mathbb{P}^1$ at the origin. The claim in \cite{Nakajima:2003pg} was that for certain theories, the partition function $\hat{Z}$ after the blowdown procedure reduces to the usual BPS partition function $Z$ on $\mathbb{C}^2$. In particular, the final partition function is independent of the background fluxes $\vec{B}$ on the $\mathbb{P}^1$. The reason for this is the following. The magnetic fluxes were supported on the $\mathbb{P}^1$ at the origin, but the $\mathbb{P}^1$ has been blown down and disappeared. Then, nothing remains to support these fluxes and moreover, there's nowhere these fluxes can flow on the flat $\mathbb{C}^2$. Therefore, we do not expect any remnant of the fluxes after the transition. 

We would like to make a remark on a subtle point in the presence of magnetic fluxes about the fermion number operator and some modifications of $Z$ associated with it. Since the partition function $\hat{Z}$ was defined with magnetic fluxes on $\hat{\mathbb{C}}^2$, the angular momentum for a state with electric charge $\sf{e}$ is shifted by ${\sf e}\cdot n$ where $n$ is the magnetic flux on $\mathbb{P}^1$ at the origin. Recall that the fermion number operator $(-1)^F$ in the index in \eqref{eq:Z} can be also defined as $(-1)^{2J_1}$. In the presence of the magnetic flux $n$, this should change as $(-1)^{2J_1} \rightarrow (-1)^{2J_1+{\sf e}\cdot n}$. This is formally equivalent to the following replacement in the index\footnote{Similar replacements $(-1)^F\rightarrow (-1)^{2J_R}$ in the superconformal index and in the holomorphic block for 3d SCFTs were discussed in \cite{Dimofte:2011py,Beem:2012mb}.}
\begin{equation}\label{eq:F-JR}
	(-1)^F \ \ \rightarrow \ \ (-1)^{2J_R} \ ,
\end{equation}
with the Cartan $J_R$ of the $SU(2)_R$ charge. This indicates that the partition function $\hat{Z}$ with magnetic fluxes on $\hat{\mathbb{C}}^2$ is in fact defined with the operator $(-1)^{2J_R}$ instead of $(-1)^F$. Moreover, since blowing down the $\mathbb{P}^1$ is a smooth transition, this definition is still valid even after the transition. Thus, the partition function (or the Witten index) in \eqref{eq:bleq} before and after the transition is defined with respect to the operator $(-1)^{2J_R}$.

One also finds that the replacement \eqref{eq:F-JR} can be implemented by a simple redefinition of the angular momentum chemical potential as $\epsilon_1\rightarrow \epsilon_1+2\pi i$. Therefore, after the transition from $\hat{\mathbb{C}}^2$ to $\mathbb{C}^2$, the partition function in the equation \eqref{eq:bleq} can be written as
\begin{align}\label{eq:hat-Z-Z}
	\hat{Z}(\phi,m;\epsilon_1,\epsilon_2) &= e^{\mathcal{E}(\phi,m;\epsilon_1,\epsilon_2)} \cdot \hat{Z}_{GV}(\phi,m;\epsilon_1,\epsilon_2) \ , \nonumber \\ 
	\hat{Z}_{GV}(\phi,m;\epsilon_1,\epsilon_2) &\equiv Z_{GV}(\phi,m;\epsilon_1+2\pi i,\epsilon_2) \ .
\end{align}
Note that the prefactor $\mathcal{E}$ in the first equation is the same as the prefactor before the replacement of $\epsilon_1$ because the redefinition doesn't affect the regularization factor in the path integral computation. Here $Z_{GV}$ is the index part of the BPS partition function, which is actually the refined Gopakumar-Vafa (GV) invariant \cite{Gopakumar:1998ii,Gopakumar:1998jq}, defined as
\begin{equation}\label{eq:GV-inv}
	Z_{GV}(\phi,m;\epsilon_1,\epsilon_2) = {\rm PE}\left[\sum_{j_l,j_r,{\bf d}}(-1)^{2(j_l+j_r)} N^{\bf d}_{j_l,j_r} \frac{\chi^{SU(2)}_{j_l}(p_1/p_2)\,\chi^{SU(2)}_{j_r}(p_1p_2)}{(p_1^{1/2}-p_1^{-1/2})(p_2^{1/2}-p_2^{-1/2})}e^{-{\bf d}\cdot {\bf m}}\right] \ ,
\end{equation}
where ${\bf d}$ denotes the charge of a BPS state, ${\bf m}$ stands for the chemical potentials (or K\"ahler parameters) $\phi,m$, and $N^{\bf d}_{j_l,j_r}$ is the degeneracy of a single-particle BPS state with spin $(j_l,j_r)$ and charge ${\bf d}$, and $\chi_j^{SU(2)}$ is the $SU(2)$ character of spin $j$. Also, $j_l\equiv \tfrac{J_1-J_2}{2}$ and $j_r\equiv \tfrac{J_1+J_2}{2}$. For example, the GV-invariants for a hypermultiplet with K\"ahler parameter $\phi$ providing a BPS state with spin $(0,0)$ are given by
\begin{align}
	Z_{GV}^{\rm hyper} &= {\rm PE}\left[\frac{\sqrt{p_1p_2}}{(1-p_1)(1-p_2)}e^{-\phi}\right] = \prod_{i,j=0}^\infty \frac{1}{1-p_1^{i+1/2}p_2^{j+1/2}e^{-\phi}} \ , \quad  \nonumber \\
	\hat{Z}_{GV}^{\rm hyper} &= Z^{\rm hyper}_{GV}(\phi;\epsilon_1+2\pi i,\epsilon_2)=\prod_{i,j=0}^\infty \frac{1}{1+p_1^{i+1/2}p_2^{j+1/2}e^{-\phi}} \ .
\end{align}

The equation \eqref{eq:bleq} with the identification \eqref{eq:hat-Z-Z} is the celebrated blowup equation for instanton partition functions on the $\Omega$-background introduced in \cite{Nakajima:2003pg,Nakajima:2005fg,Gottsche:2006bm}. See also \cite{Huang:2017mis} for a geometric generalization of the blowup formula. The blowup equation with non-trivial $\Lambda$ is called a {\it unity blowup equation}. The prefactor $\Lambda$ can also be trivial, i.e. $\Lambda=0$, for certain choices of fluxes, and the blowup equation in this case is called a {\it vanishing blowup equation} \cite{Nakajima:2005fg,Huang:2017mis}.

The purpose of this paper is to further generalize the above blowup formula such that it can cover all the 5d supersymmetric theories which have consistent UV completions. In addition, we will provide a systematic way to compute the BPS partition function $Z$ for any 5d supersymmetric theories using the blowup formula. More precisely, we propose the following conjecture:
\begin{framed}
\noindent {\bf Conjecture:} The partition function $Z$ on the $\Omega$-background in \eqref{eq:Z} for any 5d $\mathcal{N}=1$ field theory can be computed by solving the blowup equations \eqref{eq:bleq} with (i) {\it  consistent magnetic fluxes} $\vec{n}$ and $\vec{B}$, up to (ii) {\it flop transitions}.
\end{framed}

Based on this conjecture, we will present in this paper how to {\it bootstrap} BPS spectra of 5d field theories by solving the blowup equations. The seeds for this bootstrapping are the effective prepotential $\mathcal{E}$ on the $\Omega$-background and a set (or multiple sets) of consistent magnetic fluxes $\vec{n}$ and $\vec{B}$. We have already introduced how to compute the effective prepotential for every 5d field theory having either a gauge theory description in 5d or a 6d field theory origin or a geometric realization in a local CY 3-fold. We will now discuss how to choose consistent magnetic fluxes $\vec{n}$ and $\vec{B}$, basically the points (i), (ii) in the conjecture.

\subsubsection{Solving blowup equations}

Let us explain how to compute the partition function $Z$ using the blowup formula. We first remark that the index part of the partition function $Z$ of any 5d/6d SQFT must take the form of the GV-invariant  $Z_{GV}$ in \eqref{eq:GV-inv}. Now consider a power series expansion of the GV-invariant with respect to the fugacities $e^{-{\bf d}\cdot {\bf m}}$. The BPS states captured by the GV-invariant satisfy the BPS mass formula $|M|={\bf d}\cdot {\bf m}$ and at a generic point on the Coulomb branch (with mass parameters for global symmetries turned on) they have {\it positive masses} ${\bf d}\cdot {\bf m}>0$. Therefore the series expansion of $Z_{GV}$ in terms of the fugacities $e^{-{\bf d}\cdot {\bf m}}$ on the Coulomb branch is well-defined.

The blowup equation \eqref{eq:bleq} can be expressed in terms of power series in the fugacities and can be solved iteratively. Practically, we first recast the blowup equation as
\begin{align}\label{eq:bleq-GV}
	&\Lambda(m_j;\epsilon_{1},\epsilon_2) \hat{Z}_{GV}(\phi_i,m_j;\epsilon_1,\epsilon_2) = \sum_{\vec{n}}(-1)^{|\vec{n}|}e^{-V(\phi_i,m_j,\vec{n},\vec{B};\epsilon_1,\epsilon_2) } \nonumber \\
	&\times \hat{Z}_{GV}^{(N)} (\phi_i\!+\!n_i\epsilon_1,m_j\!+\!B_j\epsilon_1;\epsilon_1,\epsilon_2\!-\!\epsilon_1) \cdot \hat{Z}_{GV}^{(S)}(\phi_i\!+\!n_i\epsilon_2,m_j\!+\!B_j\epsilon_2;\epsilon_1\!-\!\epsilon_2,\epsilon_2) \ ,
\end{align}
where 
\begin{align}\label{eq:GV-V}
	&V(\phi_i,m_j,\vec{n},\vec{B};\epsilon_1,\epsilon_2) \equiv \  \mathcal{E}(\phi_i,m_j;\epsilon_1,\epsilon_2) \\
	& \qquad \quad - \mathcal{E}^{(N)}(\phi_i\!+\!n_i\epsilon_1,m_j\!+\!B_j\epsilon_1;\epsilon_1,\epsilon_2-\epsilon_1) - \mathcal{E}^{(S)}(\phi_i\!+\!n_i\epsilon_2,m_j\!+\!B_j\epsilon_2;\epsilon_1-\epsilon_2,\epsilon_2)\nonumber \ .
\end{align}
We expand both sides of the blowup equation \eqref{eq:bleq-GV} and then try to find an iterative solution of $\hat{Z}_{GV}$.

Importantly, we can use the fact that the GV-invariant should take a special form in \eqref{eq:GV-inv}. Also, spins of states at each order are bounded by the maximum spin $(j_l^{\rm max},j_r^{\rm max})$ in the series expansion, and the characters $\chi_j^{SU(2)}$ with different spins are all orthogonal to each other. Plugging the ansatz of the GV-invariant  with a finite number of trial states for a given charge ${\bf d}$ into the blowup equation and expanding it, we can iteratively solve the equations to evaluate multiplicities $N_{j_l,j_r}^{\bf d}$ of BPS states.

We conjecture that every 5d field theory enjoys enough number of independent blowup equations, enabling one to compute full BPS spectrum. It appears that a generic 5d/6d SQFT admits at least one unity blowup equation which suffices to determine all BPS degeneracies as shown in \cite{Huang:2017mis}. For instance, all 5d and 6d gauge theories with only full hypermultiplets (without any unpaired half hypermultiplets) have a number of unity blowup equations. We expect that at least one of those unity equations in this case is formulated with a set of {\it consistent magnetic fluxes}, whose definition will be given shortly, and thus the solution to such unity equation will produce the correct BPS spectrum of the gauge theory.

There are some theories having only vanishing blowup equations, though. The 6d theories involving unpaired half hypermultiplets on a circle with/without twists are such theories\footnote{On the other hand, we note that  5d gauge theories with half hypermultiplets can have unity blowup equations.} \cite{Gu:2020fem}. An analysis of small $\epsilon_{1,2}$ expansion in \cite{Huang:2017mis} suggests that a single vanishing blowup equation without other information may not be sufficient to determine all BPS degeneracies. Nevertheless, we propose that those theories in fact have enough number of vanishing blowup equations so that we can compute their BPS spectra by solving all the vanishing equations together with other supplementary consistency conditions, like the positivity of BPS degeneracies $N_{j_l,j_r}^{\bf d} \ge 0$, conformity to geometric realizations and dualities, and the KK tower structure of KK theories, etc. As a concrete example for this, in section~\ref{sec:instructive}, we will show that a single vanishing blowup equation for the pure $SU(2)_\theta$ gauge theories can be solved with the aid of additional consistency conditions so that BPS spectra  can be obtained, though the partition function for the $SU(2)_\theta$ theory can be also obtained from unity blowup equations as well. 

Though it seems trivial, the condition $N_{j_l,j_r}^{\bf d} \ge 0$ turns out to be quite powerful. While solving the blowup equations, it usually happens that degeneracies of BPS states captured in higher orders in the expansion are fixed by BPS degeneracies appearing in lower orders. Accordingly, the non-negativity of degeneracies $N_{j_l,j_r}^{\bf d} \ge 0$ for all the BPS states in higher orders puts constraints on the possible lower order BPS degeneracies. When taking into account higher expansion orders, one finds more and more additional constraints on the lower order BPS degeneracies, which would strongly restrict the allowed lower order degeneracies and hence the BPS degeneracies in a given order may be completely fixed at a certain stage in the iteration procedure\footnote{For example, the BPS spectrum of the 6d $E_7$ gauge theory with a half hypermultiplet in the fundamental representation was computed in Table 24 in \cite{Gu:2020fem} by solving vanishing blowup equations. Several undetermined BPS degeneracies (denoted by symbol `?') can actually be fixed by the condition $N_{j_l,j_r}^{\bf d} \ge 0$. For instance, from the BPS spectrum given in Table 24 in \cite{Gu:2020fem}, we could fix many degeneracies as $2(0,1)$ for $\beta=(2,1,1,0,0,1,2,1,0)$, $(0,1)$ for $\beta=(2,0,1,0,0,1,2,1,0)$, $(0,0)\oplus(0,1)$ for $\beta=(1,0,1,0,0,1,2,1,0)$, $2(0,0)\oplus(0,1)$ for $\beta=(1,0,1,0,0,2,2,1,0)$, and etc. All other undetermined degeneracies, but $\beta= (0, 3, 3, 0, 1, 0, 0, 0, 0)$, are strongly constrained and actually have only few possibilities. It seems that some higher degree computations can fix all the lower order degeneracies in the table.}.

Dualities and geometric realizations can also be useful for computation. When a theory enjoys a geometric construction or dualities, we can extract from them yet another supplementary information about the BPS states. In particular, when blowup equations have more than one distinct solutions, one can use a geometric construction or dualities to pick up the right solution for a given theory, which we will see with concrete examples below. 

Consequently, we conjecture that one can compute BPS spectra of all 5d/6d field theories using the blowup equations formulated from their geometric realizations or gauge theory descriptions, or RG-flows thereof,\footnote{We are assuming that every 5d field theory admitting a UV completion has either a geometric realization or gauge theory descriptions in 5d or in 6d on a circle possibly with twists, or can be obtained by an RG-flow from a UV complete theory.} even for the cases equipped with only vanishing blowup equations. A similar conjecture for refined BPS invariants of a local CY 3-fold was given in \cite{Huang:2017mis}. We will provide evidences for our conjecture by explicitly solving the blowup equations for all rank-2 theories and some interesting higher rank theories in sections~\ref{sec:rank2 theories} and \ref{sec:higher rank theories}.

\subsubsection{Magnetic flux quantization}\label{sec:magnetic flux quantization}

The magnetic fluxes on the $\mathbb{P}^1$ in the blowup $\hat{\mathbb{C}}^2$ cannot be arbitrary. They should satisfy suitable quantization conditions. Let us explain the quantization conditions for the magnetic fluxes in three different perspectives: the geometric perspective, the 5d gauge theoretic perspective, and the 6d gauge theoretic perspective.

\paragraph{Geometries}
In geometry, the magnetic flux $n_i$ (or $B_j$) can be turned on for each (non-)compact divisor $D_I$ in a 3-fold.
The flux quantization depends on charges and spins $(j_l,j_r)$ of wrapped M2-branes on holomorphic 2-cycles. Consider an M2-brane wrapping a primitive curve $C_i$. Here the primitive curve is a Mori cone generator and every holomorphic 2-cycle can be expressed as a linear combination of primitive curves $C_i$ as $C=\sum_i p_i C_i$ with non-negative integers $p_i$. The curve $C_i$ can consistently couple to a magnetic flux ${\sf F}$ if the following condition is satisfied \cite{Huang:2017mis}
\begin{equation}\label{eq:quantization}
	{\sf F}\cdot C_i \ \text{ is integral/half-integral when } C_i^2 \ {\rm is \ even/odd } \ ,
\end{equation}
where $C_i^2$ is the self-intersection number of $C_i$. The flux ${\sf F}$ is defined in geometry as ${\sf F}\equiv\sum_{i=1}^r n_i S_i + \sum_{j=1}^{r_F} B_j N_j$,  where $S_i, N_j\in H^{1,1}(X)$ are the basis of the compact and the non-compact surfaces inside a 3-fold $X$, respectively. The above condition \eqref{eq:quantization} is equivalent to the condition that the magnetic flux on a charged M2-brane state of spin $(j_l,j_r)$ wrapping on $C_i$ satisfies  
\begin{align}
{\sf F}\cdot C_i ~~&\text{is integral/half-integral, when $2(j_l+j_r)$ is odd/even}. 	
\end{align}
This is because the spin of a curve $C_i$ is related to the self-intersection as $2(j_l+j_r) = C_i^2+1$ mod 2.  From this we claim that the flux ${\sf F}$, or equivalently $(\vec{n},\vec{B})$, must be quantized such that all the primitive curves $C_i$ in a 3-fold satisfy the condition \eqref{eq:quantization}. In general the solution to this quantization condition is not unique. The proper choices of magnetic fluxes associated to spectrum of unitary BPS states will be discussed below.

\paragraph{5d Gauge Theories}
We can easily translate the above geometric quantization condition to physical conditions in 5d gauge theories. W-bosons and hypermultiplet states in a 5d gauge theory correspond to primitive curves with $C^2=0$ and $C^2=-1$, respectively. The charges of these elementary fields in the classical Lagrangian are all known. Based on the classical information, we can first quantize the magnetic fluxes $\vec{n}$ and $\vec{B}$ coupled to the elementary fields. The geometric quantization condition implies that the total fluxes on the W-bosons of the gauge group should be integral and those on the perturbative hypermultiplets should be half-integral. Namely,
\begin{align}\label{eq:n-B}
	\vec{n}\cdot e \in \mathbb{Z} \ , \qquad \vec{n}\cdot w_f + B_f \in \mathbb{Z}+\frac{1}{2} \ ,
\end{align}
for all roots $e$, the weights $w_f$ associated with all hypermultiplets $f$, and the fluxes $B_f$ for flavor symmetries. 

We will use the first condition of \eqref{eq:n-B} to fix the quantization of fluxes $ \vec{n}$, and use the second condition to quantize the background fluxes $B_f$.\footnote{There is one exception. A half hypermultiplet does not admit its mass parameter, so there is no corresponding $ B_f$. In this case, the second condition of \eqref{eq:n-B} associated with the half-hyper further constrains the quantization of $ \vec{n}$.} Depending on the gauge algebra, there are several possibilities of magnetic fluxes satisfying \eqref{eq:n-B}. Since a state in a given representation is obtained by subtracting roots from its highest weight state, the quantization of $ \vec{n}\cdot w_f $ is the same for every $ w_f $ in a fixed representation.

To find all possible quantizations for $\vec{n}$, it is convenient to introduce the fundamental weight $\mu_i$ which is a dual basis of the coroots $ \alpha_i^\vee $ of the gauge algebra.\footnote{For a root $ \alpha $ in Euclidean space $ E $, the coroot $\alpha^{vee}$ is a map from $ E $ to $ \mathbb{R} $, defined to be $ \inner{\alpha^\vee}{x} = 2(x, \alpha)/(\alpha, \alpha) $, where $ (\cdot, \cdot) $ is an inner product in $ E $. The fundamental weight $ \mu_i $ is dual basis of $ \alpha_i^\vee $, i.e., $ \inner{\alpha_i^\vee}{\mu_j} = \delta_{ij} $. The general weight vector can be written as a linear combination of $ \mu_i $. For example, write a simple root $ \alpha_i = \sum_k a_{ij} \mu_j$. Then $ \inner{\alpha_j^\vee}{\alpha_i} = a_{ij} $ so that $ a_{ij} $ is an element of the Cartan matrix. The details can be found in many textbooks about Lie algebras, for example, \cite{humphreys_introduction_1972} or \cite{bump_lie_2013}.} Consider a lattice $ \Lambda = \oplus \mathbb{Z}\, \mu_i $ and the corresponding root lattice $ \Lambda_r $ which is a sublattice of $ \Lambda $. The lattice $ \Lambda $ contains not only root lattice but also all possible weight vectors. Thus, the number of possible quantizations of $ \vec{n} $ with $ \vec{n} \cdot e \in \mathbb{Z} $ is counted by the number of elements in $ \Lambda / \Lambda_r $. This group is isomorphic to the center of simply connected Lie group corresponding to gauge algebra \cite{humphreys_introduction_1972, bump_lie_2013}:
\begin{align}
\begin{array}{cccccccccc}
A_r & \hspace{1ex} B_\ell & \hspace{1ex} C_\ell & \hspace{1ex} D_{\ell, \text{odd}} & \hspace{1ex} D_{\ell,\text{even}} & \hspace{1ex} E_6 & \hspace{1ex} E_7 & \hspace{1ex} E_8 & \hspace{1ex} F_4 & \hspace{1ex} G_2 \\
\mathbb{Z}_{r + 1} & \hspace{1ex} \mathbb{Z}_2 & \hspace{1ex} \mathbb{Z}_2& \hspace{1ex} \mathbb{Z}_4  & \hspace{1ex} \mathbb{Z}_2 \times \mathbb{Z}_2 & \hspace{1ex} \mathbb{Z}_3 & \hspace{1ex} \mathbb{Z}_2 & \hspace{1ex} \{1\} & \hspace{1ex} \{1\} & \hspace{1ex} \{1\}
\end{array} \ .
\end{align}

More explicitly, for the gauge algebra  of type $ A_\ell $, the possible quantization for $ \vec{n} $ is given by
\begin{align}
n_i \in \mathbb{Z} + \frac{h}{\ell+1}i \quad (1 \leq i \leq \ell)\ ,
\end{align}
where $n_i\equiv \vec{n} \cdot \mu_i$ and $ h $ is a fixed integer subject to  $0 \leq h \leq \ell$. 
For the gauge algebra of type $ B_\ell $, there are two quantizations,
\begin{align}
n_i&\in \mathbb{Z} \quad ( 1 \leq i \leq \ell-1 ) \ , \qquad
n_\ell\in \mathbb{Z} + \frac{h}{2} \ ,
\end{align}
where $ h = 0, 1 $. For type $ C_\ell $,
\begin{align}
n_i \in \mathbb{Z} + \frac{h}{2} \quad \text{for $i$ odd} \ , \qquad
n_i\in \mathbb{Z}  \quad \text{for $i$ even},
\end{align}
where $ h = 0, 1 $. The set of the quantizations of types $ B_\ell $ and $ C_\ell $ has a $ \mathbb{Z}_2 $ structure. For  
type $ D_{\ell} $ with $\ell$ odd,
\begin{align}
n_i &\in \mathbb{Z} + \frac{h}{2} \qquad (i = 1, 3, \cdots, \ell-3) \ , \quad\cr
n_i&\in \mathbb{Z}  \qquad (i=2, 4, \cdots, \ell-2) \ , \quad\\
n_{\ell-1}&\in \mathbb{Z} + \frac{h}{4}, \qquad  n_\ell \in \mathbb{Z} + \frac{h+2}{4}  \ , \nonumber
\end{align}
where $ h = 0, 1, 2, 3 $. The set of the quantization for $D_{\ell,\text{odd}}$ hence has a $ \mathbb{Z}_4 $ structure. For  
type $ D_{\ell} $ with $\ell$ even, there are four quantizations:
\begin{align}
n_i  &\in \mathbb{Z} + \frac{h_1 + h_2}{2} \qquad (i = 1, 3, \cdots, \ell-3) \ ,\cr
n_i  &\in \mathbb{Z}  \qquad (i=2, 4, \cdots, \ell-2) \ , \\
n_{\ell-1}&\in \mathbb{Z}+ \frac{h_1}{2},\qquad n_\ell \in \mathbb{Z} + \frac{h_2}{2}\ , \nonumber
\end{align}
where $ (h_1, h_2) \in \mathbb{Z}_2 \times \mathbb{Z}_2 $. For type $ E_6 $,
\begin{align}
(n_1, n_2, n_3, n_4, n_5, n_6) \in \mathbb{Z}^6 + \frac{h}{3}(2, 1, 0, 2, 1, 0)\ ,
\end{align}
with $ h = 0, 1, 2 $, which has a $\mathbb{Z}_3 $ structure. For type $ E_7 $,
\begin{align}
(n_1, n_2, n_3, n_4, n_5, n_6, n_7) \in \mathbb{Z}^7 + \frac{h}{2}(0, 0, 0, 1, 0, 1, 1) \ ,
\end{align}
with $h=0,1$, which has a $ \mathbb{Z}_2 $ structure. The remaining exceptional algebras $ E_8$, $ F_4 $ and $ G_2 $ admit only {\it canonical flux quantization} defined as
\begin{equation}\label{eq:n-B-canon}
	n_i \in \mathbb{Z} \ , \quad B_f \in \mathbb{Z}+\frac{1}{2} \ .
\end{equation}
This canonical flux quantization holds for any gauge and flavor symmetry groups except for the cases where the theory contains unpaired half hypermultiplets. Most of the examples we will discuss below use the canonical flux quantization. 

The quantization of fluxes $B_a$ for topological symmetries of gauge groups is more involved because in general we do not know spins of instanton states. Let us first discuss the quantization condition on $B_a$ when fluxes for the gauge group $G_a$ satisfy the canonical flux quantization. In this case, a single instanton state carries a unit charge under $U(1)_a$ topological symmetry for the $a$-th gauge group. The charges of the instanton state under other global symmetries can be computed by summing over contributions from the zero modes of charged hypermultiplets on the instanton background. For a hypermultiplet in a representation ${\bf r}$ of a gauge group $G_a$, the zero mode contribution to the flavor charge of a 1-instanton state on the Coulomb branch is given by its Dynkin index, i.e. $-T_a({\bf r})$. Therefore, it follows from \eqref{eq:quantization} that the ground state of a single instanton for gauge group $G_a$ should induce a quantization condition
\begin{equation}\label{eq:inst-flux}
	B_a-\sum_f B_f T_a({\bf r}_f) \  \in \ \mathbb{Z} \ \ {\rm or} \ \ \mathbb{Z}+1/2 \ ,
\end{equation} 
depending on the spin of the state. This will quantize the flux $B_a$ provided that $B_f$'s have already been quantized by \eqref{eq:n-B-canon}. Here, we denote by $f$ all hypermultiplets in the representation ${\bf r}_f$ of the gauge group $G_a$ collectively. Since the canonical gauge fluxes $n_i$ bring about only integral shifts, the gauge fluxes do not affect the above quantization condition.

When $n_i^a$'s for a gauge group $G_a$ are not canonical, we can fix $B_a$ by requiring the corresponding blowup equation to be solvable.
We first expand the blowup equation in terms of the instanton fugacity $e^{-m_a}$ for a gauge group $G_a$, assuming that it is a unity equation. The terms in the leading order in this expansion are given by a set of magnetic flux vectors $\vec{n}$ that minimize
\begin{equation}
	\partial_{m_a}V = \frac{1}{2}(\vec{n},\vec{n})^a\equiv \frac{1}{2} K^a_{ij}n^a_i n^a_j \ ,
\end{equation}
for all gauge groups, where $V$ is defined in \eqref{eq:GV-V}. We call this set ${\rm Min}(\vec{n})$. Since the GV-invariant $\hat{Z}_{GV}$ starts with $1$ in the fugacity expansion and the prefactor $\Lambda$ is independent of the Coulomb branch parameters $\phi_i$, the right-hand side in the blowup equation given in \eqref{eq:bleq-GV} must start with terms independent of $\phi_i$. This means that for a given $B_a$, there must exist at least a magnetic flux vector $\vec{n}\in {\rm Min}(\vec{n})$ such that
\begin{equation}\label{eq:Bj-cond1}
	\partial_{\phi^a_i}V = 0 \ \ {\rm at} \ (\vec{n},\vec{B}) \  {\rm for} \ {\rm all} \ i \ {\rm of} \ G_a \ .
\end{equation}
This condition then fixes the background flux $B_a$ when the gauge fluxes $n^a_i$ are non-canonical ones. We note that when the fluxes $n^a_i$ are all canonical ones, this condition is trivially satisfied since $n^a_i\in$ Min($\vec{n})$ are all zero and $V$ is independent of $\phi^a_i$ when $n^a_i=0$.

On the other hand, a vanishing blowup equation can be solved if there exists at least a pair of minimal flux vectors $\vec{n}_1,\vec{n}_2\in {\rm Min}(\vec{n})$ with $|\vec{n}_1|-|\vec{n}_2| = 1$, which is a necessary condition for that two leading terms cancel each other in the expansion, satisfying
\begin{equation}\label{eq:Bj-cond2}
	\partial_{\phi^a_i}V \ {\rm at} \ (\vec{n}_1 , \vec{B}) \quad = \quad 
	\partial_{\phi^a_i} V \ {\rm at} \ (\vec{n}_2, \vec{B}) \ ,
\end{equation}
for all $\phi_i^a$.

The conditions \eqref{eq:Bj-cond1} and \eqref{eq:Bj-cond2} will then leave only a finite number of allowed fluxes $B_a$ for the topological symmetry of $G_a$ in the unity and  vanishing blowup equations, respectively, for non-canonical magnetic fluxes $n^a_i$.

\paragraph{6d Gauge Theories with/without Twists}
Lastly, we will discuss the flux quantization conditions for the 5d KK theories in terms of the associated 6d field theory data. Upon a circle compactification, the 6d multiplets reduce to 5d vector multiplets and hypermultiplets taking representations of the 6d gauge algebra $\mathfrak{g}$, or the invariant subalgebra $\mathfrak{h}$ when $\mathfrak{g}$ is twisted. In the 5d reduction of a 6d gauge theory, the magnetic fluxes for the gauge group and those for the flavor group acting on 5d gauge and matter fields satisfy the quantization condition in \eqref{eq:n-B}. Also, the magnetic flux $B_\tau$ of the KK $U(1)$ symmetry should be an integer because all KK states from a single 6d state should have the same (half-)integrality of fluxes. So it is natural to fix
\begin{equation}
	B_\tau = 0 \ .
\end{equation}

We then need to determine the quantization conditions on the tensor fluxes $n_{\alpha}$ (or $n_{\alpha'}$ when the $\alpha$-th tensor node is twisted) for tensor symmetries. Self-dual strings are charged under the tensor symmetries. Their tensor charges are determined by the classical Green-Schwarz terms evaluated on self-dual string backgrounds. The ground state of a single string associated to the $\beta'$-th (or $\beta$-th without twist) tensor field carries charges for the $\alpha'$-th tensor symmetry as
\begin{equation}
	\Omega^{\alpha'\beta'}_S b_{a,\beta'} \ ,
\end{equation}
when the string is an instantonic string of the $a$-th gauge group. If the $\beta'$-th tensor supports no gauge algebra, then $b_{a,\beta'}=1$. 

Let us first consider the cases where magnetic fluxes for a gauge group $G_a$ coupled to the $\alpha'$-th tensor fields, so when $b_{a,\alpha'}\neq0$, satisfy the canonical flux quantization condition in \eqref{eq:n-B-canon}. The flavor and the KK charges for a single string can be computed by counting the zero mode contributions from KK states on the string background, which is the same computation we did above for instanton states in 5d gauge theories. Collecting all the zero mode contributions, we claim that the tensor flux, which we denote by $n_{\alpha'}\equiv \tilde{n}_{\alpha'}+B_{\alpha'}$ with $\tilde{n}_{\alpha'}\in \mathbb{Z}$, should satisfy the following quantization condition: given the spin $(j_l,j_r)$ of the ground state, 
\begin{align}
	-\Omega^{\alpha'\beta'}_S\!B_{\alpha'}b_{a,\beta'} &\!- \!\sum_{f_0}B_{f_0}T_a({\bf r}_{f_0}) \!+ \!\frac{1}6B_\tau\! \sum_{r}(6r(r\!-\!1)\!+\!1)\Big(T_a({\bf R}_r) \!-\! \sum_f T_a({\bf w}_{r,f})\Big) \nonumber \\
	&\in	\left\{\begin{array}{ll}
		\mathbb{Z}& ~\text{for~}2(j_l+j_r)~\text{odd}\ ,\\
		\mathbb{Z}+\frac12 & ~\text{for~}2(j_l+j_r)~\text{even}\ .
	\end{array} \right.
\end{align}
This is essentially the quantization condition for $B_{\alpha'}$.
Here, $f_0$ runs over hypermultiplets with KK-charge $0$, and $r$ runs over all fractional KK-charge shifts including $r=0$. ${\bf R}_r$ and ${\bf w}_{r,f}$ denote representations of the vector multiplets  and the hypermultiplets with fractional shift $r$ of KK charges, respectively. Note that since gauge and tensor charges of the ground state on a self-dual string are quantized to be integers and the associated fluxes $n_a$ and $\tilde{n}_{\alpha'}$ are all integers as well, they have no effect on this quantization condition.

Now consider non-canonical magnetic fluxes for a gauge group $G_a$. Let us first expand the blowup equation in terms of the fugacities $e^{-\phi_{\alpha'}}$ of tensor symmetries. The leading power of these fugacities in this expansion is determined by the minimum of
\begin{equation}
	\partial_{\phi_{\alpha'}}V=\Omega_S^{\alpha'\beta'}b_{a,\beta'}\tilde{K}_{a,ij}n^a_in^a_j  \ .
\end{equation}
This is non-zero with non-integral gauge fluxes $n^a_i$. This means that when gauge fluxes $n^a_i$ do not satisfy the canonical flux quantization condition for any $G_a$, then the blowup equation written in terms of a 6d field theory description is always a vanishing-type, i.e. $\Lambda=0$. 

As we discussed for 5d gauge theories above, the vanishing blowup equation can be solved only if we can find at least a pair of minimal flux vectors $\vec{n}_1,\vec{n}_2\in {\rm Min}(\vec{n})$ and background fluxes $\vec{B}$ subject to the condition given in \eqref{eq:Bj-cond2}, where ${\rm Min}(\vec{n})$ is a set of magnetic flux vectors minimizing $(\vec{n},\vec{n})^a$ for all $G_a$ and $\vec{B}$ here involves the tensor fluxes $B_{\alpha'}$ as well as other flavor fluxes. We can use this fact to fix the background magnetic fluxes $B_{\alpha'}$ of tensor symmetries.

We remark that if a 6d gauge theory involves unpaired half hypermultiplets and thus the background flavor flux acting on the hypermultiplet cannot be turned on, then we cannot choose the canonical flux quantization \eqref{eq:n-B-canon} for the fluxes of the gauge group $G_a$ coupled to the half hypermultiplet. Therefore, we have only vanishing blowup equations in such cases, as discussed in \cite{Gu:2020fem}.

\subsubsection{Consistent magnetic fluxes}

The above quantization conditions are a necessary condition but not a sufficient one for the BPS partition functions $Z$ of a 5d theory to satisfy the blowup equation \eqref{eq:bleq} with chosen magnetic fluxes. Among the magnetic fluxes satisfying the quantization conditions, only few of them can provide consistent blowup equations whose solution correctly produces BPS spectrum of a 5d field theory. We call such magnetic fluxes leading to the consistent blowup equations {\it consistent magnetic fluxes}. With a wrong set of fluxes, one would find an inconsistent blowup equation: the blowup equation has no solution or the solution to the blowup equation does not take the form of a GV-invariant, or the solution involves unphysical states and thus differs from the desired BPS spectrum for a 5d theory. Therefore, in order to correctly compute BPS spectrum using the blowup equations, it is crucial to know how to choose proper consistent magnetic fluxes. We now present a set of criteria for the consistent magnetic fluxes.

As discussed, all the BPS particles on the Coulomb branch must satisfy the BPS mass formula. It is expressed as $|M|= \sum_i {\sf{e}}_i \phi_i$, when we turned off mass parameters for global symmetries in strong coupling limit near the UV fixed point. Here ${\sf{e}}_i$ denotes the charge of a state under the $i$-th $U(1)$ gauge group. In geometry, these masses are identified with the volumes of holomorphic 2-cycles for the BPS states as $\sum_i {\sf{e}}_i \phi_i={\rm vol}(C)$ measured with respect to the normalizable K\"ahler parameters $\phi_i$. It then follows that the mass $\sum_i {\sf{e}}_i \phi_i$ (or ${\rm vol}(C)$) must be non-negative on the Coulomb branch, i.e. $\sum_i {\sf{e}}_i \phi_i \ge 0$ for all BPS states. In fact, the  Coulomb branch $\mathcal{C}$ is the space of the dynamical K\"ahler moduli $\phi_i$ defined as \cite{Jefferson:2017ahm,Jefferson:2018irk}
\begin{equation}\label{eq:C-Coulomb}
	\mathcal{C} = \bigg\{\phi_i,i=1,\cdots,r\, | \,{\sf{e}}\cdot \phi\equiv\sum_i {\sf{e}}_i \phi_i \ge 0\bigg\} \ ,
\end{equation}
with $\phi_i>0$ when all mass parameters for global symmetries are switched off. The Coulomb branch $\mathcal{C}$ must exist for a UV finite 5d theory equipped with dynamical K\"ahler parameters. Otherwise, unitarity of the theory will be violated at some point on the Coulomb branch and the theory cannot have a consistent UV completion.

The Coulomb branch $\mathcal{C}$ is a collection of sub-chambers $\mathcal{C}_i$ that are connected by so-called {\it flop transitions}. It is possible that a BPS hypermultiplet becomes massless $M=0$ at the boundary between two sub-chambers. If then, two sub-chambers are connected by a flop transition flipping the mass sign of the hypermultiplet as $M>0\rightarrow M<0$. The effective Chern-Simons terms in the low energy theory take different expressions in the two sub-chambers because the CS terms depend on mass signs of hypermultiplets. More precisely, the prepotential and the gauge/gravitational Chern-Simons coefficients alter their forms under a flop transition related to a hypermultiplet as
\begin{equation}
	\mathcal{F} \ \ \rightarrow \ \ \mathcal{F} + \frac16 M^3 \ , \qquad C^G_i \ \ \rightarrow \ \ C^G_i -\partial_iM \ , \qquad C_i^R \ \ \rightarrow \ \ C_i^R \ ,
\end{equation}
where $M$ here is the mass for the hypermultiplet before the flop transition. In geometry, the flop transition corresponds to a geometric transition $X \rightarrow X'$ between two CY 3-folds $X$ and $X'$  described by blowing down a $-1$ curve $C\subset X$ and blowing up a different $-1$ curve $C'\subset X'$.

The partition function $Z$ computes degeneracies of BPS states on the Coulomb branch of a  5d theory. Suppose that we solve the blowup equations for $Z$ in the fugacity expansion and the BPS states are identified up to a certain order. One can easily read off the masses for the BPS states up to that order. For a consistent 5d field theory having dynamical K\"ahler parameters, there must exist a non-vanishing Coulomb branch $\mathcal{C}$ defined in \eqref{eq:C-Coulomb}, possibly after a finite number of flop transitions, where the masses of the BPS states are all non-negative, i.e. ${\sf e}\cdot\phi\ge0$, when all non-dynamical parameters are turned off. If one cannot find a non-trivial Coulomb branch, then it implies that the 5d theory is inconsistent in UV.

Note however that the partition function $Z$ is defined and computed in a particular sub-chamber on the (extended) Coulomb branch with mass parameters $m_j$ for global symmetries turned on. In the limit $m_j\rightarrow0$, some of hypermultiplets captured in the partition function can be massless and then flop transitions should take place before their masses become negative. Since this can generically happen, we should carefully examine the existence of Coulomb branch by testing ${\sf e}\cdot\phi\ge0$ for all the states in $Z$, except for hypermultiplets (or states with spin $(j_l,j_r)=(0,0)$). We note that for hypermultiplets, either ${\sf e}\cdot\phi \ge 0$ or ${\sf e}\cdot\phi< 0$ is allowed, as flop transitions can happen. If one finds a non-trivial Coulomb branch where ${\sf e}\cdot\phi\ge0$ for all the states in $Z$ except for a hypermultiplet which is of ${\sf e}\cdot\phi < 0$, then this means that there must be a flop transition associated to the hypermultiplet as we approach the UV fixed point. In this case we should first perform flop transitions for the hypermultiplets with ${\sf e}\cdot\phi < 0$ and then test again ${\sf e}\cdot\phi\ge0$ for all states including the hypermultiplets.

If one finds a chamber with ${\sf e}\cdot\phi\ge0$ for all the states including hypermultiplets, after a sequence of such flop transitions, it is a strong indication that the theory has a UV completion with non-trivial Coulomb branch and the solution $Z$ (or $\hat{Z}$) of the blowup equations computes the BPS spectrum on the Coulomb branch around the UV fixed point. In this case, we will refer to the set of magnetic fluxes as {\it consistent magnetic fluxes} for the 5d theory on the blowup $\hat{\mathbb{C}}^2$. We conjecture that there exist enough sets of consistent magnetic fluxes for every UV finite 5d theory and therefore we can compute its BPS spectrum using the blowup equations.

Examining ${\sf e}\cdot\phi \ge 0$ for all BPS states is a formidable task as there are infinitely many BPS states but the partition function $Z$ is computed up to a certain order in the fugacity expansion. Practically we can check this non-negativity of masses of BPS states only up to a certain higher order. This however would not be very harmful from the perspective of geometry. Recall that all holomorphic 2-cycles ${C}=\sum_i n_i C_i$  can be written as a linear sum of primitive curves $C_i$ with non-negative integers $n_i$ and the number of the primitive curves is finite for a local CY 3-fold. The primitive curves (or associated BPS states) are usually captured at some lower orders in the expansion. We then need to check non-negativity of volumes only for such primitive curves in lower orders. We expect this holds also for general 5d theories. Hence one can in principle identify consistent magnetic fluxes by computing the BPS partition function up to certain leading orders.

We now have all the ingredients for solving the blowup equations and thus are ready for bootstrapping BPS spectrum of any 5d field theory. In the followings we will explicitly illustrate the bootstrap procedure with a large number of interesting examples.

%% file: sec-examples.tex
\input{sec-Rank1}
\input{sec-Rank2}
\input{sec-Rankhigh}

%% file: sec-Rank1.tex
\section{Rank 1 theories}\label{sec:rank1theories}

In this section, we will compute the BPS partition functions of rank-1 5d theories by bootstrapping with the blowup equations. All rank-1 theories can be obtained via RG flows, which integrate out massive hypermultiplets, from three KK theories that come from 6d SCFTs compactified on a circle with/without twists. When a theory admits solvable blowup equations with consistent magnetic fluxes, all the IR theories descending from this UV theory by RG flows are guaranteed to have solvable blowup equations. So we will show that the blowup equations for all the rank-1 KK theories can be solved and the solutions are consistent with the BPS spectra computed by other methods. This will prove that the BPS spectra of all rank-1 5d theories can be computed by solving the blowup equations. We will also compute the BPS partition function of a new rank-1 theory, which we call the local $\mathbb{P}^2+1\mathbf{Adj}$ theory, obtained from the $\mathcal{N}=2$ $SU(2)$ gauge theory at $\theta=\pi$ by integrating out an instantonic hypermultiplet.

\subsection{KK theories}
There are three rank-1 KK theories: the $SU(2)$ gauge theory with 8 fundamental hypermultiplets, the $\mathcal{N}=2$ $SU(2)$ gauge theory at $\theta=0,\pi$. We discuss them one by one. 

\subsubsection{\texorpdfstring{$SU(2)+8\mathbf{F}$}{SU(2) + 8F}}

The $SU(2)$ gauge theory with 8 fundamental hypermultiplets is a KK theory arising from the 6d rank-1 E-string theory compactified on a circle \cite{Witten:1995gx,Ganor:1996mu,Seiberg:1996bd}.
\begin{align}
\begin{tikzpicture}
\draw (0, 0) node {$ SU(2) + 8\mathbf{F} $}
(2.5, 0) node {$ = $}
(4.5, 0.3) node {$ \mathfrak{sp}(0)^{(1)} $}
(4.5, -0.3) node {$ 1 $};
\end{tikzpicture}
\end{align}
The BPS partition function for this theory was computed in \cite{Hwang:2014uwa,Kim:2014dza} using the ADHM instanton string construction and also in \cite{Kim:2015jba,Kim:2017jqn} based in its 5-brane webs using topological vertex. It was also shown in \cite{Gu:2019pqj} that the elliptic genus of the 6d E-string theory is consistent with the blowup equation. 

We shall here solve the blowup equation in perspective of the 5d gauge theory. The prepotential on the Coulomb branch is given by
\begin{align}
6\mathcal{F} =6m_0 \phi^2+ 8\phi^3 - \frac{1}{2} \sum_{i=1}^8 \Big((\phi + m_i)^3 + (\phi - m_i)^3\Big)\ ,
\end{align}
where $ m_0 $ is the gauge coupling and $ m_{i=1, \cdots, 8} $ are the mass parameters of fundamental hypermultiplets. Collecting the gauge/gravitational and the gauge/$SU(2)_R$ Chern-Simons terms, the effective prepotential on the $\Omega$-background is expressed as 
\begin{align}
\mathcal{E} &= \frac{1}{\epsilon_1 \epsilon_2} \qty(\mathcal{F} - \frac{\epsilon_1^2 + \epsilon_2^2}{48} \bigg(4\phi - \sum_{i=1}^8 \Big((\phi + m_i) + (\phi - m_i)\Big)\bigg) + \epsilon_+^2 \phi ) \nonumber \\
&= \frac{1}{\epsilon_1 \epsilon_2} \qty(\mathcal{F} +\frac{\epsilon_1^2+\epsilon_2^2}{4}\phi + \epsilon_+^2 \phi) \, .
\end{align}
The perturbative part of the GV-invariant is
\begin{align}
\mathcal{Z}_{\rm pert}(\phi,m_i;\epsilon_{1,2}) = {\rm PE}\left[-\frac{1+p_1p_2}{(1-p_1)(1-p_2)}e^{-2\phi}  + \frac{\sqrt{p_1p_2}}{(1-p_1)(1-p_2)}\sum_{i=1}^8e^{-(\phi\pm m_i)}\right] \, .
\end{align}

For this theory, we find a set of the consistent magnetic fluxes
\begin{align}
n \in \mathbb{Z} \, , \quad B_{m_0}=0 \, , \quad B_{m_i}=1/2 \ \ {\rm for} \ 1\leq i\leq 8 \, ,
\end{align}
which leads to a unity blowup equation. The unity blowup equation with the magnetic fluxes can be solved and the solution is summarized in Table~\ref{table:SU2_8F}. This result matches the elliptic genus of the rank-1 E-string theory computed in \cite{Kim:2014dza}.

\begin{table}
\centering
\begin{tabular}{|c|C{30ex}||c|C{30ex}|} \hline
	$ \mathbf{d} $ & $ \oplus N_{j_l, j_r}^{\mathbf{d}} (j_l, j_r) $ & $ \mathbf{d} $ & $ \oplus N_{j_l, j_r}^{\mathbf{d}} (j_l, j_r) $  \\ \hline
	$ (1, 1) $ & $ 128(0, 0) $ & $ (1, 2) $ & $ 128(0, \frac{1}{2}) $ \\ \hline
	$ (1, 3) $ & $ 128(0, 1) $ & $ (2, 1) $ & $ 576(0, 0) \oplus 16(\frac{1}{2}, \frac{1}{2}) $ \\ \hline
	$ (2, 2) $ & $ 1942(0, \frac{1}{2}) \oplus (\frac{1}{2}, 0) \oplus 121(\frac{1}{2}, 1) \oplus (1, \frac{3}{2}) $ & $ (2, 3) $ & $ 560(0, 0) \oplus 4960(0, 1) \oplus 16(\frac{1}{2}, \frac{1}{2}) \oplus 576(\frac{1}{2}, \frac{3}{2}) \oplus 16(1, 2) $ \\ \hline
\end{tabular}
\caption{BPS spectrum of $ SU(2) + 8\mathbf{F} $ theory for $ d_1 \leq 2 $ and $ d_2 \leq 3 $. Here, $ \mathbf{d} = (d_1, d_2) $ labels BPS states with charge $ d_1 m_0 + d_2 \phi $. For convenience, we set all flavor mass parameters $ m_{i=1, \cdots, 8} $ to zero.} \label{table:SU2_8F}
\end{table}

\subsubsection{\texorpdfstring{$SU(2)_0+1\mathbf{Adj}$}{SU(2)0 + 1Adj}}

The compactification of the 6d rank-1 M-string theory on a circle gives rise to the 5d $SU(2)$ gauge theory at $\theta=0$ with an adjoint hypermultiplet preserving $\mathcal{N}=2$ supersymmetry,
\begin{align}
\begin{tikzpicture}
\draw (0, 0) node {$ SU(2)_0 + 1\mathbf{Adj} $}
(2.5, 0) node {$ = $}
(4.5, 0.3) node {$ \mathfrak{su}(1)^{(1)} $}
(4.5, -0.3) node {$ 2 $};
\end{tikzpicture}
\end{align}
The validity of the blowup equation for this theory was previously checked in \cite{Gu:2019pqj}.

The cubic prepotential on the Coulomb branch where $2\phi\pm m_1\ge0$ with $\phi>0$ is
\begin{align}
6\mathcal{F} = 6m_0 \phi^2+ 8\phi^3 - \frac{1}{2} \qty((2\phi + m_1)^3 + (2\phi - m_1)^3)  \ ,
\end{align}
where $m_0$ is the gauge coupling and $ m_1 $ is the mass parameter of the adjoint hypermultiplet.
The effective prepotential on $\Omega$-background is then given by
\begin{align}\label{eq:eff-pre-su2-adj}
\mathcal{E} &= \frac{1}{\epsilon_1 \epsilon_2} \qty(\mathcal{F} - \frac{\epsilon_1^2 + \epsilon_2^2}{48} \Big(4\phi - \big((2\phi + m_1) + (2\phi - m_1)\big)\Big) + \epsilon_+^2 \phi ) \nonumber \\
&=\frac{1}{\epsilon_1 \epsilon_2} \qty(\mathcal{F} +\epsilon_+^2\phi) \, .
\end{align}
The perturbative part of the GV-invariant is
\begin{align}\label{eq:pert-su2-adj}
	\mathcal{Z}_{\rm pert}(\phi,m_1;\epsilon_{1,2}) = {\rm PE}\left[-\frac{1+p_1p_2}{(1-p_1)(1-p_2)}e^{-2\phi}  + \frac{\sqrt{p_1p_2}}{(1-p_1)(1-p_2)}e^{-(2\phi\pm m_1)}\right] \, .
\end{align}

One can formulate a unity blowup equation for this theory using the following consistent magnetic fluxes. 
\begin{align}\label{eq:flux-su2-adj}
	n\in\mathbb{Z} \, , \quad B_{m_0}=0 \, , \quad B_{m_1}=1/2 \, .
\end{align}
The solution to the blowup equation is presented in Table~\ref{table:SU2_0_1Adj}. The result indeed matches the $\mathcal{N}=2$ $SU(2)_0$  instanton partition function computed using the ADHM calculus in \cite{Hwang:2014uwa}.
\begin{table}
	\centering
	\begin{tabular}{|c|C{27ex}||c|C{27ex}|} \hline
		$ \mathbf{d} $ & $ \oplus N_{j_l, j_r}^{\mathbf{d}} (j_l, j_r) $ & $ \mathbf{d} $ & $ \oplus N_{j_l, j_r}^{\mathbf{d}} (j_l, j_r) $  \\ \hline
		$ (1, 2, -2) $ & $ (0, \frac{1}{2}) $ & $ (1, 2, -1) $ & $ (0, 0) \oplus (0, 1) \oplus (\frac{1}{2}, \frac{1}{2}) $ \\ \hline
		$ (1, 2, 0) $ & $ 2(0, \frac{1}{2}) \oplus (\frac{1}{2}, 0) \oplus (\frac{1}{2}, 1) $ & $ (1, 4, -2) $ & $ (0, \frac{3}{2}) $ \\ \hline
		$ (1, 4, -1) $ & $ (0, 1) \oplus (0, 2) \oplus (\frac{1}{2}, \frac{3}{2}) $ & $ (1, 4, 0) $ & $ 2(0, \frac{3}{2}) \oplus  (\frac{1}{2}, 1) \oplus (\frac{1}{2}, 2)$ \\ \hline
		$ (2, 2, -3) $ & $ (0, 0) $ & $ (2, 2, -2) $ & $ 2(0, \frac{1}{2}) \oplus (\frac{1}{2}, 0) \oplus (\frac{1}{2}, 1) $ \\ \hline
		$ (2, 2, -1) $ & $ 4(0,0) \oplus 2(0,1) \oplus 3(\frac{1}{2},\frac{1}{2}) \oplus (\frac{1}{2},\frac{3}{2}) \oplus (1,1) $ & $ (2, 2, 0) $ & $ 5(0,\frac{1}{2}) \oplus (0,\frac{3}{2}) \oplus 3(\frac{1}{2},0)\oplus 3(\frac{1}{2},1) \oplus (1,\frac{1}{2}) \oplus (1,\frac{3}{2}) $ \\ \hline
		$ (2, 4, -3) $ & $ (0,1) \oplus (0,2) \oplus (\frac{1}{2},\frac{3}{2}) $ & $ (2, 4, -2) $ & $ 2(0,\frac{1}{2}) \oplus 4(0,\frac{3}{2}) \oplus (0,\frac{5}{2}) \oplus 3(\frac{1}{2},1) \oplus 	3(\frac{1}{2},2) \oplus (1,\frac{3}{2}) \oplus (1,\frac{5}{2}) $ \\ \hline
		$ (2, 4, -1) $ & $ (0,0) \oplus 7(0,1) \oplus 6(0,2) \oplus 3(\frac{1}{2},\frac{1}{2}) \oplus
		8(\frac{1}{2},\frac{3}{2}) \oplus 3 (\frac{1}{2},\frac{5}{2}) \oplus 2(1,1) \oplus 3(1,2) \oplus
		(1,3) \oplus (\frac{3}{2},\frac{5}{2}) $ & $ (2, 4, 0) $ & $ 5(0,\frac{1}{2}) \oplus 10(0,\frac{3}{2}) \oplus 3(0,\frac{5}{2}) \oplus (\frac{1}{2},0) \oplus 8(\frac{1}{2},1) \oplus 8(\frac{1}{2},2) \oplus (\frac{1}{2},3) \oplus (1,\frac{1}{2}) \oplus 4(1,\frac{3}{2}) \oplus 3(1,\frac{5}{2}) \oplus (\frac{3}{2},2)\oplus (\frac{3}{2},3) $ \\ \hline
	\end{tabular}
	\caption{BPS spectrum of $ SU(2)_0 + 1\mathbf{Adj} $ theory for $ d_1 \leq 2 $ and $ d_2 \leq 4 $. Here, $ \mathbf{d} = (d_1, d_2, d_3) $ labels the BPS states with charge $ d_1 m_0 + d_2 \phi + d_3 m_1 $. The states related by the symmetry $ d_3 \leftrightarrow -d_3 $ are omitted in the table.} \label{table:SU2_0_1Adj}
\end{table}

\subsubsection{\texorpdfstring{$SU(2)_\pi+1\mathbf{Adj}$}{SU(2)pi + 1Adj}}\label{sec:su2pi+1adj}

The 5d KK-theory $ SU(2)_\pi + 1\mathbf{Adj} $ is obtained by a circle compactification of the 6d $\mathcal{N}=(2,0)$ $A_2$ theory with $ \mathbb{Z}_2 $ outer automorphism twist \cite{Tachikawa:2011ch},
\begin{align}
\begin{tikzpicture}
\draw (0, 0) node {$ SU(2)_\pi + 1\mathbf{Adj} $}
(2.5, 0) node {$ = $}
(4.5, 0.3) node {$ \mathfrak{su}(1)^{(1)} $}
(4.5, -0.3) node {$ 2 $};
\draw (4.2, -0.5) .. controls (3.8, -1.2) and (5.2, -1.2) .. (4.8, -0.5);
\end{tikzpicture}
\end{align}
This theory has no conventional geometric description \cite{Bhardwaj:2019fzv}.

The perturbative spectrum cannot distinguish whether $\theta=0$ or $\theta=\pi$ for the $SU(2)$ gauge group. Thus the effective prepotential and the perturbative GV-invariant of this theory are the same as \eqref{eq:eff-pre-su2-adj} and \eqref{eq:pert-su2-adj}, respectively, for the $SU(2)_0+1{\bf Adj}$ theory. Naively, this means that the blowup equation of this theory is the same as that of the theory at $\theta=0$. However, it turns out that there are two independent blowup equations from the same effective prepotential but distinguished by two different sets of consistent magnetic fluxes. Instead of the fluxes given in \eqref{eq:flux-su2-adj} for the theory at $\theta=0$, this theory at $\theta=\pi$ has another set of consistent magnetic fluxes quantized as
\begin{equation}
	n \in \mathbb{Z} + 1/2 \, , \quad B_{m_0}=0 \, , \quad B_{m_1}=1/2 \, .
\end{equation}

These magnetic fluxes provide a unity blowup equation. Some leading BPS invariants we computed using the blowup equation are listed in Table~\ref{table:SU2_pi_1Adj}, which are indeed different from the BPS spectrum of the theory at $\theta=0$. This result also matches the instanton partition function using the ADHM quantum mechanics in \cite{Hwang:2014uwa}. This is our first example showing that  our bootstrap approach can be applied for the BPS spectrum computation for non-geometric theories.
\begin{table}[t]
	\centering
	\begin{tabular}{|c|C{26ex}||c|C{26ex}|} \hline
		$ \mathbf{d} $ & $ \oplus N_{j_l, j_r}^{\mathbf{d}} (j_l, j_r) $ & $ \mathbf{d} $ & $ \oplus N_{j_l, j_r}^{\mathbf{d}} (j_l, j_r) $  \\ \hline
		$ (1, 1, -2) $ & $ (0, 0) $ & $ (1, 1, -1) $ & $ (0, \frac{1}{2}) \oplus (\frac{1}{2}, 0) $ \\ \hline
		$ (1, 1, 0) $ & $ 2(0, 0) \oplus (\frac{1}{2}, \frac{1}{2}) $ & $ (1, 3, -2) $ & $ (0, 1) $ \\ \hline
		$ (1, 3, -1) $ & $ (0, \frac{1}{2}) \oplus (0, \frac{3}{2}) \oplus (\frac{1}{2}, 1) $ & $ (1, 3, 0) $ & $ 2(0, 1) \oplus (\frac{1}{2}, \frac{1}{2}) \oplus (\frac{1}{2}, \frac{3}{2}) $ \\ \hline
		$ (2, 2, -2) $ & $ (0,\frac{1}{2}) \oplus (\frac{1}{2},1) $ & $ (2, 2, -1) $ & $ (0,0) \oplus 2(0,1) \oplus 2(\frac{1}{2},\frac{1}{2})\oplus (\frac{1}{2},\frac{3}{2})\oplus (1,1) $ \\ \hline
		$ (2, 2, 0) $ & $ 3(0,\frac{1}{2}) \oplus (0,\frac{3}{2}) \oplus (\frac{1}{2},0) \oplus 3(\frac{1}{2},1) \oplus	(1,\frac{1}{2})\oplus (1,\frac{3}{2}) $ & $ (2, 4, -3) $ & $ (0,0) \oplus (0,2) \oplus (\frac{1}{2},\frac{3}{2}) $ \\ \hline
		$ (2, 4, -2) $ & $ 2(0,\frac{1}{2}) \oplus 3(0,\frac{3}{2}) \oplus (0,\frac{5}{2}) \oplus (\frac{1}{2},0) \oplus 2(\frac{1}{2},1) \oplus 3(\frac{1}{2},2) \oplus (1,\frac{3}{2})\oplus (1,\frac{5}{2}) $ & $ (2, 4, -1) $ & $ 4(0,0) \oplus 4(0,1) \oplus 6(0,2) \oplus 3(\frac{1}{2},\frac{1}{2}) \oplus 7(\frac{1}{2},\frac{3}{2}) \oplus 3(\frac{1}{2},\frac{5}{2}) \oplus 2(1,1) \oplus 3(1,2) \oplus (1,3) \oplus (\frac{3}{2},\frac{5}{2}) $ \\ \hline
		$ (2, 4, 0) $ & \multicolumn{3}{C{66ex}|}{$ 5(0,\frac{1}{2}) \oplus 8(0,\frac{3}{2}) \oplus 3(0,\frac{5}{2}) \oplus 3(\frac{1}{2},0) \oplus	6(\frac{1}{2},1) \oplus 8(\frac{1}{2},2) \oplus (\frac{1}{2},3) \oplus (1,\frac{1}{2}) \oplus 4(1,\frac{3}{2}) \oplus 3(1,\frac{5}{2}) \oplus (\frac{3}{2},2) \oplus (\frac{3}{2},3) $} \\ \hline
	\end{tabular}
	\caption{BPS spectrum of $ SU(2)_\pi + 1\mathbf{Adj} $ theory for $ d_1 \leq 2 $ and $ d_2 \leq 4 $. Here, $ \mathbf{d} = (d_1, d_2, d_3) $ labels BPS states with charge $ d_1 m_0 + d_2 \phi + d_3 m_1 $. The states related by the symmetry $ d_3 \leftrightarrow -d_3 $ are omitted in the table.} \label{table:SU2_pi_1Adj}
\end{table}

\subsection{5d SCFTs: Non-Lagrangian theories}
The BPS partition functions of all the rank-1 5d SCFTs which have no gauge theory descriptions, can be computed by solving the blowup equations. To show this we will consider two simple examples: a local $\mathbb{P}^2$ theory and a local $\mathbb{P}^2 +1{\bf Adj}$ theory.

\subsubsection{\texorpdfstring{Local $\mathbb{P}^2$}{LocalP2}}

This theory is a rank-1 SCFT with no mass parameter, which is engineered by compactification of M-theory on a local $\mathbb{P}^2$ embedded in a CY 3-fold. This theory is also called the $E_0$ theory. The blowup equation for the SCFT of a local $\mathbb{P}^2$ was solved in \cite{Huang:2017mis}. We shall briefly review this.

The effective prepotential of this theory on the $\Omega$-background is simply given by
\begin{align}
\mathcal{E} = \frac{1}{\epsilon_1 \epsilon_2} \qty(9\phi^3 - \frac{(\epsilon_1^2 + \epsilon_2^2)\phi}{8} + \epsilon_+^2 \phi) \, .
\end{align}
The CY 3-fold of a local $ \mathbb{P}^2 $ contains only one primitive curve class $\ell$ with  $\ell^2= +1 $. The volume of this curve is $ \vol (\ell) = 3\phi $. Thus the magnetic fluxes that couple to the curve should be quantized as 
\begin{align}\label{eq:P2_shift}
n\in \mathbb{Z} \pm 1/6 \quad {\rm or} \quad n\in \mathbb{Z} + 1/2 \, .
\end{align}
These are all the consistent magnetic fluxes. The blowup equation with $n\in \mathbb{Z} + 1/2$ is a vanishing blowup equations, whereas the other choices of fluxes lead to unity blowup equations \cite{Huang:2017mis}, and they are all solvable.
The solution to these blowup equations is summarized in Table~\ref{table:P2}.
\begin{table}[H]
	\centering
	\begin{tabular}{|c|C{30ex}||c|C{30ex}|} \hline
		$ d $ & $ \oplus N_{j_l, j_r}^d (j_l, j_r) $ & $ d $ & $ \oplus N_{j_l, j_r}^d (j_l, j_r) $ \\ \hline
		$ 1 $ & $ (0, 1) $ & $ 2 $ & $ (0, \frac{5}{2}) $ \\ \hline
		$ 3 $ & $ (0, 3) \oplus (\frac{1}{2}, \frac{9}{2}) $  & $ 4 $ & $ \! (0, \frac{5}{2}) \oplus (0, \frac{9}{2}) \oplus (0, \frac{13}{2}) \oplus (\frac{1}{2}, 4) \oplus (\frac{1}{2}, 5) \oplus (\frac{1}{2}, 6) \oplus (1, \frac{11}{2}) \oplus (\frac{3}{2}, 7) \! $ \\ \hline
	\end{tabular}
	\caption{BPS spectrum of a local $ \mathbb{P}^2 $ for $ d \leq 4 $. Here, $ d $ labels the BPS states wrapping the degree $d$ curve in $\mathbb{P}^2$.} \label{table:P2}
\end{table}

\subsubsection{\texorpdfstring{Local $\mathbb{P}^2 + 1\mathbf{Adj}$}{Local P2 + 1Adj}}
The local $\mathbb{P}^2$ + 1$\mathbf{Adj}$ (or $``SU(2)_\pi+1\mathbf{Adj}-1\mathbf{F}"$) theory was first proposed in \cite{Bhardwaj:2019jtr}. This theory can be obtained by an RG flow from the UV $SU(2)_\pi+1{\bf Adj}$ theory after integrating out an instantonic hypermultiplet.

One way to see the existence of this theory is as follows. Recall that the SCFT of a local $\mathbb{P}^2$ (or the $E_0$ theory) can be obtained by blowing down an exceptional curve in a Hirzebruch surface $\mathbb{F}_1$. This corresponds to integrating out an instantonic hypermultiplet in the pure $SU(2)_\pi$ theory. In this regard, the $E_0$ theory can be thought of as ``$SU(2)_\pi-1\mathbf{F}$'' since we are ``subtracting'' a hypermultiplet. Likewise, the $SU(2)_\pi + 1\mathbf{Adj}$ theory we discussed in section \ref{sec:su2pi+1adj} has a hypermultiplet with charge ${\bf d}=(1,1,-2)$ at the 1-instanton sector and integrating out this hypermultiplet induces an RG flow to a consistent rank-1 SCFT with one mass parameter. We will call this IR SCFT  the local $\mathbb{P}^2+1\mathbf{Adj}$ theory. This theory is a non-Lagrangian theory. Here $+1\mathbf{Adj}$ simply means this theory contains a hypermultiplet originating from the adjoint hypermultiplet in its parent $SU(2)_\pi+1{\bf Adj}$ theory. The adjoint hypermultiplet in this SCFT can also be integrated out to give the SCFT of a local $\mathbb{P}^2$.

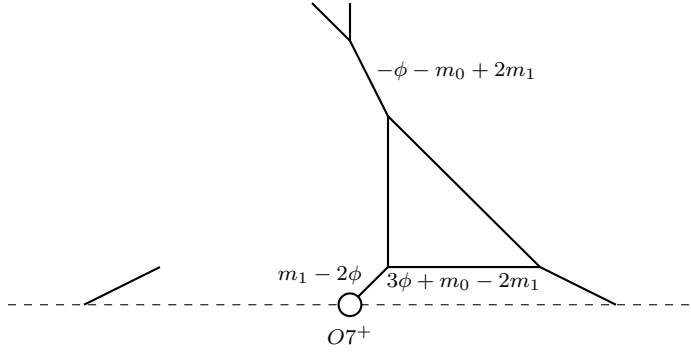
\begin{figure}
	\centering
	\begin{tikzpicture}
	\draw[thick] (0, 0) -- (2, 0) -- (0, 2) -- (0, 0)
	(0, 0) -- (-0.5, -0.5)
	(2, 0) -- (3, -0.5)
	(-3, 0) -- (-4, -0.5)
	(0, 2) -- (-0.5, 3) -- (-0.5, 3.5)
	(-0.5, 3) -- (-1, 3.5);
	\draw[dashed] (-5, -0.5) -- (4, -0.5);
	\filldraw[fill=white, thick] (-0.5, -0.5) circle (0.15);
	\draw (1, -0.2) node {\scriptsize{$ {3\phi + m_0 - 2m_1} $}}
	(0.9, 2.6) node {\scriptsize{$ {-\phi - m_0 + 2m_1} $}}
	(-0.9, -0.1) node {\scriptsize{$ {m_1 - 2\phi} $}}
	(-0.5, -0.9) node {\scriptsize{$O7^+$}}
	;
	\end{tikzpicture}
	\caption{A 5-brane web for $SU(2)_\pi+1\mathbf{Adj}$, where a flop transition associated with the $e$ curve in $\mathbb{P}^2$ is performed.} \label{fig:SU2_pi_adj_F_web}
\end{figure}

This theory is also a non-geometric theory. However, there is a 5-brane web for this theory with an $O7^+$-plane where the hypermultiplet associated with an $O7^+$-plane is in the symmetric representation of $SU(2)$ that is equivalent to the adjoint of the UV $SU(2)$ gauge symmetry. A simple way to construct the corresponding 5-brane web is to attach an $O7^+$-plane to a 5-brane web for the pure $SU(2)_\pi$ theory, as depicted in Figure \ref{fig:SU2_pi_adj_F_web}. (See also Appendix \ref{sec:appendix2} for more details of obtaining a 5-brane web for the $SU(2)_\pi + 1\mathbf{Adj}$ theory.)

The hypermultiplet which we will integrate out is the $(p,q)$-string state associated with the edge of the length $-\phi-m_0+2m_1$ in the brane web. Integrating out this hypermultiplet corresponds to taking the infinite length limit of the edge $-\phi-m_0+2m_1\rightarrow\infty$ while the lengths of other edges are kept finite. After this, we will get a 5-brane web for the local $\mathbb{P}^2+``1\mathbf{Adj}"$ theory. In this 5-brane web diagram, one can also see that by taking another limit $m_1 - 2\phi\rightarrow \infty$, after taking the limit $-\phi-m_0+2m_1\rightarrow\infty$, the diagram reduces to the brane web for a local $\mathbb{P}^2$, which is consistent with the expected RG flow for this theory.

We will now compute the partition function using the chamber for the brane web of the $SU(2)_\pi+1{\bf Adj}$ theory depicted in Figure~\ref{fig:SU2_pi_adj_F_web}. It is convenient to reparameterize the parameters in the brane web as follows:
\begin{align}
\tilde{\phi} = \phi + \frac{1}{3}m_0 - \frac{2}{3} m_1 \, , \quad
\tilde{m} = \frac{2}{3} m_0 - \frac{1}{3} m_1 \, ,
\end{align}
so that the volumes of 2-cycles in Figure~\ref{fig:SU2_pi_adj_F_web} become
\begin{align}
&3\phi + m_0 - 2m_1 = 3\tilde{\phi} \, , \quad
m_1 - 2\phi = \tilde{m} - 2\tilde{\phi} \, , \nonumber \\
&-\phi - m_0 + 2m_1 = -\tilde{\phi} + 2m_0 - 4\tilde{m} \, .
\end{align}

In order to get the IR  local $\mathbb{P}^2+1\mathbf{Adj}$ theory, we take the limit $ m_0 \to \infty $ while $ \tilde{\phi} $ and $ \tilde{m} $ are kept finite. The cubic prepotential and the mixed gravitational Chern-Simons level of the IR  theory can be obtained from those of the parent theory as
\begin{align}
\mathcal{F}_{IR}=\mathcal{F}_{UV} + \frac{1}{6}(\phi+m_0-2m_1)^3 \, , \quad C^G_{IR} =C^G_{UV} -2(\phi+m_0-2m_1) \, ,
\end{align}
whereas the mixed gauge/$SU(2)_R$ level $C^R$ remains the same along the RG flow. Thus the effective prepotential of a local  $\mathbb{P}^2+1\mathbf{Adj}$ on the extended Coulomb branch where $\tilde\phi>0$ and $\tilde{m}-2\tilde{\phi}\geq 0$ becomes 
\begin{align}
\mathcal{E} &= \frac{1}{\epsilon_1 \epsilon_2} \qty(\frac{3}{2}\tilde{\phi}^3 - \frac{(\epsilon_1^2 + \epsilon_2^2)\tilde{\phi}}{8} + \epsilon_+^2 \tilde{\phi} ) \, ,
\end{align}
up to the terms independent of $\tilde\phi$. Interestingly, this prepotential is the same as that of a local $\mathbb{P}^2$. However, they are different theories because the  $\mathbb{P}^2+1\mathbf{Adj}$ theory contains an additional hypermultiplet coming from the adjoint hypermultiplet of the parent theory, while the $E_0$ theory has no hypermultiplet. The GV-invariant for this hypermultiplet is thus a key input for solving the blowup equation
\begin{equation}
\mathcal{Z}_{\rm hyper}(\tilde{\phi},\tilde{m};\epsilon_{1,2}) = \PE \qty[\frac{\sqrt{p_1p_2}}{(1-p_1)(1-p_2)}e^{-(\tilde{m}- 2\tilde\phi)}] \, .
\end{equation}

The feasible magnetic fluxes that one can turn on are
\begin{align}
n \in \mathbb{Z}+ 1/6 \, , \quad B_{\tilde{m}} = -1/6 \, .
\end{align}
The blowup equation with this flux choice is a unity blowup equation. We solve this equation and find the BPS spectrum of the local  $\mathbb{P}^2+1\mathbf{Adj}$ theory listed in Table~\ref{table:SU2_pi_Adj-F}. As expected, this result also agrees with the RG flow from the BPS spetrum in the $SU(2)_\pi+1{\bf Adj}$ theory given in Table \ref{table:SU2_pi_1Adj} in the limit $-\tilde{\phi}+2m_0-4\tilde{m}\rightarrow \infty$ while $\tilde\phi$ and $\tilde{m}$ are kept finite.

\begin{table}
	\centering
	\begin{tabular}{|c|C{27ex}||c|C{27ex}|} \hline
		$ \mathbf{d} $ & $ \oplus N_{j_l, j_r}^{\mathbf{d}} (j_l, j_r) $ & $ \mathbf{d} $ & $ \oplus N_{j_l, j_r}^{\mathbf{d}} (j_l, j_r) $  \\ \hline
		$ (1, -2) $ & $ (0, 0) $ & $ (1, 1) $ & $ (0, \frac{1}{2}) \oplus (\frac{1}{2}, 0) $ \\ \hline
		$ (1, 4) $ & $ (0, 0) \oplus (0, 2) \oplus (\frac{1}{2}, \frac{3}{2}) $ & $ (1, 7) $ & $ (0,\frac{3}{2}) \oplus (0,\frac{5}{2}) \oplus (0,\frac{7}{2}) \oplus (\frac{1}{2},2) \oplus (\frac{1}{2},3) \oplus (\frac{1}{2},4) \oplus (1,\frac{7}{2}) $ \\ \hline
		$ (2, 2) $ & $ (0, \frac{1}{2}) \oplus (\frac{1}{2}, 1) $ & $ (2, 5) $ & $ 2(0,1) \oplus (0,2) \oplus (0,3) \oplus (\frac{1}{2},\frac{1}{2}) \oplus (\frac{1}{2},\frac{3}{2}) \oplus	2(\frac{1}{2},\frac{5}{2}) \oplus (1,2) \oplus (1,3) $ \\ \hline
		$ (2, 8) $ & \multicolumn{3}{C{63ex}|}{$ 2(0,\frac{1}{2}) \oplus 2(0,\frac{3}{2}) \oplus 6(0,\frac{5}{2}) \oplus 3(0,\frac{7}{2}) \oplus 4(0,\frac{9}{2}) \oplus 2(\frac{1}{2},1) \oplus 4(\frac{1}{2},2) \oplus 6(\frac{1}{2},3) \oplus 6(\frac{1}{2},4) \oplus 2(\frac{1}{2},5) \oplus (1,\frac{3}{2}) \oplus 3(1,\frac{5}{2}) \oplus 5(1,\frac{7}{2}) \oplus 4(1,\frac{9}{2}) \oplus (1,\frac{11}{2}) \oplus (\frac{3}{2},3) \oplus 2(\frac{3}{2},4) \oplus 2(\frac{3}{2},5) \oplus (2,\frac{9}{2}) \oplus (2,\frac{11}{2}) $} \\ \hline
	\end{tabular}
	\caption{BPS spectrum of  $\mathbb{P}^2+1\mathbf{Adj}$ theory with $ d_1 \leq 2 $ and $ d_2 \leq 9 $. Here, $ \mathbf{d} = (d_1, d_2) $ labels the BPS states with charge $d_1 \tilde{m} + d_2 \tilde{\phi} $.} \label{table:SU2_pi_Adj-F}
\end{table}

%% file: sec-Rank2.tex
\section{Rank 2 theories}\label{sec:rank2 theories}

In this section, we will compute the BPS spectra of all rank-2 KK theories by employing our bootstrap approach. The computations will then guarantee that all rank-2 5d and 6d QFTs have solvable blowup equations and their BPS spectra can be obtained by the bootstrap method. To support our claim, we will present computations of the BPS spectra of three non-geometric 5d SCFTs: $SU(3)_8$, $\mathbb{P}^2 \cup \mathbb{F}_3+1{\bf Sym}$, and $\mathbb{P}^2 \cup \mathbb{F}_6+1{\bf Sym}$.

\subsection{KK theories}
In this subsection, we solve the blowup equations for all rank-2 5d KK theories classified in \cite{Jefferson:2017ahm,Jefferson:2018irk,Bhardwaj:2019fzv}.

\subsubsection{\texorpdfstring{$Sp(2)+3\mathbf{\Lambda}^2$}{Sp(2) + 3AS}}
 
The 5d $ Sp(2)$ gauge theory with 3 anti-symmetric hypermultiplets ($Sp(2)+3\mathbf{\Lambda}^2$) is the KK-theory arising from the twisted compactification of the 6d $ SU(3) $ gauge theory with 6 fundamental hypermultiplets,
\begin{align}
\begin{tikzpicture}
\draw (0, 0) node {$ Sp(2) + 3\mathbf{\Lambda}^2 $}
(2.2, 0) node {$ = $}
(4, 0.3) node {$ \mathfrak{su}(3)^{(2)} $}
(4, -0.3) node {$ 2 $};
\end{tikzpicture}
\end{align}
This theory has a geometric description \cite{Jefferson:2018irk} as 
\begin{align}\label{eq:Sp2_3AS_geo}
\begin{tikzpicture}
\draw[thick](-3,0)--(0,0);	
\node at(-3.6,0) {$\mathbb{F}_{6}{}\big|_1$};
\node at(-2.8,0.3) {${}_e$};
\node at(0.7,0) {$\mathbb{F}_{0}^{3}{}\big|_2$};
\node at(-0.9,0.3) {${}_{2h + 4f - 2\sum x_i}$};
\end{tikzpicture}
\end{align}
and the fiber-base duality of $\mathbb{F}_0^3$ exchanges two dual descriptions: the 5d $Sp(2)$ gauge theory with 3 anti-symmetric hypers and the 6d $SU(3)$ gauge theory with 6 fundamentals on a circle with $\mathbb{Z}_2$ twist.

In 5d $Sp(2)$ gauge theory description, the cubic prepotential on the chamber described by the above geometry
 is
\begin{align}
6\mathcal{F}
&= 8\phi_1^3 + 12\phi_1^2 \phi_2 - 18\phi_1 \phi_2^2 + 8\phi_2^3 + 6m_0 (2\phi_1^2 - 2\phi_1 \phi_2 + \phi_2^2) \\
& \quad - \frac{1}{2} \sum_{i=1}^3 \qty((\phi_2 + m_i)^3 + (2\phi_1 - \phi_2 + m_i)^3 + (-2\phi_1 + \phi_2 + m_i)^3 + (\phi_2 - m_i)^3) \, , \nonumber
\end{align}
where $ m_0 $ is the gauge coupling and $ m_{i=1, 2, 3} $ are mass parameters for the 3 antisymmetric hypermultiplets. The full effective prepotential on the $\Omega$-background is
\begin{align}
\mathcal{E}
= \frac{1}{\epsilon_1 \epsilon_2} \left(\mathcal{F} - \frac{\epsilon_1^2 + \epsilon_2^2}{48} (4\phi_1 - 2\phi_2) + \epsilon_+^2 (\phi_1 + \phi_2) \right) \, .
\end{align}

The volumes of the primitive 2-cycles in the geometry \eqref{eq:Sp2_3AS_geo} in terms of the $Sp(2)$ gauge theory parameters can be written as
\begin{align}
&\vol (f_1) = 2\phi_1 - \phi_2 \, , \quad
&&\vol (f_2) = -2\phi_1 + 2\phi_2 \, , \nonumber \\
&\vol (e_2) = -4\phi_1 + 2\phi_2 + m_0 \, , \quad
&&\vol (x_i) = -2\phi_1 + \phi_2 + m_i \quad (i = 1, 2, 3) \, .
\end{align}
From this information, we can choose a set of magnetic fluxes as
\begin{align}
n_i \in \mathbb{Z} \, , \quad
B_{m_0} = 0 \, , \quad
B_{m_i} = 1/2 \quad (i = 1, 2, 3) \, ,
\end{align}
which gives rise to a unity blowup equation. The unity blowup equation can be easily solved and the resulting BPS spectrum is summarized in Table~\ref{table:Sp2_3AS}.

\begin{table}
	\centering
	\begin{tabular}{|c|C{25ex}||c|C{25ex}|} \hline
		$ \mathbf{d} $ & $ \oplus N_{j_l, j_r}^{\mathbf{d}} (j_l, j_r) $ & $ \mathbf{d} $ & $ \oplus N_{j_l, j_r}^{\mathbf{d}} (j_l, j_r) $ \\ \hline
		$ (1, 0, 0, -1) $ & $ 3(0, 0) $ & $ (1, 0, 0, 0) $ & $ (0, \frac{1}{2}) $ \\ \hline
		$ (1, 0, 1, -3) $ & $ (0, 0) $ & $ (1, 0, 1, -2) $ & $ 3(0, \frac{1}{2}) $ \\ \hline
		$ (1, 0, 1, -1) $ & $ 3(0, 1) $ & $ (1, 0, 1, 0) $ & $ (0, \frac{3}{2}) $ \\ \hline
		$ (1, 1, 0, 0) $ & $ (0, \frac{1}{2}) $ & $ (1, 1, 1, -2) $ & $ 3(0, \frac{1}{2}) $ \\ \hline
		$ (1, 1, 1, -1) $ & $ 3(0, 0) \oplus 3(0, 1) $ & $ (1, 1, 1, 0) $ & $ (0, \frac{1}{2}) \oplus (0, \frac{3}{2}) $ \\ \hline
		$ (2, 0, 1, -3) $ & $ (0, 1) $ & $ (2, 0, 1, -2) $ & $ 3(0, \frac{3}{2}) $ \\ \hline
		$ (2, 0, 1, -1) $ & $ 3(0, 2) $ & $ (2, 0, 1, 0) $ & $ (0, \frac{5}{2}) $ \\ \hline
		$ (2, 1, 1, -3) $ & $ (0, 0) \oplus (0, 1) $ & $ (2, 1, 1, -2) $ & $ 3(0, \frac{1}{2}) \oplus 3(0, \frac{3}{2}) $ \\ \hline
		$ (2, 1, 1, -1) $ & $ 3(0, 1) \oplus 3(0, 2) $ & $ (2, 1, 1, 0) $ & $ (0, \frac{3}{2}) \oplus (0, \frac{5}{2}) $ \\ \hline
	\end{tabular}
	\caption{BPS spectrum of $ Sp(2) + 3\mathbf{\Lambda}^2 $ for $ d_1 \leq 2 $, $ d_2 \leq 1 $, $ d_3 \leq 1 $. Here, $ \mathbf{d} = (d_1, d_2, d_3, d_4) $ labels the BPS states with charge $ d_1 e_2 + d_2 f_1 + d_3 f_2 + d_4 x_i $ and $d_4$ counts collective degrees of all anti-symmetric hypermultiplets.} \label{table:Sp2_3AS}
\end{table}

Now we consider the 6d $SU(3)$ gauge theory. Under the $\mathbb{Z}_2$ twist, the vector multiplet and the hypermultiplets in 6d are decomposed into the representations of the invariant subalgebra $\mathfrak{su}(2)$ as
\begin{align}
\mathbf{8} \text{ of } \mathfrak{su}(3) \ &\to \ \mathbf{3}_0 \oplus \mathbf{2}_{1/4} \oplus \mathbf{2}_{3/4} \oplus \mathbf{1}_{1/2} \text{ of } \mathfrak{su}(2)\ , \nonumber \\
\mathbf{3} \oplus \bar{\mathbf{3}} \text{ of } \mathfrak{su}(3) \ &\to \ \mathbf{2}_0 \oplus \mathbf{1}_{1/4} \oplus \mathbf{2}_{1/2} \oplus \mathbf{1}_{3/2} \text{ of } \mathfrak{su}(2) \ ,
\end{align}
where the subscripts denote the shifted KK charges due to the $\mathbb{Z}_2$ twist. Collecting the 1-loop contributions from these KK states and the classical Green-Schwarz contributions, we obtain the full effective prepotential on the $\Omega$-background as
\begin{align}
\mathcal{E} \label{eq:su3_2_E}
&=\frac{1}{\epsilon_1 \epsilon_2} \bigg( \mathcal{E}_{\mathrm{tree}} + \mathcal{F}_{\text{1-loop}} + \frac{\epsilon_1^2 + \epsilon_2^2}{24}\phi_1 + \epsilon_+^2 \phi_1 \bigg) \ , \nonumber \\
\mathcal{E}_{\mathrm{tree}}
&= \tau \phi_0^2 + 2\phi_0 \qty( \phi_1^2 - \frac{1}{2}\sum_{i=1}^3 m_i^2 + \frac{3}{2} \epsilon_+^2 )\ , \nonumber  \\
\mathcal{F}_{\text{1-loop}}
&= \frac{5}{6}\phi_1^3 - \frac{3}{16}\tau \phi_1^2 - \frac{1}{2} \sum_{i=1}^3 m_i^2 \phi_1\ . 
\end{align}
Here, the K\"ahler parameters in the 6d gauge theory description can be converted to the parameters in the 5d $Sp(2)$ gauge theory description simply by the following reparametrization:
\begin{align}
&\phi_0 \to \phi_1 + \frac{m_0}{16} - \frac{1}{4} \sum_{i=1}^3 m_i\,, \quad
\phi_1 \to \phi_2 - 2\phi_1 + \frac{m_0}{2}\,, \quad
\tau \to 2m_0\,,  \quad
m_i \to m_i - \frac{m_0}{2} \,.
\end{align}
One can easily check that the effective prepotential of the 6d theory with this reparametrization reproduces that of the 5d theory given above up to terms independent of $\phi_i$.

It is convenient in the 6d perspective to use the 6d parameters shifted as
\begin{align}
\phi_0 \to \phi_0 - \frac{5}{32}\tau - \frac{1}{4}\sum_{i=1}^3 m_i\, , \quad
\phi_1 \to \phi_1 - 2\phi_0 \, .
\end{align}
The volumes of 2-cycles in the geometry \eqref{eq:Sp2_3AS_geo} can be written in terms of the shifted 6d parameters as
\begin{align}
&\vol (f_1) = 2\phi_0 - \phi_1 + \frac{\tau}{4}\,, \quad
&&\vol (f_2) = -4\phi_0 + 2\phi_1\,, \nonumber \\
&\vol (e_2) = -2\phi_0 + 2\phi_1 - \frac{\tau}{2}\,, \quad
&&\vol (x_i) = -2\phi_0 + \phi_1 + m_i \, .
\end{align}
One finds a pair of consistent magnetic fluxes as
\begin{align}
	n_1 \in \mathbb{Z} \, , \quad B_\tau = 0 \,, \quad B_{m_i}=1/2 \,, \quad n_0\in \mathbb{Z}+B_{0} \quad {\rm with} \quad B_0 = 0,1/2  \, ,
\end{align}
which lead to unity blowup equations. We checked that the solutions to the blowup equations match the result in Table~\ref{table:Sp2_3AS} from the 5d $Sp(2)$ gauge theory.

\subsubsection{\texorpdfstring{$SU(3)_4+6\mathbf{F}$, $Sp(2)+2\mathbf{\Lambda}^2+4\mathbf{F}$, $G_2+6\mathbf{F}$}{SU(3)4 + 6F}}

This theory is a KK theory coming from the 6d $ SU(3)$ gauge theory with 12 fundamental hypermultiplets on a circle with $ \mathbb{Z}_2 $ twist. It has three different 5d gauge theory descriptions: $ SU(3)_4 + 6\mathbf{F} $, $ Sp(2) + 2\mathbf{\Lambda}^2 + 4\mathbf{F} $, and  $G_2+6\mathbf{F}$,
\begin{align}
\begin{tikzpicture}
\draw (0, 0) node {$ SU(3)_4 + 6\mathbf{F} $}
(1.7, 0) node {$ = $}
(3, 0) node {$ G_2 + 6\mathbf{F} $}
(4.3, 0) node {$ = $}
(6.5, 0) node {$ Sp(2) + 2\mathbf{\Lambda}^2 + 4\mathbf{F} $}
(8.7, 0) node {$ = $}
(10, 0.3) node {$ \mathfrak{su}(3)^{(2)} $}
(10, -0.3) node {$ 1 $};
\end{tikzpicture}
\end{align}
This theory is geometrically engineered by gluing $ \mathbb{F}_2 $ and $ \mathbb{F}_0^6 $ as \cite{Jefferson:2018irk}
\begin{align}
\begin{tikzpicture}
\draw[thick](-3,0)--(0,0);	
\node at(-3.6,0) {$\mathbb{F}_{2}{}\big|_1$};
\node at(-2.8,0.3) {${}_e$};
\node at(0.7,0) {$\mathbb{F}_{0}^{6}{}\big|_2$};
\node at(-0.9,0.3) {${}_{e + 2f - \sum_{i=1}^4 x_i}$};
\end{tikzpicture}
\end{align}

We shall solve the blowup equations by using the $ SU(3) $ and $ G_2 $ frames. We first compute the cubic prepotential of the $SU(3)$ gauge theory on the chamber $\phi_2\ge \phi_1>0$ as
\begin{align}
6\mathcal{F}
&= 8\phi_1^3 - 3\phi_1^2 \phi_2 - 3\phi_1 \phi_2^2 + 8\phi_2^2 + 12\phi_1 \phi_2(\phi_1 - \phi_2) \\
& \quad - \frac{1}{2} \sum_{i=1}^6 \Big((\phi_2 - m_i)^3 + (\phi_2 - \phi_1 + m_i)^3 + (\phi_1 + m_i)^3\Big) + 6m_0(\phi_1^2 - \phi_1 \phi_2 + \phi_2^2) \, ,  \nonumber
\end{align}
where $ m_0 $ is the $ SU(3) $ gauge coupling and $ m_{i=1, \cdots, 6} $ are mass parameters of 6 fundamentals. The full effective prepotential is
\begin{align}
\mathcal{E}
= \frac{1}{\epsilon_1 \epsilon_2} \qty(\mathcal{F} - \frac{\epsilon_1^2 + \epsilon_2^2}{48}(4\phi_1 - 8\phi_2 ) + \epsilon_+^2 (\phi_1 + \phi_2) ) \, .
\end{align}
A choice of magnetic fluxes
\begin{align}
n_i \in \mathbb{Z} \, , \quad
B_{m_0} = -1/2 \, , \quad
B_{m_i} = 1/2 \quad (1 \leq i \leq 6) \, 
\end{align}
provides a solvable unity blowup equation. We solve the blowup equation and find the BPS spectrum of the $SU(3)_4+6{\bf F}$ theory listed in Table~\ref{table:SU3_4_6F}.
\begin{table}
	\centering
	\begin{tabular}{|c|C{25ex}||c|C{25ex}|} \hline
		$ \mathbf{d} $ & $ \oplus N_{j_l, j_r}^{\mathbf{d}} (j_l, j_r) $ & $ \mathbf{d} $ & $ \oplus N_{j_l, j_r}^{\mathbf{d}} (j_l, j_r) $ \\ \hline
		$ (1, -\frac{4}{3}, \frac{1}{3}) $ & $ (0, \frac{1}{2}) $ & $ (1, -\frac{4}{3}, \frac{4}{3}) $ & $ (0, \frac{3}{2}) $ \\ \hline
		$ (1, -1, 0) $ & $ 6(0, 0) $ & $ (1, -1, 1) $ & $ 6(0, 1) $ \\ \hline
		$ (1, -1, 2) $ & $ 6(0, 2) $ & $ (1, -\frac{2}{3}, \frac{2}{3}) $ & $ 15(0, \frac{1}{2}) $ \\ \hline
		$ (1, -\frac{2}{3}, \frac{5}{3}) $ & $ 15(0, \frac{3}{2}) $ & $ (1, -\frac{1}{3}, \frac{1}{3}) $ & $ 20(0, 0) \oplus (0, \frac{1}{2}) $ \\ \hline
		$ (1, -\frac{1}{3}, \frac{4}{3}) $ & $ (0,\frac{1}{2}) \oplus 20(0,1) \oplus (0,\frac{3}{2}) $ & $ (1, 0, 1) $ & $ 6(0,0) \oplus 15(0,\frac{1}{2}) \oplus 6(0,1) $ \\ \hline
		$ (1, 0, 2) $ & $ 6(0,1) \oplus 15(0,\frac{3}{2}) \oplus 6(0,2) $ & $ (1, \frac{1}{3}, \frac{2}{3}) $ & $ 6(0,0) \oplus 15(0,\frac{1}{2}) $ \\ \hline
		$ (1, \frac{1}{3}, \frac{5}{3}) $ & $ \! 15(0,\frac{1}{2}) \oplus 6(0,1) \oplus 15(0,\frac{3}{2}) \! $ & $ (1, \frac{2}{3}, \frac{1}{3}) $ & $ 20(0,0) \oplus (0,\frac{1}{2}) $ \\ \hline
		$ (1, \frac{2}{3}, \frac{4}{3}) $ & $ 20(0,0) \oplus 2(0,\frac{1}{2}) \oplus 20(0,1) \oplus (0,\frac{3}{2}) $ & $ (1, 1, 0) $ & $ 6(0, 0) $ \\ \hline
		$ (1, 1, 1) $ & $ 6(0,0) \oplus 15(0,\frac{1}{2}) \oplus 6(0,1) $ & $ (1, 1, 2) $ & $ 6(0,0) \oplus 15(0,\frac{1}{2}) \oplus 6(0,1) \oplus 15(0,\frac{3}{2}) \oplus 6(0,2) $ \\ \hline
		$ (2, -2, 1) $ & $ 15(0, \frac{3}{2}) $ & $ (2, -\frac{5}{3}, \frac{2}{3}) $ & $ 20(0, 1) $  \\ \hline
		$ (2, -\frac{4}{3}, \frac{1}{3}) $ & $ 15(0, \frac{1}{2}) $ & $ (2, -1, 0) $ & $ 6(0, 0) $ \\ \hline
		$ (2, -1, 1) $ & $ 6(0,0) \oplus 15(0,\frac{1}{2}) \oplus 96(0,1) \oplus 15(0,\frac{3}{2}) \oplus 6(\frac{1}{2},\frac{3}{2}) $ & $ (2, -\frac{2}{3}, \frac{2}{3}) $ & $ 20(0,0) \oplus 37(0,\frac{1}{2}) \oplus 20(0,1) \oplus	(\frac{1}{2},1) $  \\ \hline
		$ (2, -\frac{1}{3}, \frac{1}{3}) $ & $ 12(0,0) \oplus 15(0,\frac{1}{2}) $ & $ (2, 0, 1) $ & $ 102(0,0) \oplus 66(0,\frac{1}{2}) \oplus 102(0,1) \oplus 15(0,\frac{3}{2}) \oplus 6(\frac{1}{2},\frac{1}{2}) \oplus 6(\frac{1}{2},\frac{3}{2}) $  \\ \hline
	\end{tabular}
	\caption{BPS spectrum of the $ SU(3)_4 + 6\mathbf{F} $ theory for $ (d_1 = 1, d_2 \leq 1, d_3 \leq 2) $ and $ (d_1 = 2, d_2 \leq 0, d_3 \leq 1) $ where $ \mathbf{d} = (d_1, d_2, d_3) $ labels the BPS states with charge $ d_1 m_0 + d_2 \alpha_1 + d_3 \alpha_2 $ for simple roots $ \alpha_1 = 2\phi_1 - \phi_2 $, $ \alpha_2 = -\phi_1 + 2\phi_2 $ of $ \mathfrak{su}(3) $ algebra, and 6 flavor charges are blindly summed over.} \label{table:SU3_4_6F}
\end{table}

On the other hand, the full effective prepotential in the $ G_2 $ gauge theory in the same chamber is
\begin{align}
\mathcal{E}
&= \frac{1}{\epsilon_1 \epsilon_2} \qty(\mathcal{F} - \frac{\epsilon_1^2 + \epsilon_2^2}{48} (4\phi_1 - 8\phi_2) + \epsilon_+^2 (\phi_1 + \phi_2) ) \, , \nonumber \\
6\mathcal{F}
&= 8\phi_1^3 + 18\phi_1^2 \phi_2 - 24\phi_1 \phi_2^2 + 8\phi_2^3 + 6m_0(3\phi_1^2 - 3\phi_1 \phi_2 + \phi_2^2) \nonumber \\
& \quad - \frac{1}{2} \sum_{i=1}^6 \Big( (\phi_1 + m_i)^3 + (-\phi_1 + \phi_2 + m_i)^3 + (2\phi_1 - \phi_2 + m_i)^3 \nonumber \\
& \qquad \qquad + (-2\phi_1 + \phi_2 + m_i)^3 + (-\phi_1 + \phi_2 - m_i)^3 + (\phi_1 - m_i)^3 + m_i^3 \Big) \, .
\end{align}
Here, $ m_0 $ is the $ G_2 $ gauge coupling and $ m_{i=1, \cdots, 6} $ are mass parameters of the $G_2$ fundamental hypers.

Under the duality between the $SU(3)$ and $G_2$ descriptions, the K\"ahler parameters in two theories have a natural map \cite{Hayashi:2018lyv} given by 
\begin{align}
\phi_1^{SU} &= \phi_1^{G_2} + \frac{1}{3} m_0^{G_2} - \frac{1}{3} \sum_{j=1}^6 m_j^{G_2}  \,, \quad \phi_2^{SU} = \phi_2^{G_2} + \frac{2}{3} m_0^{G_2} - \frac{2}{3} \sum_{j=1}^6 m_j^{G_2}\ ,  \\
m_0^{SU} &= \frac{2}{3} m_0^{G_2} - \frac{1}{6} \sum_{j=1}^6 m_j^{G_2} \,, \quad
m_i^{SU} = -\frac{1}{3} m_0^{G_2} - m_i^{G_2} + \frac{1}{3} \sum_{j=1}^6 m_j^{G_2} \quad (1 \leq i \leq 6) \, . \nonumber
\end{align}

The magnetic flux quantization is satisfied by
\begin{align}
n_i \in \mathbb{Z}\,, \quad
B_{m_0} = 0\,, \quad
B_{m_i} = 1/2 \quad (1 \leq i \leq 6) \, .
\end{align}
This choice of magnetic fluxes provides a solvable unity blowup equation. We check that the solution to the equation matches the BPS spectrum of the $SU(3)$ theory in Table~\ref{table:SU3_4_6F} under the above parameter map.

\subsubsection{\texorpdfstring{$SU(3)_{\frac32}+9\mathbf{F}$,  $Sp(2)+1\mathbf{\Lambda}^2+8\mathbf{F}$}{SU(3)3/2 + 9F}}

This theory comes from the circle compactification of the 6d rank-2 E-string theory. This theory can also be described by two 5d gauge theories: $ SU(3)_{3/2} + 9\mathbf{F} $ and $ Sp(2) + 1\mathbf{\Lambda}^2 + 8\mathbf{F} $,
\begin{align}
\begin{tikzpicture}
\draw (0, 0) node {$ SU(3)_{3/2} + 9\mathbf{F} $}
(2, 0) node {$ = $}
(4.5, 0) node {$ Sp(2) + 1\mathbf{\Lambda}^2 + 8\mathbf{F} $}
(7, 0) node {$ = $}
(8.5, 0.3) node {$ \mathfrak{sp}(0)^{(1)} $}
(8.5, -0.3) node {$ 1 $}
(10.5, 0.3) node {$ \mathfrak{su}(1)^{(1)} $}
(10.5, -0.3) node {$ 2 $}
(9, -0.3) -- (10, -0.3)
;
\end{tikzpicture}
\end{align}
The geometric construction for this theory is given by
\begin{align}
\begin{tikzpicture}
\draw[thick](-3,0)--(0,0);	
\node at(-3.6,0) {$\mathbb{F}_{0}{}\big|_1$};
\node at(-2.6,0.3) {${}_{f+e}$};
\node at(0.7,0) {$\mathbb{F}_{4}^{9}{}\big|_2$};
\node at(-0.3,0.3) {${}_{e}$};
\end{tikzpicture}
\end{align}

The validity of the blowup equations for the 6d rank-2 E-string was checked in \cite{Gu:2019pqj}. Here, we will solve the blowup equations from the perspective of two dual 5d gauge theories. First, the full effective prepotential of the $ SU(3) $ theory on the Coulomb branch where $\phi_2\ge \phi_1>0$ is
\begin{align}
\mathcal{E} & = \frac{1}{\epsilon_1 \epsilon_2} \qty(\mathcal{F} - \frac{\epsilon_1^2 + \epsilon_2^2}{48} \qty(4\phi_1 - 14\phi_2 ) + \epsilon_+^2 (\phi_1 + \phi_2) ) \,, \nonumber \\
6\mathcal{F}
&= 8\phi_1^3 - 3\phi_1^2 \phi_2 - 3\phi_1 \phi_2^2 + 8\phi_2^2 + \frac{9}{2} \phi_1 \phi_2 (\phi_1 - \phi_2) + 6m_0 (\phi_1^2 - \phi_1 \phi_2 + \phi_2^2) \nonumber \\
& \quad - \frac{1}{2} \sum_{i=1}^9 \qty((\phi_2 - m_i)^3 + (\phi_2 - \phi_1 + m_i)^3 + (\phi_1 + m_i)^3) \, ,
\end{align}
where $ m_0 $ is the $SU(3)$ gauge coupling and $ m_{i=1, \cdots, 9} $ are mass parameters of the fundamental hypermultiplets. A set of consistent magnetic fluxes
\begin{align}
n_i \in \mathbb{Z} \, , \quad
B_{m_0} = -3/4 \, , \quad
B_{m_i} = 1/2 \quad (1 \leq i \leq 9) \, .
\end{align}
provides a solvable unity blowup equation. By solving the blowup equation, we obtain the BPS spectrum given in Table~\ref{table:SU3_3/2_9F}.

\begin{table}
	\centering
	\begin{tabular}{|c|C{25ex}||c|C{25ex}|} \hline
		$ \mathbf{d} $ & $ \oplus N_{j_l, j_r}^{\mathbf{d}} (j_l, j_r) $ & $ \mathbf{d} $ & $ \oplus N_{j_l, j_r}^{\mathbf{d}} (j_l, j_r) $ \\ \hline
		$ (1, -1, 0) $ & $ (0, 0) $ & $ (1, -1, 1) $ & $ (0, 1) $ \\ \hline
		$ (1, -1, 2) $ & $ (0, 2) $ & $ (1, -\frac{2}{3}, \frac{2}{3}) $ & $ 9(0, \frac{1}{2}) $ \\ \hline
		$ (1, -\frac{2}{3}, \frac{5}{3}) $ & $ 9(0, \frac{3}{2}) $ & $ (1, -\frac{1}{3}, \frac{1}{3}) $ & $ 36(0, 0) $ \\ \hline
		$ (1, -\frac{1}{3}, \frac{4}{3}) $ & $ 36(0, 1) $ & $ (1, 0, 1) $ & $ (0,0) \oplus 84(0,\frac{1}{2}) \oplus (0,1) $ \\ \hline
		$ (1, 0, 2) $ & $ (0,1) \oplus 84(0,\frac{3}{2}) \oplus (0,2) $ & $ (1, \frac{1}{3}, \frac{2}{3}) $ & $ 126(0,0) \oplus 9(0,\frac{1}{2}) $ \\ \hline
		$ (1, \frac{1}{3}, \frac{5}{3}) $ & $ 9(0,\frac{1}{2}) \oplus 126(0,1) \oplus 9(0,\frac{3}{2}) $ & $ (1, \frac{2}{3}, \frac{1}{3}) $ & $ 45(0,0) $ \\ \hline
		$ (1, \frac{2}{3}, \frac{4}{3}) $ & $ 36(0,0) \oplus 126(0,\frac{1}{2}) \oplus 36(0,1) $ & $ (1, 1, 0) $ & $ (0,0) \oplus (0,\frac{1}{2}) $ \\ \hline
		$ (1, 1, 1) $ & $ 85(0,0) \oplus 85(0,\frac{1}{2})\oplus (0,1) $ & $ (1, 1, 2) $ & $ (0,0) \oplus 84(0,\frac{1}{2}) \oplus 85(0,1) \oplus 84(0,\frac{3}{2}) \oplus (0,2) $ \\ \hline
		$ (2, -1, 1) $ & $ 84(0, 1) $ & $ (2, -\frac{2}{3}, \frac{2}{3}) $ & $ 126(0, \frac{1}{2}) $ \\ \hline
		$ (2, -\frac{1}{3}, \frac{1}{3}) $ & $ 126(0, 0) $ & $ (2, 0, 1) $ & $ 85(0, 0) \oplus 1219(0, \frac{1}{2}) \oplus 85(0, 1) \oplus 84(\frac{1}{2}, 1) $ \\ \hline
		$ (2, \frac{1}{3}, \frac{2}{3}) $ & $ 801(0, 0) \oplus 135(0, \frac{1}{2}) \oplus 36(\frac{1}{2}, \frac{1}{2}) $ & $ (2, \frac{2}{3}, \frac{1}{3}) $ & $ \! 162(0, 0) \oplus 9(0, \frac{1}{2}) \oplus 9(\frac{1}{2}, 0) \! $ \\ \hline
		$ (2, 1, 0) $ & $ (0, 0) \oplus (0, 1) \oplus (\frac{1}{2}, \frac{1}{2}) $ & $ (2, 1, 1) $ & $ 1465(0, 0) \oplus 1387(0, \frac{1}{2}) \oplus 168(0, 1) \oplus (0, 2) \oplus 84(\frac{1}{2}, 0) \oplus 83(\frac{1}{2}, \frac{1}{2}) \oplus 84(\frac{1}{2}, 1) \oplus (\frac{1}{2}, \frac{3}{2}) \oplus (1, 1) $ \\ \hline
	\end{tabular}
	\caption{BPS spectrum of the $ SU(3)_{3/2} + 9\mathbf{F} $ theory for $ d_1 = 1 $, $ d_2 \leq 1 $, $ d_3 \leq 2 $ and $ d_1 = 2 $, $ d_2, d_3 \leq 1 $. All flavor mass parameters $ m_{i=1, \cdots, 9} $ are turned off and $ \mathbf{d} = (d_1, d_2, d_3) $ labels the BPS states with charge $ d_1 m_0 + d_2 \alpha_1 + d_3 \alpha_2 $ for the simple roots $ \alpha_1 $ and $ \alpha_2 $ of $ \mathfrak{su}(3) $ algebra.} \label{table:SU3_3/2_9F}
\end{table}

On the other hand, in the $ Sp(2) $ theory, the full effective prepotential in the same chamber is
\begin{align}
\mathcal{E}& = \frac{1}{\epsilon_1 \epsilon_2} \qty(\mathcal{F} - \frac{\epsilon_1^2 + \epsilon_2^2}{48} \qty(4\phi_1 - 14\phi_2) +  \epsilon_+^2 (\phi_1 + \phi_2) ) \,,  \\
6\mathcal{F}
&= 8\phi_1^3 + 12\phi_1^2 \phi_2 - 18\phi_1 \phi_2^2 + 8\phi_2^3 + m_0 (2\phi_1^2 - 2\phi_1 \phi_2 + \phi_2^2)  \nonumber \\
& \quad - \frac{1}{2} \sum_{i=1}^8 \qty((\phi_1 + m_i)^3 + (-\phi_1 + \phi_2 + m_i)^3 + (-\phi_1 + \phi_2 - m_i)^3 + (\phi_1 - m_i)^3) \nonumber \\
& \quad - \frac{1}{2} \Big((\phi_2 + m_9)^3 + (2\phi_1 - \phi_2 + m_9)^3 + (-2\phi_1 + \phi_2 + m_9)^3 + (\phi_2 - m_9)^3\Big) \nonumber \, ,
\end{align}
where $ m_0 $ is the $Sp(2)$ gauge coupling, $ m_{i=1, \cdots, 8} $ are mass parameters of the $Sp(2)$ fundamentals and $ m_9 $ denotes the mass parameter of the antisymmetric hyper.

The K\"ahler parameters in the $Sp(2)$ theory are mapped to those in the $ SU(3) $ theory as
\begin{align}
\phi_1^{Sp} &= \phi_1^{SU} + \frac{1}{2} m_0^{SU} + \frac{1}{4} \sum_{j=1}^8 m_j^{SU} - \frac{1}{4} m_9^{SU} , \
\phi_2^{Sp} = \phi_2^{SU} + m_0^{SU} + \frac{1}{2} \sum_{j=1}^8 m_j^{SU} - \frac{1}{2} m_9^{SU}\ , \nonumber \\
m_0^{Sp} &= \frac{3}{2} m_0^{SU} + \frac{1}{4} \sum_{i=1}^9 m_i^{SU} , \ 
m_i^{Sp} = \frac{1}{2} m_0^{SU} - m_i^{SU} + \frac{1}{4} \sum_{j=1}^8 m_j^{SU} - \frac{1}{4} m_9^{SU} \ (1 \leq i \leq 7)\ ,  \nonumber \\
m_8^{Sp} &= -\frac{1}{2} m_0^{SU} + m_8^{SU} - \frac{1}{4} \sum_{j=1}^8 m_j^{SU} + \frac{1}{4} m_9^{SU} , \ \
m_9^{Sp} = m_0^{SU} + \frac{1}{2} \sum_{j=1}^9 m^{SU}_j \, .
\end{align}

In the $Sp(2)$ frame, we find a set of consistent magnetic fluxes
\begin{align}
n_i \in \mathbb{Z} \, , \quad
B_{m_0} = 0 \, , \quad
B_{m_i} = 1/2 \quad  (1 \leq i \leq 9) \, .
\end{align}
This leads to a solvable unity blowup equation. We have checked that the solution to the blowup equation agrees with the BPS spectrum of the dual $SU(3)$ gauge theory given in Table~\ref{table:SU3_3/2_9F}.

The instanton partition function of the $ Sp(2) + 1\mathbf{\Lambda}^2 + 8\mathbf{F} $ theory can also be calculated using the ADHM construction introduced in \cite{Aharony:1997pm, Kim:2012gu, Hwang:2014uwa}. We checked that our result from the blowup formula is in perfect agreement with the ADHM result in the K\"ahler parameter expansion.

\subsubsection{\texorpdfstring{$SU(3)_0+10\mathbf{F}$, $Sp(2)+10\mathbf{F}$}{SU(3)0 + 10F}}

The 5d $ SU(3)_0 + 10\mathbf{F} $ theory and the $ Sp(2) + 10\mathbf{F} $ theory are two dual 5d descriptions for the KK theory arising from the circle compactification of the 6d $ Sp(1) $ gauge theory with 10 fundamental hypermultiplets,
\begin{align}
\begin{tikzpicture}
\draw (0, 0) node {$ SU(3)_{0} + 10\mathbf{F} $}
(2, 0) node {$ = $}
(4, 0) node {$ Sp(2) + 10\mathbf{F} $}
(6, 0) node {$ = $}
(7.7, 0.3) node {$ \mathfrak{sp}(1)^{(1)} $}
(7.7, -0.3) node {$ 1 $}
;
\end{tikzpicture}
\end{align}
This theory is geometrically engineered by gluing $ \mathbb{F}_6^{10} $ and $ \mathbb{F}_0 $ \cite{Jefferson:2018irk}.
\begin{align}
\begin{tikzpicture}
\draw[thick](-3,0)--(0,0);	
\node at(-3.6,0) {$\mathbb{F}_{0}{}\big|_1$};
\node at(-2.6,0.3) {${}_{e+2f}$};
\node at(0.7,0) {$\mathbb{F}_{6}^{10}{}\big|_2$};
\node at(-0.2,0.3) {${}_{e}$};
\end{tikzpicture}
\end{align}
The blowup equation for this theory from the 6d gauge theory perspective was checked to be satisfied by substituting the known one-string elliptic genus into the blowup equation  in \cite{Gu:2020fem}.

We will now compute the BPS spectrum of this theory using the blowup equations in the 5d gauge theory descriptions. First, the full prepotential of the $SU(3)$ gauge theory on the chamber $\phi_2\ge \phi_1>0$ is
\begin{align}
\mathcal{E} &= \frac{1}{\epsilon_1 \epsilon_2} \qty(\mathcal{F} - \frac{\epsilon_1^2 + \epsilon_2^2}{48} \qty(4\phi_1 - 16\phi_2 ) + \epsilon_+^2 (\phi_1 + \phi_2) ) \,, \nonumber \\
6\mathcal{F}
&= 8\phi_1^3 - 3\phi_1^2 \phi_2 - 3\phi_1 \phi_2^2 + 8\phi_2^2 + 6m_0 (\phi_1^2 - \phi_1 \phi_2 + \phi_2^2) \nonumber \\
& \quad - \frac{1}{2} \sum_{i=1}^{10} \qty((\phi_2 - m_i)^3 + (\phi_2 - \phi_1 + m_i)^3 + (\phi_1 + m_i)^3) \, .
\end{align}
We find a set of consistent magnetic fluxes
\begin{align}
n_i \in \mathbb{Z} \, , \ \
B_{m_0} = 0 \, , \ \
B_{m_i} = 1/2 \ \ (1 \leq i \leq 5) \, , \ \ 
B_{m_i} = -1/2 \ \ (6 \leq i \leq 10) \ ,
\end{align}
which provides a solvable unity blowup equation.

In the $ Sp(2) $ gauge theory, on the other hand, the full effective prepotential on the same chamber takes the form of
\begin{align}
\mathcal{E}
&= \frac{1}{\epsilon_1 \epsilon_2} \qty(\mathcal{F} - \frac{\epsilon_1^2 + \epsilon_2^2}{48} (4\phi_1 - 16\phi_2) + \epsilon_+^2 (\phi_1 + \phi_2)) \,, \nonumber \\
6\mathcal{F}
&= 8\phi_1^3 + 12\phi_1^2 \phi_2 - 18\phi_1 \phi_2^2 + 8\phi_2^3 + m_0(2\phi_1^2 - 2\phi_1 \phi_2 + \phi_2^2)  \nonumber\\
& \quad - \frac{1}{2}\sum_{i=1}^{10} \qty((\phi_1 \pm m_i)^3 + (-\phi_1 + \phi_2 \pm m_i)^3 )\,,  
\end{align}
where $ m_0 $ is the $ Sp(2) $ gauge coupling and $ m_{i=1, \cdots, 10} $ are mass parameters of the fundamental hypers. 
One possible choice of consistent magnetic fluxes in the $Sp(2)$ theory is
\begin{align}
n_i \in \mathbb{Z} \, , \quad
B_{m_0} = 0 \, , \quad
B_{m_i} = 1/2 \quad (1 \leq i \leq 10) \ .
\end{align}
The K\"ahler parameters in two 5d theories are mapped to each other by the relations \cite{Hayashi:2016abm}
\begin{align}
\phi_1^{Sp} &= \phi_1^{SU} + \frac{1}{2}m_0^{SU} + \frac{1}{4}\sum_{i=1}^{10} m_i^{SU} \,, \ \
\phi_2^{Sp} = \phi_2^{SU} + m_0^{SU} + \frac{1}{2}\sum_{i=1}^{10} m_i^{SU} \,, \ \ m_0^{Sp} = m_0^{SU} \,, \nonumber \\
m_i^{Sp} &= \frac{1}{2} m_0^{SU} - m_j^{SU} + \frac{1}{4}\sum_{j=1}^{10} m_j^{SU} \quad (1 \leq i \leq 9) \,, \ \
m_{10}^{Sp} = -\frac{1}{2} m_0^{SU} - \frac{1}{4}\sum_{j=1}^{9} m_j^{SU} + \frac{3}{4} m_{10}^{SU} .
\end{align} 

We list in Table~\ref{table:sp2_10F} some leading BPS states computed by solving the blowup equation in the $ Sp(2) + 10\mathbf{F} $ theory.
We checked that this solution agrees with the result from the  $ SU(3)_0 + 10\mathbf{F} $ description and also agree with the ADHM result for the $ Sp(2) + 10\mathbf{F} $ theory computed in \cite{Yun:2016yzw} as well as topological vertex \cite{Hayashi:2016abm} in series expansion of K\"ahler parameters.
\begin{table}
	\centering
	\begin{tabular}{|c|C{25ex}||c|C{25ex}|} \hline
		$ \mathbf{d} $ & $ \oplus N_{j_l, j_r}^{\mathbf{d}} (j_l, j_r) $ & $ \mathbf{d} $ & $ \oplus N_{j_l, j_r}^{\mathbf{d}} (j_l, j_r) $ \\ \hline
		$ (1, 1, 1) $ & $ 512(0, 0) $ & $ (1, 1, \frac{3}{2}) $ & $ 512(0, \frac{1}{2}) $ \\ \hline
		$ (1, 1, 2) $ & $ 512(0, 1) $ & $ (1, 2, \frac{3}{2}) $ & $ 512(0, \frac{1}{2}) $ \\ \hline
		$ (1, 2, 2) $ & $ 512(0, 0) \oplus 512(0, 1) $ & $ (2, 0, -1) $ & $ (0, \frac{1}{2}) $ \\ \hline
		$ (2, 0, -\frac{1}{2}) $ & $ 20(0, 0) $ & $ (2, 0, \frac{1}{2}) $ & $ 20(0, 0) $ \\ \hline
		$ (2, 0, 1) $ & $ (0, \frac{1}{2}) $ & $ (2, 1, -1) $ & $ (0, \frac{3}{2}) $ \\ \hline
		$ (2, 1, -\frac{1}{2}) $ & $ 20(0, 1) $ & $ (2, 1, 0) $ & $ 192(0, \frac{1}{2}) \oplus (0, \frac{3}{2}) \oplus (\frac{1}{2}, 0) \oplus (\frac{1}{2}, 1) $ \\ \hline
		$ (2, 1, \frac{1}{2}) $ & $ 1180(0, 0) \oplus 20(0, 1) \oplus 20(\frac{1}{2}, \frac{1}{2}) $ & $ (2, 1, 1) $ & $ 192(0, \frac{1}{2}) \oplus (0, \frac{3}{2}) \oplus (\frac{1}{2}, 0) \oplus (\frac{1}{2}, 1) $ \\ \hline
		$ (2, 2, -1) $ & $ (0, \frac{5}{2}) $ & $ (2, 2, -\frac{1}{2}) $ & $ 20(0, 2) $ \\ \hline
		$ (2, 2, 0) $ & $ (0, \frac{1}{2}) \oplus 192(0, \frac{3}{2}) \oplus (0, \frac{5}{2}) \oplus (\frac{1}{2}, 1) \oplus (\frac{1}{2}, 2) $ & $ (2, 1, \frac{1}{2}) $ & $ 20(0, 0) \oplus 1180(0, 1) \oplus 20(0, 2) \oplus 20(\frac{1}{2}, \frac{1}{2}) \oplus 20(\frac{1}{2}, \frac{3}{2}) $ \\ \hline
		$ (2, 2, 1) $ & \multicolumn{3}{C{65ex}|}{$ 5230(0, \frac{1}{2}) \oplus 192(0, \frac{3}{2}) \oplus (0, \frac{5}{2}) \oplus 192(\frac{1}{2}, 0) \oplus 193(\frac{1}{2}, 1) \oplus (\frac{1}{2}, 2) \oplus (1, \frac{1}{2}) \oplus (1, \frac{3}{2}) $} \\ \hline
	\end{tabular}
	\caption{BPS spectrum of the $ Sp(2) + 10\mathbf{F} $ theory for $ d_1 = 1 $, $ d_2, d_3 \leq 2 $ and $ d_1 = 2 $, $ d_2 \leq 2 $, $ d_3 \leq 1 $. Here, $ \mathbf{d} = (d_1, d_2, d_3) $ labels BPS states with charge $ d_1 m_0 + d_2 \alpha_1 + d_3 \alpha_2 $, where $ \alpha_1 = 2\phi_1 - \phi_2 $, $ \alpha_2 = -2\phi_1 + 2\phi_2 $ are the simple roots of $ \mathfrak{sp}(2) $ algebra, and 10 flavor charges are blindly summed over.} \label{table:sp2_10F}
\end{table}

\subsubsection{\texorpdfstring{$SU(2)\times SU(2)+2\,\mathbf{bi\text{-}F}$}{SU(2)*SU(2) + 2bifund}}

We will now give an example for 5d quiver gauge theories. The 5d theory with $SU(2)\times SU(2)$ gauge group and 2 bi-fundamental hypermultiplets ($\mathbf{bi\text{-}F}$) is the KK theory obtained from a circle compactification of the 6d $SU(2)$ gauge theory with 4 fundamental hypermultiplets.

The full effective prepotential of the $ SU(2) \times SU(2) $ gauge theory on the Coulomb branch where $\phi_2\ge\phi_1>0$ is given by
\begin{align}
\mathcal{E}
&= \frac{1}{\epsilon_1 \epsilon_2} \qty(\mathcal{F} - \frac{\epsilon_1^2 + \epsilon_2^2}{12} \qty(\phi_1 - \phi_2) + \epsilon_+^2 (\phi_1 + \phi_2) ) \,,  \\
6\mathcal{F}
&= 8\phi_1^3 + 8\phi_2^3 + 6m_1 \phi_1^2 + 6m_2 \phi_2^2 - \frac{1}{2}\sum_{i=3}^4\Big((\phi_1 + \phi_2 \pm m_i)^3 + (-\phi_1 + \phi_2 \pm m_i)^3 \Big) \ , \nonumber
\end{align}
where $ m_1 $ and $ m_2 $ are the gauge couplings for two $ SU(2) $ gauge groups respectively, while $ m_3$ and $m_4 $ are the mass parameters of two bi-fundamentals. 

A unity blowup equation can be formulated with a set of the consistent magnetic fluxes given by
\begin{align}
n_i \in \mathbb{Z} \ , \quad
B_{m_1} = 0 \ , \quad
B_{m_2} = 0 \ , \quad
B_{m_{3,4}} = 1/2 \ .
\end{align}
The solution to the blowup equation is given in Table~\ref{table:SU2_bifund}. We also computed the partition function of this theory using the ADHM method and confirmed that the result agrees with the BPS spectrum in Table~\ref{table:SU2_bifund}. 

\begin{table}
	\centering
	\begin{tabular}{|c|C{21ex}||c|C{21ex}|} \hline
		$ \mathbf{d} $ & $ \oplus N_{j_l, j_r}^{\mathbf{d}} (j_l, j_r) $ & $ \mathbf{d} $ & $ \oplus N_{j_l, j_r}^{\mathbf{d}} (j_l, j_r) $ \\ \hline
		$ (1, 0, 1, -1, -1) $ & $ 2(0, 0) $ & $ (1, 0, 1, 1, -1) $ & $ 2(0, 0) $ \\ \hline
		$ (1, 0, 2, -2, 0) $ & $ (0, \frac{1}{2}) $ & $ (1, 0, 2, 0, -2) $ & $ (0, \frac{1}{2}) $ \\ \hline
		$ (1, 0, 2, 0, 0) $ & $ 4(0, \frac{1}{2}) $ & $ (1, 0, 2, 2, 0) $ & $ (0, \frac{1}{2}) $ \\ \hline
		$ (1, 1, -1, 1, -1) $ & $ 2(0, 0) $ & $ (1, 1, 0, 2, -2) $ & $ 4(0, \frac{1}{2}) $ \\ \hline
		$ (1, 1, 0, 2, 0) $ & $ 10(0, \frac{1}{2}) \oplus (0, \frac{3}{2}) \oplus (\frac{1}{2}, 0) \oplus (\frac{1}{2}, 1) $ & $ (1, 1, 1, -1, -1) $ & $ 2(0, 0) $ \\ \hline
		$ (1, 1, 1, 1, -3) $ & $ 2(0, 0) $ & $ (1, 1, 1, 1, -1) $ & $ 14(0, 0) \oplus 4(0, 1) \oplus 2(\frac{1}{2}, \frac{1}{2}) $ \\ \hline
		$ (1, 1, 2, 0, -2) $ & $ 4(0, \frac{1}{2}) $ & $ (1, 1, 2, 0, 0) $ & $ 10(0, \frac{1}{2}) \oplus (0, \frac{3}{2}) \oplus (\frac{1}{2}, 0) \oplus (\frac{1}{2}, 1) $ \\ \hline
		$ (1, 1, 2, 2, -2) $ & $ 10(0, \frac{1}{2}) \oplus 5(0, \frac{3}{2}) \oplus (\frac{1}{2}, 0) \oplus (\frac{1}{2}, 1) $ & $ (1, 1, 2, 2, 0) $ & $ 25(0, \frac{1}{2}) \oplus 14(0, \frac{3}{2}) \oplus (0, \frac{5}{2}) \oplus 4(\frac{1}{2}, 0) \oplus 5(\frac{1}{2}, 1) \oplus (\frac{1}{2}, 2) $ \\ \hline
	\end{tabular}
	\caption{BPS spectrum of the $ SU(2) \times SU(2) $ theory with $ 2 $ bifundamentals. Here, $ \mathbf{d}=(d_1, d_2, d_3, d_4, d_5) $ labels the BPS states with charge $ d_1 m_1 + d_2 m_2 + d_3 \phi_1 + d_4 \phi_2 + d_5 m_i $ where $d_5$ counts collective degrees of two bi-fundamental matters with masses $m_{i=3,4}$. There is a symmetry exchanging $ (d_1, d_2, d_3, d_4, d_5) \leftrightarrow (d_2, d_1, d_4, d_3, d_5) $ and $ d_5 \leftrightarrow -d_5 $, so we list BPS states up to $ (d_3, d_4) \leq (2, 2) $ and $ d_5 \leq 0 $ for $ (1, 0) $ and $ (1, 1) $ instantons.} \label{table:SU2_bifund}
\end{table}

\subsubsection{\texorpdfstring{$ SU(3)_0 + 1\mathbf{Adj} $}{SU(3)0 + 1Adj}}

The 5d $ SU(3)_0 $ gauge theory with one adjoint hypermultiplet is equivalent to a circle reduction of the 6d $ \mathcal{N}=(2, 0) $ $ A_2$ theory \cite{Bhardwaj:2019fzv},
\begin{align}
\begin{tikzpicture}
\draw (0, 0) node {$ SU(3)_0 + 1\mathbf{Adj} $}
(2.3, 0) node {$ = $}
(4, 0.3) node {$ \mathfrak{su}(2)^{(1)} $}
(4, -0.3) node {$ 2 $}
(6, 0.3) node {$ \mathfrak{su}(2)^{(1)} $}
(6, -0.3) node {$ 2 $}
(4.3, -0.3) -- (5.7, -0.3);
\end{tikzpicture}
\end{align}
The cubic prepotential of this theory on the chamber where $2\phi_2>\phi_1 > \frac{1}{2}\phi_2$ is
\begin{align}
6\mathcal{F}
&= 8\phi_1^3 - 3\phi_1^2 \phi_2 - 3\phi_1 \phi_2^2 + 8\phi_2^3 + 6m_0 (\phi_1^2 - \phi_1 \phi_2 + \phi_2^2) \nonumber \\
& \quad - \frac{1}{2}\Big((\phi_1 + \phi_2 \pm m_1)^3 + (-\phi_1 + 2\phi_2 \pm m_1)^3 + (2\phi_1 - \phi_2 \pm m_1)^3 \Big),
\end{align}
where $ m_0 $ is the gauge coupling and $ m_1 $ is the mass parameter of the adjoint matter. The effective prepotential is
\begin{align}
\mathcal{E}
= \frac{1}{\epsilon_1 \epsilon_2} \qty(\mathcal{F} + \epsilon_+^2 (\phi_1 + \phi_2)) \ .
\end{align}

We find that a set of consistent magnetic fluxes
\begin{align}
n_i \in \mathbb{Z} \ , \quad
B_{m_0} = 0 \ , \quad
B_{m_1} = 1/2 \ 
\end{align}
gives rise to a solvable unity blowup equation.
The BPS spectrum obtained by solving the blowup equation is listed in Table~\ref{table:su3_adj}. We also computed the instanton partition function for $ SU(3)_0 + 1\mathbf{Adj}$ independently using the ADHM construction and confirmed that our result agrees with the result from the ADHM method.

\begin{table}
	\centering
	\begin{tabular}{|c|C{25ex}||c|C{25ex}|} \hline
		$ \mathbf{d} $ & $ \oplus N_{j_l, j_r}^{\mathbf{d}} (j_l, j_r) $ & $ \mathbf{d} $ & $ \oplus N_{j_l, j_r}^{\mathbf{d}} (j_l, j_r) $ \\ \hline
		$ (1, 0, 1, -2) $ & $ (0, \frac{1}{2}) $ & $ (1, 0, 1, -1) $ & $ (0, 0) \oplus (0, 1) \oplus (\frac{1}{2}, \frac{1}{2}) $ \\ \hline
		$ (1, 0, 1, 0) $ & $ 2(0, \frac{1}{2}) \oplus (\frac{1}{2}, 0) \oplus (\frac{1}{2}, 1) $ & $ (1, 0, 2, -2) $ & $ (0, \frac{3}{2}) $ \\ \hline
		$ (1, 0, 2, -1) $ & $ (0, 1) \oplus (0, 2) \oplus (\frac{1}{2}, \frac{3}{2}) $ & $ (1, 0, 2, 0) $ & $ 2(0, \frac{3}{2}) \oplus (\frac{1}{2}, 1) \oplus (\frac{1}{2}, 2) $ \\ \hline
		$ (1, 1, 1, -3) $ & $ (0, 0) $ & $ (1, 1, 1, -2) $ & $ 3(0, \frac{1}{2}) \oplus (\frac{1}{2}, 0) $ \\ \hline
		$ (1, 1, 1, -1) $ & $ 5(0, 0) \oplus 2(0, 1) \oplus 3(\frac{1}{2}, \frac{1}{2}) $ & $ (1, 1, 1, 0) $ & $ 6(0, \frac{1}{2}) \oplus 4(\frac{1}{2}, 0) \oplus 2(\frac{1}{2}, 1) $ \\ \hline
		$ (1, 1, 2, -3) $ & $ (0, 1) $ & $ (1, 1, 2, -2) $ & $ 2(0, \frac{1}{2}) \oplus 2(0, \frac{3}{2}) \oplus (\frac{1}{2}, 1) $ \\ \hline
		$ (1, 1, 2, -1) $ & $ (0,0) \oplus 5(0,1) \oplus (0,2) \oplus 2(\frac{1}{2},\frac{1}{2}) \oplus 2(\frac{1}{2},\frac{3}{2}) $ & $ (1, 1, 2, 0) $ & $ 4(0,\frac{1}{2}) \oplus 4(0,\frac{3}{2}) \oplus (\frac{1}{2},0) \oplus 4(\frac{1}{2},1) \oplus (\frac{1}{2},2) $ \\ \hline
		$ (1, 2, 2, -3) $ & $ (0, 0) \oplus (0, 1) $ & $ (1, 2, 2, -2) $ & $ 4(0,\frac{1}{2}) \oplus 3(0,\frac{3}{2}) \oplus (\frac{1}{2},0) \oplus (\frac{1}{2},1) $ \\ \hline
		$ (1, 2, 2, -1) $ & $ 5(0,0) \oplus 7(0,1) \oplus 2(0,2) \oplus 4(\frac{1}{2},\frac{1}{2}) \oplus 3(\frac{1}{2},\frac{3}{2}) $ & $ (1, 2, 2, 0) $ & $ 8(0,\frac{1}{2}) \oplus 6(0,\frac{3}{2}) \oplus 4(\frac{1}{2},0) \oplus 6(\frac{1}{2},1) \oplus 2(\frac{1}{2},2) $ \\ \hline
		$ (2, 0, 1, -3) $ & $ (0, 0) $ & $ (2, 0, 1, -2) $ & $ 2(0, \frac{1}{2}) \oplus (\frac{1}{2}, 0) \oplus (\frac{1}{2}, 1) $ \\ \hline
		$ (2, 0, 1, -1) $ & $ 4(0, 0) \oplus 2(0, 1) \oplus 3(\frac{1}{2}, \frac{1}{2}) \oplus (\frac{1}{2}, \frac{3}{2}) \oplus (1, 1) $ & $ (2, 0, 1, 0) $ & $ 5(0,\frac{1}{2}) \oplus (0,\frac{3}{2}) \oplus 3(\frac{1}{2},0) \oplus 3(\frac{1}{2},1) \oplus (1,\frac{1}{2}) \oplus (1,\frac{3}{2}) $ \\ \hline
		$ (2, 1, 1, -3) $ & $ 5(0,0) \oplus 2(0,1) \oplus 3(\frac{1}{2},\frac{1}{2}) $ & $ (2, 1, 1, -2) $ & $ 14(0,\frac{1}{2}) \oplus (0,\frac{3}{2}) \oplus 9(\frac{1}{2},0) \oplus 7(\frac{1}{2},1) \oplus 2(1,\frac{1}{2}) $ \\ \hline
		$ (2, 1, 1, -1) $ & $ 22(0,0) \oplus 14(0,1) \oplus 22(\frac{1}{2},\frac{1}{2}) \oplus 4(\frac{1}{2},\frac{3}{2}) \oplus 4(1,0) \oplus 5(1,1) $ & $ (2, 1, 1, 0) $ & $ 31(0,\frac{1}{2}) \oplus 5(0,\frac{3}{2}) \oplus 21(\frac{1}{2},0) \oplus 17(\frac{1}{2},1) \oplus 9(1,\frac{1}{2}) \oplus 3(1,\frac{3}{2}) $ \\ \hline
	\end{tabular}
	\caption{BPS spectrum of $ SU(3)_0 + 1\mathbf{Adj} $ theory for $ d_1 = 1 $, $ d_2, d_3 \leq 2 $ and $ d_1 = 2 $, $ d_2, d_3 \leq 1 $. Here $ \mathbf{d} = (d_1, d_2, d_3, d_4) $ labels the BPS states with charge $ d_1 m_0 + d_2 \alpha_1 + d_3 \alpha_2 + d_4 m_1 $, where $ \alpha_1 $ and $ \alpha_2 $ are simple roots of $ \mathfrak{su}(3) $ algebra. The states related by the symmetries $ d_2 \leftrightarrow d_3 $ and $ d_4 \leftrightarrow -d_4 $ are omitted.} \label{table:su3_adj}
\end{table}

\subsubsection{\texorpdfstring{$Sp(2)_0+1\mathbf{Adj}$}{Sp(2)0 + 1Adj}}

There are two $Sp(2)_\theta$ gauge theories with an adjoint hypermultiplet in 5d distinguished by two distinct theta angles $\theta=0$ or $\theta=\pi$. In this subsection, we discuss the case with $\theta= 0$.

The 5d $ Sp(2)_0 + 1\mathbf{Adj} $ theory is the KK-theory obtained by the $ \mathbb{Z}_2 $ twisted compactification of the 6d $ \mathcal{N} = (2, 0) $ $A_3$ theory \cite{Tachikawa:2011ch},
\begin{align}
\begin{tikzpicture}
\draw (0, 0) node {$ Sp(2)_0 + 1\mathbf{Adj} $}
(2.3, 0) node {$ = $}
(4, 0.3) node {$ \mathfrak{su}(1)^{(1)} $}
(4, -0.3) node {$ 2 $}
(6, 0.3) node {$ \mathfrak{su}(1)^{(1)} $}
(6, -0.3) node {$ 2 $}
(4.3, -0.3) -- (4.7, -0.3)
(5, -0.3) node {$ _2 $}
;
\draw [->] (5.25, -0.3) -- (5.7, -0.3);
\end{tikzpicture}
\end{align}
This is a non-geometric theory. This theory on the other hand has a 5-brane web realization with an $O7^+$ plane (or frozen singularity) which will be discussed in Appendix~\ref{sec:appendix2}. 

The BPS spectrum of this theory can be obtained by solving the blowup equation as follows. The full effective prepotential of this theory on the chamber where $2\phi_1>\phi_2>\phi_1>0$ is
\begin{align}\label{eq:Sp2_0_adj_F}
\mathcal{E} & = \frac{1}{\epsilon_1 \epsilon_2} \qty(\mathcal{F} - \frac{\epsilon_1^2 + \epsilon_2^2}{48} \qty(4\phi_1 - 2\phi_2) + \epsilon_+^2 (\phi_1 + \phi_2) ) \,, \nonumber \\
6\mathcal{F}
&= 8\phi_1^3 + 12\phi_1^2 \phi_2 - 18\phi_1 \phi_2^2 + 8\phi_2^3 + m_0 (2\phi_1^2 - 2\phi_1 \phi_2 + \phi_2^2) \\
& \quad -\frac{1}{2} \qty((2\phi_1 \pm m_1)^3 + (\phi_2 \pm m_1)^3 + (-2\phi_1 + 2\phi_2 \pm m_1)^3 + (\pm(2\phi_1 - \phi_2) + m_1)^3), \nonumber
\end{align}
where $ m_0 $ is the gauge coupling and $ m_1 $ is the adjoint mass parameter. One can choose the consistent magnetic fluxes as
\begin{align} \label{eq:Sp2_0_adj_shift}
n_i \in \mathbb{Z} \ , \quad
B_{m_0} = 0 \ , \quad
B_{m_1} = 1/2 \ ,
\end{align}
and formulate a unity blowup equation. The solution to the blowup equation is given in Table~\ref{table:Sp2_0_1Adj}. We checked that this result matches the instanton partition function of the 5d $\mathcal{N}=2$ $Sp(2)_0$ gauge theory computed in \cite{Nekrasov:2004vw,Shadchin:2004yx} using the ADHM construction.
\begin{table}
	\centering
	\begin{tabular}{|c|C{25ex}||c|C{25ex}|} \hline
		$ \mathbf{d} $ & $ \oplus N_{j_l, j_r}^{\mathbf{d}} (j_l, j_r) $ & $ \mathbf{d} $ & $ \oplus N_{j_l, j_r}^{\mathbf{d}} (j_l, j_r) $ \\ \hline
		$ (1, 0, 1, -2) $ & $ (0, \frac{1}{2}) $ & $ (1, 0, 1, -1) $ & $ (0,0) \oplus (0,1) \oplus (\frac{1}{2},\frac{1}{2}) $ \\ \hline
		$ (1, 0, 1, 0) $ & $ 2(0,\frac{1}{2}) \oplus (\frac{1}{2},0) \oplus	(\frac{1}{2},1) $ & $ (1, 0, 2, -2) $ & $ (0, \frac{3}{2}) $ \\ \hline
		$ (1, 0, 2, -1) $ & $ (0,1) \oplus (0,2) \oplus (\frac{1}{2},\frac{3}{2}) $ & $ (1, 0, 2, 0) $ & $ 2(0,\frac{3}{2}) \oplus (\frac{1}{2},1) \oplus (\frac{1}{2},2) $ \\ \hline
		$ (1, 1, 1, -3) $ & $ (0, 0) $ & $ (1, 1, 1, -2) $ & $ 2(0,\frac{1}{2}) \oplus (\frac{1}{2},0) $ \\ \hline
		$ (1, 1, 1, -1) $ & $ 4(0,0) \oplus (0,1) \oplus 2(\frac{1}{2},\frac{1}{2}) $ & $ (1, 1, 1, 0) $ & $ 4(0,\frac{1}{2}) \oplus 3(\frac{1}{2},0) \oplus	(\frac{1}{2},1) $ \\ \hline
		$ (2, 0, 1, -2) $ & $ 2(0,\frac{1}{2}) \oplus (\frac{1}{2},0) \oplus (\frac{1}{2},1) $ & $ (2, 0, 1, -1) $ & $ 4(0,0) \oplus 2(0,1) \oplus 3(\frac{1}{2},\frac{1}{2}) \oplus (\frac{1}{2},\frac{3}{2}) \oplus (1,1) $ \\ \hline
		$ (2, 0, 1, 0) $ & $ 5(0,\frac{1}{2}) \oplus (0,\frac{3}{2}) \oplus	3(\frac{1}{2},0) \oplus 3(\frac{1}{2},1) \oplus	(1,\frac{1}{2}) \oplus (1,\frac{3}{2}) $ & $ (2, 1, 0, -2) $ & $ (0, \frac{1}{2}) $ \\ \hline
		$ (2, 1, 0, -1) $ & $ (0,0) \oplus (0,1) \oplus (\frac{1}{2},\frac{1}{2}) $ & $ (2, 1, 0, 0) $ & $ 2(0,\frac{1}{2}) \oplus (\frac{1}{2},0) \oplus (\frac{1}{2},1) $ \\ \hline
		$ (2, 1, 1, -3) $ & $ 2(0,0) \oplus (\frac{1}{2},\frac{1}{2}) $ & $ (2, 1, 1, -2) $ & $ 6(0,\frac{1}{2}) \oplus 4(\frac{1}{2},0) \oplus	2(\frac{1}{2},1) \oplus (1,\frac{1}{2}) $ \\ \hline
		$ (2, 1, 1, -1) $ & $ 11(0,0) \oplus 5(0,1) \oplus 10(\frac{1}{2},\frac{1}{2}) \oplus (\frac{1}{2},\frac{3}{2}) \oplus 2(1,0) \oplus 2(1,1) $ & $ (2, 1, 1, 0) $ & $ 14(0,\frac{1}{2}) \oplus (0,\frac{3}{2}) \oplus 11(\frac{1}{2},0) \oplus 7(\frac{1}{2},1) \oplus 4(1,\frac{1}{2}) \oplus (1,\frac{3}{2}) $ \\ \hline
	\end{tabular}
	\caption{BPS spectrum of  the $ Sp(2)_{0} + 1\mathbf{Adj} $ theory for $ d_1 = 1, 2 $ and $ d_2 \leq 1 $, $ d_3 \leq 1 $. Here, $ \mathbf{d} = (d_1, d_2, d_3, d_4) $ labels the BPS states with charge $ d_1 m_0 + d_2 \alpha_1 + d_3 \alpha_2 + d_4 m_1 $ where $ \alpha_1 = 2\phi_1 - \phi_2 $, $ \alpha_2 = -2\phi_1 + 2\phi_2 $ are simple roots of $ \mathfrak{sp}(2) $ algebra. The theory has a symmetry exchanging $ d_4 \leftrightarrow -d_4 $ which provides BPS states with flipped charge $d_4\rightarrow -d_4$.} \label{table:Sp2_0_1Adj}
\end{table}

\subsubsection{\texorpdfstring{$SU(3)_{\frac{3}{2}}+1\mathbf{Sym}$, $Sp(2)_\pi+1\mathbf{Adj}$}{SU(3)3/2 + 1Sym}}

The 5d $ Sp(2)_\pi + 1\mathbf{Adj} $ theory is the KK-theory obtained by $ \mathbb{Z}_2 $ twisted compactification of the 6d $ \mathcal{N}=(2, 0) $ $ A_4 $ theory. This theory is also dual to the $ SU(3)_{3/2} + 1\mathbf{Sym} $ theory \cite{Jefferson:2017ahm},
\begin{align}
\begin{tikzpicture}
\draw (0, 0) node {$ SU(3)_{3/2} + 1\mathbf{Sym} $}
(2.1, 0) node {$ = $}
(4, 0) node {$ Sp(2)_\pi + 1\mathbf{Adj} $}
(6, 0) node {$ = $}
(7.3, 0.3) node {$ \mathfrak{su}(1)^{(1)} $}
(7.3, -0.3) node {$ 2 $}
(9.3, 0.3) node {$ \mathfrak{su}(1)^{(1)} $}
(9.3, -0.3) node {$ 2 $}
(7.6, -0.3) -- (9.0, -0.3);
\draw (7.0, -0.5) .. controls (6.6, -1.2) and (8.0, -1.2) .. (7.6, -0.5);
\end{tikzpicture}
\end{align}
This theory has no geometric construction, but it can be realized by a 5-brane web with an $O7^+$ plane as discussed in \cite{Hayashi:2018lyv}. See also Appendix~\ref{sec:appendix2} for more details. 

In the $ SU(3) $ description, the effective prepotential on the Coulomb branch where $\phi_2\ge \phi_1>0$ is given by
\begin{align}
\mathcal{E} & = \frac{1}{\epsilon_1 \epsilon_2} \qty(\mathcal{F} - \frac{\epsilon_1^2 + \epsilon_2^2}{48} \qty(4\phi_1 - 2\phi_2) + \epsilon_+^2 (\phi_1 + \phi_2) ) \,, \nonumber \\
6\mathcal{F}
&= 8\phi_1^3 - 3\phi_1^2 \phi_2 - 3\phi_1 \phi_2^2 + 8\phi_2^2 + \frac{9}{2} \phi_1 \phi_2 (\phi_1 - \phi_2) + m_0 (\phi_1^2 - \phi_1 \phi_2 + \phi_2^2) \nonumber \\
& \quad - \frac{1}{2} \big( (2\phi_1 + m_1)^3 + (\phi_2 + m_1)^3 + (-2\phi_1 + 2\phi_2 + m_1)^3 \nonumber \\
& \qquad \qquad + (-\phi_1 + \phi_2 - m_1)^3 + (\phi_1 - m_1)^3 + (2\phi_2 - m_1)^3 \big) \ ,
\end{align}
where $ m_0 $ is the gauge coupling and $ m_1 $ is the mass parameter of the symmetric matter. The following magnetic fluxes give a solvable unity blowup equation for this theory, 
\begin{align}\label{eq:su3_sym_flux}
n_i \in \mathbb{Z} \, , \quad
B_{m_0} = -1/4 \, , \quad
B_{m_1} = 1/2 \ .
\end{align}

On the other hand, the blowup equation from the $Sp(2)$ theory perspective is rather subtle. The effective prepotentials and the 1-loop GV-invariant of $ Sp(2)_\pi + 1\mathbf{Adj} $ theory are the same as those of the $ \theta = 0 $ case. Therefore the form of the blowup equation for the $Sp(2)_\theta$ theory at $\theta=\pi$ is expected to be determined by another choice of magnetic fluxes different from the fluxes used for the theory at $\theta=0$ in the previous subsection. Indeed, there exists another set of consistent magnetic fluxes given by
\begin{align}\label{eq:Sp2-adj-pi}
n_1 \in \mathbb{Z} + 1/2 \, , \quad
n_2 \in \mathbb{Z} \, , \quad
B_{m_0} = 0 \, , \quad
B_{m_1} = 1/2 \ .
\end{align}

Rather surprisingly, we notice that the blowup equation coming from this set of fluxes has two distinct solutions. While solving the blowup equation, we always find two independent solutions at each order in the expansion. For example, $ N_{(0, 0)}^{(1, 1, 1, -3)} $ at 1-instanton order has two solutions, either $1$ or $0$. Interestingly, it turns out that these two solutions of the single blowup equation correspond to the $\theta=0$ and $\theta=\pi$ cases respectively. Indeed, we checked up to 3-instantons that the instanton partition functions of the $\mathcal{N}=2\ Sp(2)_\theta$ gauge theories both at $\theta=0,\pi$ computed using their ADHM constructions satisfy the same blowup equation formulated with the fluxes in \eqref{eq:Sp2-adj-pi}. This example tells us that two $Sp(N)$ gauge theories with different theta angles can be distinguished by flux choices leading to different blowup equations, or by distinct solutions of a single blowup equation.

The map between  K\"ahler parameters  of the $SU(3)$ and $Sp(2)$ theories is
\begin{align}
\phi_1^{SU} &= \phi_1^{Sp} + \frac{1}{3} \qty(m_0^{Sp} - 3m_1^{Sp}) \,, \ \ \phi_2^{SU} = \phi_2^{Sp} + \frac{2}{3} \qty(m_0^{Sp} - 3m_1^{Sp}) \,,\nonumber \\
m_0^{SU} &= m_0^{Sp} - \frac{1}{2} m_1^{Sp} \,, \ \ m_1^{SU} = -\frac{2}{3} m_0^{Sp} + m_1^{Sp}.
\end{align}
Under this map, the BPS spectrum from the solution to the blowup equation in the $SU(3)$ description agrees with that from the dual $Sp(2)$ description. Some leading BPS states of the $SU(3)$ theory are listed in Table~\ref{table:SU3_1Sym}. We also confirmed that this result agrees with the ADHM result for the $Sp(2)_\pi + 1{\bf Adj}$ theory.
\begin{table}
	\centering
	\begin{tabular}{|c|C{23ex}||c|C{23ex}|} \hline
		$ \mathbf{d} $ & $ \oplus N_{j_l, j_r}^{\mathbf{d}} (j_l, j_r) $ & $ \mathbf{d} $ & $ \oplus N_{j_l, j_r}^{\mathbf{d}} (j_l, j_r) $ \\ \hline
		$ (1, -1, 0, \frac{3}{2}) $ & $ (0, 0) $ & $ (1, -1, 1, \frac{3}{2}) $ & $ (0, 1) $ \\ \hline
		$ (1, -\frac{2}{3}, \frac{2}{3}, \frac{5}{2}) $ & $ (0, \frac{1}{2}) $ & $ (1, -\frac{1}{3}, \frac{1}{3}, \frac{1}{2}) $ & $ (0,\frac{1}{2}) \oplus (\frac{1}{2},0) $ \\ \hline
		$ (1, 0, 1, \frac{3}{2}) $ & $ 2(0,0) \oplus 2(0,1) \oplus (\frac{1}{2},\frac{1}{2}) $ & $ (1, \frac{1}{3}, \frac{2}{3}, -\frac{1}{2}) $ & $ 2(0,0) \oplus (\frac{1}{2},\frac{1}{2}) $ \\ \hline
		$ (1, \frac{1}{3}, \frac{2}{3}, \frac{5}{2}) $ & $ (0, \frac{1}{2}) $ & $ (1, \frac{2}{3}, \frac{1}{3}, -\frac{5}{2}) $ & $ (0, 0) $ \\ \hline
		$ (1, \frac{2}{3}, \frac{1}{3}, \frac{1}{2}) $ & $ (0,\frac{1}{2}) \oplus (\frac{1}{2},0) $ & $ (1, 1, 0, -\frac{3}{2}) $ & $ (0, \frac{1}{2}) $ \\ \hline
		$ (1, 1, 0, \frac{3}{2}) $ & $ (0, 0) $ & $ (1, 1, 1, -\frac{3}{2}) $ & $ 2(0,\frac{1}{2}) \oplus (\frac{1}{2},0) $ \\ \hline
		$ (1, 1, 1, \frac{3}{2}) $ & $ 2(0,0) \oplus 2(0,1) \oplus (\frac{1}{2},\frac{1}{2}) $ & $ (2, -1, 1, 3) $ & $ (0, \frac{1}{2}) \oplus (0, \frac{3}{2}) \oplus (\frac{1}{2}, 1) $ \\ \hline
		$ (2, -\frac{2}{3}, \frac{2}{3}, 1) $ & $ (0, \frac{1}{2}) \oplus (\frac{1}{2}, 1) $ & $ (2, -\frac{2}{3}, \frac{2}{3}, 4) $ & $ (0, 0) $ \\ \hline
		$ (2, -\frac{1}{3}, \frac{1}{3}, 2) $ & $ 2(0, 0) \oplus (\frac{1}{2}, \frac{1}{2}) $ & $ (2, 0, 1, 0) $ & $ 2(0,0) \oplus 3(0,1) \oplus 2(\frac{1}{2},\frac{1}{2}) \oplus (\frac{1}{2},\frac{3}{2}) \oplus (1,1) $ \\ \hline
		$ (2, 0, 1, 3) $ & $ 5(0,\frac{1}{2}) \oplus (0,\frac{3}{2}) \oplus 2(\frac{1}{2},0) \oplus 2(\frac{1}{2},1) $ & $ (2, \frac{1}{3}, \frac{2}{3}, 1) $ & $ 6(0,\frac{1}{2}) \oplus 4(\frac{1}{2},0) \oplus 2(\frac{1}{2},1) \oplus (1,\frac{1}{2}) $ \\ \hline
		$ (2, \frac{1}{3}, \frac{2}{3}, 4) $ & $ (0, 0) $ & $ (2, \frac{2}{3}, \frac{1}{3}, -1) $ & $ 2(0, \frac{1}{2}) \oplus 2(\frac{1}{2}, 0) $ \\ \hline
		$ (2, \frac{2}{3}, \frac{1}{3}, 2) $ & $ 2(0,0) \oplus (\frac{1}{2},\frac{1}{2}) $ & $ (2, 1, 0, 0) $ & $ (0, 0) \oplus (0, 1) \oplus (\frac{1}{2}, \frac{1}{2}) $ \\ \hline
		$ (2, 1, 1, 0) $ & $ 11(0,0) \oplus 9(0,1) \oplus (0,2) \oplus 10(\frac{1}{2},\frac{1}{2}) \oplus 2(\frac{1}{2},\frac{3}{2}) \oplus 2(1,0) \oplus 2(1,1) $ & $ (2, 1, 1, 3) $ & $ 5(0,\frac{1}{2}) \oplus (0,\frac{3}{2}) \oplus 2(\frac{1}{2},0) \oplus 2(\frac{1}{2},1) $ \\ \hline
	\end{tabular}
	\caption{BPS spectrum of the $ SU(3)_{3/2} + 1\mathbf{Sym} $ for $ d_1 \leq 2 $ and $ d_2, d_3 \leq 1 $. Here, $ \mathbf{d} = (d_1, d_2, d_3, d_4) $ labels the BPS states with charge $ d_1 m_0 + d_2 \alpha_1 + d_3 \alpha_2 + d_4 m_1 $ for simple roots $ \alpha_1 $ and $ \alpha_2 $ of $ \mathfrak{su}(3) $ algebra.} \label{table:SU3_1Sym}
\end{table}

\subsubsection{\texorpdfstring{$ SU(3)_0 + 1\mathbf{Sym} + 1\mathbf{F} $}{SU(3)0 + 1Sym + 1F}}

The 5d $ SU(3) $ gauge theory at CS-level $ 0 $ with a symmetric and a fundamental hypermultiplets is the KK-theory obtained by a twisted compactification of the 6d rank-2 $(A_1,A_1)$ conformal matter theory introduced in \cite{DelZotto:2014hpa},
\begin{align}
\begin{tikzpicture}
\draw (0, 0) node {$ SU(3)_0 + 1\mathbf{Sym} + 1\mathbf{F} $}
(3.1, 0) node {$ = $}
(5, 0.3) node {$ \mathfrak{su}(2)^{(1)} $}
(5, -0.3) node {$ 2 $}
;
\draw (4.7, -0.5) .. controls (4.3, -1.2) and (5.7, -1.2) .. (5.3, -0.5);
\end{tikzpicture}
\end{align}
This is another non-geometric rank-2 theory that can be realized by a brane web with $O7^+$ plane. The associated brane web will be given in Appendix~\ref{sec:appendix2}.

We can compute the BPS spectrum of this theory using the blowup formula.
The effective prepotential of this theory on the chamber $\phi_2\ge\phi_1>0$  is given by
\begin{align}
\mathcal{E} & = \frac{1}{\epsilon_1 \epsilon_2} \bigg( \mathcal{F} - \frac{\epsilon_1^2 + \epsilon_2^2}{48} (4\phi_1 - 4\phi_2) + \Big(\frac{\epsilon_1 + \epsilon_2}{2}\Big)^2 (\phi_1 + \phi_2)\bigg) \,, \nonumber \\
6\mathcal{F}
&= 8\phi_1^3 - 3\phi_1^2 \phi_2 - 3\phi_1 \phi_2^2 + 8\phi_2^3 + 6m_0(\phi_1^2 - \phi_1 \phi_2 + \phi_2^2) \nonumber \\
& \quad - \frac{1}{2} \big((2\phi_1\! + \!m_1)^3 + (\phi_2 \!+\! m_1)^3 + (-2\phi_1 \!+\! 2\phi_2 + m_1)^3 + (-\phi_1 \!+\! \phi_2 - m_1)^3 + (\phi_1 \!-\! m_1)^3\big) \nonumber \\
& \quad - \frac{1}{2} \big( (\phi_1 + m_2)^3 + (-\phi_1 + \phi_2 + m_2)^3 + (\phi_2 - m_2)^3 \big),
\end{align}
where $ m_0 $ is the gauge coupling, $ m_1 $ and $ m_2 $ are mass parameters of the symmetric and fundamental hypers, respectively. 
One finds a set of consistent magnetic fluxes
\begin{align}
n_i \in \mathbb{Z} \, , \quad
B_{m_0} = 0 \, , \quad
B_{m_1} = 1/2 \, , \quad
B_{m_2} = 1/2 \ ,
\end{align}
which provides a solvable unity blowup equation. The BPS spectrum from the solution of the blowup equation is given in Table~\ref{table:SU3_sym_F}.
\begin{table}
	\centering
	\begin{tabular}{|c|C{19ex}||c|C{19ex}|} \hline
		$ \mathbf{d} $ & $ \oplus N_{j_l, j_r}^{\mathbf{d}} (j_l, j_r) $ & $ \mathbf{d} $ & $ \oplus N_{j_l, j_r}^{\mathbf{d}} (j_l, j_r) $ \\ \hline
		$ (1, -\frac{2}{3}, \frac{2}{3}, \frac{3}{2}, \frac{1}{2}) $ & $ (0, \frac{1}{2}) $ & $ (1, -\frac{1}{3}, \frac{1}{3}, \frac{3}{2}, -\frac{1}{2}) $ & $ (0, 0) $ \\ \hline
		$ (1, -\frac{1}{3}, \frac{1}{3}, \frac{5}{2}, \frac{1}{2}) $ & $ (0, 0) $ & $ (1, 0, 1, \frac{1}{2}, \frac{1}{2}) $ & $ \! (0, 0) \oplus (0, 1) \oplus (\frac{1}{2}, \frac{1}{2}) \! $ \\ \hline
		$ (1, 0, 1, \frac{5}{2}, -\frac{1}{2}) $ & $ (0, \frac{1}{2}) $ & $ (1, \frac{1}{3}, -\frac{1}{3}, -\frac{5}{2}, -\frac{1}{2}) $ & $ (0, 0) $ \\ \hline
		$ (1, \frac{1}{3}, -\frac{1}{3}, -\frac{3}{2}, \frac{1}{2}) $ & $ (0, 0) $ & $ (1, \frac{1}{3}, \frac{2}{3}, -\frac{5}{2}, -\frac{1}{2}) $ & $ (0, 0) $ \\ \hline
		$ (1, \frac{1}{3}, \frac{2}{3}, -\frac{3}{2}, \frac{1}{2}) $ & $ (0, 0) $ & $ (1, \frac{1}{3}, \frac{2}{3}, \frac{1}{2}, -\frac{1}{2}) $ & $ (0, \frac{1}{2}) \oplus (\frac{1}{2}, 0) $ \\ \hline
		$ (1, \frac{1}{3}, \frac{2}{3}, \frac{3}{2}, \frac{1}{2}) $ & $ 2(0, \frac{1}{2}) \oplus (\frac{1}{2}, 0) $ & $ (1, \frac{2}{3}, -\frac{2}{3}, -\frac{3}{2}, -\frac{1}{2}) $ & $ (0, \frac{1}{2}) $ \\ \hline
		$ (1, \frac{2}{3}, \frac{1}{3}, -\frac{3}{2}, -\frac{1}{2}) $ & $ 2(0, \frac{1}{2}) \oplus (\frac{1}{2}, 0) $ & $ (1, \frac{2}{3}, \frac{1}{3}, -\frac{1}{2}, \frac{1}{2}) $ & $ (0, \frac{1}{2}) \oplus (\frac{1}{2}, 0) $ \\ \hline
		$ (1, \frac{2}{3}, \frac{1}{3}, \frac{3}{2}, -\frac{1}{2}) $ & $ (0, 0) $ & $ (1, \frac{2}{3}, \frac{1}{3}, \frac{5}{2}, \frac{1}{2}) $ & $ (0, 0) $ \\ \hline
		$ (1, 1, 0, -\frac{5}{2}, \frac{1}{2}) $ & $ (0, \frac{1}{2}) $ & $ (1, 1, 0, -\frac{1}{2}, -\frac{1}{2}) $ & $ \! (0, 0) \oplus (0, 1) \oplus (\frac{1}{2}, \frac{1}{2}) \! $ \\ \hline
		$ (1, 1, 1, -\frac{5}{2}, \frac{1}{2}) $ & $ (0, \frac{1}{2}) $ & $ (1, 1, 1, -\frac{1}{2}, -\frac{1}{2}) $ & $ 3(0, 0) \oplus (0, 1) \oplus 2(\frac{1}{2}, \frac{1}{2}) $ \\ \hline
		$ (1, 1, 1, \frac{1}{2}, \frac{1}{2}) $ & $ 3(0, 0) \oplus (0, 1) \oplus 2(\frac{1}{2}, \frac{1}{2}) $ & $ (1, 1, 1, \frac{5}{2}, -\frac{1}{2}) $ & $ (0, \frac{1}{2}) $ \\ \hline
	\end{tabular}
	\caption{BPS spectrum of the $SU(3)_0 + 1\mathbf{Sym} + 1\mathbf{F}$ theory. Here, $ \mathbf{d} = (d_1, d_2, d_3, d_4, d_5) $ labels the BPS states with charge $ d_1 m_0 + d_2 \alpha_1 + d_3 \alpha_2 + d_4 m_1 + d_5 m_2 $ where $ \alpha_1 $ and $ \alpha_2 $ are simple roots of $ \mathfrak{su}(3) $. } \label{table:SU3_sym_F}
\end{table}

\subsubsection{\texorpdfstring{$SU(3)_{\frac{15}{2}}+1\mathbf{F}$, $G_2+1\mathbf{Adj}$}{G2 + 1Adj}}

The last rank-2 KK theory is the theory obtained by $ \mathbb{Z}_3 $ twisted compactification of the 6d $\mathcal{N}=(2,0)$ $ D_4$ theory. This theory has two 5d gauge theory descriptions: one is the $G_2$ gauge theory with an adjoint hypermultiplet \cite{Tachikawa:2011ch} and another one is the $ SU(3)_{15/2} + 1\mathbf{F} $ theory \cite{Bhardwaj:2019jtr},
\begin{align}
\begin{tikzpicture}
\draw (0, 0) node {$ G_2 + 1\mathbf{Adj} $}
(1.8, 0) node {$ = $}
(4, 0) node {$ SU(3)_{15/2} + 1\mathbf{F} $}
(6.2, 0) node {$ = $}
(7.7, 0.3) node {$ \mathfrak{su}(1)^{(1)} $}
(7.7, -0.3) node {$ 2 $}
(9.7, 0.3) node {$ \mathfrak{su}(1)^{(1)} $}
(9.7, -0.3) node {$ 2 $}
(8.15, -0.3) -- (8.5, -0.3)
(8.7, -0.3) node {$ _3 $}
;
\draw [->] (8.9, -0.3) -- (9.3, -0.3);
\end{tikzpicture}
\end{align}
It can be geometrically described by 
\begin{align}\label{eq:G2_adj_geo}
\begin{tikzpicture}
\draw[thick](-3,0)--(0,0);	
\node at(-3.6,0) {$\mathbb{F}_{8}{}\big|_1$};
\node at(-2.8,0.3) {${}_e$};
\node at(0.5,0) {$\mathbb{F}^1_{0}{}\big|_2$};
\node at(-0.45,0.3) {${}_{f+3h}$};
\end{tikzpicture} \ .
\end{align}
However, this geometry is not shrinkable and there is no known shrinkable geometric phase for this theory \cite{Jefferson:2018irk}. Nonetheless, it is expected that a series of flop transitions give rise to a phase for the unitary KK theory \cite{Bhardwaj:2019jtr}. As we will see below, this expectation is consistent with the BPS spectrum which we can compute by solving the blowup equations. 

We shall solve the blowup equations from both the $G_2$ and  $SU(3)$ theory perspectives. The effective prepotential for the $SU(3)$ theory on the Coulomb branch where $\phi_2\ge \phi_1>0$ is given by
\begin{align}
\mathcal{E} &= \frac{1}{\epsilon_1 \epsilon_2} \qty(\mathcal{F} - \frac{\epsilon_1^2 + \epsilon_2^2}{48} \big(4\phi_1 + 2\phi_2\big) + \epsilon_+^2 \qty(\phi_1 + \phi_2) ) \,, \nonumber \\
6\mathcal{F}
&= 8\phi_1^3 - 3\phi_1^2 \phi_2 - 3\phi_1 \phi_2^2 + 8\phi_2^3 + \frac{45}{2} \phi_1 \phi_2(\phi_1 - \phi_2) + 6m_0 (\phi_1^2 - \phi_1 \phi_2 + \phi_2^2) \nonumber \\
& \quad - \frac{1}{2} \Big((\phi_2 - m_1)^3 + (\phi_2 - \phi_1 + m_1)^3 + (\phi_1 + m_1)^3\Big) \,,
\end{align}
where $m_0$ is the $SU(3)$ gauge coupling and $m_1$ is the mass parameter of the fundamental hyper.

In the $G_2$ gauge theory, the effective prepotential in the same chamber is
\begin{align}
\mathcal{E} &= \frac{1}{\epsilon_1 \epsilon_2} \bigg( \mathcal{F} - \frac{\epsilon_1^2 + \epsilon_2^2}{48} (4\phi_1 + 2\phi_2) + \epsilon_+^2 (\phi_1 + \phi_2)\bigg) \,, \nonumber \\
6\mathcal{F}
&= 8\phi_1^3 + 18\phi_1^2 \phi_2 - 24\phi_1 \phi_2^2 + 8\phi_2^3 + 6m_0 (3\phi_1^2 - 3\phi_1 \phi_2 + \phi_2^2) \nonumber \\
& \quad -\frac{1}{2} \big((\phi_2 \pm m_1)^3 + (\pm(3\phi_1 - \phi_2) + m_1)^3 + (\pm \phi_1 + m_1)^3 + (\pm(\phi_2 - \phi_1) + m_1)^3 \nonumber \\
& \qquad \qquad + (\pm(-3\phi_1 + 2\phi_2) + m_1)^3 + (\pm(2\phi_1 - \phi_2) + m_1)^3) \big),
\end{align}
where $m_0$ is the $G_2$ gauge coupling and $m_1$ is the mass parameter of the adjoint matter. 

Under the duality, the K\"ahler parameters in the $ G_2 + 1\mathbf{Adj} $ and in the $ SU(3)_{15/2} + 1\mathbf{F} $ theories are mapped to each other as follows:
\begin{align}\label{eq:G2adj-su3+1f_map}
\phi_1^{G_2} &= \phi_1^{SU} + m_0^{SU} - \frac{1}{2} m_1^{SU} \,, \ \ \phi_2^{G_2} = \phi_2^{SU} + 2m_0^{SU(3)} - m_1^{SU} \,, \cr 
m_0^{G_2} &= 5m_0^{SU} + \frac{3}{2}m_1^{SU} \,, \ \ m_1^{G_2} = 2m_0^{SU}.
\end{align}
One can check with this map that the effective prepotentials of two dual gauge theories match well up to constant terms.

We first solve the blowup equation from the perspective of $ G_2 $ gauge theory. A set of consistent magnetic fluxes 
\begin{align}\label{eq:G2_adj_flux}
n_i \in \mathbb{Z} \, , \quad
B_{m_0} = 0 \, , \quad
B_{m_1} = 1/2 \ 
\end{align}
provides a solvable blowup equation.
By solving the blowup equation, we find the BPS spectrum of the $G_2$ gauge theory given in Table~\ref{table:G2_adj}.

\begin{table}
	\centering
	\begin{tabular}{|c|C{26ex}||c|C{26ex}|} \hline
		$ \mathbf{d} $ & $ \oplus N_{j_l, j_r}^{\mathbf{d}} (j_l, j_r) $ & $ \mathbf{d}$ & $\oplus N_{j_l, j_r}^{\mathbf{d}} (j_l, j_r) $ \\ \hline
		$ (1, 0, 1, 0) $ & $ 2(0,\frac{1}{2}) \oplus (\frac{1}{2},0) \oplus (\frac{1}{2},1) $ & $ (1, 0, 1, 1) $ & $ (0,0) \oplus (0,1) \oplus (\frac{1}{2},\frac{1}{2}) $ \\ \hline
		$ (1, 0, 1, 2) $ & $ (0, \frac{1}{2}) $ & $ (1, 0, 2, 0) $ & $ 2(0,\frac{3}{2}) \oplus (\frac{1}{2},1) \oplus (\frac{1}{2},2) $ \\ \hline
		$ (1, 0, 2, 1) $ & $ (0,1) \oplus (0,2) \oplus (\frac{1}{2},\frac{3}{2}) $ & $ (1, 0, 2, 2) $ & $ (0, \frac{3}{2}) $ \\ \hline
		$ (1, 1, 1, 0) $ & $ 4(0,\frac{1}{2}) \oplus 3(\frac{1}{2},0) \oplus (\frac{1}{2},1) $ & $ (1, 1, 1, 1) $ & $ 4(0,0) \oplus (0,1) \oplus 2(\frac{1}{2},\frac{1}{2}) $ \\ \hline
		$ (1, 1, 1, 2) $ & $ 2(0,\frac{1}{2}) \oplus (\frac{1}{2},0) $ & $ (1, 1, 1, 3) $ & $ (0, 0) $  \\ \hline
		$ (1, 1, 2, 0) $ & $ 4(0,\frac{1}{2}) \oplus 4(0,\frac{3}{2}) \oplus (\frac{1}{2},0) \oplus 4(\frac{1}{2},1) \oplus (\frac{1}{2},2) $ & $ (1, 1, 2, 1) $ & $ (0,0) \oplus 5(0,1) \oplus (0,2) \oplus 2(\frac{1}{2},\frac{1}{2}) \oplus 2(\frac{1}{2},\frac{3}{2}) $ \\ \hline
		$ (1, 1, 2, 2) $ & $ 2(0,\frac{1}{2}) \oplus 2(0,\frac{3}{2}) \oplus (\frac{1}{2},1) $ & $ (1, 1, 2, 3) $ & $ (0, 1) $ \\ \hline
		$ (1, 2, 1, 0) $ & $ 4(0,\frac{1}{2}) \oplus 3(\frac{1}{2},0) \oplus (\frac{1}{2},1) $ & $ (1, 2, 1, 1) $ & $ 4(0,0) \oplus (0,1) \oplus 2(\frac{1}{2},\frac{1}{2}) $ \\ \hline
		$ (1, 2, 1, 2) $ & $ 2(0,\frac{1}{2}) \oplus (\frac{1}{2},0) $ & $ (1, 2, 1, 3) $ & $ (0, 0) $  \\ \hline
		$ (1, 2, 2, 0) $ & $ 8(0,\frac{1}{2}) \oplus 4(0,\frac{3}{2}) \oplus 4(\frac{1}{2},0) \oplus 5(\frac{1}{2},1) \oplus (\frac{1}{2},2) $ & $ (1, 2, 2, 1) $ & $ 5(0,0) \oplus 6(0,1) \oplus (0,2) \oplus 4(\frac{1}{2},\frac{1}{2}) \oplus 2(\frac{1}{2},\frac{3}{2}) $ \\ \hline
		$ (1, 2, 2, 2) $ & $ 4(0,\frac{1}{2}) \oplus 2(0,\frac{3}{2}) \oplus (\frac{1}{2},0) \oplus (\frac{1}{2},1) $ & $ (1, 2, 2, 3) $ & $ (0,0) \oplus (0,1) $ \\ \hline
		$ (2, 0, 1, 0) $ & $ 5(0,\frac{1}{2}) \oplus 1(0,\frac{3}{2}) \oplus 3(\frac{1}{2},0) \oplus 3(\frac{1}{2},1) \oplus (1,\frac{1}{2}) \oplus (1,\frac{3}{2}) $ & $ (2, 0, 1, 1) $ & $ 4(0,0) \oplus 2(0,1) \oplus 3(\frac{1}{2},\frac{1}{2}) \oplus (\frac{1}{2},\frac{3}{2}) \oplus (1,1) $ \\ \hline
		$ (2, 0, 1, 2) $ & $ 2(0,\frac{1}{2}) \oplus (\frac{1}{2},0) \oplus (\frac{1}{2},1) $ & $ (2, 0, 1, 3) $ & $ (0, 0) $ \\ \hline
		$ (2, 1, 1, 0) $ & $ \! 12(0,\frac{1}{2}) \oplus (0,\frac{3}{2}) \oplus 10(\frac{1}{2},0) \oplus 6(\frac{1}{2},1) \oplus 4(1,\frac{1}{2}) \oplus (1,\frac{3}{2}) \! $ & $ (2, 1, 1, 1) $ & $ \! 10(0,0) \oplus 4(0,1) \oplus 9(\frac{1}{2},\frac{1}{2}) \oplus (\frac{1}{2},\frac{3}{2}) \oplus 2(1,0) \oplus 2(1,1) \! $ \\ \hline
		$ (2, 1, 1, 2) $ & $ 5(0,\frac{1}{2}) \oplus 4(\frac{1}{2},0) \oplus 2(\frac{1}{2},1) \oplus (1,\frac{1}{2}) $ & $ (2, 1, 1, 3) $ & $ 2(0,0) \oplus (\frac{1}{2},\frac{1}{2}) $ \\ \hline
	\end{tabular}
	\caption{BPS spectrum of the $ G_2 + 1\mathbf{Adj} $ for $ d_1 = 1 $, $ d_2, d_3 \leq 2 $ and $ d_1 = 2 $, $ d_2, d_3 \leq 1 $. Here, $ \mathbf{d} = (d_1, d_2, d_3, d_4) $ labels the BPS states with charge $ d_1 m_0 + d_2 \alpha_1 + d_3 \alpha_2 + d_4 m_1 $, where $ \alpha_1 = 2\phi_1 - \phi_2 $ and $ \alpha_2 = -3\phi_1 + 2\phi_2 $ are simple roots of $ G_2 $. The theory has a symmetry exchanging $ d_4 \leftrightarrow -d_4 $.} \label{table:G2_adj}
\end{table}

In the $ SU(3) $ description, one can use the same magnetic fluxes to formulate a unity blowup equation. The duality map \eqref{eq:G2adj-su3+1f_map} implies that the above fluxes in the $G_2$ perspective are converted into the magnetic fluxes in the $SU(3)$ gauge theory given by 
\begin{align}
n_1 \in \mathbb{Z} + 1/3 \, , \quad
n_2 \in \mathbb{Z} + 2/3 \, , \quad
B_{m_0} = 1/4 \, , \quad
B_{m_1} = -5/6 \ .
\end{align}
We checked that the blowup equation with these fluxes in the $SU(3)$ perspective can be solved and the solution perfectly agrees with the result from the $ G_2 + 1\mathbf{Adj} $ description.

The BPS spectrum tells us how to move on to the unitary phase where all the BPS states have non-negative masses in the UV limit. First note that the geometric phase \eqref{eq:G2_adj_geo} is non-shrinkable due to an exceptional curve $f_2-x$ whose volume cannot be non-negative while keeping volumes of all other curves non-negative at the same time. This implies that we need to flop the $f_2-x$ curve in the UV limit. Unfortunately, this transition leads to a phase which cannot be described by a smooth CY 3-fold.

However, we are able to trace all the phase transitions toward the unitary phase relying on the BPS spectrum we computed. To move on  to the unitary phase, we need to  flop 4 curves associated with 4 hypermultiplets with the masses, consecutively as,
\begin{equation}
	m_1-3\phi_1+\phi_2 \ \ \rightarrow \ \ m_1 -\phi_1 \ \  \rightarrow \ \ m_1 + \phi_1-\phi_2 \ \ \rightarrow \ \ m_1 +3\phi_1-2\phi_2 
\end{equation}
in the $G_2$ description. In the $SU(3)$ description, these hypermultiplets are all instantonic states. In the geometry \eqref{eq:G2_adj_geo}, the flop transitions performed on the exceptional curves as
\begin{equation}\label{eq:G2-flop}
	f_2 -x \ \ \rightarrow \ \ f_1+f_2 -x \ \ \rightarrow \ \ 2f_1+f_2 -x \ \rightarrow \ \ 3f_1+f_2 -x \ ,
\end{equation}
will lead to the unitary phase smoothly connected to the UV fixed point. One can check that in the final phase all the BPS states have non-negative masses. This result provides a strong evidence that the flop transitions in \eqref{eq:G2-flop} in the geometric description \eqref{eq:G2_adj_geo} are physically well-established transitions, though the phases after the flop transitions are non-geometrical.

\subsection{Rank 2 5d SCFTs}\label{sec:rank2SCFTs}

\subsubsection{\texorpdfstring{$ \mathbb{P}^2 \cup \mathbb{F}_3 $}{P2 U F3}}

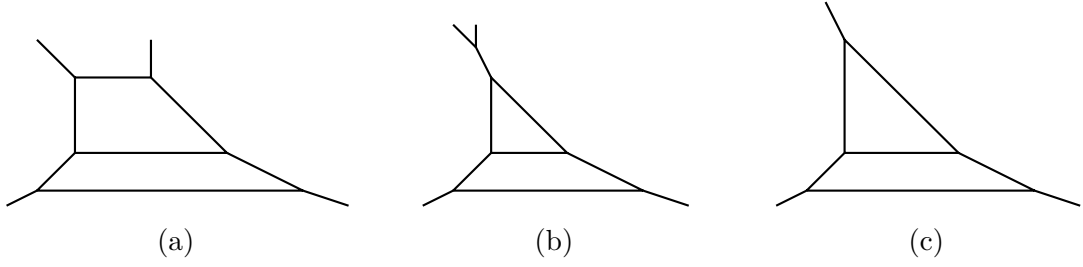
\begin{figure}
	\centering
	\begin{subfigure}[b]{0.33\textwidth}
		\centering
		\begin{tikzpicture}
		\draw[thick] (0.5, 0.5) -- (1, 1) -- (1, 2) -- (2, 2) -- (3, 1) -- (1, 1)
		(1, 2) -- (0.5, 2.5)
		(2, 2) -- (2, 2.5)
		(3, 1) -- (4, 0.5)
		(0.1, 0.3) -- (0.5, 0.5) -- (4, 0.5) -- (4.6, 0.3);
		\end{tikzpicture}
		\caption{}
	\end{subfigure}
	\begin{subfigure}[b]{0.31\textwidth}
		\centering
		\begin{tikzpicture}
		\draw[thick] (0.5, 0.5) -- (1, 1) -- (1, 2) -- (2, 1) -- (1, 1)
		(1, 2) -- (0.8, 2.4) -- (0.8, 2.7)
		(0.8, 2.4) -- (0.5, 2.7)
		(2, 1) -- (3, 0.5)
		(0.1, 0.3) -- (0.5, 0.5) -- (3, 0.5) -- (3.6, 0.3);
		\end{tikzpicture}
		\caption{}
	\end{subfigure}
	\begin{subfigure}[b]{0.32\textwidth}
		\centering
		\begin{tikzpicture}
		\draw[thick] (0.5, 0.5) -- (1, 1) -- (1, 2.5) -- (2.5, 1) -- (1, 1)
		(1, 2.5) -- (0.75, 3)
		(2.5, 1) -- (3.5, 0.5)
		(0.1, 0.3) -- (0.5, 0.5) -- (3.5, 0.5) -- (4.1, 0.3);
		\end{tikzpicture}
		\caption{}
	\end{subfigure}
	\caption{A 5-brane web construction for $ \mathbb{P}^2 \cup \mathbb{F}_3 $. Starting from the web diagram of $ SU(3)_2 $ theory (a), flop transition for an instantonic hypermultiplet gives (b). Integrating out the hypermultiplet gives a web diagram of $ \mathbb{P}^2 \cup \mathbb{F}_3 $ depicted in (c).} \label{fig:P2-F3}
\end{figure}

The local $\mathbb{P}^2 \cup \mathbb{F}_3 $ theory is a rank-2 analog of the SCFT engineered by a local $\mathbb{P}^2$. It is a non-Lagrangian theory with no mass parameter. Its geometric construction is represented by
\begin{align}
\begin{tikzpicture}
\draw[thick](-2.5,0)--(0,0);	
\node at(-3.1,0) {$\eval{\mathbb{P}^2}_1$};
\node at(-2.3,0.3) {${}_\ell$};
\node at(0.6,0) {$\eval{\mathbb{F}_{3}}_2$};
\node at(-0.2,0.3) {${}_{e}$};
\end{tikzpicture} \ .
\end{align}
The volumes of two primitive 2-cycles are given by
\begin{align}
\vol(\ell) = 3\phi_1 - \phi_2 \ , \quad
\vol(f_2) = -\phi_1 + 2\phi_2\ ,
\end{align}
where $\ell$ is a curve with $\ell^2=1$ in $\mathbb{P}^2$ and $ f_2 $ is the fiber in $ \mathbb{F}_3 $. The geometry can be obtained by blowing down an exceptional cycle of $ \mathbb{F}_1 $ in geometry $ \mathbb{F}_1 - \mathbb{F}_3 $, which implies that this theory can be obtained by an RG flow from  the pure $ SU(3)_2 $ theory by integrating out an instantonic hypermultiplet. The web diagram for this theory is depicted in Figure~\ref{fig:P2-F3}.

The effective prepotential on the $\Omega$-background is
\begin{align}
\mathcal{E} &= \frac{1}{\epsilon_1 \epsilon_2} \qty(\mathcal{F} - \frac{\epsilon_1^2 + \epsilon_2^2}{48}(6\phi_1 + 4\phi_2) + \epsilon_+^2 (\phi_1 + \phi_2)) \ , \nonumber \\
6\mathcal{F} &= 9\phi_1^3 - 9\phi_1^2 \phi_2 + 3\phi_1 \phi_2^2 + 8\phi_2^3 \ .
\end{align}
For a unity blowup equation, we choose the magnetic fluxes as
\begin{align}
n_1 \in \mathbb{Z} + 1/5 \ , \quad
n_2 \in \mathbb{Z} + 1/10 \ ,
\end{align}
which assign half-integral fluxes for $ \ell $ and integral fluxes for $ f_2 $. The BPS spectrum obtained by solving the blowup equation is given in Table~\ref{table:P2-F3}. We checked that the result matches the spectrum after taking an RG flow from the BPS spectrum of the pure $ SU(3)_2 $ theory by integrating out the instantonic hypermultiplet.

\begin{table}
	\centering
	\begin{tabular}{|c|C{29ex}||c|C{29ex}|} \hline
		$\mathbf{d}$ & $\oplus N_{j_l, j_r}^{\mathbf{d}} (j_l, j_r)$ & $\mathbf{d}$ & $\oplus N_{j_l, j_r}^{\mathbf{d}} (j_l, j_r)$ \\ \hline
		$ (0, 1) $ & $ (0, \frac{1}{2}) $ & $ (1, 0) $ & $ (0, 1) $ \\ \hline
		$ (1, 1) $ & $ (0, 0) \oplus (0, 1) $ & $ (1, 2) $ & $ (0, 1) $ \\ \hline
		$ (1, 3) $ & $ (0, 2) $ & $ (2, 0) $ & $ (0, \frac{5}{2}) $ \\ \hline
		$ (2, 1) $ & $ (0, \frac{3}{2}) \oplus (0, \frac{5}{2}) $ & $ (2, 2) $ & $ (0, \frac{1}{2}) \oplus (0, \frac{3}{2}) \oplus (0, \frac{5}{2}) $ \\ \hline
		$ (2, 3) $ & $ (0, \frac{1}{2}) \oplus (0, \frac{3}{2}) \oplus (0, \frac{5}{2}) $ & $ (3, 0) $ & $ (0, 3) \oplus (\frac{1}{2}, \frac{9}{2}) $ \\ \hline
		$ (3, 1) $ & $ (0,2) \oplus 2(0,3) \oplus (0,4) \oplus (\frac{1}{2},\frac{7}{2}) \oplus (\frac{1}{2},\frac{9}{2}) $ & $ (3, 2) $ & $ (0,1) \oplus 2(0,2) \oplus 3(0,3) \oplus (0,4) \oplus (\frac{1}{2},\frac{5}{2}) \oplus (\frac{1}{2},\frac{7}{2}) \oplus (\frac{1}{2},\frac{9}{2}) $ \\ \hline
		$ (3, 3) $ & \multicolumn{3}{C{67ex}|}{$ \! (0,0) \oplus 2(0,1) \oplus 3(0,2) \oplus 3(0,3) \oplus (0,4) \oplus (\frac{1}{2},\frac{3}{2}) \oplus (\frac{1}{2},\frac{5}{2}) \oplus (\frac{1}{2},\frac{7}{2}) \oplus (\frac{1}{2},\frac{9}{2}) \! $} \\ \hline
	\end{tabular}
	\caption{BPS spectrum of $ \mathbb{P}^2 \cup \mathbb{F}_3 $ for $ d_i \leq 3 $. Here, $ \mathbf{d} = (d_1, d_2) $ labels the BPS states with charge $ d_1 \ell + d_2 f_2 $.}\label{table:P2-F3}
\end{table}

\subsubsection{\texorpdfstring{$ \mathbb{P}^2 \cup \mathbb{F}_6 $}{P2 U F6}}

\begin{figure}
	\centering
	\begin{subfigure}[b]{0.32\textwidth}
		\centering
		\begin{tikzpicture}
		\draw[thick] (0.5, 0.5) -- (1, 1) -- (1, 2) -- (2, 2) -- (3, 1)
		(1, 1) -- (3, 1)
		(1, 2) -- (0.5, 2.5) -- (0.5, 0.5) -- (4, 0.5) -- (3, 1)
		(0.5, 2.5) -- (0.25, 3)
		(2, 2) -- (2, 3)
		(4, 0.5) -- (4.75, 0.25);
		\draw[thick, double] (0, 0) -- (0.5, 0.5);
		\end{tikzpicture}
		\caption{}
	\end{subfigure}
	\begin{subfigure}[b]{0.32\textwidth}
		\centering
		\begin{tikzpicture}
		\draw[thick] (0.5, 0.5) -- (1, 1) -- (1, 2.5) -- (2.5, 1) -- (3.5, 0.5) -- (4.25, 0.25)
		(1, 1) -- (2.5, 1)
		(3.5, 0.5) -- (0.5, 0.5) -- (0.5, 3.5) -- (1, 2.5)
		(0.5, 3.5) -- (0.333, 4) -- (0.333, 4.5)
		(0.333, 4) -- (0.083, 4.5);
		\draw[thick, double] (0, 0) -- (0.5, 0.5);
		\end{tikzpicture}
		\caption{}
	\end{subfigure}
	\begin{subfigure}[b]{0.32\textwidth}
		\centering
		\begin{tikzpicture}
		\draw[thick] (0.5, 0.5) -- (1, 1) -- (1, 2.5) -- (2.5, 1) -- (3.5, 0.5) -- (4.25, 0.25)
		(1, 1) -- (2.5, 1)
		(3.5, 0.5) -- (0.5, 0.5) -- (0.5, 3.5) -- (1, 2.5)
		(0.5, 3.5) -- (0.333, 4);
		\draw[thick, double] (0, 0) -- (0.5, 0.5);
		\end{tikzpicture}
		\caption{}
	\end{subfigure}
	\caption{A 5-brane web construction for $ \mathbb{P}^2 \cup \mathbb{F}_6 $. Starting from the web diagram of $ SU(3)_4 $ theory (a), flop transition of two instantonic hypermultiplets gives (b). Integrating out the hypermultiplet gives a web diagram of $ \mathbb{P}^2 \cup \mathbb{F}_6 $ depicted in (c).} \label{fig:P2-F6}
\end{figure}
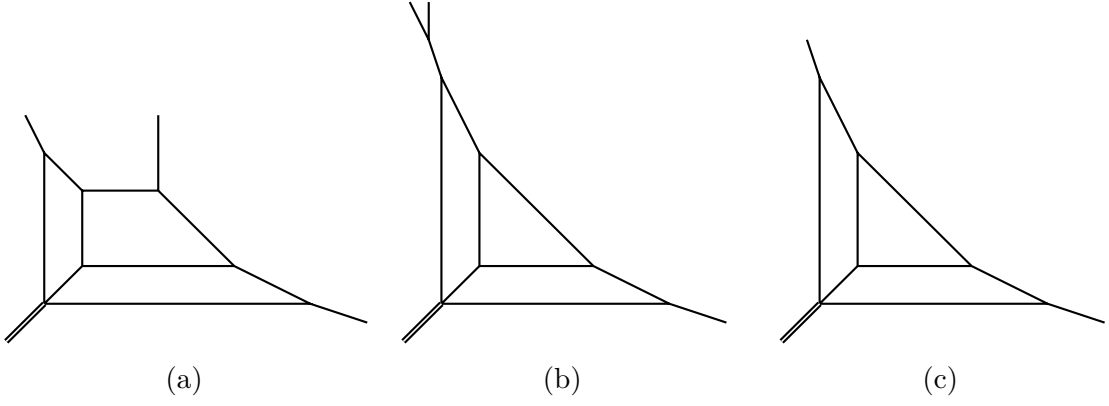

The local $ \mathbb{P}^2 \cup \mathbb{F}_6 $ theory is another non-Lagrangian rank-2 theory with no mass parameter. Its geometric construction is given by
\begin{align}
\begin{tikzpicture}
\draw[thick](-2.5,0)--(0,0);	
\node at(-3.1,0) {$\eval{\mathbb{P}^2}_1$};
\node at(-2.3,0.3) {${}_{2\ell}$};
\node at(0.6,0) {$\eval{\mathbb{F}_{6}}_2$};
\node at(-0.2,0.3) {${}_{e}$};
\end{tikzpicture} \ .
\end{align}
The volumes of two primitive 2-cycles are
\begin{align}
\vol(\ell) = 3\phi_1 - 2\phi_2 \, , \quad
\vol(f_2) = -\phi_1 + 2\phi_2
\end{align}
where $ \ell $ is the curve class in $ \mathbb{P}^2 $ and $ f_2 $ is the fiber in $ \mathbb{F}_6 $. This theory can be obtained by an integrating out an instantonic hypermultiplet of the pure $ SU(3)_4 $ theory. The web diagram and the RG flow of this theory are given in Figure~\ref{fig:P2-F6}.

The effective prepotential is
\begin{align}
\mathcal{E} &= \frac{1}{\epsilon_1 \epsilon_2} \qty( \mathcal{F} - \frac{\epsilon_1^2 + \epsilon_2^2}{48}(6\phi_1 + 4\phi_2) + \epsilon_+^2 (\phi_1 + \phi_2) ) \, ,  \nonumber \\
6\mathcal{F} &= 9\phi_1^3 - 18\phi_1^2 \phi_2 + 12\phi_1 \phi_2^2 + 8\phi_2^3 \ .
\end{align}
We choose the magnetic fluxes
\begin{align}
n_1 \in \mathbb{Z} + 1/4 \, , \quad
n_2 \in \mathbb{Z} + 1/8\ ,
\end{align}
to get a solvable unity blowup equation. The BPS spectrum obtained by solving the blowup equation is given in Table~\ref{table:P2-F6}.  We checked that the result matches the spectrum after taking an RG flow from the BPS spectrum of the pure $ SU(3)_4 $ theory given in Table \ref{table:SU(3)_4} by integrating out the instantonic hypermultiplet.

\begin{table}
	\centering
	\begin{tabular}{|c|C{29ex}||c|C{29ex}|} \hline
		$\mathbf{d}$ & $\oplus N_{j_l, j_r}^{\mathbf{d}} (j_l, j_r)$ & $\mathbf{d}$ & $\oplus N_{j_l, j_r}^{\mathbf{d}} (j_l, j_r)$ \\ \hline
		$ (0, 1) $ & $ (0, \frac{1}{2}) $ & $ (1, 0) $ & $ (0, 1) $ \\ \hline
		$ (1, 1) $ & $ (0, 0) \oplus (0, 1) $ & $ (1, 2) $ & $ (0, 1) $ \\ \hline
		$ (2, 0) $ & $ (0, \frac{5}{2}) $ & $ (2, 1) $ & $ (0, \frac{3}{2}) \oplus (0, \frac{5}{2}) $ \\ \hline
		$ (2, 2) $ & $ (0,\frac{1}{2}) \oplus (0,\frac{3}{2}) \oplus (0,\frac{5}{2}) $ & $ (2, 3) $ & $ (0, \frac{1}{2}) \oplus (0, \frac{3}{2}) \oplus (0, \frac{5}{2}) $ \\ \hline
		$ (3, 0) $ & $ (0, 3) \oplus (\frac{1}{2}, \frac{9}{2}) $ & $ (3, 1) $ & $ (0,2) \oplus 2(0,3) \oplus (0,4) \oplus (\frac{1}{2},\frac{7}{2})\oplus (\frac{1}{2},\frac{9}{2}) $ \\ \hline
		$ (3, 2) $ & $ (0,1) \oplus 2(0,2) \oplus 3(0,3) \oplus (0,4) \oplus (\frac{1}{2},\frac{5}{2}) \oplus (\frac{1}{2},\frac{7}{2}) \oplus (\frac{1}{2},\frac{9}{2}) $ & $ (3, 3) $ & $ (0,0) \oplus 2(0,1) \oplus 3(0,2) \oplus 3(0,3) \oplus (0,4) \oplus (\frac{1}{2},\frac{3}{2}) \oplus (\frac{1}{2},\frac{5}{2}) \oplus (\frac{1}{2},\frac{7}{2}) \oplus (\frac{1}{2},\frac{9}{2}) $ \\ \hline
	\end{tabular}
	\caption{BPS spectrum of $ \mathbb{P}^2 \cup \mathbb{F}_6 $ for $ d_i \leq 3 $. Here, $ \mathbf{d} = (d_1, d_2) $ labels the BPS state with charge $ d_1 \ell + d_2 f_2 $.} \label{table:P2-F6}
\end{table}

\subsubsection{\texorpdfstring{$ \mathbb{P}^2 \cup \mathbb{F}_3 + 1\mathbf{Sym} $}{P2 U F3 + 1Sym}}

The $\mathbb{P}^2 \cup \mathbb{F}_3 + 1\mathbf{Sym}$ theory is a rank-2 analog of the local $\mathbb{P}^2+{\bf 1Adj}$ theory. As proposed in \cite{Bhardwaj:2019jtr}, we can obtain this theory from the UV $ Sp(2)_\pi + 1\mathbf{Adj} $ theory or $ SU(3)_{3/2} + 1\mathbf{Sym} $ theory by integrating out an instantonic hypermultiplet. See Appendix~\ref{sec:app-P2 U F3+1Sym} for their 5-brane constructions. 

\begin{figure}
	\centering
	\begin{tikzpicture}
	\draw[thick] (0, 0) -- (3, 1) -- (5, 2) -- (6, 3) -- (6, 2) -- (7, 1) -- (9, 0)
	(5, 2) -- (6, 2)
	(3, 1) -- (7, 1)
	(6, 3) -- (6.5, 4) -- (6.5, 4.5)
	(6.5, 4) -- (7, 4.5)
	(11, 0) -- (10, 1);
	\draw[dashed] (-1, 0) -- (12, 0);
	\filldraw[fill=white, thick] (9, 0) circle (0.15);
	\draw (7.5, 3.5) node {\scriptsize{$ -m_0 - \phi_1 + \frac{5}{2}m_1 $}}
	(7.6, 2.5) node {\scriptsize{$ m_0 + 3\phi_1 - \phi_2 - \frac{5}{2}m_1 $}}
	(3, 1.7) node {\scriptsize{$ -\phi_1 + 2\phi_2 $}}
	(6.9, 0.4) node {\scriptsize{$ m_1 - 2\phi_2 $}}
	(9, -0.4) node {\scriptsize{$ O7^+ $}};
	\end{tikzpicture}
	\caption{A 5-brane web for $ SU(3)_{3/2} + 1\mathbf{Sym} $ and $Sp(2)_\pi + 1{\bf Adj}$, where a flop transition for the edge with length $\phi_1+m_0-\frac{5}{2}m_1$ is performed.} \label{fig:P2-F3-Sym}
\end{figure}
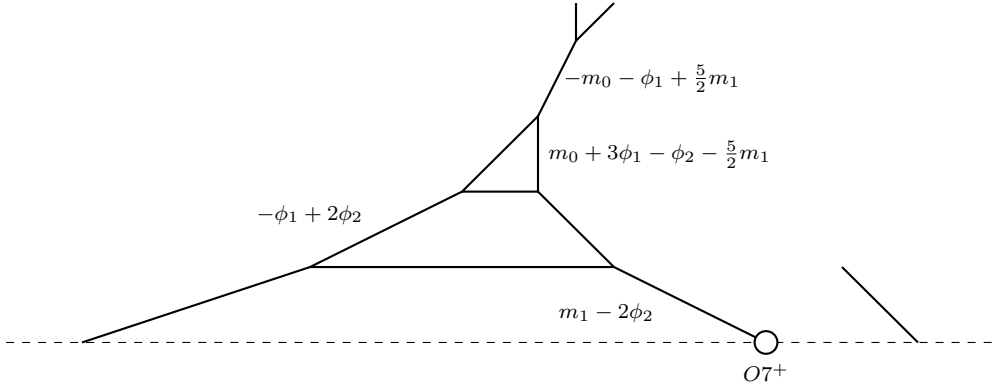

More precisely, there is a hypermultiplet with $ \mathbf{d} = (1, \frac{2}{3}, \frac{1}{3}, -\frac{5}{2}) $ which has mass $ m_0 + \phi_1 - \frac{5}{2}m_1 $ in the UV parent theory, as listed in Table~\ref{table:SU3_1Sym}. We flop this hypermultiplet and integrate it out to get the $ \mathbb{P}^2 \cup \mathbb{F}_3 + 1\mathbf{Sym} $ theory in IR. This RG flow is realized in the 5-brane web in Figure~\ref{fig:P2-F3-Sym} as taking a limit $ -m_0 - \phi_1 + \frac{5}{2}m_1 \to \infty $ while the lengths of other edges are kept finite.

To compute the effective prepotential and the BPS partition function of this theory, it is convenient to redefine the parameters in the web diagram in Figure~\ref{fig:P2-F3-Sym} as
\begin{align}\label{eq:P2-F3_var_redef}
\tilde{\phi}_1 = \phi_1 + \frac{2}{5}m_0 - m_1 \, , \quad
\tilde{\phi}_2 = \phi_2 + \frac{m_0}{5} - \frac{m_1}{2} \, , \quad
\tilde{m} = \frac{2}{5} m_0 \ .
\end{align}
Then the volumes of 2-cycles in Figure~\ref{fig:P2-F3-Sym} become
\begin{align}
&-m_0 -\phi_1 + \frac{5}{2}m_1 = -\tilde{\phi}_1 - \frac{3}{2}\tilde{m} + \frac{3}{2}m_1 \, , \quad -\phi_1 + 2\phi_2 = -\tilde{\phi}_1 + 2\tilde{\phi}_2 \,,  \nonumber \\
&m_0 + 3\phi_1 - \phi_2 - \frac{5}{2}m_1 = 3\tilde{\phi}_1 - \tilde{\phi}_2 \, , \quad m_1 - 2\phi_2 = \tilde{m} - 2\tilde{\phi}_2 \ . 
\end{align}
Hence, the RG flow corresponds to the limit $ m_1 \to \infty $,  while keeping $ \tilde{\phi}_i $ and $ \tilde{m} $ finite.

The cubic prepotential and mixed gravitational Chern-Simons term in the IR $ \mathbb{P}^2 \cup \mathbb{F}_3 + 1\mathbf{Sym} $ theory can be obtained from those of the UV $ SU(3)_{3/2} + 1\mathbf{Sym} $ theory as follows:
\begin{align}
\mathcal{F}_{IR} = \mathcal{F}_{UV} + \frac{1}{6} \qty(m_0 + \phi_1 - \frac{5}{2}m_1)^3 \, , \quad
C_{IR} = C_{UV} - 2\qty(m_0 + \phi_1 - \frac{5}{2}m_1) \, ,
\end{align}
whereas the mixed gauge/$ SU(2)_R $ Chern-Simons term remains the same. In terms of the parameters in \eqref{eq:P2-F3_var_redef}, the effective prepotential in the IR theory takes the form
\begin{align}
\mathcal{E} &= \frac{1}{\epsilon_1 \epsilon_2} \qty(\mathcal{F}_{IR} - \frac{\epsilon_1^2 + \epsilon_2^2}{48} (6 \tilde{\phi}_1 + 4 \tilde{\phi}_2) + \epsilon_+^2 (\tilde{\phi}_1 + \tilde{\phi}_2) ) \, , \nonumber \\
6\mathcal{F}_{IR} &= 9\tilde{\phi}_1^3 - 9\tilde{\phi}_1^2 \tilde{\phi}_2 + 3\tilde{\phi}_1 \tilde{\phi}_2^2 + 8\tilde{\phi}_2^3 \ ,
\end{align}
up to constant terms. One may notice that the prepotential is the same as that of $ \mathbb{P}^2 \cup \mathbb{F}_3 $. However, this theory has an additional hypermultiplet coming from the symmetric matter of the parent theory. The GV-invariant for this hypermultiplet
\begin{align}
\mathcal{Z}_{\mathrm{hyper}}(\tilde{\phi}_i, \tilde{m}; \epsilon_{1,2})
= \PE\qty[ \frac{(p_1 p_2)^{1/2}}{(1-p_1)(1-p_2)} e^{-(\tilde{m} - 2\tilde{\phi}_2)}  ] 
\end{align}
is an additional input for the blowup equation of the IR theory.

We find a set of  consistent magnetic fluxes
\begin{align}
\tilde{n}_1 \in \mathbb{Z} - 3/5 \, , \quad
\tilde{n}_2 \in \mathbb{Z} - 3/10 \, , \quad
B_{\tilde{m}} = -1/10
\end{align}
which leads to a unity blowup equation. The solution of the blowup equation provides the BPS spectrum of the theory, summarized in Table~\ref{table:P2-F3+1Sym}. This solution matches the spectrum that we can obtain indirectly from the RG flow of the BPS spectrum in $ SU(3)_{3/2} + 1\mathbf{Sym} $ in Table~\ref{table:SU3_1Sym} after integrating out the hypermultiplet.

\begin{table}
	\centering
	\begin{tabular}{|c|C{28ex}||c|C{28ex}|} \hline
		$\mathbf{d}$ & $\oplus N_{j_l, j_r}^{\mathbf{d}} (j_l, j_r)$ & $\mathbf{d}$ & $\oplus N_{j_l, j_r}^{\mathbf{d}} (j_l, j_r)$ \\ \hline
		$ (1, 0, 0) $ & $ (0, 0) $ & $ (1, 0, 1) $ & $ (0, 0) $ \\ \hline
		$ (1, 0, 2) $ & $ (0, 0) $ & $ (1, 1, 1) $ & $ (0, \frac{1}{2}) $ \\ \hline
		$ (1, 1, 2) $ & $ 2(0, \frac{1}{2}) \oplus (\frac{1}{2}, 0) $ & $ (1, 1, 3) $ & $ (0, \frac{1}{2}) \oplus (0, \frac{3}{2}) \oplus (\frac{1}{2}, 1) $ \\ \hline
		$ (1, 2, 1) $ & $ (0, 2) $ & $ (1, 2, 2) $ & $ \! (0, 0) \oplus (0, 1) \oplus 2(0, 2) \oplus (\frac{1}{2}, \frac{3}{2}) \! $ \\ \hline
		$ (1, 2, 3) $ & $ 2(0,0) \oplus 3(0,1) \oplus 2(0,2) \oplus (\frac{1}{2},\frac{1}{2}) \oplus (\frac{1}{2},\frac{3}{2}) $ & $ (1, 3, 1) $ & $ (0,\frac{5}{2}) \oplus (0,\frac{7}{2}) \oplus (\frac{1}{2},4) $ \\ \hline
		$ (1, 3, 2) $ & $ 2(0,\frac{3}{2}) \oplus 4(0,\frac{5}{2}) \oplus 3(0,\frac{7}{2}) \oplus (\frac{1}{2},2) \oplus 2(\frac{1}{2},3) \oplus 2(\frac{1}{2},4) \oplus (1,\frac{7}{2}) $ & $ (1, 3, 3) $ & $ 2(0,\frac{1}{2}) \oplus 6(0,\frac{3}{2}) \oplus 8(0,\frac{5}{2}) \oplus 4(0,\frac{7}{2}) \oplus (\frac{1}{2},1) \oplus 4(\frac{1}{2},2) \oplus 4(\frac{1}{2},3) \oplus 2(\frac{1}{2},4) \oplus (1,\frac{5}{2}) \oplus (1,\frac{7}{2}) $ \\ \hline
		$ (2, 1, 3) $ & $ 2(0, 0) \oplus (\frac{1}{2}, \frac{1}{2}) $ & $ (2, 2, 3) $ & $ (0,\frac{1}{2}) \oplus (0,\frac{3}{2}) \oplus (\frac{1}{2},1) $ \\ \hline
		$ (2, 3, 2) $ & $ (0, 3) $ & $ (2, 3, 3) $ & $ (0,1) \oplus 3(0,2) \oplus 3(0,3) \oplus (\frac{1}{2},\frac{3}{2}) \oplus 2(\frac{1}{2},\frac{5}{2}) \oplus (\frac{1}{2},\frac{7}{2}) \oplus (1,3) $ \\ \hline
	\end{tabular}
	\caption{BPS spectrum of $ \mathbb{P}^2 \cup \mathbb{F}_3 + 1\mathbf{Sym} $ for $ d_1 \leq 2 $, $ d_2, d_3 \leq 3 $. Here, $ \mathbf{d} = (d_1, d_2, d_3) $ labels the BPS state with charge $ d_1(\tilde{m}-2\tilde{\phi}_2) + d_2(3\tilde{\phi}_1 - \tilde{\phi}_2) + d_3(-\tilde{\phi}_1 + 2\tilde{\phi}_2) $. The $ d_1 = 0 $ sector is the same with that of $ \mathbb{P}^2 \cup \mathbb{F}_3 $.} \label{table:P2-F3+1Sym}
\end{table}

\subsubsection{\texorpdfstring{$ \mathbb{P}^2 \cup \mathbb{F}_6 + 1\mathbf{Sym} $}{P2 U F6 + 1Sym}}

\begin{figure}
	\centering
	\begin{tikzpicture}
	\draw[thick] (0.5, 0.5) -- (1, 1) -- (1, 2.5) -- (2.5, 1) -- (3.5, 0.5) -- (5, 0)
	(1, 1) -- (2.5, 1)
	(3.5, 0.5) -- (0.5, 0.5) -- (0.5, 3.5) -- (1, 2.5)
	(0.5, 3.5) -- (0.333, 4) -- (0.333, 4.5)
	(0.333, 4) -- (0.083, 4.5);
	\draw[thick, double] (0, 0) -- (0.5, 0.5)
	(7.5, 0) -- (6, 0.5);
	\draw[dashed] (-1, 0) -- (8, 0);
	\filldraw[fill=white, thick] (5, 0) circle (0.15);
	\draw (5, -0.4) node {\scriptsize{$ O7^+ $}}
	(1.7, 3.8) node {\scriptsize{$ -m_0 - \phi_2 + \frac{5}{2}m_1 $}}
	(1.7, 3) node {\scriptsize{$ -\phi_1 + 2\phi_2 $}}
	(3.4, 1.9) node {\scriptsize{$ m_0 + 3\phi_1 - 2\phi_2 - \frac{5}{2}m_1 $}}
	(5, 0.45) node {\scriptsize{$ m_1 - 2\phi_2 $}};
	\end{tikzpicture}
	\caption{A 5-brane web for $ SU(3)_{-1/2} + 1\mathbf{Sym} $, after flopping the edges with lengths $ m_0 + \phi_1 - \phi_2 - \frac{5}{2}m_1 $ and $ m_0 + \phi_2 -\frac{5}{2}m_1 $.} \label{fig:P2-F6-Sym}
\end{figure}
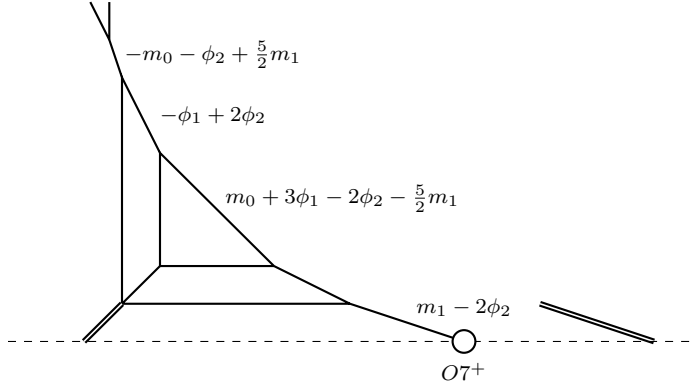

The theory $ \mathbb{P}^2 \cup \mathbb{F}_6 + 1\mathbf{Sym} $ is a non-Lagrangian theory obtained from the KK-theory $ SU(3)_0 + 1\mathbf{Sym} + 1\mathbf{F} $ by integrating out the fundamental hypermultiplet and an instantonic hypermultiplet together \cite{Bhardwaj:2019jtr}. The corresponding 5-brane web is given in Figure \ref{fig:P2-F6-Sym}. This theory can also be obtained from a non-perturbative Higgsing on  $SU(4)_0+1\mathbf{Sym}$. See Appendix~\ref{sec:app-P^2F_6+1Sym} for 5-brane construction and generalizations.

From $ SU(3)_0 + 1\mathbf{Sym} + 1\mathbf{F} $, we can first integrate out the fundamental matter. The IR theory then becomes the $ SU(3)_{-1/2} + 1\mathbf{Sym} $ theory, whose effective prepotential is given by
\begin{align}
\mathcal{E}_{SU(3)} & = \frac{1}{\epsilon_1 \epsilon_2} \qty( \mathcal{F}_{SU(3)} - \frac{\epsilon_1^2 + \epsilon_2^2}{24}(2\phi_1 + 2\phi_2 - 3m_1) + \epsilon_+^2 (\phi_1 + \phi_2))
 \,, \nonumber \\
6\mathcal{F}_{SU(3)}
&= 8\phi_1^3 - 3\phi_1^2 \phi_2 - 3\phi_1 \phi_2^2 + 8\phi_2^3 - \frac{3}{2} \phi_1 \phi_2 (\phi_1 - \phi_2) + 6m_0 (\phi_1^2 - \phi_1 \phi_2 + \phi_2^2) \nonumber \\
& \quad - \frac{1}{2} \Big( (2\phi_1 + m_1)^3 + (\phi_2 + m_1)^3 + (-2\phi_1 + 2\phi_2 + m_1)^3 \nonumber \\
& \qquad \quad + (\phi_1 - \phi_2 + m_1)^3 + (-\phi_1 + m_1)^3 + (-2\phi_1 + m_1)^3 \Big)\ ,
\end{align}
where $ m_0 $ is the gauge coupling and $ m_1 $ is the mass parameter of the symmetric matter. 

This theory has a set of consistent magnetic fluxes
\begin{align}
n_i \in \mathbb{Z} \, , \quad
B_{m_0} = -1/4 \, , \quad
B_{m_1} = 1/2 \ ,
\end{align}
which gives a unity blowup equation. Solving the blowup equation yields the BPS spectrum of this theory.

Among the BPS states in $ SU(3)_{-1/2} + 1\mathbf{Sym} $ theory, there are instantonic hypermultiplets whose masses are $ m_0 + \phi_1 - \phi_2 - \frac{5}{2}m_1 $ and $ m_0 + \phi_2 - \frac{5}{2}m_1 $. In terms of $ SU(3)_0 + 1\mathbf{Sym} + 1\mathbf{F} $ theory given in Table~\ref{table:SU3_sym_F}, they correspond to $ \mathbf{d} = (1, \frac{1}{3}, -\frac{1}{3}, -\frac{5}{2}, -\frac{1}{2}) $ and $ \mathbf{d} = (1, \frac{1}{3}, \frac{2}{3}, -\frac{5}{2}, -\frac{1}{2}) $, respectively. We can perform flop transitions for these two hypermultiplets to get a 5-brane web shown in Figure~\ref{fig:P2-F6-Sym}.

The $ SU(3)_{-1/2} + 1\mathbf{Sym} $ theory has an RG flow to the $ \mathbb{P}^2 \cup \mathbb{F}_6 + 1\mathbf{Sym} $ theory. The RG flow is generated by integrating out the instantonic hypermultiplet corresponding to the limit $ -m_0 - \phi_2 + \frac{5}{2}m_1 \to \infty $, while keeping the lengths of other edges finite, in the brane web. To see this, it is convenient to use the parameters as
\begin{align}\label{eq:P2-F6-var-redef}
\tilde{\phi}_1 = \phi_1 + \frac{m_0}{2} - \frac{5}{4}m_1 \, , \quad
\tilde{\phi}_2 = \phi_2 + \frac{m_0}{4} - \frac{5}{8}m_1 \, , \quad
\tilde{m} = \frac{m_0}{2} - \frac{m_1}{4} \ .
\end{align}
Then volumes of 2-cycles in Figure~\ref{fig:P2-F6-Sym} become
\begin{align}
&-m_0 - \phi_2 + \frac{5}{2}m_1 = -\tilde{\phi}_2 + 3m_0 - \frac{15}{2}\tilde{m} \, , \quad
m_0 + 3\phi_1 - 2\phi_2 - \frac{5}{2}m_1 = 3\tilde{\phi}_1 - 2\tilde{\phi}_2 \, , \nonumber \\
&-\phi_1 + 2\phi_2 = -\tilde{\phi}_1 + 2\tilde{\phi}_2 \, , \quad
m_1 - 2\phi_2 = \tilde{m} - 2\tilde{\phi}_2 \ .
\end{align}
The RG-flow amounts to the limit $ m_0 \to \infty $ while $ \tilde{\phi}_i $ and $ \tilde{m} $ are kept finite.

The cubic prepotential and mixed gravitational Chern-Simons term of the $ \mathbb{P}^2 \cup \mathbb{F}_6 + 1\mathbf{Sym} $ theory after the RG flow can be obtained from those of the UV $ SU(3)_{-1/2} + 1\mathbf{Sym} $ theory as follows:
\begin{align}
\mathcal{F}_{IR} &= \mathcal{F}_{SU(3)} + \frac{1}{6} \qty(m_0 + \phi_1 - \phi_2 - \frac{5}{2}m_1)^3 + \frac{1}{6} \qty(m_0 + \phi_2 - \frac{5}{2}m_1)^3 \, , \\
C_{IR} &= C_{SU(3)} - 2 \qty(m_0 + \phi_1 - \phi_2 - \frac{5}{2}m_1) - 2 \qty(m_0 + \phi_2 - \frac{5}{2}m_1) \ .
\end{align}
The mixed gauge/$ SU(2)_R $ Chern-Simons term is unchanged along the RG flow. So, in terms of the parameters in \eqref{eq:P2-F6-var-redef}, the effective prepotential of the IR theory is
\begin{align}
\mathcal{E}_{IR} &= \frac{1}{\epsilon_1 \epsilon_2} \qty(\mathcal{F}_{IR} - \frac{\epsilon_1^2 + \epsilon_2^2}{48}(6\tilde{\phi}_1 + 4\tilde{\phi}_2) + \epsilon_+^2 (\tilde{\phi}_1 + \tilde{\phi}_2)) \, , \nonumber \\
6\mathcal{F}_{IR} &= 9\tilde{\phi}_1^3 - 18\tilde{\phi}_1^2 \tilde{\phi}_2 + 12\tilde{\phi}_1 \tilde{\phi}_2^2 + 8\tilde{\phi}_2^3 \ .
\end{align}
This prepotential is the same as that of the $ \mathbb{P}^2 \cup \mathbb{F}_6 $ theory, but one should remember that this theory has an additional hypermultiplet coming from the symmetric matter of the parent theory whose GV-invariant is given by
\begin{align}
\mathcal{Z}_{\mathrm{hyper}}(\tilde{\phi}_i, \tilde{m}; \epsilon_{1,2})
= \PE \qty[\frac{(p_1 p_2)^{1/2}}{(1-p_1)(1-p_2)} e^{-(\tilde{m}-2\tilde{\phi}_2)}] \ .
\end{align}

We find a unity blowup equation for this SCFT with the magnetic fluxes
\begin{align}
n_1 \in \mathbb{Z} - 3/4 \, , \quad
n_2 \in \mathbb{Z} - 3/8 \, , \quad
B_{\tilde{m}} = -1/4 \, .
\end{align}
The solution to the blowup equation is listed in Table~\ref{table:P2-F6+1Sym}. We checked that this result matches the BPS spectrum that can be obtained from the RG flow of the $ SU(3)_{-1/2} + 1\mathbf{Sym} $ spectrum after integrating out an instantonic hypermultiplet.

\begin{table}
	\centering
	\begin{tabular}{|c|C{28ex}||c|C{28ex}|} \hline
		$\mathbf{d}$ & $\oplus N_{j_l, j_r}^{\mathbf{d}} (j_l, j_r)$ & $\mathbf{d}$ & $\oplus N_{j_l, j_r}^{\mathbf{d}} (j_l, j_r)$ \\ \hline
		$ (1, 0, 0) $ & $ (0, 0) $ & $ (1, 0, 1) $ & $ (0, 0) $ \\ \hline
		$ (1, 0, 2) $ & $ (0, 0) $ & $ (1, 1, 1) $ & $ (0, \frac{1}{2}) $ \\ \hline
		$ (1, 1, 2) $ & $ 2(0, \frac{1}{2}) \oplus (0, \frac{1}{2}) $ & $ (1, 1, 3) $ & $ (0, \frac{1}{2}) $ \\ \hline
		$ (1, 2, 1) $ & $ (0, 2) $ & $ (1, 2, 2) $ & $ \! (0,0) \oplus (0,1) \oplus 2(0,2) \oplus (\frac{1}{2},\frac{3}{2}) \! $ \\ \hline
		$ (1, 2, 3) $ & $ 2(0,0) \oplus 3(0,1) \oplus 2(0,2) \oplus (\frac{1}{2},\frac{1}{2}) \oplus (\frac{1}{2},\frac{3}{2}) $ & $ (1, 3, 1) $ & $ (0,\frac{5}{2}) \oplus (0,\frac{7}{2}) \oplus (\frac{1}{2},4) $ \\ \hline
		$ (1, 3, 2) $ & $ 2(0,\frac{3}{2}) \oplus 4(0,\frac{5}{2}) \oplus 3(0,\frac{7}{2}) \oplus (\frac{1}{2},2) \oplus 2(\frac{1}{2},3) \oplus 2(\frac{1}{2},4) \oplus (1,\frac{7}{2}) $ & $ (1, 3, 3) $ & $ 2(0,\frac{1}{2}) \oplus 6(0,\frac{3}{2}) \oplus 8(0,\frac{5}{2}) \oplus 4(0,\frac{7}{2}) \oplus (\frac{1}{2},1) \oplus 4(\frac{1}{2},2) \oplus 4(\frac{1}{2},3) \oplus 2(\frac{1}{2},4) \oplus (1,\frac{5}{2}) \oplus (1,\frac{7}{2}) $ \\ \hline
		$ (2, 2, 3) $ & $ (0, \frac{1}{2}) \oplus (0, \frac{3}{2}) \oplus (\frac{1}{2}, 1) $ & $ (2, 3, 2) $ & $ (0, 3) $ \\ \hline
		$ (2, 3, 3) $ & \multicolumn{3}{c|}{$ (0,1) \oplus 3(0,2) \oplus 3(0,3) \oplus (\frac{1}{2},\frac{3}{2}) \oplus 2(\frac{1}{2},\frac{5}{2}) \oplus (\frac{1}{2},\frac{7}{2}) \oplus (1,3) $} \\ \hline
	\end{tabular}
	\caption{BPS spectrum of $ \mathbb{P}^2 \cup \mathbb{F}_6 + 1\mathbf{Sym} $ for $ d_1 \leq 2 $ and $ d_2, d_3 \leq 3 $. Here, $ \mathbf{d} = (d_1, d_2, d_3) $ labels the BPS state with charge $ d_1(\tilde{m}-2\tilde{\phi}_2) + d_2(3\tilde{\phi}_1 - 2\tilde{\phi}_2) + d_3(-\tilde{\phi}_1 + 2\tilde{\phi}_2) $. The $ d_1 = 0 $ sector is the same with that of $ \mathbb{P}^2 \cup \mathbb{F}_6 $.} \label{table:P2-F6+1Sym}
\end{table}

\subsubsection{\texorpdfstring{$SU(3)_8$}{SU(3)8}}\label{sec:SU3_8}

The $SU(3)$ gauge theory with the CS-level $ 8 $ is geometrically realized as
\begin{align}\label{fig:SU3_8}
\begin{tikzpicture}
\draw[thick](-3,0)--(0,0);	
\node at(-3.6,0) {$\mathbb{F}_{9}{}\big|_1$};
\node at(-2.8,0.3) {${}_e$};
\node at(0.5,0) {$\mathbb{F}_{1}{}\big|_2$};
\node at(-0.5,0.3) {${}_{h+3f}$};
\end{tikzpicture}
\end{align}
This geometry is non-shrinkable because the volumes of  the primitive curves
\begin{align}\label{eq:SU3_8_vol}
\vol (f_1) = 2\phi_1 - \phi_2 \, , \quad
\vol (f_2) = -\phi_1 + 2\phi_2 \, , \quad
\vol (e_2) = -3\phi_1 + \phi_2 + m\ ,
\end{align}
cannot all be non-negative at the same time in the limit $m\rightarrow0$, and thus this theory cannot have a UV completion in the geometric phase \cite{Jefferson:2018irk}. However, it was pointed out in \cite{Bhardwaj:2019jtr} that this theory can be obtained from an RG flow of the UV $SU(3)_{15/2}+1{\bf F}$ theory and the UV theory is dual to the $\mathcal{N}=2$ $G_2$ gauge theory (or $G_2+1\mathbf{Adj}$). So the duality of its parent theory ensures that the $SU(3)_8$ theory has a consistent UV completion though it has no unitary geometric realization having non-trivial Coulomb branch in the UV limit.

We will now compute the BPS spectrum of this theory by solving blowup equations in the geometric (and also gauge theory) phase and show that this theory has a unitary phase with non-trivial Coulomb branch which can be directly connected to the UV fixed point. The effective prepotential is given by
\begin{align}
\mathcal{E} &= \frac{1}{\epsilon_1 \epsilon_2} \qty(\mathcal{F} - \frac{\epsilon_1^2 + \epsilon_2^2}{12}(\phi_1 + \phi_2) + \epsilon_+^2 (\phi_1 + \phi_2)) \, , \nonumber \\
6\mathcal{F} &= 8\phi_1^3 + 21\phi_1^2 \phi_2 - 27\phi_1 \phi_2^2 + 8\phi_2^3 + 6m (\phi_1^2 - \phi_1 \phi_2 + \phi_2^2) \ .
\end{align}
We find a set of consistent magnetic fluxes given by
\begin{align}
n_1 \in \mathbb{Z} + 1/3 \, , \quad
n_2 \in \mathbb{Z} + 2/3 \, , \quad
B_{m} = -1/6 \ .
\end{align}
This set can be used to formulate a unity blowup equation. The solution to the blowup equation is listed in Table~\ref{table:SU(3)_8}. Here, we have fixed $N^{(1,1,0)}_{0,0}$ using the RG flow from the spectrum of the $SU(3)_{15/2}+1{\bf F}$ theory since it remains undetermined until $e^{-3m}$ order, though it will possibly be fixed in higher order computations. All other states are fixed by solving the blowup equation and the result is consistent with the expected RG flow from the spectrum of the  parent theories, the $SU(3)_{15/2}+1{\bf F}$ theory and the $\mathcal{N}=2$ $G_2$ theory.

\begin{table}
	\centering
	\begin{tabular}{|c|C{28ex}||c|C{28ex}|} \hline
		$ \mathbf{d} $ & $ \oplus N_{j_l, j_r}^{\mathbf{d}} (j_l, j_r) $ & $ \mathbf{d}$ & $\oplus N_{j_l, j_r}^{\mathbf{d}} (j_l, j_r) $ \\ \hline
		$ (1, 0, 0) $ & $ (0, 0) $ & $ (1, 0, 1) $ & $ (0, 1) $ \\ \hline
		$ (1, 0, 2) $ & $ (0, 2) $ & $ (1, 0, 3) $ & $ (0, 3) $ \\ \hline
		$(1, 1, 0)$ & $(0, 0)$ & $(1, 1, 1)$ & $(0, 0) \oplus (0, 1)$ \\ \hline
		$(1, 1, 2)$ & $(0, 1) \oplus (0, 2)$ & $ (1, 1, 3) $ & $ (0, 2) \oplus (0, 3) $ \\ \hline
		$(1, 2, 0)$ & $(0, 0)$ & $(1, 2, 1)$ & $(0, 0) \oplus (0, 1)$ \\ \hline
		$(1, 2, 2)$ & $(0, 0) \oplus (0, 1) \oplus (0, 2)$ & $ (1, 2, 3) $ & $ (0, 1) \oplus (0, 2) \oplus (0, 3) $ \\ \hline
		$ (1, 3, 0) $ & $ (0, 0) $ & $ (1, 3, 1) $ & $ (0, 0) \oplus (0, 1) $ \\ \hline
		$ (1, 3, 2) $ & $ (0, 0) \oplus (0, 1) \oplus (0, 2) $ & $ (1, 3, 3) $ & $ (0, 0) \oplus (0, 1) \oplus (0, 2) \oplus (0, 3) $ \\ \hline
		$(2, 0, 2)$ & $(0, \frac{5}{2})$ & $ (2, 0, 3) $ & $ (0, \frac{5}{2}) \oplus (0, \frac{7}{2}) \oplus (\frac{1}{2}, 4) $ \\ \hline
		$(2, 1, 2)$ & $(0, \frac{3}{2}) \oplus (0, \frac{5}{2})$ & $ (2, 1, 3) $ & $ (0, \frac{3}{2}) \oplus 3(0, \frac{5}{2}) \oplus 2(0, \frac{7}{2}) \oplus (\frac{1}{2}, 3) \oplus (\frac{1}{2}, 4) $ \\ \hline
		$(2, 2, 1)$ & $(0, \frac{1}{2})$ & $(2, 2, 2)$ & $(0, \frac{1}{2}) \oplus 2(0, \frac{3}{2}) \oplus (0, \frac{5}{2})$ \\ \hline
		$ (2, 2, 3) $ & $ (0, \frac{1}{2}) \oplus 3(0, \frac{3}{2}) \oplus 5(0, \frac{5}{2}) \oplus 2(0, \frac{7}{2}) \oplus (\frac{1}{2}, 2) \oplus (\frac{1}{2}, 3) \oplus (\frac{1}{2}, 4) $ & $ (2, 3, 1) $ & $ 2(0, \frac{1}{2}) \oplus (\frac{1}{2}, 0) $ \\ \hline
		$ (2, 3, 2) $ & $ 3(0, \frac{1}{2}) \oplus 3(0, \frac{3}{2}) \oplus (0, \frac{5}{2}) \oplus (\frac{1}{2}, 1) $ & $ (2, 3, 3) $ & $ 3(0, \frac{1}{2}) \oplus 6(0, \frac{3}{2}) \oplus 6(0, \frac{5}{2}) \oplus 2(0, \frac{7}{2}) \oplus (\frac{1}{2}, 1) \oplus 2(\frac{1}{2}, 2) \oplus (\frac{1}{2}, 3) \oplus (\frac{1}{2}, 4) $ \\ \hline
	\end{tabular}
	\caption{BPS spectrum of $SU(3)_8$ for $d_1 \leq 2$ and $ d_2, d_3 \leq 3 $. Here, $\mathbf{d} = (d_1, d_2, d_3)$ labels the state wrapping curve $d_1 e_2 + d_2 f_1 + d_3 f_2$.} \label{table:SU(3)_8}
\end{table}

From the BPS spectrum, one finds that three hypermultiplets with degree $(1,0,0)$, $(1,1,0)$ and $ (1, 2, 0) $ have negative masses, in the limit $m\rightarrow0$, when we take masses of all other BPS states to be non-negative. This implies that once we flop these three hypermultiplets, we can attain a unitary chamber where all BPS states including the three flopped hypermultiplets have non-negative masses. This suggests that the $SU(3)_8$ theory actually has a consistent UV fixed point, and the unitary chamber we obtained by flop transitions describes the Coulomb branch of the moduli space of the CFT fixed point (without mass deformations). Thus, our computation of the BPS spectrum strongly supports the existence of the UV completion for the $SU(3)_8$ theory.

%% file: sec-Rankhigh.tex
\section{Theories of higher ranks}\label{sec:higher rank theories}

In this section, we apply our bootstrap approach for some interesting higher rank theories and compute their BPS spectra.

\subsection{\texorpdfstring{$SU(4)_8$}{SU(4)8}}\label{sec:SU(4)8}

The first example is the 5d $ SU(4) $ gauge theory with the CS level 8. This theory can be obtained from $\mathbb{Z}_3$ automorphism twist of the 6d minimal $SO(8)$ SCFT
\begin{align}
\begin{tikzpicture}
\draw (0, 0) node {$ SU(4)_8 $}
(1.5, 0) node {$ = $}
(3, 0.3) node {$ \mathfrak{so}(8)^{(3)} $}
(3, -0.3) node {$ 4 $}
;
\end{tikzpicture}
\end{align}
It has a geometric realization as \cite{Bhardwaj:2019fzv}
\begin{align}\label{eq:su4_8_geo}
\begin{tikzpicture}
\draw (0, 0) node {$ \eval{\mathbb{F}_{10}}_1 $}
(4, 0) node {$ \eval{\mathbb{F}_8}_2 $}
(8, 0) node {$ \eval{\mathbb{F}_0}_3 $};
\draw [thick](0.5, 0) -- (3.5, 0);
\draw [thick] (4.5, 0) -- (7.5, 0);
\draw (0.7, 0.3) node {$ _{e} $}
(3.3, 0.3) node {$ _h $}
(4.7, 0.3) node {$ _e $}
(7, 0.3) node {$ _{h + 3f} $}
;
\end{tikzpicture}
\end{align}
It is convenient to use the volumes of primitive 2-cycles in the geometry \eqref{eq:su4_8_geo} as the basis in the computation below,
\begin{align}\label{eq:su4_8_vol}
&\vol (f_1) = 2\phi_1 - \phi_2, 
&&\vol (f_2) = -\phi_1 + 2\phi_2 - \phi_3, \nonumber \\
&\vol (f_3) = -\phi_2 + 2\phi_3,
&&\vol (e_3) = -3\phi_2 + 2\phi_3 + m_0.
\end{align}

We will now solve the blowup equations in both the 5d $SU(4)$ and 6d $SO(8)$ descriptions. In the 5d description, the effective prepotential on the $\Omega$-background is given by
\begin{align}\label{eq:su4_8_prepotential}
\mathcal{E} = & ~\frac{1}{\epsilon_1 \epsilon_2} \qty( \mathcal{F} - \frac{\epsilon_1^2 + \epsilon_2^2}{12} \qty(\phi_1 +\phi_2 +\phi_3) + \epsilon_+^2 (\phi_1 + \phi_2+ \phi_3) )\,, \nonumber \\
6\mathcal{F}
=&~ 8\phi_1^3 + 24\phi_1^2 \phi_2 - 30\phi_1 \phi_2^2 + 8\phi_2^3 + 18\phi_2^2 \phi_3 - 24\phi_2 \phi_3^2 + 8\phi_3^3 \nonumber \\
& + m_0 (\phi_1^2 - \phi_1 \phi_2 + \phi_2^2 - \phi_2 \phi_3 + \phi_3^2)\ ,
\end{align}
where $ m_0 $ is the $SU(4)$ gauge coupling. We find a set of consistent magnetic fluxes as
\begin{align}
n_i \in \mathbb{Z} \ , \quad
B_{m_0} = 0 \ ,
\end{align}
which provides a solvable unity blowup equation. The solution to the blowup equation is summarized in Table~\ref{table:SU4_8}.

We now turn to the 6d description. The subalgebra $G_2$ of $\mathfrak{so}(8)$ is invariant under the $\mathbb{Z}_3$ twist. So in the 5d reduction the 6d $SO(8)$ vector multiplet is decomposed into a $ G_2 $ adjoint with KK charge 0 and two $ G_2 $ fundamentals with KK charges $\tau/3,2\tau/3 $. The perturbative contribution to the GV-invariant is thus given by
\begin{align}
\mathcal{Z}_{\text{1-loop}}
= \PE \Bigg[-\frac{1+p_1 p_2}{(1-p_1)(1-p_2)} \frac{1}{1-q} &\bigg(\sum_{e \in \mathbf{R}^+} e^{-e\cdot \phi} + q^{1/3} \sum_{w \in \mathbf{F}} e^{-w\cdot \phi} \nonumber \\
& + q^{2/3} \sum_{w \in \mathbf{F}} e^{-w\cdot \phi} +  q\sum_{e\in \mathbf{R}^+} e^{e\cdot\phi}\bigg) \Bigg]\,,
\end{align}
where $ \mathbf{R}^+ $ denotes the positive roots and $ \mathbf{F} $ denotes the weights in the fundamental representation of $ G_2 $ algebra. See Appendix~\ref{appendix:1-loop} for more details.

The effective prepotential in the 6d description reads
\begin{align}
\mathcal{E} &= \frac{1}{\epsilon_1 \epsilon_2} \qty( \mathcal{F}_{\mathrm{tree}} + \mathcal{F}_{\text{1-loop}} - \frac{\epsilon_1^2 + \epsilon_2^2}{12}(\phi_1 + \phi_2) + \epsilon_+^2 (\phi_1 + \phi_2 ) )\ , \nonumber \\
\mathcal{F}_{\mathrm{tree}} &= 2\tau \phi_0^2 + \phi_0 \qty(12\phi_1^2 - 12\phi_1 \phi_2 + 4\phi_2^2 -\frac{\epsilon_1^2 + \epsilon_2^2}{2} + 6\epsilon_+^2)\ , \nonumber \\
\mathcal{F}_{\text{1-loop}} &= \frac{4}{3}\phi_1^3 + 3\phi_1^2 \phi_2 - 4\phi_1 \phi_2^2 + \frac{4}{3}\phi_2^3 - \frac{5}{9}\tau (3\phi_1^2 - 3\phi_1 \phi_2 + \phi_2^2) \ . \label{eq:1-loopSO8Aut}
\end{align}
One can check that after taking the shifts $ \phi_1 \to \phi_1 - 2\phi_0 $, $ \phi_2 \to \phi_2 - 3\phi_0 $, where the coefficients of $ \phi_0 $ are the dual Coxeter labels of affine $ D_4^{(3)} $ algebra, the cubic prepotential $\mathcal{F}_{\text{1-loop}}$ reproduces the triple intersection numbers of compact 4-cycles in the geometry \eqref{eq:su4_8_geo}. Also, the 5d effective prepotential in \eqref{eq:su4_8_prepotential} agrees with this 6d effective prepotential after redefining the 5d parameters as 
\begin{align}
\phi_1 \to \phi_0 + \frac{\tau}{9}, \quad
\phi_2 \to \phi_1 + 2\phi_0 - \frac{\tau}{9}, \quad
\phi_3 \to \phi_2 + 3\phi_0 - \frac{\tau}{3}, \quad
m_0 \to \frac{\tau}{3},
\end{align}
up to constant terms.

To compute the BPS spectrum (or the elliptic genus), we can use three unity blowup equations with three sets of consistent magnetic fluxes given by
\begin{align}
n_{1, 2} \in \mathbb{Z} \ , \quad
B_\tau = 0 \ , \quad
n_0 \in \mathbb{Z} + B_0 \quad (B_0 = -1/4,\ 0,\ 1/4) \ ,
\end{align}
which preserve the affine $ D_4^{(3)} $ structure. We can expand the three blowup equations in terms of the instanton string number and solve them at each order to find a closed expression for the elliptic genus of $k$-instanton strings, which is a similar procedure we did for the 6d $ SU(3) $ gauge theory with $ \mathbb{Z}_2 $ twist in section~\ref{subsubsec:SU3/Z2}.
\begin{table}
	\centering
	\begin{tabular}{|c|C{25ex}||c|C{25ex}|} \hline
		$\mathbf{d}$ & $\oplus N_{j_l, j_r}^{\mathbf{d}} (j_l, j_r)$ & $\mathbf{d}$ & $\oplus N_{j_l, j_r}^{\mathbf{d}} (j_l, j_r)$ \\ \hline
		$ (1, 0, 0, 0) $ & $ (0, \frac{1}{2}) $ & $ (1, 0, 0, 1) $ & $ (0, \frac{3}{2}) $ \\ \hline
		$ (1, 0, 0, 2) $ & $ (0, \frac{5}{2}) $ & $ (1, 0, 1, 0) $ & $ (0, \frac{1}{2}) $ \\ \hline
		$ (1, 0, 1, 1) $ & $ (0, \frac{1}{2}) \oplus (0, \frac{3}{2}) $ & $ (1, 0, 1, 2) $ & $ (0, \frac{3}{2}) \oplus (0, \frac{5}{2}) $ \\ \hline
		$ (1, 0, 2, 0) $ & $ (0, \frac{1}{2}) $ & $ (1, 0, 2, 1) $ & $ (0, \frac{1}{2}) \oplus (0, \frac{3}{2}) $ \\ \hline
		$ (1, 0, 2, 2) $ & $ (0, \frac{1}{2}) \oplus (0, \frac{3}{2}) \oplus (0, \frac{5}{2}) $ & $ (1, 1, 1, 0) $ & $ (0, \frac{1}{2}) $ \\ \hline
		$ (1, 1, 1, 1) $ & $ (0, \frac{1}{2}) \oplus (0, \frac{3}{2}) $ & $ (1, 1, 1, 2) $ & $ (0, \frac{3}{2}) \oplus (0, \frac{5}{2}) $ \\ \hline
		$ (1, 1, 2, 1) $ & $ 2(0, \frac{1}{2}) \oplus (0, \frac{3}{2}) $ & $ (1, 1, 2, 2) $ & $ (0, \frac{1}{2}) \oplus 2(0, \frac{3}{2}) \oplus (0, \frac{5}{2}) $ \\ \hline
		$ (1, 2, 2, 0) $ & $ (0, \frac{1}{2}) $ & $ (1, 2, 2, 1) $ & $ (0, \frac{1}{2}) \oplus (0, \frac{3}{2}) $ \\ \hline
		$ (1, 2, 2, 2) $ & $ (0, \frac{1}{2}) \oplus (0, \frac{3}{2}) \oplus (0, \frac{5}{2}) $ & $ (2, 0, 0, 1) $ & $ (0, \frac{5}{2}) $ \\ \hline
		$ (2, 0, 0, 2) $ & $ (0, \frac{5}{2}) \oplus (0, \frac{7}{2}) \oplus (\frac{1}{2}, 4) $ & $ (2, 0, 1, 1) $ & $ (0, \frac{3}{2}) \oplus (0, \frac{5}{2}) $ \\ \hline
		$ (2, 0, 1, 2) $ & $ (0,\frac{3}{2}) \oplus 3(0,\frac{5}{2}) \oplus 2(0,\frac{7}{2}) \oplus 1(\frac{1}{2},3) \oplus (\frac{1}{2},4) $ & $ (2, 0, 2, 1) $ & $ (0,\frac{1}{2}) \oplus (0,\frac{3}{2}) \oplus (0,\frac{5}{2}) $ \\ \hline
		$ (2, 0, 2, 2) $ & $ (0,\frac{1}{2}) \oplus 3(0,\frac{3}{2}) \oplus 4(0,\frac{5}{2}) \oplus 2(0,\frac{7}{2}) \oplus (\frac{1}{2},2) \oplus (\frac{1}{2},3) \oplus (\frac{1}{2},4) $ & $ (2, 1, 1, 1) $ & $ (0, \frac{3}{2}) \oplus (0, \frac{5}{2}) $ \\ \hline
		$ (2, 1, 2, 1) $ & $ (0, \frac{1}{2}) \oplus 2(0, \frac{3}{2}) \oplus (0, \frac{5}{2}) $ & $ (2, 1, 2, 2) $ & $ (0,\frac{1}{2}) \oplus 5(0,\frac{3}{2}) \oplus 7(0,\frac{5}{2}) \oplus 3(0,\frac{7}{2}) \oplus (\frac{1}{2},2) \oplus 2(\frac{1}{2},3) \oplus (\frac{1}{2},4) $ \\ \hline
		$ (2, 2, 2, 1) $ & $ (0, \frac{1}{2}) \oplus (0, \frac{3}{2}) \oplus (0, \frac{5}{2}) $ & $ (2, 2, 2, 2) $ & $ (0,\frac{1}{2}) \oplus 3(0,\frac{3}{2}) \oplus 4(0,\frac{5}{2}) \oplus 2(0,\frac{7}{2}) \oplus (\frac{1}{2},2) \oplus (\frac{1}{2},3) \oplus (\frac{1}{2},4) $ \\ \hline
	\end{tabular}
	\caption{BPS spectrum of the pure $ SU(4)_8 $ theory for $ d_i \leq 2 $. Here, $ \mathbf{d} = (d_1, d_2, d_3, d_4) $ labels the state with charge $ d_1 e_3 + d_2 f_1 + d_3 f_2 + d_4 f_3 $.} \label{table:SU4_8}
\end{table}
We checked in the K\"ahler parameter expansion that the 6d solution perfectly agrees with the 5d result in Table~\ref{table:SU4_8}.

\subsection{\texorpdfstring{6d $ SO(8) $ gauge theory with $ \mathbb{Z}_2 $ twist}{6d SO(8) gauge theory with Z2 twist}}

We can also consider $ \mathbb{Z}_2 $ twisted compactification of the 6d minimal $ SO(8) $ SCFT. This theory has no 5d gauge theory description, but it has a geometric realization given by \cite{Bhardwaj:2019fzv}
\begin{align}
\begin{tikzpicture}\label{eq:so8/z2_geo}
\draw (0, 0) node {$ \eval{\mathbb{F}_{6}}_0 $}
(3, 0) node {$ \eval{\mathbb{F}_1}_1 $}
(6, 0) node {$ \eval{\mathbb{F}_1}_2 $}
(9, 0) node {$ \eval{\mathbb{F}_6}_3 $};
\draw [thick](0.5, 0) -- (2.5, 0)
(3.5, 0) -- (5.5, 0)
(6.5, 0) -- (8.5, 0);
\draw (0.7, 0.3) node {$ _{e} $}
(2.3, 0.3) node {$ _{2h} $}
(3.7, 0.3) node {$ _e $}
(5.3, 0.3) node {$ _e $}
(6.7, 0.3) node {$ _{2h} $}
(8.3, 0.3) node {$ _{e} $}
;
\end{tikzpicture}
\end{align}

Upon this compactification, the 6d $SO(8)$ gauge field is decomposed into a 5d gauge field in the adjoint representation and another 5d field carrying KK-charge $\frac{1}{2}$ in the fundamental representation of the invariant subalgebra $ \mathfrak{so}(7)$. The perturbative part of the GV-invariant in this theory can then be written as
\begin{align}
Z_{\mathrm{1-loop}}
= \PE \qty[-\frac{1+p_1 p_2}{(1-p_1)(1-p_2)} \frac{1}{1-q} \qty( \sum_{e \in \mathbf{R}^+} e^{-w\cdot \phi} + q^{1/2} \sum_{w \in \mathbf{F}} e^{-w\cdot \phi} + q \sum_{e \in \mathbf{R}^+} e^{w\cdot \phi} ) ] \ ,
\end{align}
where  $ \mathbf{R}^+ $ denotes the positive roots and $ \mathbf{F} $ denotes the fundamental weights of the $ \mathfrak{so}(7) $ algebra.

The effective prepotential on the $\Omega$-background reads
\begin{align}
\mathcal{E} &= \frac{1}{\epsilon_1 \epsilon_2} \qty( \mathcal{F}_{\mathrm{tree}} + \mathcal{F}_{\text{1-loop}} - \frac{\epsilon_1^2 + \epsilon_2^2}{12}(\phi_1 + \phi_2 + \phi_3) +\epsilon_+^2 (\phi_1 + \phi_2 + \phi_3)  ) \ , \nonumber \\
\mathcal{F}_{\mathrm{tree}} &= 2\tau \phi_0^2 + \phi_0 \qty(4\phi_1^2 - 4\phi_1 \phi_2 + 4\phi_2^2 - 8\phi_2 \phi_3 + 8\phi_3^2 -\frac{\epsilon_1^2 + \epsilon_2^2}{2} + 6\epsilon_+^2)\ , \nonumber \\
\mathcal{F}_{\text{1-loop}} &= \frac{4}{3}\phi_1^3 - \frac{1}{2} \phi_1^2 \phi_2 - \frac{1}{2} \phi_1 \phi_2^2 + \frac{4}{3}\phi_2^3 - 3\phi_2^2 \phi_3 + 2\phi_2 \phi_3^2 + \frac{4}{3}\phi_3^3 \nonumber \\
& \quad  - \frac{3}{4}\tau (\phi_1^2 - \phi_1 \phi_2 + \phi_2^2 - 2\phi_2 \phi_3 + 2\phi_3^2)\ .
\end{align}
One can easily check that the cubic prepotential reproduces the triple intersection numbers of compact 4-cycles in the geometry \eqref{eq:so8/z2_geo} by shifting $ \phi_1 \to \phi_1 - 2\phi_0 $, $ \phi_2 \to \phi_2 - 2\phi_0 $ and $ \phi_3 \to \phi_3 - \phi_0 $.

We find three unity blowup equations from the 3 sets of consistent magnetic fluxes:
\begin{align}
n_{1, 2, 3} \in \mathbb{Z}\ , \quad
B_\tau = 0 \ , \quad
n_0 \in \mathbb{Z} + B_0 \quad (B_0 = -1/8,\ 1/8,\ 3/8) \ .
\end{align}
One can solve these three blowup equations together at each string order and compute a closed expression of the elliptic genus of the 6d $ SO(8) $ gauge theory with $ \mathbb{Z}_2 $ twist. The solution is shown in Table~\ref{table:SO8/Z2}. We checked that this result agrees with the BPS spectrum computed using topological vertex as well as the ADHM calculations in \cite{Kim:2021cua}.
\begin{table}
	\centering
	\begin{tabular}{|c|C{23ex}||c|C{23ex}|} \hline
		$\mathbf{d}$ & $\oplus N_{j_l, j_r}^{\mathbf{d}} (j_l, j_r)$ & $\mathbf{d}$ & $\oplus N_{j_l, j_r}^{\mathbf{d}} (j_l, j_r)$ \\ \hline
		$ (1, 0, 2, 3, 3) $ & $ (0, 0) $ & $ (1, 0, 3, 3, 3) $ & $ (0, 1) $ \\ \hline
		$ (1, \frac{1}{2}, 2, 2, 2) $ & $ (0, 0) \oplus (0, 1) $ & $ (1, \frac{1}{2}, 2, 3, 2) $ & $ (0, 0) \oplus (0, 1) $ \\ \hline
		$ (1, \frac{1}{2}, 2, 3, 3) $ & $ (0, 0) \oplus (0, 1) $ & $ (1, \frac{1}{2}, 3, 2, 2) $ & $ (0, 1) \oplus (0, 2) $ \\ \hline
		$ (1, \frac{1}{2}, 3, 3, 2) $ & $ (0, 0) \oplus 2(0, 1) \oplus (0, 2) $ & $ (1, \frac{1}{2}, 3, 3, 3) $ & $ (0, 0) \oplus 2(0, 1) \oplus (0, 2) $ \\ \hline
		$ (1, 1, 1, 1, 1) $ & $ (0, 1) $ & $ (1, 1, 1, 2, 1) $ & $ (0, 0) \oplus (0, 1) $ \\ \hline
		$ (1, 1, 1, 2, 2) $ & $ (0, 0) \oplus (0, 1) $ & $ (1, 1, 1, 2, 3) $ & $ (0, 0) \oplus (0, 1) $ \\ \hline
		$ (1, 1, 1, 3, 1) $ & $ (0, 1) \oplus (0, 2) $ & $ (1, 1, 1, 3, 2) $ & $ (0, 0) \oplus 2(0, 1) \oplus (0, 2) $ \\ \hline
		$ (1, 1, 1, 3, 3) $ & $ 2(0, 0) \oplus 3(0, 1) \oplus (0, 2) $ & $ (1, 1, 2, 1, 1) $ & $ (0, 0) \oplus (0, 1) \oplus (0, 2) $ \\ \hline
		$ (1, 1, 2, 2, 1) $ & $ 2(0, 0) \oplus 3(0, 1) \oplus (0, 2) $ & $ (1, 1, 2, 2, 2) $ & $ 2(0, 0) \oplus 3(0, 1) \oplus (0, 2) $ \\ \hline
		$ (1, 1, 2, 2, 3) $ & $ 2(0, 0) \oplus 3(0, 1) \oplus (0, 2) $ & $ (1, 1, 2, 3, 1) $ & $ \! 2(0, 0) \oplus 3(0, 1) \oplus 2(0, 2) \! $ \\ \hline
		$ (1, 1, 2, 3, 2) $ & $ \! 4(0, 0) \oplus 6(0, 1) \oplus 2(0, 2) \! $ & $ (1, 1, 2, 3, 3) $ & $ 9(0, 0) \oplus 11(0, 1) \oplus 3(0, 2) \oplus (\frac{1}{2}, \frac{1}{2}) $ \\ \hline
		$ (1, 1, 3, 1, 1) $ & $ (0, 1) \oplus (0, 2) \oplus (0, 3) $ & $ (1, 1, 3, 2, 1) $ & $ (0, 0) \oplus 3(0, 1) \oplus 3(0, 2) \oplus (0, 3) $ \\ \hline
		$ (1, 1, 3, 2, 2) $ & $ (0, 0) \oplus 3(0, 1) \oplus 3(0, 2) \oplus (0, 3) $ & $ (1, 1, 3, 2, 3) $ & $ (0, 0) \oplus 3(0, 1) \oplus 3(0, 2) \oplus (0, 3) $ \\ \hline
		$ (1, 1, 3, 3, 1) $ & $ 2(0, 0) \oplus 5(0, 1) \oplus 3(0, 2) \oplus (0, 3) $ & $ (1, 1, 3, 3, 2) $ & $ 4(0, 0) \oplus 8(0, 1) \oplus 5(0, 2) \oplus (0, 3) $ \\ \hline
		$ (1, 1, 3, 3, 3) $ & \multicolumn{3}{c|}{$ 8(0, 0) \oplus 17(0, 1) \oplus 10(0, 2) \oplus 2(0, 3) \oplus (\frac{1}{2}, \frac{1}{2}) \oplus (\frac{1}{2}, \frac{3}{2}) $} \\ \hline
	\end{tabular}
	\caption{BPS spectrum of $ SO(8) $ gauge theory with $ \mathbb{Z}_2 $ twist for $ d_1 = 1 $, $ d_2 \leq 1 $ and $ d_{3,4, 5} \leq 3 $. Here, $ \mathbf{d} = (d_1, d_2, d_3, d_4, d_5) $ labels BPS states with charge $ d_1 \Phi + d_2 \tau + d_3 \alpha_1 + d_4 \alpha_2 + d_5 \alpha_3 $, where $ \alpha_1 = 2\phi_1-\phi_2 $, $ \alpha_2 = -\phi_1 + 2\phi_2 - 2\phi_3 $, $ \alpha_3 = -\phi_2 + 2\phi_3 $ are simple roots of $ \mathfrak{so}(7) $ algebra.} \label{table:SO8/Z2}
\end{table}

\subsection{\texorpdfstring{$ SU(5)_8 $: undetermined}{SU(5)8}}

Our last example is the 5d $ SU(5) $ gauge theory with the CS level $ 8 $. This theory was classified as an `undetermined theory' in \cite{Bhardwaj:2020gyu} since its UV completion has not been identified. Following our bootstrapping procedure, we found that this theory may have no consistent UV completion. We now explain it below.

To begin with, the geometric realization of this theory is given by 
\begin{align}\label{eq:su5_8}
\begin{tikzpicture}
\draw (0, 0) node {$ \eval{\mathbb{F}_{11}}_1 $}
(3, 0) node {$ \eval{\mathbb{F}_9}_2 $}
(6, 0) node {$ \eval{\mathbb{F}_7}_3 $}
(9, 0) node {$ \eval{\mathbb{F}_1}_4 $};
\draw [thick](0.5, 0) -- (2.5, 0)
(3.5, 0) -- (5.5, 0)
(6.5, 0) -- (8.5, 0);
\draw (0.7, 0.3) node {$ _{e} $}
(2.3, 0.3) node {$ _{h} $}
(3.7, 0.3) node {$ _e $}
(5.3, 0.3) node {$ _h $}
(6.7, 0.3) node {$ _{e} $}
(8.1, 0.3) node {$ _{h+2f} $}
;
\end{tikzpicture} \, .
\end{align}
However, this geometry is not shrinkable because it does not have  non-trivial Coulomb branch where the volumes of primitive 2-cycles 
\begin{align}\label{eq:su5_8_vol}
\vol (f_1) & = 2\phi_1 - \phi_2, \quad
&\vol (f_2) &= -\phi_1 + 2\phi_2 - \phi_3, \quad
&\vol (f_3) &= -\phi_2 + 2\phi_3 - \phi_4, \nonumber \\
\vol (f_4) &= -\phi_3 + 2\phi_4, \quad
&\vol (e_4) &= -2\phi_3 + \phi_4 + m_0,&&
\end{align}
are all non-negative simultaneously in the UV limit where $ m_0 \to 0 $. Thus, from the geometric perspective, it is unclear whether this theory is UV completable or not.

\begin{table}
	\centering
	\begin{tabular}{|c|C{25ex}||c|C{25ex}|} \hline
		$\mathbf{d}$ & $\oplus N_{j_l, j_r}^{\mathbf{d}} (j_l, j_r)$ & $\mathbf{d}$ & $\oplus N_{j_l, j_r}^{\mathbf{d}} (j_l, j_r)$ \\ \hline
			$ (1, 0, 0, 0, 0) $ & $ (0, 0) $ & $ (1, 0, 0, 0, 1) $ & $ (0, 1) $ \\ \hline
			$ (1, 0, 0, 0, 2) $ & $ (0, 2) $ & $ (1, 0, 0, 1, 0) $ & $ (0, 0) $ \\ \hline
			$ (1, 0, 0, 1, 1) $ & $ (0, 0) \oplus (0, 1) $ & $ (1, 0, 0, 1, 2) $ & $ (0, 1) \oplus (0, 2) $ \\ \hline
			$ (1, 0, 1, 1, 0) $ & $ (0, 0) $ & $ (1, 0, 1, 1, 1) $ & $ (0, 0) \oplus (0, 1) $  \\ \hline
			$ (1, 0, 1, 1, 2) $ & $ (0, 1) \oplus (0, 2) $ & $ (1, 1, 1, 1, 0) $ & $ (0, 0) $ \\ \hline
			$ (1, 1, 1, 1, 1) $ & $ (0, 0) \oplus (0, 1) $ & $ (2, 0, 0, 0, 2) $ & $ (0, \frac{5}{2}) $ \\ \hline
			$ (2, 0, 0, 1, 2) $ & $ (0, \frac{3}{2}) \oplus (0, \frac{5}{2}) $ & $ (2, 0, 1, 1, 2) $ & $ (0, \frac{3}{2}) \oplus (0, \frac{5}{2}) $ \\ \hline
			$ (2, 1, 1, 1, 2) $ & $ (0, \frac{3}{2}) \oplus (0, \frac{5}{2}) $ & $ \vdots$ & $ \vdots $ \\ \hline
			$ \vdots $ & $ \vdots $ & $ (6, 1, 4, 8, 2) $ & $ (0, \frac{1}{2}) $ \\ \hline
	\end{tabular}
	\caption{BPS spectrum of the pure $ SU(5)_8 $ theory up to $ d_1, d_5 \leq 2 $, $ d_2, d_3, d_4 \leq 1$ together with the negative volume state of $ \mathbf{d} = (6, 1, 4, 8, 2) $. Here, $ \mathbf{d} = (d_1, d_2, d_3, d_4, d_5) $ labels state with charge $ d_1 e_4 + d_2 f_1 + d_3 f_2 + d_4 f_3 + d_5 f_4 $.} \label{table:SU5_8}
\end{table}

Let us try to solve blowup equations and compute the BPS spectrum of this theory. The effective prepotential for the theory is given by
\begin{align}
\mathcal{E} 
&= \frac{1}{\epsilon_1 \epsilon_2} \qty( \mathcal{F} -\frac{\epsilon_1^2 + \epsilon_2^2}{12}(\phi_1 + \phi_2 + \phi_3 + \phi_4) + \epsilon_+^2 (\phi_1 + \phi_2 + \phi_3 + \phi_4)) \, ,
\nonumber \\
6\mathcal{F}
&= 8\phi_1^3 + 27\phi_1^2 \phi_2 - 33\phi_1 \phi_2^2 + 8\phi_2^3 + 21\phi_2^2 \phi_3 - 27\phi_2 \phi_3^2 + 8\phi_3^3 + 15\phi_3^2 \phi_4 \nonumber \\
& \quad - 21\phi_3 \phi_4^2 + 8\phi_4^3 + 6m_0 (\phi_1^2 - \phi_1 \phi_2 + \phi_2^2 - \phi_2 \phi_3 + \phi_3^2 - \phi_3 \phi_4 + \phi_4^2) \, .
\end{align}
We can set a (trial) unity blowup equation using the following set of magnetic fluxes:
\begin{align}
n_i \in \mathbb{Z} \, , \quad
B_{m_0} = 1/2 \, .
\end{align}
The solution to the blowup equation is summarized in Table \ref{table:SU5_8}.
The solution  involves many negative volume hypermultiplet states at one- and three-instanton orders. As explained, we should flop such states to move on to a unitary chamber where all the BPS states have non-negative masses. For instance, the instantonic hypermultiplets with charges $e_4$, $e_4+f_3$, and $e_4+f_2+f_3$ in Table \ref{table:SU5_8} should be flopped.

After a series of flop transitions, we find that all the BPS states up to certain higher orders can be written as non-negative linear combinations of basis curves given by
\begin{align}\label{eq:SU58floppedNewBasis}
\{ &-e_4 - 2f_1 - 2f_2 - 2f_3,\ -3e_4 - 2f_2 - 4f_3 - f_4,\  -3e_4 - f_1 - f_2 - 4f_3 -f_4, \nonumber \\
&3e_4 + 2f_2 + 5f_3 + f_4,\   6e_4 + f_1 + 4f_2 + 8f_3 + 2f_4  \} \, .
\end{align}
Here the minus signs in the first three curves indicate that the associated hypermultiplets, which exist in the spectrum, are flopped. The spectrum also involves a BPS hypermultiplet wrapping the fourth curve $ 3e_4 + 2f_1 + 5f_3 + f_4 $.

The higher order solution shows that there exists a vector multiplet wrapping the last curve $ 6e_4 + f_1 + 4f_2 + 8f_3 + 2f_4 $. One can then check that the volume of this curve cannot be non-negative while satisfying the non-negative volume conditions $ \vol(f_i) \geq 0 $ for the perturbative vector multiplets. Unfortunately we cannot flop vector multiplets. This implies that the theory with the spectrum we obtained from the blowup equation has no unitary chamber where masses of all the BPS states are non-negative. One may wonder if the $SU(5)_8$ theory has other choices of magnetic fluxes leading to physically consistent BPS spectrum, but we could not find any other solvable blowup equation for this theory. Thus our computation suggests that the 5d $ SU(5)_8 $ theory may have no UV completion.

%% file: sec-conclusion.tex
\section{Conclusion}\label{sec:Conclusion}

In this paper, we proposed a systematic bootstrap method for BPS spectra of 5d $\mathcal{N} = 1$ field theories including KK theories, based on the Nakajima-Yoshioka's blowup equation. As the main input, we introduced the effective prepotential $\mathcal{E}$ and the consistent magnetic fluxes. The effective prepotential incorporates the usual cubic prepotential, the mixed gauge/gravitational, and the mixed gauge/$SU(2)_R$ Chern-Simons terms, which can be readily obtained for every 5d SQFT, as discussed in section~\ref{sec:5dthyOmega}. The consistent magnetic fluxes that one can turn on should satisfy the quantization condition. We discussed possible quantization conditions in section~\ref{sec:BlowupEqReview}. Equipped with these inputs, one can formulate blowup equations and solve them recursively to obtain BPS spectrum for a 5d QFT. We conjecture that our method applies to all the 5d $\mathcal{N}=1$ theories and 6d theories on a circle with/without a twist. To support this we explicitly showed how to bootstrap for all rank-1 and rank-2 theories as well as some of interesting higher rank theories. In particular, we computed BPS spectra of the theories whose partition functions still remain as challenges from other methods such as the ADHM or topological vertex method. For instance, $SU(3)_8$, $SU(4)_8$, and some non-Lagrangian theories.

There are some open questions that beg to be resolved. The first question concerns our main conjecture which asserts that BPS spectra of all UV finite theories in 5d and 6d can be obtained by solving blowup equations. The bootstrapping method proposed in this paper allows us to build a collection of blowup equations for any arbitrary supersymmetric field theory in 5d or 6d. However, physical or mathematical proofs for our conjecture that the blowup equations must be solved to compute the correct BPS spectrum of a QFT are currently lacking. The proof may require a physical derivation of blowup equations and further studies on the structure of blowup equations for general 5d and 6d field theories. We would like to provide more discussions about this in the future.

A related question is whether the blowup equations can always be solved. As shown in \cite{Huang:2017mis}, the blowup equations are solvable if there exists at least one unity blowup equation. There is however a class of theories only admitting blowup equations of the vanishing type, for example, 6d SCFTs with half hypermultiplets. Unfortunately, it is not proven yet whether the vanishing blowup equations are enough to compute full BPS spectra of those theories. Nevertheless, we claimed without proof that though vanishing blowup equations may not be solved by themselves, we can still compute full BPS spectra for those theories from them provided that they are supplemented with other additional constraints such as non-negativity of BPS degeneracies and KK tower structure for 6d theories as well as dualities, geometric descriptions. We have checked this claim for several examples including the 6d minimal $E_7$ SCFT with a half hypermultiplet up to certain lower orders, but we leave the proof of solvability of vanishing blowup equations to future study.

In section~\ref{sec:higher rank theories}, we saw that the $SU(5)_8$ theory has a solvable blowup equation, but its solution shows that this theory has inconsistent BPS spectrum, by which we mean that the theory does not have physical Coulomb branch where all BPS states have non-negative masses in UV. This theory was classified as an `undetermined' theory in \cite{Bhardwaj:2020gyu} as its existence couldn't be proved or disproved yet. Our computation for this theory supports that this theory has no UV completion. Similarly, one can try to compute BPS spectra for other `undetermined' theories and see if they can have non-trivial Coulomb branches. If not, such theories are inconsistent in UV with singularities on their Coulomb branches. Thus we suggest that BPS spectra computed by our bootstrap method for such `undetermined' theories can be used to distinguish UV completable theories.

Partition functions on other supersymmetric backgrounds are also important observables in 5d/6d field theories. Many of them can be factorized into a product of partition functions on the $\Omega$-background localized at fixed points of spatial Lorentz rotations under supersymmetric localization. For instance, a superconformal index which counts BPS operators in a 5d CFT can be understood as a partition function on $S^4\times S^1$ and a supersymmetric localization factorizes it into a product of two  partition functions on $\mathbb{R}^4\times S^1$, each of which we can compute by solving the blowup equations, around the north and the south poles of $S^4$ \cite{Kim:2012gu}. This implies that in principle the blowup equations can be used to compute the superconformal index as well. We notice however that the superconformal index is usually expressed as a power series expansion of the fugacity $x\equiv e^{-\epsilon_+}$ related to conformal dimensions of local operators, and thus it is calculated in the parameter regime where the $\Omega$-parameter $\epsilon_+$ is much bigger than K\"ahler parameters $\phi_i$ and $m_j$ in the theory. On the other hand, we have solved the blowup equations in a different parameter regime where all the K\"ahler parameters are bigger than the $\Omega$-deformation parameters. It would be interesting to see if the blowup equations can be solved in the other parameter regimes, in particular where $\epsilon_+\gg \phi_i,m_j$, so that we can also compute superconformal indices and other observables having similar factorization structures by applying the blowup method.

Lastly, it is intriguing to ask whether the blowup equations can capture asymptotics of BPS partition functions in the large $N$ limit (or in the large rank limit) of 5d and 6d field theories. Asymptotic behavior of supersymmetric partition functions at a large number of degrees of freedom has been playing a crucial role in studying holographic dual theories. Holographic interpretation of the blowup formula in the large $N$ limit, if well-defined, can give us a new implication on physics of dual gravity theories.

\acknowledgments
We would like to thank Lakshya Bhardwaj, Hirotaka Hayashi, Chiung Hwang, Seok Kim, Kimyeong Lee, Houri-Christina Tarazi and Futoshi Yagi for valuable discussions and comments. SSK thanks APCTP, KIAS and POSTECH for his visit where part of this work is done. The research of HK and MK is supported by the POSCO Science Fellowship of POSCO TJ Park Foundation and the National Research Foundation of Korea (NRF) Grant 2018R1D1A1B07042934.

%% file: sec-appendix.tex
\appendix

\section{1-loop prepotentials of 6d SCFTs on a circle with twist} \label{appendix:1-loop}

\subsection{Affine root system and 1-loop prepotential of W-bosons}\label{sec:App-affineroots}

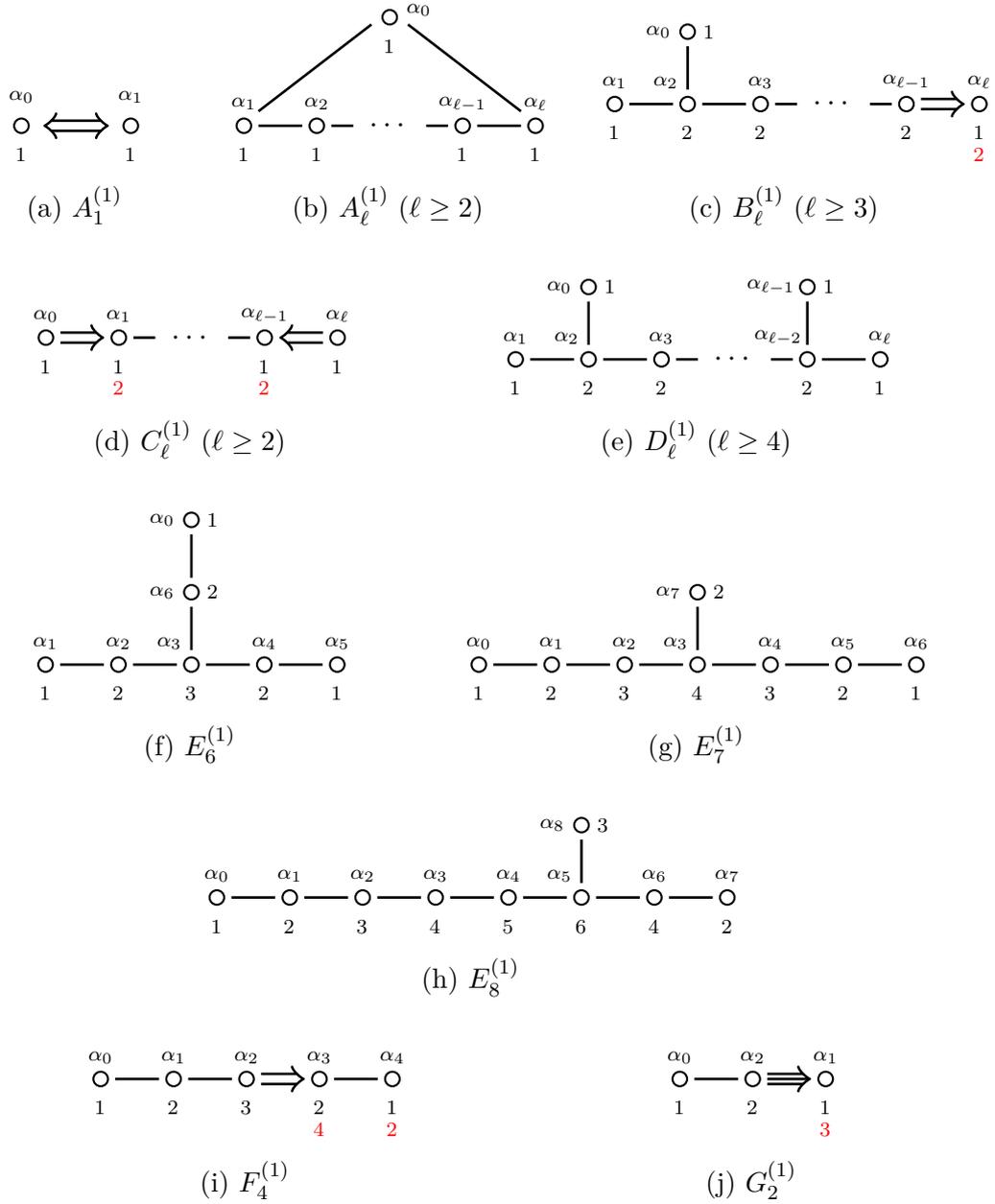
\begin{figure}[t]
	\centering
	\begin{subfigure}[b]{0.2\textwidth}
		\centering
		\begin{tikzpicture}
		\draw[thick] (0, 0) circle (0.1)
		(1.5, 0) circle (0.1);
		\draw[white, decoration={markings,mark=at position 1 with {\arrow[scale=3, black]{>}}},postaction={decorate}] (1,0) -- (1.2,0);
		\draw[white, decoration={markings,mark=at position 1 with {\arrow[scale=3, black]{>}}},postaction={decorate}] (1,0) -- (0.3,0);
		\draw[line width=1pt] (0.41, 0.07) -- (1.09, 0.07)
		(0.41, -0.07) -- (1.09, -0.07);
		\draw (0, -0.4) node {$ _1 $}
		(1.5, -0.4) node {$ _1 $};
		\draw (0, 0.4) node {$ \scriptstyle \alpha_0 $}
		(1.5, 0.4) node {$ \scriptstyle \alpha_1 $};
		\end{tikzpicture}
		\caption{$ A_1^{(1)} $}
		\vspace{3ex}
	\end{subfigure}
	\begin{subfigure}[b]{0.35\textwidth}
		\centering
		\begin{tikzpicture}
		\draw[thick] (0, 0) circle (0.1)
		(1, 0) circle (0.1)
		(2, 0) node {$ \cdots $}
		(3, 0) circle (0.1)
		(4, 0) circle(0.1)
		(2, 1.5) circle (0.1);
		\draw[line width=1pt] (0.2, 0) -- (0.8, 0)
		(1.2, 0) -- (1.5, 0)
		(2.5, 0) -- (2.8, 0)
		(3.2, 0) -- (3.8, 0)
		(0.2, 0.2) -- (1.75, 1.4)
		(3.8, 0.2) -- (2.25, 1.4);
		\draw (0, -0.4) node {$ _1 $}
		(1, -0.4) node {$ _1 $}
		(3, -0.4) node {$ _1 $}
		(4, -0.4) node {$ _1 $}
		(2, 1.1) node {$ _1 $};
		\draw (2.4, 1.6) node {$ \scriptstyle \alpha_0 $}
		(0, 0.3) node {$ \scriptstyle \alpha_1 $}
		(1, 0.3) node {$ \scriptstyle \alpha_2 $}
		(3, 0.3) node {$ \scriptstyle \alpha_{\ell-1} $}
		(4, 0.3) node {$ \scriptstyle \alpha_{\ell} $};
		\end{tikzpicture}
		\caption{$ A_{\ell}^{(1)} \ (\ell \geq 2) $}
		\vspace{3ex}
	\end{subfigure}
	\begin{subfigure}[b]{0.35\textwidth}
		\centering
		\begin{tikzpicture}
		\draw[thick] (0, 0) circle (0.1)
		(1, 0) circle (0.1)
		(2, 0) circle (0.1)
		(3, 0) node {$ \cdots $}
		(4, 0) circle (0.1)
		(5, 0) circle (0.1)
		(1, 1) circle (0.1);
		\draw[line width=1pt] (0.2, 0) -- (0.8, 0)
		(1.2, 0) -- (1.8, 0)
		(2.2, 0) -- (2.5, 0)
		(3.5, 0) -- (3.8, 0)
		(1, 0.2) -- (1, 0.8);
		\draw[white, decoration={markings,mark=at position 1 with {\arrow[scale=3, black]{>}}},postaction={decorate}] (4.5,0) -- (4.8,0);
		\draw[line width=1pt] (4.2, 0.07) -- (4.69, 0.07)
		(4.2, -0.07) -- (4.69, -0.07);
		\draw (0, -0.4) node {$ _1 $}
		(1, -0.4) node {$ _2 $}
		(2, -0.4) node {$ _2 $}
		(4, -0.4) node {$ _2 $}
		(5, -0.4) node {$ _1 $}
		(1.3, 1) node {$ _1 $}
		(5, -0.7) node {\textcolor{red}{$ _2 $}};
		\draw (0.6, 1) node {$ \scriptstyle \alpha_0 $}
		(0, 0.3) node {$ \scriptstyle \alpha_1 $}
		(0.7, 0.3) node {$ \scriptstyle \alpha_2 $}
		(2, 0.3) node {$ \scriptstyle \alpha_3 $}
		(4, 0.3) node {$ \scriptstyle \alpha_{\ell-1} $}
		(5, 0.3) node {$ \scriptstyle \alpha_{\ell} $};
		\end{tikzpicture}
		\caption{$ B_{\ell}^{(1)} \ (\ell \geq 3) $}
		\vspace{3ex}
	\end{subfigure}
	\hfill
	\begin{subfigure}[b]{0.4\textwidth}
		\centering
		\begin{tikzpicture}
		\draw[thick] (0, 0) circle (0.1)
		(1, 0) circle (0.1)
		(2, 0) node {$ \cdots $}
		(3, 0) circle (0.1)
		(4, 0) circle (0.1);
		\draw[line width=1pt] (1.2, 0) -- (1.5, 0)
		(2.5, 0) -- (2.8, 0);
		\draw[white, decoration={markings,mark=at position 1 with {\arrow[scale=3, black]{>}}},postaction={decorate}] (3.5, 0) -- (3.2, 0);
		\draw[white, decoration={markings,mark=at position 1 with {\arrow[scale=3, black]{>}}},postaction={decorate}] (0.5, 0) -- (0.8, 0);
		\draw[line width=1pt] (3.31, 0.07) -- (3.8, 0.07)
		(3.31, -0.07) -- (3.8, -0.07)
		(0.2, 0.07) -- (0.69, 0.07)
		(0.2, -0.07) -- (0.69, -0.07);
		\draw (0, -0.4) node {$ _1 $}
		(1, -0.4) node {$ _1 $}
		(3, -0.4) node {$ _1 $}
		(4, -0.4) node {$ _1 $}
		(1, -0.7) node {\textcolor{red}{$ _2 $}}
		(3, -0.7) node {\textcolor{red}{$ _2 $}};
		\draw (0, 0.3) node {$ \scriptstyle \alpha_{0} $}
		(1, 0.3) node {$ \scriptstyle \alpha_{1} $}
		(3, 0.3) node {$ \scriptstyle \alpha_{\ell-1} $}
		(4, 0.3) node {$ \scriptstyle \alpha_{\ell} $};
		\end{tikzpicture}
		\caption{$ C_{\ell}^{(1)} \ (\ell \geq 2) $}
		\vspace{3ex}
	\end{subfigure}
	\begin{subfigure}[b]{0.5\textwidth}
		\centering
		\begin{tikzpicture}
		\draw[thick] (0, 0) circle (0.1)
		(1, 0) circle (0.1)
		(2, 0) circle (0.1)
		(3, 0) node {$ \cdots $}
		(4, 0) circle (0.1)
		(5, 0) circle (0.1)
		(1, 1) circle (0.1)
		(4, 1) circle (0.1);
		\draw[line width=1pt] (0.2, 0) -- (0.8, 0)
		(1.2, 0) -- (1.8, 0)
		(2.2, 0) -- (2.5, 0)
		(3.5, 0) -- (3.8, 0)
		(4.2, 0) -- (4.8, 0)
		(1, 0.2) -- (1, 0.8)
		(4, 0.2) -- (4, 0.8);
		\draw (0, -0.4) node {$ _1 $}
		(1, -0.4) node {$ _2 $}
		(2, -0.4) node {$ _2 $}
		(4, -0.4) node {$ _2 $}
		(5, -0.4) node {$ _1 $}
		(1.3, 1) node {$ _1 $}
		(4.3, 1) node {$ _1 $};
		\draw (0.6, 1) node {$ \scriptstyle \alpha_{0} $}
		(0, 0.3) node {$ \scriptstyle \alpha_{1} $}
		(0.7, 0.3) node {$ \scriptstyle \alpha_{2} $}
		(2, 0.3) node {$ \scriptstyle \alpha_{3} $}
		(3.6, 0.3) node {$ \scriptstyle \alpha_{\ell-2} $}
		(3.5, 1) node {$ \scriptstyle \alpha_{\ell-1} $}
		(5, 0.3) node {$ \scriptstyle \alpha_{\ell} $};
		\end{tikzpicture}
		\caption{$ D_{\ell}^{(1)} \ (\ell \geq 4) $}
		\vspace{3ex}
	\end{subfigure}
	\hfill
	\begin{subfigure}[b]{0.4\textwidth}
		\centering
		\begin{tikzpicture}
		\draw[thick] (0, 0) circle (0.1)
		(1, 0) circle (0.1)
		(2, 0) circle (0.1)
		(3, 0) circle (0.1)
		(4, 0) circle (0.1)
		(2, 1) circle (0.1)
		(2, 2) circle (0.1);
		\draw[line width=1pt] (0.2, 0) -- (0.8, 0)
		(1.2, 0) -- (1.8, 0)
		(2.2, 0) -- (2.8, 0)
		(3.2, 0) -- (3.8, 0)
		(2, 0.2) -- (2, 0.8)
		(2, 1.2) -- (2, 1.8);
		\draw (0, -0.4) node {$ _1 $}
		(1, -0.4) node {$ _2 $}
		(2, -0.4) node {$ _3 $}
		(3, -0.4) node {$ _2 $}
		(4, -0.4) node {$ _1 $}
		(2.3, 1) node {$ _2 $}
		(2.3, 2) node {$ _1 $};
		\draw (1.6, 2) node {$ \scriptstyle \alpha_{0} $}
		(0, 0.3) node {$ \scriptstyle \alpha_{1} $}
		(1, 0.3) node {$ \scriptstyle \alpha_{2} $}
		(1.7, 0.3) node {$ \scriptstyle \alpha_{3} $}
		(3, 0.3) node {$ \scriptstyle \alpha_{4} $}
		(4, 0.3) node {$ \scriptstyle \alpha_{5} $}
		(1.6, 1) node {$ \scriptstyle \alpha_{6} $};
		\end{tikzpicture}
		\caption{$ E_{6}^{(1)} $}
		\vspace{3ex}
	\end{subfigure}
	\begin{subfigure}[b]{0.5\textwidth}
		\centering
		\begin{tikzpicture}
		\draw[thick] (0, 0) circle (0.1)
		(1, 0) circle (0.1)
		(2, 0) circle (0.1)
		(3, 0) circle (0.1)
		(4, 0) circle (0.1)
		(5, 0) circle (0.1)
		(6, 0) circle (0.1)
		(3, 1) circle (0.1);
		\draw[line width=1pt] (0.2, 0) -- (0.8, 0)
		(1.2, 0) -- (1.8, 0)
		(2.2, 0) -- (2.8, 0)
		(3.2, 0) -- (3.8, 0)
		(4.2, 0) -- (4.8, 0)
		(5.2, 0) -- (5.8, 0)
		(3, 0.2) -- (3, 0.8);
		\draw (0, -0.4) node {$ _1 $}
		(1, -0.4) node {$ _2 $}
		(2, -0.4) node {$ _3 $}
		(3, -0.4) node {$ _4 $}
		(4, -0.4) node {$ _3 $}
		(5, -0.4) node {$ _2 $}
		(6, -0.4) node {$ _1 $}
		(3.3, 1) node {$ _2 $};
		\draw (0, 0.3) node {$ \scriptstyle \alpha_{0} $}
		(1, 0.3) node {$ \scriptstyle \alpha_{1} $}
		(2, 0.3) node {$ \scriptstyle \alpha_{2} $}
		(2.7, 0.3) node {$ \scriptstyle \alpha_{3} $}
		(4, 0.3) node {$ \scriptstyle \alpha_{4} $}
		(5, 0.3) node {$ \scriptstyle \alpha_{5} $}
		(6, 0.3) node {$ \scriptstyle \alpha_{6} $}
		(2.6, 1) node {$ \scriptstyle \alpha_{7} $};
		\end{tikzpicture}
		\caption{$ E_{7}^{(1)} $}
		\vspace{3ex}
	\end{subfigure}
	\hfill
	\begin{subfigure}[b]{0.9\textwidth}
		\centering
		\begin{tikzpicture}
		\draw[thick] (0, 0) circle (0.1)
		(1, 0) circle (0.1)
		(2, 0) circle (0.1)
		(3, 0) circle (0.1)
		(4, 0) circle (0.1)
		(5, 0) circle (0.1)
		(6, 0) circle (0.1)
		(7, 0) circle (0.1)
		(5, 1) circle (0.1);
		\draw[line width=1pt] (0.2, 0) -- (0.8, 0)
		(1.2, 0) -- (1.8, 0)
		(2.2, 0) -- (2.8, 0)
		(3.2, 0) -- (3.8, 0)
		(4.2, 0) -- (4.8, 0)
		(5.2, 0) -- (5.8, 0)
		(6.2, 0) -- (6.8, 0)
		(5, 0.2) -- (5, 0.8);
		\draw (0, -0.4) node {$ _1 $}
		(1, -0.4) node {$ _2 $}
		(2, -0.4) node {$ _3 $}
		(3, -0.4) node {$ _4 $}
		(4, -0.4) node {$ _5 $}
		(5, -0.4) node {$ _6 $}
		(6, -0.4) node {$ _4 $}
		(7, -0.4) node {$ _2 $}
		(5.3, 1) node {$ _3 $};
		\draw (0, 0.3) node {$ \scriptstyle \alpha_{0} $}
		(1, 0.3) node {$ \scriptstyle \alpha_{1} $}
		(2, 0.3) node {$ \scriptstyle \alpha_{2} $}
		(3, 0.3) node {$ \scriptstyle \alpha_{3} $}
		(4, 0.3) node {$ \scriptstyle \alpha_{4} $}
		(4.7, 0.3) node {$ \scriptstyle \alpha_{5} $}
		(6, 0.3) node {$ \scriptstyle \alpha_{6} $}
		(7, 0.3) node {$ \scriptstyle \alpha_{7} $}
		(4.6, 1) node {$ \scriptstyle \alpha_{8} $};
		\end{tikzpicture}
		\caption{$ E_{8}^{(1)} $}
		\vspace{3ex}
	\end{subfigure}
	\hfill
	\begin{subfigure}[b]{0.5\textwidth}
		\centering
		\begin{tikzpicture}
		\draw[thick] (0, 0) circle (0.1)
		(1, 0) circle (0.1)
		(2, 0) circle (0.1)
		(3, 0) circle (0.1)
		(4, 0) circle (0.1);
		\draw[line width=1pt] (0.2, 0) -- (0.8, 0)
		(1.2, 0) -- (1.8, 0)
		(3.2, 0) -- (3.8, 0);
		\draw[white, decoration={markings,mark=at position 1 with {\arrow[scale=3, black]{>}}},postaction={decorate}] (2.5, 0) -- (2.8, 0);
		\draw[line width=1pt] (2.2, 0.07) -- (2.69, 0.07)
		(2.2, -0.07) -- (2.69, -0.07);
		\draw (0, -0.4) node {$ _1 $}
		(1, -0.4) node {$ _2 $}
		(2, -0.4) node {$ _3 $}
		(3, -0.4) node {$ _2 $}
		(4, -0.4) node {$ _1 $}
		(3, -0.7) node {\textcolor{red}{$ _4 $}}
		(4, -0.7) node {\textcolor{red}{$ _2 $}};
		\draw (0, 0.3) node {$ \scriptstyle \alpha_{0} $}
		(1, 0.3) node {$ \scriptstyle \alpha_{1} $}
		(2, 0.3) node {$ \scriptstyle \alpha_{2} $}
		(3, 0.3) node {$ \scriptstyle \alpha_{3} $}
		(4, 0.3) node {$ \scriptstyle \alpha_{4} $};
		\end{tikzpicture}
		\caption{$ F_4^{(1)} $}
	\end{subfigure}
	\begin{subfigure}[b]{0.4\textwidth}
		\centering
		\begin{tikzpicture}
		\draw[thick] (0, 0) circle (0.1)
		(1, 0) circle (0.1)
		(2, 0) circle (0.1);
		\draw[line width=1pt] (0.2, 0) -- (0.8, 0);
		\draw[white, decoration={markings,mark=at position 1 with {\arrow[scale=3, black]{>}}},postaction={decorate}] (1.5, 0) -- (1.8, 0);
		\draw[line width=1pt] (1.2, 0.07) -- (1.69, 0.07)
		(1.2, 0) -- (1.78, 0)
		(1.2, -0.07) -- (1.69, -0.07);
		\draw (0, -0.4) node {$ _1 $}
		(1, -0.4) node {$ _2 $}
		(2, -0.4) node {$ _1 $}
		(2, -0.7) node {\textcolor{red}{$ _3 $}};
		\draw (0, 0.3) node {$ \scriptstyle \alpha_{0} $}
		(1, 0.3) node {$ \scriptstyle \alpha_{2} $}
		(2, 0.3) node {$ \scriptstyle \alpha_{1} $};
		\end{tikzpicture}
		\caption{$ G_2^{(1)} $}
	\end{subfigure}
	\caption{Dynkin diagrams of untwisted affine Lie algebras $ X_\ell^{(1)} $. The number for each node is the dual Coxeter label (comark) $ d_i^\vee $. If a Coxeter label (mark) $ d_i $ differs from $ d_i^\vee $, it is represented as a red number. Dynkin diagrams without the affine node $ \alpha_0 $ reduce to the Dynkin diagrams of simple Lie algebras $ X_\ell $.} 
	\label{fig:dynkin_untwist}
\end{figure}

In this appendix, we determine KK-momentum shifts for perturbative states in 6d theories. When a 6d gauge theory is compactified on a circle, the periodic boundary condition imposed on the gauge algebra $ \mathfrak{g} $ defines a map from $ S^1 $ to $ \mathfrak{g} $. The space of such map is generated by $ T_n^a =  T^a \otimes z^n $ for $ n \in \mathbb{Z} $ where $ T^a $ are the generators of $ \mathfrak{g} $ and $ z $ is the coordinate along $ S^1 $. The natural commutation relation is given by
\begin{align}
\comm*{T_n^a}{T_m^b} = \comm{T^a}{T^b} \otimes z^{n+m} = f^{abc} \,T_{n+m}^c\ ,
\end{align}
where $ f^{abc} $ is the structure constant of $ \mathfrak{g} $. This defines a loop algebra of $ \mathfrak{g} $, and its central extension is called an untwisted affine Lie algebra. Thus, a 6d gauge theory compactified on $ S^1 $ naturally has an affine Lie algebra structure. The untwisted affine Lie algebra for a simple Lie algebra of type $ X_\ell $ is denoted by $ X_\ell^{(1)}$. Dynkin diagrams of the untwisted affine Lie algebras $ X_\ell^{(1)}$ are drawn in Figure~\ref{fig:dynkin_untwist}, where the nodes of the Dynkin diagrams are labeled by the simple roots $\alpha_i$ ($i=1, \cdots, \ell$), while the affine node is labeled by $\alpha_0$. 

There is another type of affine Lie algebras. Lie algebras of types $ A_\ell $, $ D_\ell $ and $ E_6 $ have nontrivial outer automorphisms. Here, an outer automorphism on a Lie algebra $\mathfrak{g}$ is an automorphism which is not an inner automorphism, i.e., conjugation. An outer automorphism of Lie algebras can be viewed as a symmetry of their Dynkin diagrams. For Lie algebras of type $ A_\ell $, their Dynkin diagrams have a $ \mathbb{Z}_2 $ outer automorphism which exchanges the simple roots $ \alpha_i $ and $ \alpha_{\ell-i+1} $. For Lie algebras of type $ D_\ell $, Dynkin diagrams have a $ \mathbb{Z}_2 $ outer automorphism exchanging the simple roots $ \alpha_{\ell-1} $ and $ \alpha_{\ell} $. In particular, the Dynkin diagram of $ D_4 $ has a triality and thus $ D_4 $ algebra additionally has a $ \mathbb{Z}_3 $ outer automorphism whose action on the simple roots is given by $ \alpha_1 \to \alpha_2 \to \alpha_3 \to \alpha_1 $. For the Lie algebra of type $ E_6 $, its Dynkin diagram has a $ \mathbb{Z}_2 $ outer automorphism exchanging  $ \alpha_1 \leftrightarrow \alpha_5 $ and $ \alpha_2 \leftrightarrow \alpha_4 $.

\begin{figure}
\centering
\begin{subfigure}[b]{0.3\textwidth}
	\centering
	\begin{tikzpicture}
	\draw[thick] (0, 0) circle (0.1)
	(1.5, 0) circle (0.1);
	\draw[white, decoration={markings,mark=at position 1 with {\arrow[scale=3, black]{>}}},postaction={decorate}] (1,0) -- (0.3,0);
	\draw[line width=1pt] (0.42, 0.09) -- (1.09, 0.09)
	(0.37, 0.03) -- (1.09, 0.03)
	(0.37, -0.03) -- (1.09, -0.03)
	(0.42, -0.09) -- (1.09, -0.09);
	\draw (0, -0.4) node {$ _1 $}
	(1.5, -0.4) node {$ _2 $}
	(0, -0.7) node {\textcolor{red}{$ _2 $}}
	(1.5, -0.7) node {\textcolor{red}{$ _1 $}};
	\draw (0, 0.4) node {$ \scriptstyle \alpha_0 $}
	(1.5, 0.4) node {$ \scriptstyle \alpha_1 $};
	\end{tikzpicture}
	\caption{$ A_2^{(2)} $}
	\vspace{3ex}
\end{subfigure}
\begin{subfigure}[b]{0.6\textwidth}
	\centering
	\begin{tikzpicture}
	\draw[thick] (0, 0) circle (0.1)
	(1, 0) circle (0.1)
	(2, 0) node {$ \cdots $}
	(3, 0) circle (0.1)
	(4, 0) circle (0.1);
	\draw[line width=1pt] (1.2, 0) -- (1.5, 0)
	(2.5, 0) -- (2.8, 0);
	\draw[white, decoration={markings,mark=at position 1 with {\arrow[scale=3, black]{>}}},postaction={decorate}] (3.5, 0) -- (3.2, 0);
	\draw[white, decoration={markings,mark=at position 1 with {\arrow[scale=3, black]{>}}},postaction={decorate}] (0.5, 0) -- (0.2, 0);
	\draw[line width=1pt] (3.31, 0.07) -- (3.8, 0.07)
	(3.31, -0.07) -- (3.8, -0.07)
	(0.31, 0.07) -- (0.8, 0.07)
	(0.31, -0.07) -- (0.8, -0.07);
	\draw (0, -0.4) node {$ _1 $}
	(1, -0.4) node {$ _2 $}
	(3, -0.4) node {$ _2 $}
	(4, -0.4) node {$ _2 $}
	(0, -0.7) node {\textcolor{red}{$ _2 $}}
	(4, -0.7) node {\textcolor{red}{$ _1 $}};
	\draw (0, 0.3) node {$ \scriptstyle \alpha_{0} $}
	(1, 0.3) node {$ \scriptstyle \alpha_{1} $}
	(3, 0.3) node {$ \scriptstyle \alpha_{\ell-1} $}
	(4, 0.3) node {$ \scriptstyle \alpha_{\ell} $};
	\end{tikzpicture}
	\caption{$ A_{2\ell}^{(2)} \ ( \ell \geq 2) $}
	\vspace{3ex}
\end{subfigure}
\hfill
\begin{subfigure}[b]{0.45\textwidth}
	\centering
	\begin{tikzpicture}
	\draw[thick] (0, 0) circle (0.1)
	(1, 0) circle (0.1)
	(2, 0) circle (0.1)
	(3, 0) node {$ \cdots $}
	(4, 0) circle (0.1)
	(5, 0) circle (0.1)
	(1, 1) circle (0.1);
	\draw[line width=1pt] (0.2, 0) -- (0.8, 0)
	(1.2, 0) -- (1.8, 0)
	(2.2, 0) -- (2.5, 0)
	(3.5, 0) -- (3.8, 0)
	(1, 0.2) -- (1, 0.8);
	\draw[white, decoration={markings,mark=at position 1 with {\arrow[scale=3, black]{>}}},postaction={decorate}] (4.5,0) -- (4.2,0);
	\draw[line width=1pt] (4.31, 0.07) -- (4.8, 0.07)
	(4.31, -0.07) -- (4.8, -0.07);
	\draw (0, -0.4) node {$ _1 $}
	(1, -0.4) node {$ _2 $}
	(2, -0.4) node {$ _2 $}
	(4, -0.4) node {$ _2 $}
	(5, -0.4) node {$ _2 $}
	(1.3, 1) node {$ _1 $}
	(5, -0.7) node {\textcolor{red}{$ _1 $}};
	\draw (0.6, 1) node {$ \scriptstyle \alpha_0 $}
	(0, 0.3) node {$ \scriptstyle \alpha_1 $}
	(0.7, 0.3) node {$ \scriptstyle \alpha_2 $}
	(2, 0.3) node {$ \scriptstyle \alpha_3 $}
	(4, 0.3) node {$ \scriptstyle \alpha_{\ell-1} $}
	(5, 0.3) node {$ \scriptstyle \alpha_{\ell} $};
	\end{tikzpicture}
	\caption{$ A_{2\ell-1}^{(2)} \ (\ell \geq 3) $}
	\vspace{3ex}
\end{subfigure}
\begin{subfigure}[b]{0.45\textwidth}
	\centering
	\begin{tikzpicture}
	\draw[thick] (0, 0) circle (0.1)
	(1, 0) circle (0.1)
	(2, 0) node {$ \cdots $}
	(3, 0) circle (0.1)
	(4, 0) circle (0.1);
	\draw[line width=1pt] (1.2, 0) -- (1.5, 0)
	(2.5, 0) -- (2.8, 0);
	\draw[white, decoration={markings,mark=at position 1 with {\arrow[scale=3, black]{>}}},postaction={decorate}] (3.5, 0) -- (3.8, 0);
	\draw[white, decoration={markings,mark=at position 1 with {\arrow[scale=3, black]{>}}},postaction={decorate}] (0.5, 0) -- (0.2, 0);
	\draw[line width=1pt] (3.2, 0.07) -- (3.69, 0.07)
	(3.2, -0.07) -- (3.69, -0.07)
	(0.31, 0.07) -- (0.8, 0.07)
	(0.31, -0.07) -- (0.8, -0.07);
	\draw (0, -0.4) node {$ _1 $}
	(1, -0.4) node {$ _2 $}
	(3, -0.4) node {$ _2 $}
	(4, -0.4) node {$ _1 $}
	(1, -0.7) node {\textcolor{red}{$ _1 $}}
	(3, -0.7) node {\textcolor{red}{$ _1 $}};
	\draw (0, 0.3) node {$ \scriptstyle \alpha_{0} $}
	(1, 0.3) node {$ \scriptstyle \alpha_{1} $}
	(3, 0.3) node {$ \scriptstyle \alpha_{\ell-1} $}
	(4, 0.3) node {$ \scriptstyle \alpha_{\ell} $};
	\end{tikzpicture}
	\caption{$ D_{\ell+1}^{(2)} \ ( \ell \geq 2) $}
	\vspace{3ex}
\end{subfigure}
\hfill
\begin{subfigure}[b]{0.45\textwidth}
	\centering
	\begin{tikzpicture}
	\draw[thick] (0, 0) circle (0.1)
	(1, 0) circle (0.1)
	(2, 0) circle (0.1)
	(3, 0) circle (0.1)
	(4, 0) circle (0.1);
	\draw[line width=1pt] (0.2, 0) -- (0.8, 0)
	(1.2, 0) -- (1.8, 0)
	(3.2, 0) -- (3.8, 0);
	\draw[white, decoration={markings,mark=at position 1 with {\arrow[scale=3, black]{>}}},postaction={decorate}] (2.5, 0) -- (2.2, 0);
	\draw[line width=1pt] (2.31, 0.07) -- (2.8, 0.07)
	(2.31, -0.07) -- (2.8, -0.07);
	\draw (0, -0.4) node {$ _1 $}
	(1, -0.4) node {$ _2 $}
	(2, -0.4) node {$ _3 $}
	(3, -0.4) node {$ _4 $}
	(4, -0.4) node {$ _2 $}
	(3, -0.7) node {\textcolor{red}{$ _2 $}}
	(4, -0.7) node {\textcolor{red}{$ _1 $}};
	\draw (0, 0.3) node {$ \scriptstyle \alpha_{0} $}
	(1, 0.3) node {$ \scriptstyle \alpha_{1} $}
	(2, 0.3) node {$ \scriptstyle \alpha_{2} $}
	(3, 0.3) node {$ \scriptstyle \alpha_{3} $}
	(4, 0.3) node {$ \scriptstyle \alpha_{4} $};
	\end{tikzpicture}
	\caption{$ E_6^{(2)} $}
\end{subfigure}
\begin{subfigure}[b]{0.45\textwidth}
	\centering
	\begin{tikzpicture}
	\draw[thick] (0, 0) circle (0.1)
	(1, 0) circle (0.1)
	(2, 0) circle (0.1);
	\draw[line width=1pt] (0.2, 0) -- (0.8, 0);
	\draw[white, decoration={markings,mark=at position 1 with {\arrow[scale=3, black]{>}}},postaction={decorate}] (1.5, 0) -- (1.2, 0);
	\draw[line width=1pt] (1.31, 0.07) -- (1.8, 0.07)
	(1.22, 0) -- (1.8, 0)
	(1.31, -0.07) -- (1.8, -0.07);
	\draw (0, -0.4) node {$ _1 $}
	(1, -0.4) node {$ _2 $}
	(2, -0.4) node {$ _3 $}
	(2, -0.7) node {\textcolor{red}{$ _1 $}};
	\draw (0, 0.3) node {$ \scriptstyle \alpha_{0} $}
	(1, 0.3) node {$ \scriptstyle \alpha_{1} $}
	(2, 0.3) node {$ \scriptstyle \alpha_{2} $};
	\end{tikzpicture}
	\caption{$ D_4^{(3)} $}
\end{subfigure}
\caption{Dynkin diagrams of twisted affine Lie algebras $ X_\ell^{(r=2, 3)} $. The number assigned each node is the dual Coxeter label (comark) $ d_i^\vee $. If a Coxeter label (mark) $ d_i $ is different from $ d_i^\vee $, it is represented as a red number.} \label{fig:dynkin_twist}
\end{figure}
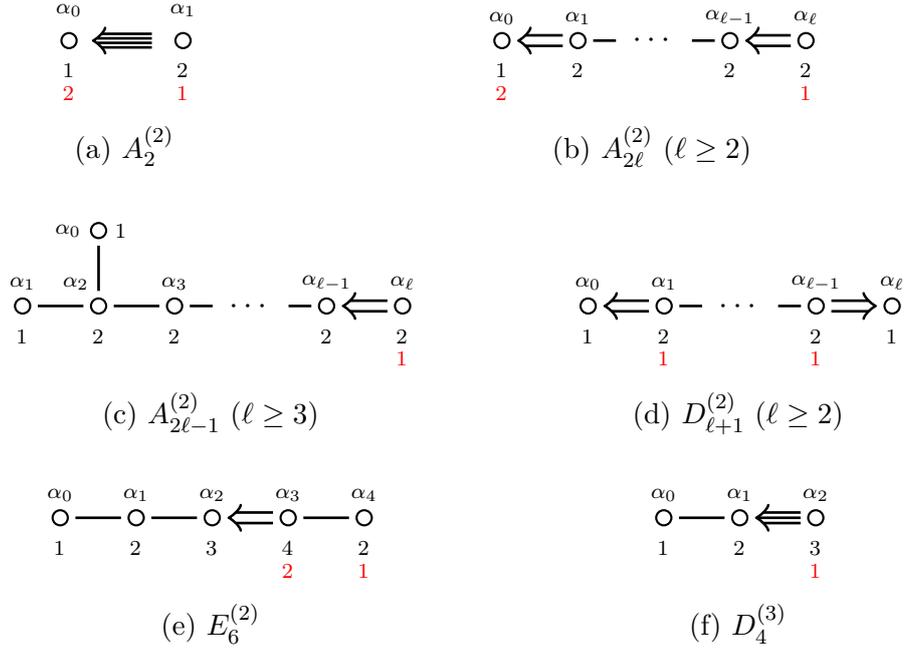

Instead of imposing periodic boundary condition along $S^1$, one can impose twisted boundary condition using an outer automorphism of a simple Lie algebra $\mathfrak{g}$ associated with a gauge group of a 6d theory. That is to say, instead of considering a function $ f : \mathbb{R} \to \mathfrak{g} $ with $ f(x + 2\pi) = f(x) $, we introduce $ f(x + 2\pi) = \sigma(f(x)) $, where $ \sigma : \mathfrak{g} \to \mathfrak{g} $ is an automorphism. When $ \sigma $ is a non-trivial outer automorphism, the resultant loop algebra with central extension differs from untwisted affine Lie algebra. Such affine algebra is called a twisted affine Lie algebra, denoted by $ X_\ell^{(r)} $, where $ r = 2, 3 $ is the order of the automorphism in the Dynkin diagram. Dynkin diagrams of twisted affine Lie algebras are drawn in Figure~\ref{fig:dynkin_twist}.

The 1-loop contribution to the prepotential of a 6d gauge theory on a circle is determined by matter representations of (twisted-)affine Lie algebras. Let us discuss the affine root systems. A root $ \alpha $ is called {\it real} if there exists an element $ w $ in the Weyl group such that $ w(\alpha) $ is a simple root. A root which is not real is called an {\it imaginary} root. It is known that there is no imaginary root for a simple Lie algebra \cite{kac_infinite_2003}. To describe a root system of affine Lie algebras, let $ \hat{\mathfrak{g}} $ be an affine Lie algebra of type $ X_\ell^{(r)} $ and $ \hat{\Delta} $ be its root system. We also define a simple Lie algebra $ \mathfrak{h} $ obtained by removing the affine node $ \alpha_0 $ in the Dynkin diagram of $ \hat{\mathfrak{g}} $ and its root system $ \Delta $. When $ \hat{\mathfrak{g}} = X_\ell^{(1)} $ is untwisted, $ \mathfrak{h} = \mathfrak{g} = X_\ell $, but if $ \hat{\mathfrak{g}} = X_\ell^{(r)} $ is twisted, then $ \mathfrak{h} $ is different from $ \mathfrak{g} = X_\ell $. The root system of $ \hat{\mathfrak{g}} $ can be written in terms of the roots of $ \mathfrak{h} $ \cite{kac_infinite_2003}. First, define
\begin{align}
\delta = \sum_{i=0}^{\ell} d_i \alpha_i\ ,
\end{align}
where $ d_i $ are the Coxeter labels of $ \hat{\mathfrak{g}} $ algebra. The set of imaginary roots of $ \hat{\mathfrak{g}} $ is
\begin{align}
\hat{\Delta}^{\mathrm{im}} = \{ n \delta \mid n \in \mathbb{Z} \setminus \{0\} \} \ .
\end{align}
The multiplicities of imaginary roots are as follows: 
\begin{align}\label{eq:multiplicities-imaginary}
\hat{\mathfrak{g}} = X_\ell^{(1)} &\quad \text{multiplicity of } n\delta \text{ is } \ell\nonumber \\
\hat{\mathfrak{g}} = A_{2\ell}^{(2)} &\quad \text{multiplicity of } n\delta \text{ is } \ell\nonumber \\
\hat{\mathfrak{g}} = A_{2\ell-1}^{(2)} &\quad \text{multiplicity of } 2n\delta \text{ is } \ell \text{, } (2n+1)\delta \text{ is } \ell-1 \nonumber\\
\hat{\mathfrak{g}} = D_{\ell+1}^{(2)} &\quad \text{multiplicity of } 2n\delta \text{ is } \ell \text{, } (2n+1)\delta \text{ is } 1 \nonumber\\ 
\hat{\mathfrak{g}} = E_{6}^{(2)} &\quad \text{multiplicity of } 2n\delta \text{ is } 4 \text{, } (2n+1)\delta \text{ is } 2 \nonumber\\
\hat{\mathfrak{g}} = D_{4}^{(3)} &\quad \text{multiplicity of } 3n\delta \text{ is } 2 \text{, } (3n\pm 1)\delta \text{ are } 1\ .
\end{align}
On the other hand, sets of real roots are given as follows: if $ \hat{\mathfrak{g}} = X_\ell^{(1)} $, then
\begin{align}\label{eq:untwist_real_root}
\hat{\Delta}^{\mathrm{re}} = \{ \alpha + n \delta \mid \alpha \in \Delta,\ n \in \mathbb{Z} \} \, .
\end{align}
If $ \hat{\mathfrak{g}} = X_\ell^{(r = 2, 3)} $ and $ \hat{\mathfrak{g}} \neq A_{2\ell}^{(2)} $, then
\begin{align}\label{eq:twist_real_root}
\hat{\Delta}^{\mathrm{re}} = \{ \alpha + n \delta \mid \alpha \in \Delta_s,\ n \in \mathbb{Z} \}  \cup \{\alpha + nr\delta \mid \alpha \in \Delta_l,\ n \in \mathbb{Z} \}\ ,
\end{align}
where $ \Delta_l $ and $ \Delta_s $ denote the long and the short roots of $ \mathfrak{h} $, respectively. If $ \hat{\mathfrak{g}} = A_{2\ell}^{(2)} $, then
\begin{align}\label{eq:twist_real_root2}
\hat{\Delta}^{\mathrm{re}}
&= \Big\{ \frac{1}{2} \big(\alpha + (2n-1)\delta \big)\ \Big\vert\  \alpha \in \Delta_l,\ n \in \mathbb{Z} \Big\} \nonumber \\
& \quad \cup \{\alpha + n\delta \mid \alpha \in \Delta_s,\ n \in \mathbb{Z} \} \cup \{\alpha + 2n\delta \mid \alpha \in \Delta_l,\ n \in \mathbb{Z} \} \, .
\end{align}
The multiplicities of real roots are $ 1 $. The total root system of an affine Lie algebra $ \hat{\mathfrak{g}} $ is then given by $ \hat{\Delta}^{\mathrm{re}} \cup \hat{\Delta}^{\mathrm{im}} $.

Using the affine root system we can compute the 1-loop contributions of 6d vector multiplets to the effective prepotential. We first identify $ r\delta $ as a KK-momentum for $ \hat{\mathfrak{g}} = X_\ell^{(r)} $. The reason for $ r $ factor is as follows: from the real root system \eqref{eq:untwist_real_root}-\eqref{eq:twist_real_root2}, the total root system $ \Delta $ of the underlying simple Lie algebra $ \mathfrak{h} $ is repeated for every $ r \delta $. This reflects that $ r \delta $ is a KK-momentum.

For an untwisted affine algebra $ \hat{\mathfrak{g}} = X_\ell^{(1)} $, \eqref{eq:multiplicities-imaginary} and \eqref{eq:untwist_real_root} implies that the 1-loop prepotential is given by
\begin{align}
\mathcal{F}_{\text{1-loop}} = \frac{1}{12} \sum_{n \in \mathbb{Z}} \sum_{\alpha \in \Delta} \abs{\alpha \cdot \phi + n \tau}^3\ ,
\end{align}
where $ \Delta $ is the root system of $ X_\ell $ algebra and $\phi$ denotes gauge holonomies and $\tau$ is the inverse of 6d circle radius. The $\ell$ imaginary roots can be understood as Cartan elements in the adjoint representation of $X_\ell$ algebra and their contribution to the prepotential are discarded because they only carry KK-charges and thus they provide only constant shifts to the prepotential.

Let us now discuss the twisted cases. For $ \hat{\mathfrak{g}} = X_\ell^{(2)} $ and $ \hat{\mathfrak{g}} \neq A_{2\ell}^{(2)} $, one can read from \eqref{eq:twist_real_root} that
\begin{align}
\mathcal{F}_{\text{1-loop}}
&= \frac{1}{12} \sum_{n \in \mathbb{Z}} \qty( \sum_{\alpha \in \Delta_s} \abs{\alpha\cdot \phi + \frac{n \tau}{2}}^3 + \sum_{\alpha \in \Delta_l} \abs{\alpha\cdot \phi + n \tau}^3 ) \nonumber \\
&= \frac{1}{12} \sum_{n \in \mathbb{Z}} \qty( \sum_{\alpha \in \Delta} \abs{\alpha\cdot \phi + n \tau}^3 + \sum_{\alpha \in \Delta_s} \abs{\alpha \cdot \phi+ \qty(n + \frac{1}{2})\tau}^3 )\ ,
\end{align}
where $ \Delta $, $ \Delta_l $ and $ \Delta_s $ are sets of all the roots, the long roots and the short roots of $ \mathfrak{h} $ algebra respectively. The first term carrying integral KK-charges forms the root system of $\mathfrak{h}$ algebra. The second term carrying half-integral KK-charges amounts to the rank-2 antisymmetric representation of $C_\ell$ for $A^{(2)}_{2\ell-1}$ and the fundamental representation of $B_\ell$ and $F_4$ for $D^{(2)}_{\ell+1}$ and $E^{(2)}_6$, respectively. The imaginary roots correspond to Cartan elements in each representation. This situation can be interpreted as
\begin{align}
\begin{array}{rlrlrl}
A_{2\ell-1} &\to C_{\ell} & \quad D_{\ell+1} & \to B_\ell & \qquad E_6 &\to F_4  \\
\mathbf{Adj} &\to \mathbf{Adj}_0 \oplus \mathbf{\Lambda}^2_{1/2} \, , & \mathbf{Adj} &\to \mathbf{Adj}_0 \oplus \mathbf{F}_{1/2} \, , & \mathbf{Adj} &\to \mathbf{Adj}_0 \oplus \mathbf{F}_{1/2}\ ,
\end{array}
\end{align}
which are branching rules of the adjoint representation of $ \mathfrak{g} $ to $ \mathfrak{h} $. Here, the subscript in a representation ${\bf r}$ stands for KK-momentum shift. For instance, the vector multiplet contribution in the $ \mathbb{Z}_2 $ twisted compactification of $ \mathfrak{su}(4) = \mathfrak{so}(6) $ gauge theory is given by
\begin{align}
\mathcal{F}_{\text{1-loop}}
= \frac{1}{12} \sum_{n \in \mathbb{Z}} \sum_{e \in \mathbf{R}} \abs{n \tau + e\cdot \phi}^3 + \frac{1}{12} \sum_{n \in \mathbb{Z}} \sum_{w \in \mathbf{\Lambda}^2} \abs{\qty(n+\frac{1}{2})\tau + w\cdot \phi}^3 \, ,
\end{align}
where $ \mathbf{R} $ and $ \mathbf{\Lambda}^2 $ are the root system and the rank-2 anti-symmetric representation of $ \mathfrak{sp}(2) $ algebra, respectively. By evaluating the 1-loop contribution explicitly using the zeta function regularization, one can compute
\begin{align}
\mathcal{F}_{\text{1-loop}}
= \frac{4}{3}\phi_1^3 + 2\phi_1^2 \phi_2 - 3\phi_1 \phi_2^2 + \frac{4}{3}\phi_2^3 - \frac{11}{24}\tau (2\phi_1^2 - 2\phi_1 \phi_2 + \phi_2^2)\, ,
\end{align}
up to constant terms. 

For $\hat{\mathfrak{g}} = D_4^{(3)}$, the l-loop contribution can be obtained in a similar fashion with $r=3$:
\begin{align}\label{eq:D4(3)_1-loop}
\mathcal{F}_{\text{1-loop}}
&= \frac{1}{12} \sum_{n \in \mathbb{Z}} \qty(\sum_{\alpha \in \Delta_s} \abs{\alpha\cdot \phi + \frac{n\tau}{3}}^3 + \sum_{\alpha \in \Delta_l} \abs{\alpha\cdot \phi + n\tau}^3 ) \\
& =\frac{1}{12} \sum_{n \in \mathbb{Z}} \qty[\sum_{\alpha \in \Delta} \abs{\alpha\cdot \phi + n \tau}^3 + \sum_{\alpha \in \Delta_s} \qty( \abs{\alpha\cdot \phi \!+\! \qty(n \!+\! \tfrac{1}{3})\tau}^3 + \abs{\alpha\cdot \phi \!+\! \qty(n \!+\! \tfrac{2}{3})\tau}^3 ) ] \, . \nonumber
\end{align}
The first term carrying integer KK-charge forms the root system of $ \mathfrak{h} = G_2 $ algebra, while the second and third terms amount to the fundamental representation of $ G_2 $. This can be summarized as
\begin{align}
D_4 &\to G_2 \nonumber \\
\mathbf{Adj} &\to \mathbf{Adj}_0 + \mathbf{F}_{1/3} + \mathbf{F}_{2/3} \, .
\end{align}
The prepotential in \eqref{eq:D4(3)_1-loop} after performing the zeta function regularization reduces to
\begin{align}
\mathcal{F}_{\text{1-loop}} &= \frac{4}{3}\phi_1^3 + 3\phi_1^2 \phi_2 - 4\phi_1 \phi_2^2 + \frac{4}{3}\phi_2^3 - \frac{5}{9}\tau (3\phi_1^2 - 3\phi_1 \phi_2 + \phi_2^2) \, ,
\end{align}
which is used in subsection~\ref{sec:SU(4)8}.

For $ \hat{\mathfrak{g}} = A_{2\ell}^{(2)} $, one can read from the root system \eqref{eq:twist_real_root2} that
\begin{align}
\mathcal{F}_{\text{1-loop}}
&= \frac{1}{12} \sum_{n \in \mathbb{Z}} \qty(\sum_{\alpha \in \Delta_l} \abs{\frac{\alpha\cdot \phi}{2} + \frac{2n\!-\!1}{4}\tau}^3 + \sum_{\alpha \in \Delta_s} \abs{\alpha\cdot \phi + \frac{n \tau}{2}}^3 + \sum_{\alpha \in \Delta_l} \abs{\alpha\cdot \phi + n \tau}^3 ) \nonumber \\
&= \frac{1}{12} \sum_{n \in \mathbb{Z}} \left(\sum_{\alpha \in \Delta} \abs{\alpha\cdot \phi+ n \tau}^3 + \sum_{\alpha \in \Delta_l} \abs{\frac{\alpha\cdot \phi}{2} + \qty(n+\frac{1}{4})\tau}^3 \right. \nonumber \\
& \qquad \qquad \left. +\sum_{\alpha \in \Delta_l} \abs{\frac{\alpha\cdot \phi}{2} + \qty(n+\frac{3}{4})\tau}^3+\sum_{\alpha \in \Delta_s} \abs{\alpha\cdot \phi + \qty(n+\frac{1}{2})\tau}^3\right) \, .
\end{align}
The first term in the second line forms the root system of $ C_\ell $ algebra. The second and third terms carrying $\tau/4 $ and $3\tau/4$ are half of the long roots of $ C_\ell $, so they amount to the fundamental representation of $ C_\ell $ algebra. The last term carrying half-integral KK-charges amounts to the rank-2 antisymmetric representation. There is one more singlet carrying half-integral KK-momentum which arises from the imaginary root, so we can understand the situation as a branching of $A_{2\ell}$ algebra into $C_\ell$ algebra
\begin{align}\label{eq:A_2l_branching}
A_{2\ell} &\to C_\ell \nonumber \\
\mathbf{Adj} &\to \mathbf{Adj}_0 \oplus \mathbf{F}_{1/4} \oplus \mathbf{F}_{3/4} \oplus \mathbf{\Lambda}^2_{1/2} \oplus \mathbf{1}_{1/2} \, .
\end{align}
As an example, the vector multiplet contribution in $ \mathbb{Z}_2 $ twist compactification of a $SU(3)$ gauge field is given by
\begin{align}
\mathcal{F}_{\text{1-loop}}
& = \frac{1}{12} \sum_{n \in \mathbb{Z}} \sum_{e \in \mathbf{R}} \abs{e\cdot \phi+n \tau }^3 + \frac{1}{12} \sum_{n \in \mathbb{Z}} \sum_{w \in \mathbf{F}} \abs{w\cdot \phi+\qty(n + \frac{1}{4})\tau}^3  \nonumber \\
& \qquad + \frac{1}{12} \sum_{n \in \mathbb{Z}} \sum_{w \in \mathbf{F}} \abs{w\cdot \phi+\qty(n + \frac{3}{4})\tau}^3 \nonumber \\
&= \frac{4}{3}\phi_1^3 - \frac{5}{16}\tau \phi_1^2 \ .
\end{align}
This is used in subsection~\ref{subsubsec:SU3/Z2}.

\subsection{Graded representations and 1-loop prepotential }

To find a KK-momentum shift for a hypermultiplet, we first reinterpret the KK-momentum shifts for W-bosons we discussed in the previous subsection. Let $ \mathfrak{g} = X_\ell $ be a simple Lie algebra, $ r = 2, 3 $ be an order of automorphism in a Dynkin diagram, $ s = (s_0, s_1, \cdots, s_\ell) $ be non-negative relatively prime integers, and $ m = r \sum_{i=0}^\ell d_i s_i $ for the Coxeter labels $ d_i $. For a given $ s $, there is a unique outer automorphism $ \sigma_{s, r} $ of $ \mathfrak{g} $ up to conjugation such that $ \sigma_{s,r}^m = 1 $ \cite{kac_infinite_2003}. $ r $ is the smallest integer for which $ \sigma_{s,r}^r $ is an inner automorphism. An order $ m $ outer automorphism $ \sigma_{s, r} $ naturally induces the grading
\begin{align}\label{eq:graded_alg}
\mathfrak{g} = \bigoplus_{j \in \mathbb{Z}_m} \mathfrak{g}_j \ ,
\end{align}
with $ \comm*{\mathfrak{g}_i}{\mathfrak{g}_j} \subset \mathfrak{g}_{i+j}$, where an element of $ \mathfrak{g}_j $ has an eigenvalue $ e^{2\pi i j/m} $ under $ \sigma_{s, r} $. Thus, under twisted compactification of a 6d theory with gauge algebra $ \mathfrak{g} $, the W-boson states in $ \mathfrak{g}_j $ acquire additional fractional KK-charge $ j /m $. In particular, W-boson states in the invariant subalgebra $ \mathfrak{g}_0 $ of $ \sigma_{s, r} $ have integral KK-charges. The affine root system demands $ \mathfrak{g}_0 $ to be $ \mathfrak{h} $ defined in  Appendix~\ref{sec:App-affineroots}. It is known that when $ s = (1, 0, 0, \cdots, 0) $, the invariant subalgebra $ \mathfrak{g}_0 $ is isomorphic to $ \mathfrak{h} $, and $ \mathfrak{g}_0 $-module $ \mathfrak{g}_1 $ is isomorphic to an irreducible module with the highest weight $ -\alpha_0 $ \cite{kac_infinite_2003}.  

We now reproduce the branching rules with KK-momentum shifts of the adjoint representations in Appendix \ref{sec:App-affineroots}, using the automorphism $ \sigma_{s,r} $ and grading \eqref{eq:graded_alg}. As a representative example, consider $ \mathfrak{g} = A_{2\ell - 1} $ with $ s = (1, 0, \cdots, 0) $, $ r = 2 $. Under the corresponding outer automorphism $ \sigma_{s,r} $, $ \mathfrak{g} $ is graded by
\begin{align}\label{eq:su(2l)_grading}
\mathfrak{g} = \mathfrak{g}_0 \oplus \mathfrak{g}_1\ ,
\end{align}
where $ \mathfrak{g}_0 $ is $ C_\ell $ algebra and $ \mathfrak{g}_1 $ corresponds to the rank-2 antisymmetric representation of $ C_\ell $ algebra. We can also explicitly construct $ \mathfrak{g}_0 $ and $ \mathfrak{g}_1 $. Let $ E_{\alpha} $ be a ladder operator of $ A_{2\ell - 1} $ algebra associated with a root $ \alpha $. We write $ E_i = E_{\alpha_i} $ for a simple root $ \alpha_i $. The Cartan generators $ H_i $ are given as $ H_i = \alpha_i^\vee = \comm*{E_{\alpha_i}}{E_{-\alpha_i}} $. Note that $ \comm{H_i}{E_j} = \inner{\alpha_j}{\alpha_i^\vee} E_j = a_{ij} E_j $, where $ a_{ij} $ is a Cartan matrix element. An automorphism $ \mu $ in a Dynkin diagram acts on the simple roots by $ \mu(\alpha_i) = \alpha_{2\ell-i} $, so we can use the outer automorphism $ \sigma_{s,r} $ as the induced map $ \sigma_{s,r}(H_i) = H_{2\ell-i} $ and $ \sigma_{s,r}(E_i) = E_{2\ell-i} $ by uniqueness. The invariant combinations of the Cartan generators under $ \sigma_{s,r} $ are 
\begin{align}\label{eq:su(2l)_g0_cartan}
\tilde{H}_i = H_i + H_{2n-i} \ (1 \leq i \leq \ell-1), \quad
\tilde{H}_\ell = H_\ell \ ,
\end{align}
and they become the Cartan generators of $ \mathfrak{g}_0 $. In the case of the ladder operators corresponding to the simple roots, invariant and non-invariant combinations under $ \sigma_{s,r} $ are given as follows:
\begin{align}
\{ E_i + E_{2\ell-i}, E_\ell \mid 1 \leq i \leq \ell-1 \} \subset \mathfrak{g}_0 \, , \quad
\{ E_i - E_{2\ell-i} \mid 1 \leq i \leq \ell-1 \} \subset \mathfrak{g}_1 \, .
\end{align}
The eigenvalues of such elements in $ \mathfrak{g}_0 $ under the Cartan generators $ \tilde{H}_i $ amount to the simple root of $ \mathfrak{g}_0 = C_\ell $ algebra, and hence the collection of charges makes the Cartan matrix of $\mathfrak{g}_0$, in this case $\mathfrak{sp(\ell)}$, as shown in Table~\ref{table:A_2l-1_root_class}.
\begin{table}
	\centering
	\begin{tabular}{ccc}
		$\mathfrak{g}_0$ & $\mathfrak{g}_1$ & Charge \\ \hline
		$E_1 + E_{2\ell}$ & $E_1 - E_{2\ell}$ & $(2, -1, 0, 0, \cdots, 0)$ \\
		$E_2 + E_{2\ell-1}$ & $E_2 - E_{2\ell-1}$ & $(-1, 2, -1, 0, \cdots, 0)$ \\
		$\vdots$ & $\vdots$ & $\vdots$ \\
		$E_{\ell-1} + E_{\ell+1}$ & $E_{\ell-1} - E_{\ell+1}$ & $(0, \cdots, 0, -1, 2, -1)$ \\
		$E_\ell$ & & $(0, 0, \cdots, 0, -2, 2)$
	\end{tabular}
	\caption{Classification of the ladder operators corresponding to the simple roots of $ A_{2\ell-1} $ under order 2 outer automorphism $ \sigma $. The charge implies the eigenvalues under the Cartan generators $ \tilde{H}_i $ of $ \mathfrak{g}_0 $.} \label{table:A_2l-1_root_class}
\end{table}
For a general root $ \alpha $ of $ A_{2\ell-1} $, $ \sigma_{s,r}(E_\alpha) = u_\alpha E_{\mu(\alpha)} $, where $ u_\alpha = \pm 1 $. If $ \alpha \neq \mu(\alpha) $, then
\begin{align}
E_\alpha + u_\alpha E_{\mu(\alpha)} \in \mathfrak{g}_0, \quad
E_\alpha - u_\alpha E_{\mu(\alpha)} \in \mathfrak{g}_1 \, .
\end{align}
When $ \alpha = \mu(\alpha) $, then $ E_\alpha \in \mathfrak{g}_0 $ and there is no corresponding element in $ \mathfrak{g}_1 $. By considering the eigenvalues of all such elements in $ \mathfrak{g}_0 $ and $ \mathfrak{g}_1 $ under $ \tilde{H}_i $ defined in \eqref{eq:su(2l)_g0_cartan}, one can confirm that charges of the elements in $ \mathfrak{g}_0 $ and $ \mathfrak{g}_1 $ form the weight systems of the adjoint representation and the rank-2 antisymmetric representation, respectively. For $ \mathfrak{g} = D_{\ell+1}$ and $E_6 $, one can treat them in a similar way as the $ A_{2\ell-1} $ case and obtain the corresponding branching rules.

As discussed for the affine root system in Appendix \ref{sec:App-affineroots}, the $\mathfrak{g} = A_{2\ell}$ case is rather involved. By setting $ s = (1, 0, \cdots, 0) $ to construct an outer automorphism that leaves $ \mathfrak{h} = C_\ell $ invariant, we get order 4 automorphism $ \sigma_{s,r} $ since the Coxeter label $ d_0 = 2 $. This map $ \sigma_{s,r} $ induces the grading
\begin{align}\label{eq:su(2l+1)_grading}
\mathfrak{g} = \mathfrak{g}_0 \oplus \mathfrak{g}_1 \oplus \mathfrak{g}_2 \oplus \mathfrak{g}_3\ ,
\end{align}
where $ \mathfrak{g}_0 $ is $ C_\ell $ algebra. To find $ \mathfrak{g}_{1,2,3} $, we consider the following choice of $ \sigma_{s,r} $ \cite{Tachikawa:2011ch}:
\begin{align}\label{eq:A_2l_sigma}
\sigma_{s,r}(x) = -\Omega\, x^T \Omega^{-1} \, ,
\end{align}
where $ x $ is a traceless $ (2\ell+1) \times (2\ell+1) $ matrix and
\begin{align}
\Omega = \mqty(1 & 0 & 0 \\ 0 & 0 & \mathbf{I}_{2\ell} \\ 0 & -\mathbf{I}_{2\ell} & 0)\ ,
\end{align}
with a $ 2\ell \times 2\ell $ identity matrix $ \mathbf{I}_{2\ell} $. Explicitly, the action of $\sigma_{s,r}$ on  $x$ in the matrix representation is as follows: 
\begin{align}\label{eq:ABCD}
\left(
\begin{array}{c|ccccc}
x_{11} & \phantom{x} & \vec{y}_1 & \phantom{x} & \vec{y}_2 & \phantom{x} \\ \hline
\vec{z}_1{}^T & & A & & B & \\
\vec{z}_2{}^T & & C & & D & 
\end{array}\right)
\overset{\sigma_{s,r}}{\longmapsto}
\left(
\begin{array}{c|ccccc}
-x_{11} & \phantom{x} & -\vec{z}_2 & \phantom{x} & \vec{z}_1 & \phantom{x} \\ \hline
-\vec{y}_2{}^T & & -D^T & & B^T & \\
\vec{y}_1{}^T & & C^T & & -A^T &
\end{array}\right)
\overset{\sigma_{s,r}}{\longmapsto}
\left(
\begin{array}{c|ccccc}
x_{11} & \phantom{x} & -\vec{y}_1 & \phantom{x} & -\vec{y}_2 & \phantom{x} \\ \hline
-\vec{z}_1{}^T & & A & & B & \\
-\vec{z}_2{}^T & & C & & D & 
\end{array}\right)
\end{align}
where $ \vec{y}_1 = (x_{12}, \cdots, x_{1,\ell+1}) $, $ \vec{y}_2 = (x_{1,\ell+2}, \cdots, x_{1,2\ell+1}) $, $ \vec{z}_1 = (x_{21}, \cdots, x_{\ell+1,1}) $, $ \vec{z}_2 = (x_{\ell+2,1}, \cdots, x_{2\ell+1, 1}) $ and $ A, B, C, D $ are $ \ell \times \ell $ matrices. The invariant subalgebra $ \mathfrak{g}_0 $ lies in the $ 2\ell \times 2 \ell $ matrix (the lower right part of \eqref{eq:ABCD}) which consists of $A, B, C, D $, satisfying $ D = -A^T $, $ B = B^T $ and $ C = C^T $. 
This naturally identifies with $ C_\ell $ algebra. Suitable linear combinations of $ (\vec{y}_1, \vec{y}_2) $ and $ (\vec{z}_1, \vec{z}_2) $ are in $ \mathfrak{g}_1 $ and $ \mathfrak{g}_3 $, and they form two fundamental representations of $ C_\ell $. The remaining $ \mathfrak{g}_2 $ is identified with the rank-2 anti-symmetric representation of $ C_\ell $, because the lower right $ 2\ell \times 2\ell $ matrix is nothing but an embedding of $ A_{2\ell-1} $ algebra and it is decomposed to the adjoint and rank-2 anti-symmetric representations of $ C_\ell $ algebra. As the adjoint representation belongs to $\mathfrak{g}_0$, this reproduces \eqref{eq:A_2l_branching}.

We extend the above argument to general representations. When a Lie algebra is graded as \eqref{eq:graded_alg}, the representation $ \pi : \mathfrak{g} \to \mathfrak{gl}(V) $ is said to be {\it compatible with the grading}, if the vector space $ V $ is decomposed as
\begin{align}
V = \bigoplus_{j \in \mathbb{Z}_m} V_j\ ,
\end{align}
such that
\begin{align}
\pi(x_i) V_j \subset V_{i+j} \ ,
\end{align}
for each $ i, j \in \mathbb{Z}_m $ and any $ x_i \in \mathfrak{g}_i $ \cite{havlicek_representations_2009}. Consequently, for matter fields, the states in a representation space $ V_j $ acquire additional fractional KK-charge $ j/m $, similar to the adjoint representation. As an example, consider a twisted compactification of $ \mathfrak{g} = A_{2\ell-1} $ type gauge theory coupled with matter in the (anti-)fundamental representation. Let $ \pi : \mathfrak{g} \to \mathfrak{gl}(\mathbb{C}^{2\ell}) $ be the fundamental representation and $ \{\ket{\mu_i} \mid 1 \leq i \leq 2\ell \} $ be the basis of the representation space $ \mathbb{C}^{2\ell} $ whose charges under the Cartan generators $ \pi(H_i) $ form the weight system of the fundamental representation:
\begin{align}\label{eq:su(2l)_fund_weight}
&\pi(H_i) \ket{\mu_1} = \delta_{i,1} \ket{\mu_1} \, , \quad
\nonumber \\
&\pi(H_i) \ket{\mu_j} = (-\delta_{i,j-1} + \delta_{i,j}) \ket{\mu_j} \quad (1 < j < 2\ell) \, , \nonumber\\
&\pi(H_i) \ket{\mu_{2\ell}} = -\delta_{i,2\ell-1} \ket{\mu_{2\ell}}\ .
\end{align}
We note that $ \ket{\mu_{i+1}} = E_{-\alpha_{i}} \ket{\mu_i} $. An outer automorphism $ \sigma_{s,r} $ induced by Dynkin diagram automorphism acts on the Cartan generators as $ \sigma_{s,r}(H_i) = H_{2\ell-i} $, so the charges of basis vectors $ \ket{\mu_i} $ under the Cartan generators $ \pi^* (H_i) := \pi \circ \sigma_{s,r} (H_i)= \pi(\sigma_{s,r}(H_i)) $ form the weight system of antifundamental representation:
\begin{align}\label{eq:su(2l)_antifund_weight}
&\pi^*(H_i) \ket{\mu_1} = \delta_{i,2\ell-1} \ket{\mu_1}, \nonumber\\
&\pi^*(H_i) \ket{\mu_j} = (\delta_{i,2\ell-j} - \delta_{i,2\ell+1-j}) \ket{\mu_j} \ (1 < j < 2\ell) \ ,\nonumber \\
&\pi^*(H_i) \ket{\mu_{2\ell}} = -\delta_{i,1} \ket{\mu_{2\ell}}.
\end{align}
Hence, the representation $ \pi^* $ is naturally identified to the anti-fundamental representation. The representation $ \pi \oplus \pi^* $ which is of the representation space $ \mathbb{C}^{4\ell} $ is then compatible with the grading \eqref{eq:su(2l)_grading}. To see it, let $ \ket{\mu_i} $ be an embedding of the basis of $ \pi $ representation space $ \mathbb{C}^{2\ell} $ into $ \mathbb{C}^{4\ell} $ satisfying \eqref{eq:su(2l)_fund_weight}. Similarly, let $ \ket{\nu_i} $ be an embedding of the basis of $ \pi^* $ representation space $ \mathbb{C}^{2\ell} $ into $ \mathbb{C}^{4\ell} $ satisfying \eqref{eq:su(2l)_antifund_weight}. Then we find that charges of $ \ket{\mu_i} \pm \ket{\nu_i} $ under the Cartan generators $ \pi \oplus \pi^*(\tilde{H}_i) $ of the invariant subalgebra $ \mathfrak{g}_0 $ are given as
\begin{align}
(\pi \oplus \pi^*(\tilde{H}_i)) (\ket{\mu_1} \pm \ket{\nu_1}) &= \delta_{i,1} (\ket{\mu_1} \pm \ket{\nu_1}) \, , \nonumber \\
(\pi \oplus \pi^*(\tilde{H}_i)) (\ket{\mu_j} \pm \ket{\nu_j}) &= (-\delta_{i,j-1} + \delta_{i,j}) (\ket{\mu_j} \pm \ket{\nu_j}) \quad (1 < j \leq \ell) \, , \nonumber \\
(\pi \oplus \pi^*(\tilde{H}_i)) (\ket{\mu_j} \pm \ket{\nu_j}) &= (\delta_{i,2\ell-j} - \delta_{i,2\ell+1-j}) (\ket{\mu_j} \pm \ket{\nu_j}) \quad (\ell < j < 2\ell) \, , \nonumber \\
(\pi \oplus \pi^*(\tilde{H}_i)) (\ket{\mu_{2\ell}} \pm \ket{\nu_{2\ell}}) &= -\delta_{i,1} (\ket{\mu_{2\ell}} \pm \ket{\nu_{2\ell}}) \, . 
\end{align}
Here, we see that the representation space $ \mathbb{C}^{4\ell} $ of $ \pi \oplus \pi^* $ is decomposed into $ \mathbb{C}^{2\ell} \oplus \mathbb{C}^{2\ell} $ spanned by $ \{\ket{\mu_i} + \ket{\nu_i} \} $ and $ \{\ket{\mu_i} - \ket{\nu_i} \} $, respectively. 
Therefore, the eigenvalues of the basis vectors form the weight system of the fundamental representation of the invariant subalgebra $ \mathfrak{g}_0 = C_\ell $. Moreover, if one applies element $ \pi \oplus \pi^*(E_{-\alpha_i} - E_{-\alpha_{2\ell-i}}) $ in $ \mathfrak{g}_1 $ to $ \ket{\mu_i} \pm \ket{\nu_i} $, then
\begin{align}
(\pi \oplus \pi^*(E_{-\alpha_i} - E_{-\alpha_{2\ell-i}}))(\ket{\mu_i} \pm \ket{\nu_i}) = \ket{\mu_{i+1}} \mp \ket{\nu_{i+1}} \quad (1 \leq i < \ell) \, .
\end{align}
Thus, the representation $ \pi \oplus \pi^* $ is compatible with the grading. If we identify one fundamental representation of $ C_\ell $ to integer KK-momentum modes, then the other one corresponds to half-integer KK-momentum modes,
\begin{align}
A_{2\ell-1} &\to C_\ell \nonumber \\
\mathbf{F} \oplus \overline{\mathbf{F}} &\to \mathbf{F}_0 \oplus \mathbf{F}_{1/2} \, .
\end{align}

Next, consider the twisted compactification of $ \mathfrak{g} = A_{2\ell} $ type gauge algebra coupled with (anti-)fundamental matter. To use the explicit definition of $ \sigma_{s,r} $ \eqref{eq:A_2l_sigma}, we note that a matrix representation of $ H_i $ of $ \mathfrak{g} $ is given by
\begin{align}
(H_i)_{\alpha \beta} &= \delta_{i \alpha} \delta_{i \beta} - \delta_{i+1,\alpha} \delta_{i+1,\beta}\ ,
\end{align}
where $ 1 \leq \alpha, \beta \leq 2\ell+1 $. The automorphism $ \sigma_{s,r} $ acts on them by
\begin{align}
&\sigma_{s,r}(H_1) = -\sum_{i=1}^{\ell+1} H_i \, , \quad
\sigma_{s,r}(H_j) = -H_{\ell+j} \ (2 \leq j \leq \ell) \, , \nonumber \\
&\sigma_{s,r}(H_{\ell+1}) = \sum_{i=2}^{2\ell} H_i \, , \quad
\sigma_{s,r}(H_j) = -H_{j-\ell} \ (\ell+2 \leq j \leq 2\ell) \, .
\end{align}
The Cartan generators of the invariant subalgebra $ \mathfrak{g}_0 = C_\ell $ are
\begin{align}
\tilde{H}_i = H_{i+1} - H_{\ell + i + 1} \ (1 \leq i \leq \ell-1)\, , \quad
\tilde{H}_\ell = \sum_{i = \ell+1}^{2\ell} H_i \, .
\end{align}
Now, let $ \pi : \mathfrak{g} \to \mathfrak{gl}(\mathbb{C}^{2\ell+1}) $, $ \pi(x) = x $ be the fundamental representation and $ \{\ket{\mu_i} \mid 1 \leq i \leq 2\ell+1 \} $ be a basis of $ \mathbb{C}^{2\ell+1} $ whose eigenvalues form  the weight system of the representation
\begin{align}\label{eq:su(2l+1)_fund_weight}
&\pi(H_i) \ket{\mu_1} = \delta_{i,1} \ket{\mu_1} \, ,  \nonumber \\
&\pi(H_i) \ket{\mu_j} = (-\delta_{i,j-1} + \delta_{i,j}) \ket{\mu_j} \quad (1 < j < 2\ell+1) \, ,\nonumber\\
&\pi(H_i) \ket{\mu_{2\ell+1}} = -\delta_{i,2\ell} \ket{\mu_{2\ell+1}} \, .
\end{align}
Explicitly, $ (\ket{\mu_i})_\alpha = \delta_{i\alpha} $ for $ 1 \leq \alpha \leq 2\ell+1 $. The composition $ \pi \circ \sigma_{s,r} $ is the anti-fundamental representation:
\begin{align}\label{eq:su(2l+1)_antifund_weight}
&\pi(\sigma(H_i)) \ket{\mu_1} = -\delta_{i,1} \ket{\mu_1} \, ,  \nonumber \\
&\pi(\sigma(H_i)) \ket{\mu_j} = (\delta_{i,\ell+j-1} - \delta_{i,\ell+j}) \ket{\mu_j} \quad (2 \leq j \leq \ell) \, , \nonumber \\
&
\pi(\sigma(H_i)) \ket{\mu_{\ell+1}} = \delta_{i,2\ell} \ket{\mu_{\ell+1}} \, , \nonumber\\
&\pi(\sigma(H_i)) \ket{\mu_{j}} = (\delta_{i,j-\ell-1} - \delta_{i,j-\ell}) \ket{\mu_{j}} \quad (\ell+2 \leq j \leq 2\ell) \, .
\end{align}
Similar to the $ A_{2\ell-1} $ case, the direct sum representation $ \pi \oplus \pi^* $ is compatible with the grading \eqref{eq:su(2l+1)_grading}. To see it, let $ \ket{\mu_i} $ be an embedding of the basis of the fundamental representation space $ \mathbb{C}^{2\ell+1} $ into the direct sum representation space $ \mathbb{C}^{4\ell+2} $ satisfying \eqref{eq:su(2l+1)_fund_weight}. Likewise, let $ \ket{\nu_i} $ be an embedding of the basis of the anti-fundamental representation space into $ \mathbb{C}^{4\ell+2} $ satisfying \eqref{eq:su(2l+1)_antifund_weight}. Following \cite{havlicek_representations_2009}, define a $ (4\ell + 2) \times (4\ell + 2) $ matrix
\begin{align}
R = \mqty(0 & \mathbf{I}_{2\ell+1} \\ \Omega^2 & 0)\ ,
\end{align}
which satisfies
\begin{align}\label{eq:compatible_grading_R}
(\pi \oplus \pi^* (\sigma(x))) = R (\pi \oplus \pi^*(x)) R^{-1}\ ,
\end{align}
where
\begin{align}
\pi \oplus \pi^*(x) = \mqty(x & 0 \\ 0 & -\Omega x^T \Omega^{-1}) \in \mathfrak{gl}(\mathbb{C}^{4\ell+2}) \, .
\end{align}
Since $ R^4 $ is an identity matrix, 
 it decomposes $ \mathbb{C}^{4\ell+2} $ as
\begin{align}\label{eq:su(2l+1)_fund_decomp}
\mathbb{C}^{4\ell+2} = \bigoplus_{j \in \mathbb{Z}_4} V_j\ ,
\end{align}
where an element of $ V_j $ has an eigenvalue $ e^{2\pi i (n+j)/4} $ for some fixed integer $ n $ under $ R $. Moreover, for $ x_j \in \mathfrak{g}_j $ and $ v_k \in V_k $,
\begin{align}
\Big(\pi \oplus \pi^*\big(\sigma(x_j)\big)\Big)R v_k
= e^{2\pi i (n+j+k)/4} \big(\pi \oplus \pi^*(x_j)\big) v_k = R \big(\pi \oplus \pi^*(x_j)\big) v_k\ ,
\end{align}
from \eqref{eq:compatible_grading_R}. This shows that $ (\pi \oplus \pi^*(x_j)) v_k \in V_{j+k} $, so the grading on representation space is compatible with Lie algebra grading. The remaining thing is to identify each $ V_j $ into a representation of the invariant subalgebra. The states $ \ket{\mu_1} \pm \ket{\nu_1} $ have eigenvalues $ \pm 1 $ under $ R $, and $ \ket{\mu_j} \pm i\ket{\nu_j} $ have eigenvalues $ \pm i $ for $ j > 1 $. The charges of these states under the Cartan generators of the invariant subalgebra are
\begin{align}
&(\pi \oplus \pi^*(\tilde{H}_k)) (\ket{\mu_1} \pm \ket{\nu_1}) = 0 \, , \\
&(\pi \oplus \pi^*(\tilde{H}_k)) (\ket{\mu_2} \pm i\ket{\nu_2}) = \delta_{k,1} (\ket{\mu_2} \pm i\ket{\nu_2}) \, , \nonumber \\
&(\pi \oplus \pi^*(\tilde{H}_k)) (\ket{\mu_j} \pm i\ket{\nu_j}) = (-\delta_{k,j-2} + \delta_{k,j-1}) (\ket{\mu_j} \pm i\ket{\nu_j}) \ (2 \leq j \leq \ell+1) \, , \nonumber \\
&(\pi \oplus \pi^*(\tilde{H}_k)) (\ket{\mu_{\ell+2}} \pm i\ket{\nu_{\ell+2}}) = -\delta_{k,1} (\ket{\mu_{\ell+2}} \pm i\ket{\nu_{\ell+2}}) \, , \nonumber \\
&(\pi \oplus \pi^*(\tilde{H}_k)) (\ket{\mu_j} \pm i\ket{\mu_j}) = (\delta_{k,j-\ell-2} - \delta_{k,j-\ell-1}) (\ket{\mu_j} \pm i\ket{\nu_j}) \ (\ell+3 \leq j \leq 2\ell+1) \, .\nonumber
\end{align}
Thus, the representation space spanned by $ \{\ket{\mu_1} + \ket{\nu_1} \} $ and $ \{\ket{\mu_1} - \ket{\nu_1} \} $ correspond to two singlets, while $ \{\ket{\mu_j} + i\ket{\nu_j} \} $ and $ \{\ket{\mu_j} - i\ket{\nu_j} \} $ correspond to two fundamental representations of $ C_\ell $. We choose $ n = -1 $ in the eigenvalue $e^{2\pi i (n+j)/4}$ of the elements $V_j$ in \eqref{eq:su(2l+1)_fund_decomp} so that $ V_0 $ and $ V_2 $ correspond to fundamental matter, while $ V_1 $ and $ V_3 $ become singlets. This is because the 5d reduction of the theory contains fundamental matter. In other words, under twisted compactification,
\begin{align}
A_{2\ell} &\to C_\ell \nonumber \\
\mathbf{F} \oplus \overline{\mathbf{F}} &\to \mathbf{F}_0 \oplus \mathbf{1}_{1/4} \oplus \mathbf{F}_{1/2} \oplus \mathbf{1}_{3/4} \, .
\end{align}
We summarize such rules for twisted compactifications in Table~\ref{table:affine_KK}.
\begin{table}
	\centering
	\begin{tabular}{|c|c|rcl|} \hline
		$ \hat{\mathfrak{g}} $ & $ \mathfrak{h} $ & $ \mathfrak{R}_{\mathfrak{g}} $ & $ \to $ & $ \mathfrak{R}_{\mathfrak{h}} $  \\ \hline
		\multirow{2}{*}{$ A_{2\ell}^{(2)} $} & \multirow{2}{*}{$ C_\ell $} & $ \mathbf{Adj} $ & $ \to $ & $ \mathbf{Adj}_0 \oplus \mathbf{F}_{1/4} \oplus \mathbf{\Lambda}^2_{1/2} \oplus \mathbf{1}_{1/2} \oplus \mathbf{F}_{3/4}  $ \\ 
		& & $ \mathbf{F} \oplus \overline{\mathbf{F}} $ & $ \to $ & $ \mathbf{F}_0 \oplus \mathbf{1}_{1/4} \oplus \mathbf{F}_{1/2} \oplus \mathbf{1}_{3/4} $ \\ \hline
		\multirow{3}{*}{$ A_{2\ell-1}^{(2)} $} & \multirow{3}{*}{$ C_\ell $} & $ \mathbf{Adj} $ & $ \to $ & $ \mathbf{Adj}_0 \oplus \mathbf{\Lambda}^2_{1/2} $ \\
		& & $ \mathbf{F} \oplus \overline{\mathbf{F}} $ & $ \to $ & $ \mathbf{F}_{0} \oplus \mathbf{F}_{1/2} $ \\
		& & $ \mathbf{\Lambda}^2 \oplus \overline{\mathbf{\Lambda}^2} $ & $ \to $ & $ \mathbf{\Lambda}^2_0 \oplus \mathbf{\Lambda}^2_{1/2} \oplus \mathbf{1}_0 \oplus \mathbf{1}_{1/2} $ \\ \hline
		\multirow{3}{*}{$ D_{\ell+1}^{(2)} $} & \multirow{3}{*}{$ B_\ell $} & $ \mathbf{Adj} $ & $ \to $ & $ \mathbf{Adj}_0 \oplus \mathbf{F}_{1/2} $ \\
		& & $ \mathbf{F} $ & $ \to $ &$ \mathbf{F}_0 \oplus \mathbf{1}_{1/2} $ \\
		& & $ \mathbf{S} \oplus \mathbf{C} $ & $ \to $ & $ \mathbf{S}_0 \oplus \mathbf{S}_{1/2} $ \\ \hline
		\multirow{2}{*}{$ E_6^{(2)} $} & \multirow{2}{*}{$ F_4 $} & $ \mathbf{Adj} $ & $ \to $ & $ \mathbf{Adj}_0 \oplus \mathbf{F}_{1/2} $ \\
		& & $ \mathbf{F} \oplus \overline{\mathbf{F}} $ & $ \to $ & $ \mathbf{F}_0 \oplus \mathbf{F}_{1/2} \oplus \mathbf{1}_0 \oplus \mathbf{1}_{1/2} $ \\ \hline
		\multirow{2}{*}{$ D_4^{(3)} $} & \multirow{2}{*}{$ G_2 $} & $ \mathbf{Adj} $ & $ \to $ & $ \mathbf{Adj}_0 \oplus \mathbf{F}_{1/3} \oplus \mathbf{F}_{2/3} $ \\
		& & $ \mathbf{F} \oplus \mathbf{S} \oplus \mathbf{C} $ & $ \to $ & $ \mathbf{F}_0 \oplus \mathbf{1}_0 \oplus \mathbf{F}_{1/3} \oplus \mathbf{1}_{1/3} \oplus \mathbf{F}_{2/3} \oplus  \mathbf{1}_{2/3} $ \\ \hline
	\end{tabular}
	\caption{Summary of KK-momentum shifts on each representation under twisted compactification. When $ \hat{\mathfrak{g}} = X_\ell^{(r)} $, $ \mathfrak{R}_\mathfrak{g} $ and $ \mathfrak{R}_\mathfrak{h} $ denote representation in $ \mathfrak{g} = X_\ell $ and $ \mathfrak{h} $, respectively. The subscript in the representations of $ \mathfrak{R}_{\mathfrak{h}} $ denotes KK-momentum shift.}\label{table:affine_KK}
\end{table}

%% file: sec-appendix2.tex
\section{5-brane webs for theories of frozen singularities}\label{sec:appendix2}

In this section, we present 5-brane webs for some theories of frozen singularities, which contain an $O7^+$-plane. 

\subsection{\texorpdfstring{5-brane webs for $SU(2)_\pi+1\mathbf{Adj}$ and local $\mathbb{P}^2+``1\mathbf{Adj}"$}{5-brane webs for SU(2)pi + 1Adj and local P2 + 1Adj}}

\begin{figure}
	\includegraphics[width=5cm]{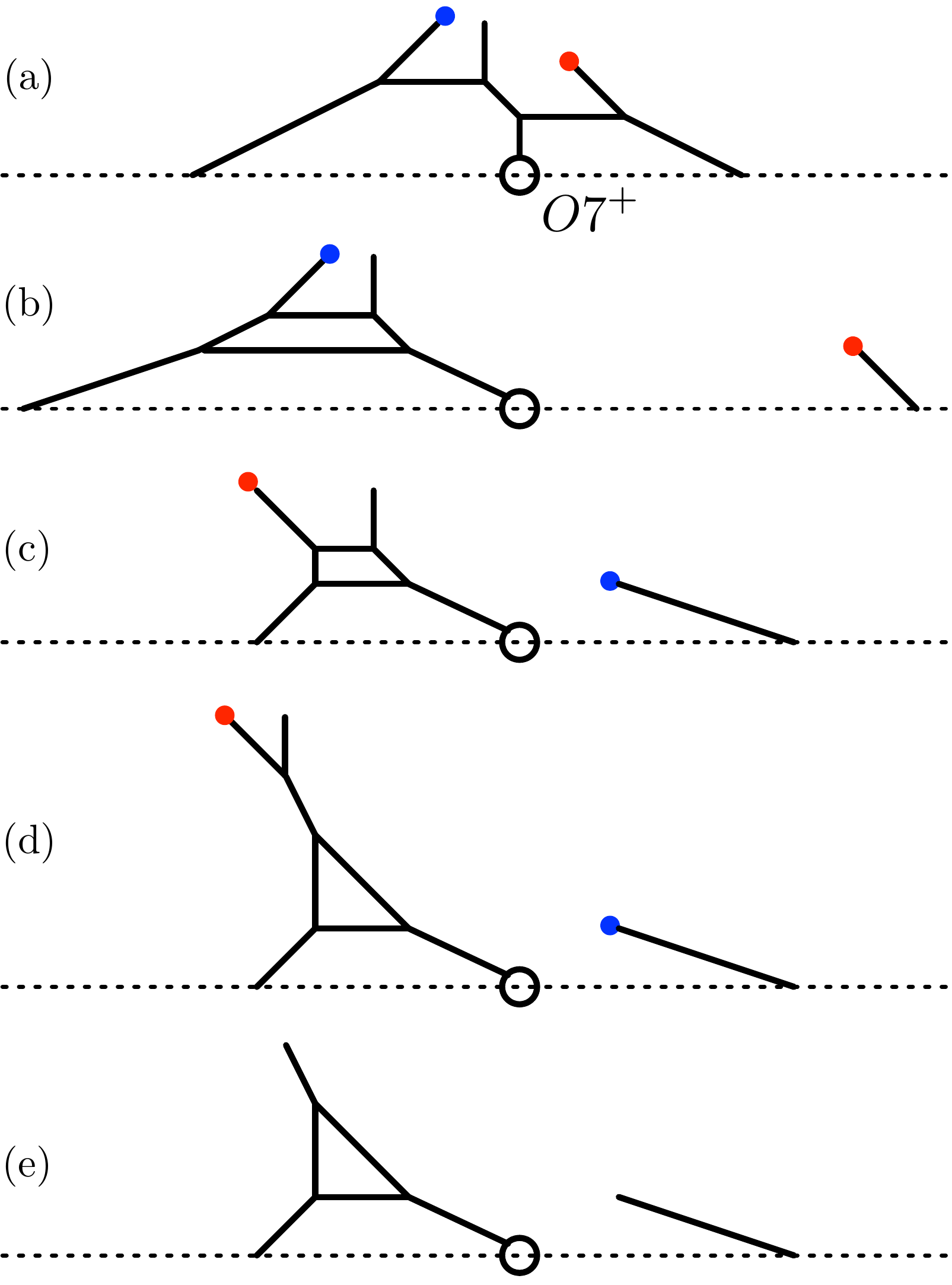}
	\centering	
	\caption{(a) A 5-brane web for $SU(2)_\pi+1\mathbf{Adj}$. (b) An equivalent 5-brane web for $SU(2)_\pi+1\mathbf{Adj}$ where the D5 brane located on the right hand side of an $O7^+$-plane in figure (a) is moved to the left hand side. (c) Hanany-Witten transition associated with the (1,1) 7-brane painted in blue in figure (b), going through the cut of the $O7^+$-plane. (d) A flop transition. (e) Decoupling the flopped `instantonic' hypermultiplet, giving rise to a 5-brane web for the local $\mathbb{P}^2+``1\mathbf{Adj}"$ theory. Clearly putting an $O7^+$-plane far away leads to a local $\mathbb{P}^2$ which corresponds to decoupling ``$1\mathbf{Adj}$". }
	\label{fig:su2+1adj}
\end{figure}

A 5-brane configuration for $SU(2)_\pi+1\mathbf{Adj}$ is depicted in Figure~\ref{fig:su2+1adj}(a). A little bit of deformation of this 5-brane web leads to a 5-brane web given in Figure~\ref{fig:su2+1adj}(b). By taking the $(1,-1)$ 7-brane painted in red through the cut of an $O7^+$-plane, one gets a 5-brane web in Figure~\ref{fig:su2+1adj}(c). Notice here that on the left hand side, the 5-brane configuration looks like ${\rm d}\mathbb{P}_1$ geometry locally. Hence, if one decouples the adjoint hypermultiplet by taking an $O7^+$ far away, then one obtains a 5-brane web for the pure $SU(2)_\pi$ theory as expected. On the other hand, recalling that ${\rm d}\mathbb{P}_1$ has an $\mathcal{O}(-1)$ curve, we can flop this $-1$ curve, which gives a 5-brane web given in Figure~\ref{fig:su2+1adj}(d). By taking this flopped part away, one finds that the remaining part is a local $\mathbb{P}^2$ with the adjoint hypermultiplet inherited from $SU(2)_\pi+1\mathbf{Adj}$, as depicted in Figure \ref{fig:su2+1adj}(e). which we have referred to as $\mathbb{P}^2+``1\mathbf{Adj}"$. One can also view this decoupling as decoupling the `instantonic' hypermultiplet as discussed in the main text.

\subsection{\texorpdfstring{5-brane web for $SU(2)_0+1\mathbf{Adj}$}{5-brane web for SU(2)0 + 1Adj}}

\begin{figure}[t]
\includegraphics[width=8cm]{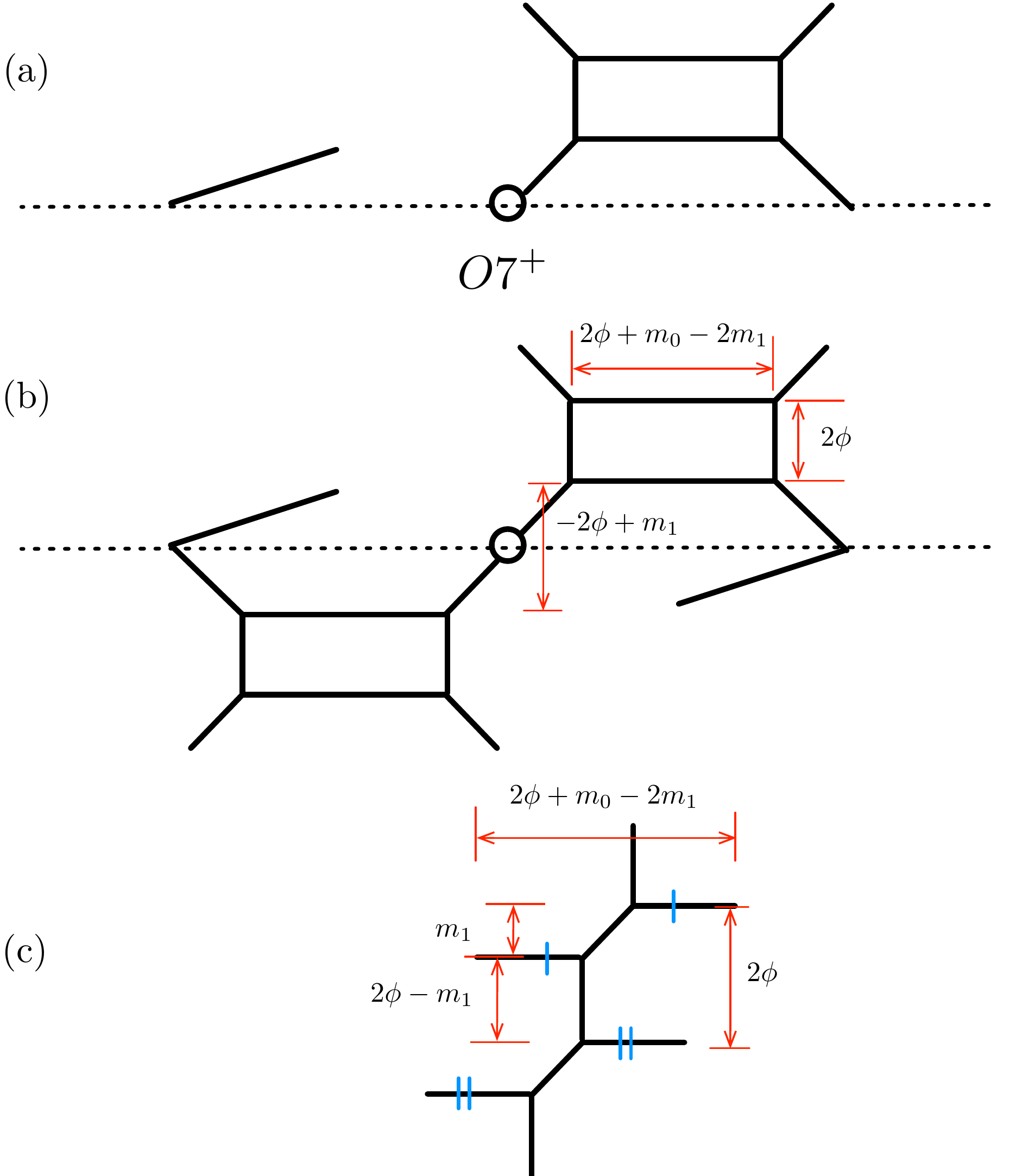}
\centering	
\caption{(a) A 5-brane web for $SU(2)_0+1\mathbf{Adj}$. (b) The reflected image of (a) is included.  (c) A 5-brane configuration for the M-string.}
\label{fig:su2_0+1adj}
\end{figure}

$SU(2)_0+1\mathbf{Adj}$ is a KK theory which is also referred to as the M-string. The corresponding web, depicted in Figure~\ref{fig:su2_0+1adj}(c), was discussed in \cite{Haghighat:2013gba}. We note that one also depicts it with an $O7^+$-plane as given in Figure~\ref{fig:su2_0+1adj}(a). It is not difficult to see that the K\"ahler parameters of these two 5-brane webs are equivalent as shown in Figure~\ref{fig:su2_0+1adj}(b) and Figure \ref{fig:su2_0+1adj}(c).

\subsection{\texorpdfstring{5-brane web for $Sp(2)_{0}+1\mathbf{Adj}$}{5-brane web for Sp(2)0 + 1Adj}}

\begin{figure}[t]
	\includegraphics[width=7cm]{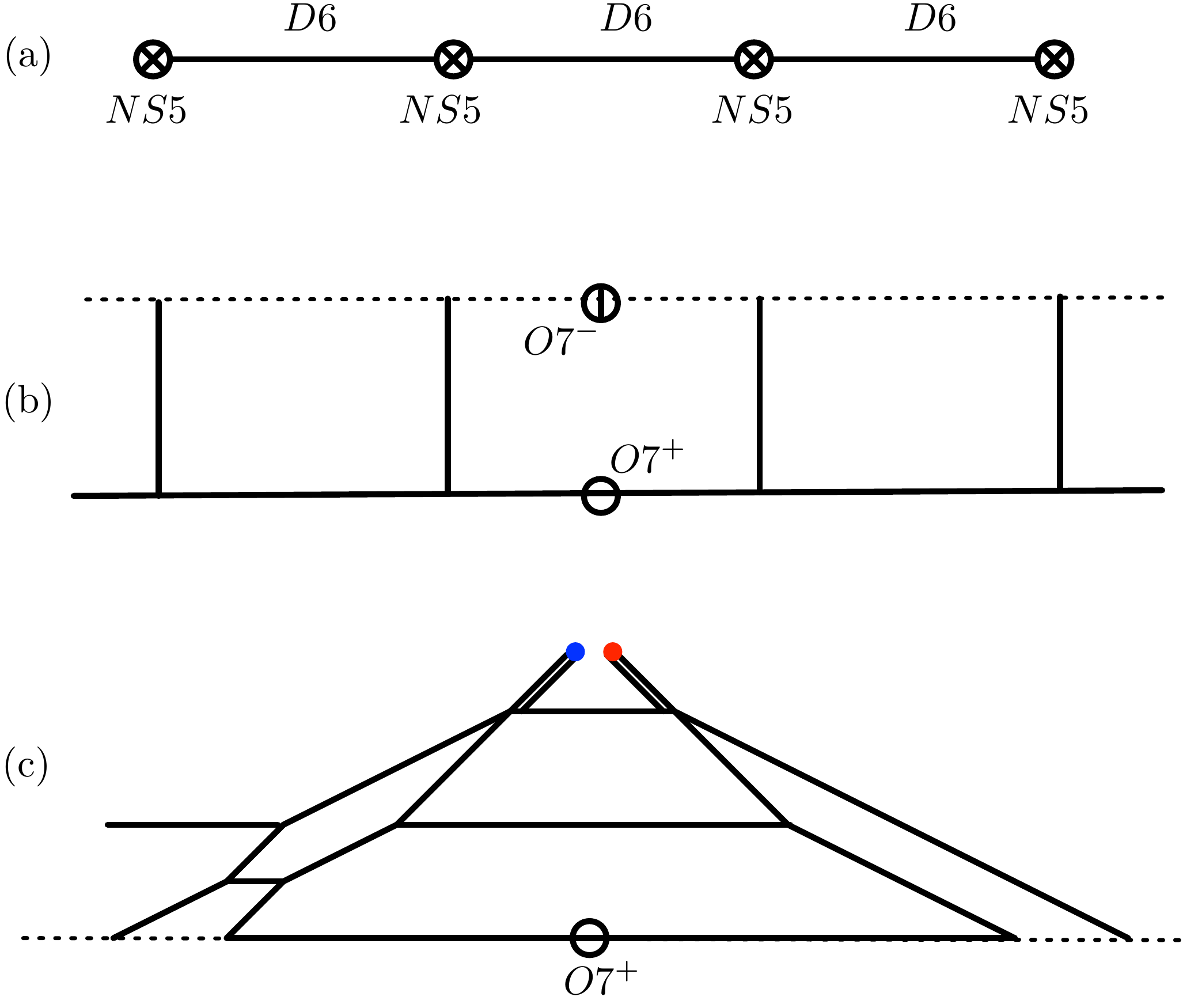}
	\centering	
	\caption{(a) Type IIA configuration for 6d $SU(1)-SU(1)-SU(1)$ theory or equivalently $SU(4)$ theory. (b) A $\mathbb{Z}_2$ twisted compactification of (a), where 1/2 $D5$-branes are stuck along the cut of an $O7^+$-plane. (c) Type IIB 5-brane configuration after resolving an $O7^-$ into two 7-branes of charge $(1,1)$ and $(1, -1)$. Here a blue dot refers to a $(1,1)$ 7-brane while a red dot refers to a $(1,-1)$ 7-brane.}
	\label{fig:Sp2_0+1Adj}
\end{figure}

A 5-brane web for 5d $Sp(2)_0+1\mathbf{Adj}$ can be obtained by a $\mathbb{Z}_2$ twisting of the 6d $\mathcal{N}=(2,0)$ $A_3$ theory whose brane configuration can be realized as a $D6$ brane suspended between 4 $NS5$ branes, as depicted in Figure~\ref{fig:Sp2_0+1Adj}(a).
The procedure of a $\mathbb{Z}_2$ twisting of the 6d $\mathcal{N}=(2,0)$ $A_N$ theory is already considered in \cite{Hayashi:2015vhy}. Such twisted compactification on 5-brane gives rise to a pair of $O7^-$- and $O7^+$-planes and the number of $D6$ branes in Type IIA brane configuration is halved to yield half the number of $D5$-branes. For this $A_3$ case, the resulting 5-brane configuration is given in Figure~\ref{fig:Sp2_0+1Adj}(b), where a half $D5$ is stuck along the cut of an $O7^+$-plane. By resolving an $O7^-$ into two 7-branes of charges $(1,1)$ and $(1, -1)$, one gets a 5-brane configuration for $Sp(2)_0+1\mathbf{Adj}$ as depicted in Figure~\ref{fig:Sp2_0+1Adj}(c). We note that in Figure~\ref{fig:Sp2_0+1Adj}(c), we moved two half $D5$-branes on the right hand side of an $O7^+$ to the left to form a full $D5$-brane so that they then can be away from the $O7^+$ cut at the bottom, which allows us to give the mass of the adjoint hypermultiplet.

\subsection{\texorpdfstring{5-brane web for $Sp(2)_\pi+1\mathbf{Adj},~ SU(3)_{\frac32}+1\mathbf{Sym}$}{5-brane web for Sp(2)pi + 1Adj}}

\begin{figure}
\centering	
\includegraphics[width=6cm]{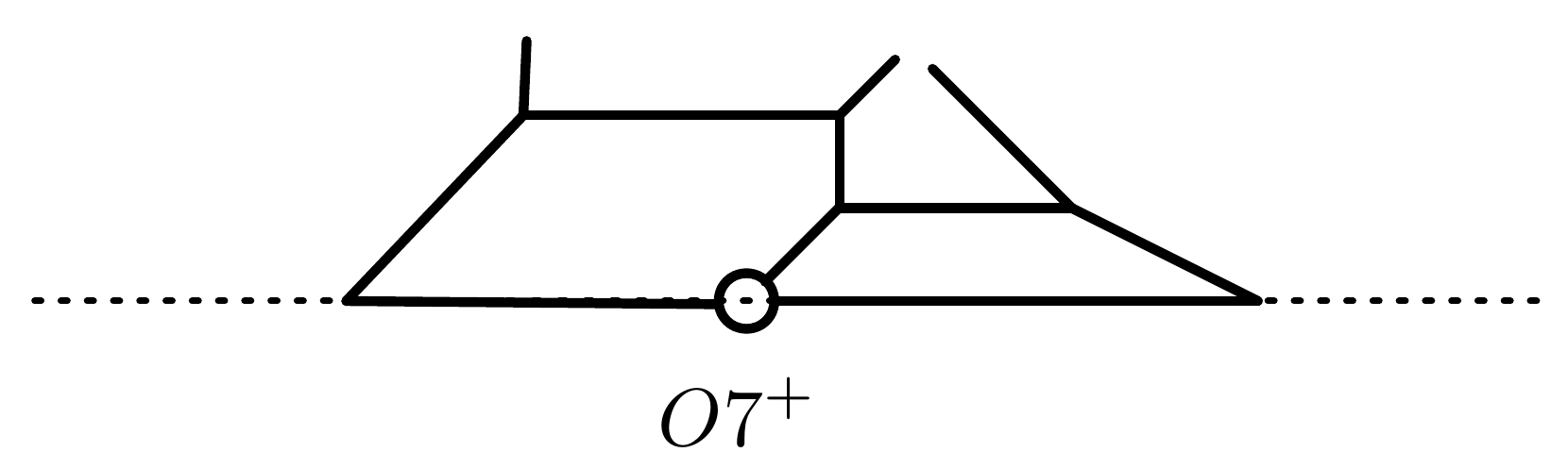}
\caption{A 5-brane web for $Sp(2)_\pi + 1\mathbf{Adj}$ or $SU(3)_\frac32+1\mathbf{Sym}$.}
\label{fig:Sp2_pi+1Adj}
\end{figure}

\begin{figure}[t]
	\includegraphics[width=7cm]{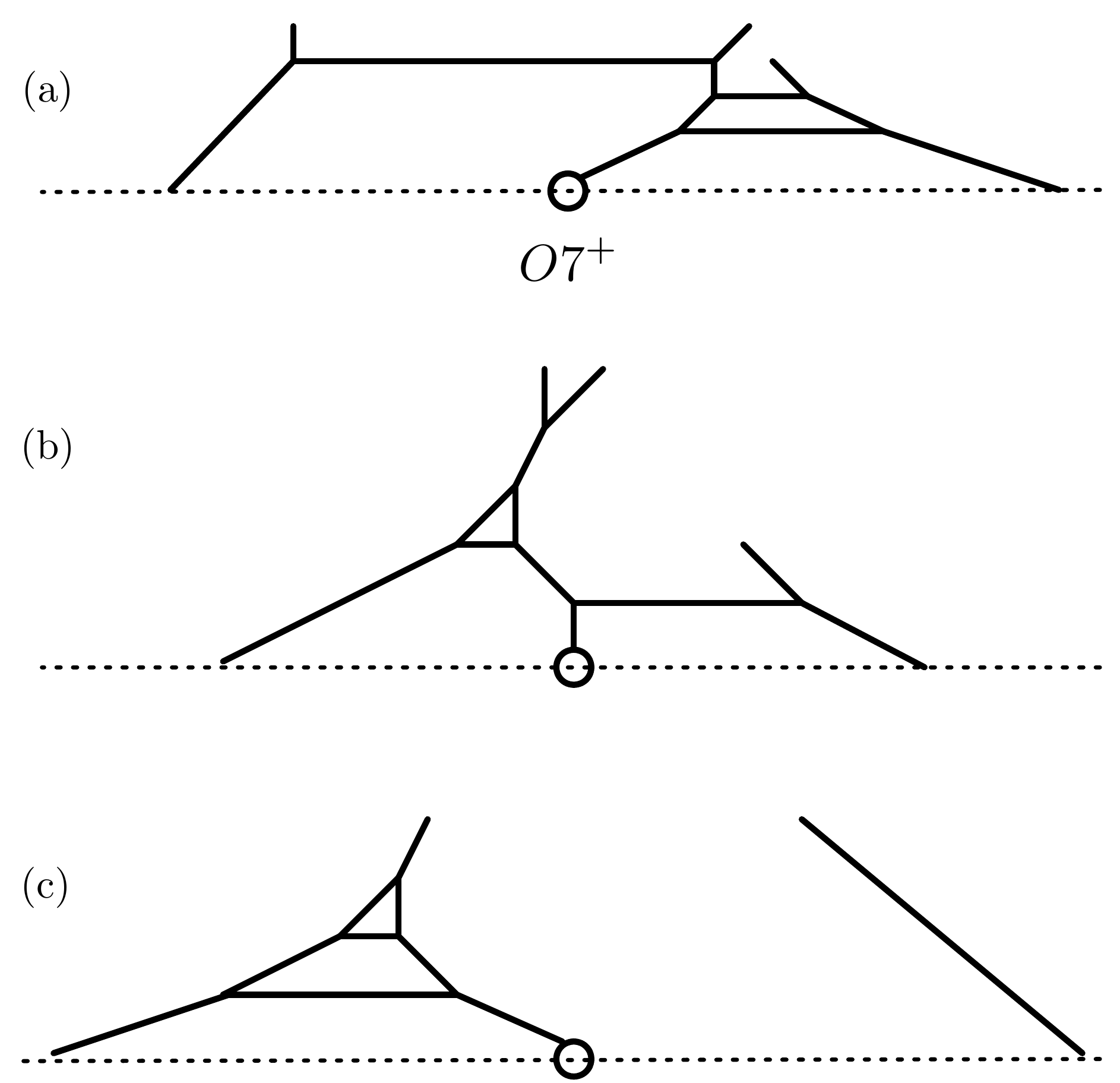}
	\centering	
	\caption{(a) A 5-brane web for $SU(3)_\frac32+1\mathbf{Sym}$ which has a $-1$ curve on the left hand side of an $O7^+$-plane. (b) A flop transition and allocating a D5-brane from the right to the left. (c)  Decoupling the `instantonic' hypermultiplet, which leads to a 5-brane configuration for a non-Lagrangian theory, $\mathbb{P}^2\cup\mathbb{F}_3+``1\mathbf{Sym}."$}
	\label{fig:F3+P2+1Sym}
\end{figure}

A 5-brane web for $Sp(2)_\pi + 1\mathbf{Adj}$ can be obtained a $\mathbb{Z}_2$ twisted compactification of the 6d $\mathcal{N}=(2,0)$ $A_4$ theory. The resulting web is given in Figure~\ref{fig:Sp2_pi+1Adj}, where two half $D5$ branes are suspended either between an $O7^+$ and $(1,1)$ 5-brane or between an $O7^+$ and $(-2, 1)$ 5-brane. 
As before, two half $D5$-branes can be put together to form a full $D5$-brane. It can be then easily recognized that the resulting 5-brane configuration is nothing but that for $SU(3)_\frac32+1\mathbf{Sym}$ as given in Figure~\ref{fig:F3+P2+1Sym}(a).

\subsection{\texorpdfstring{5-brane web for $\mathbb{P}^2\cup\mathbb{F}_3+``1\mathbf{Sym}"$ or $ ``SU(3)_{\frac32}+1\mathbf{Sym}-1\mathbf{F}"$}{5-brane web for P2 U F3 + 1Sym}}\label{sec:app-P2 U F3+1Sym}
In a similar way as done for the 5-brane configuration for $\mathbb{P}^2+1\mathbf{Adj}$, one can easily get a 5-brane configuration for $\mathbb{P}^2\cup\mathbb{F}_3+``1\mathbf{Sym}"$ or $ ``SU(3)_{\frac32}+1\mathbf{Sym}-1\mathbf{F}"$ as depicted in Figure~\ref{fig:F3+P2+1Sym}. Given a 5-brane configuration for $SU(3)_\frac32+1\mathbf{Sym}$ in Figure~\ref{fig:F3+P2+1Sym}(a), which has a $-1$ curve, one performs a flop transition to yield 5-brane web in Figure~\ref{fig:F3+P2+1Sym}(b). Decoupling this `instantonic' hypermultiplet gives rise to $\mathbb{P}^2\cup\mathbb{F}_3+``1\mathbf{Sym}"$ whose 5-brane web is given in Figure~\ref{fig:F3+P2+1Sym}(c), which we may be referred to as $ ``SU(3)_{\frac32}+1\mathbf{Sym}-1\mathbf{F}."$

\subsection{5-brane web for \texorpdfstring{$SU(3)_{0}+1\mathbf{Sym}+1\mathbf{F}$}{SU(3)0 + 1Sym + 1F}}
\begin{figure}[t]
\includegraphics[width=6cm]{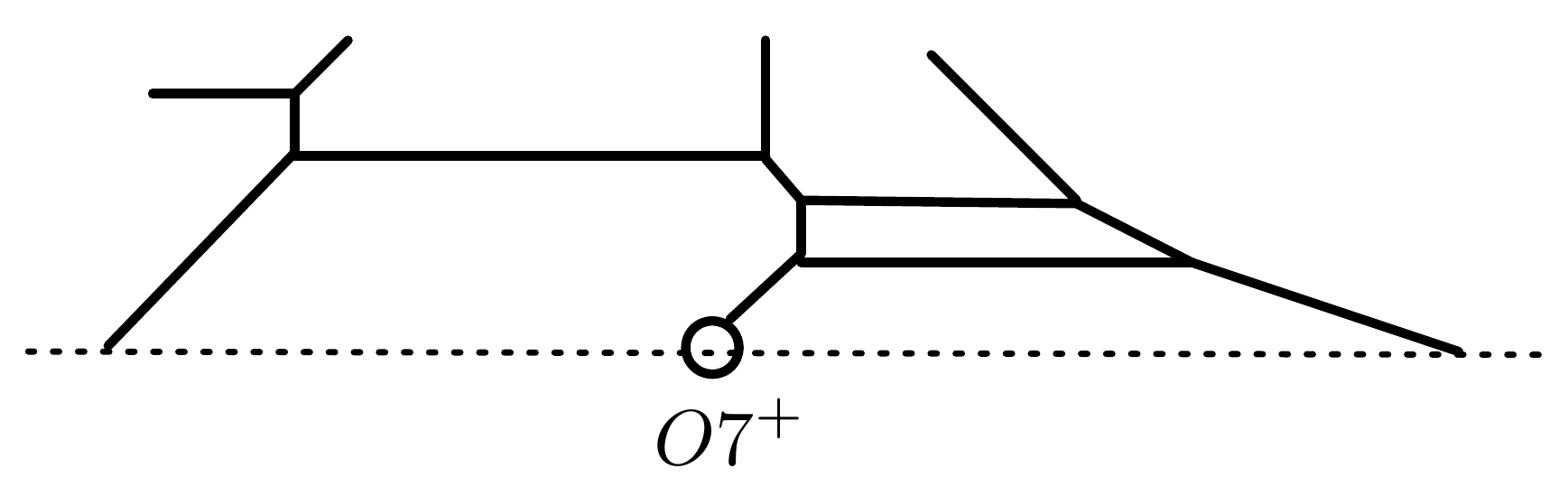}
\centering	
\caption{A 5-brane configuration for $SU(3)_0+1\mathbf{Sym}+1\mathbf{F}$.}
\label{fig:SU3_0+1Sym+1F}
\end{figure}

A 5-brane configuration for $SU(3)_{0}+1\mathbf{Sym}+1\mathbf{F}$ is depicted in Figure~\ref{fig:SU3_0+1Sym+1F}, which is discussed in detail in \cite{Hayashi:2018lyv}.

\subsection{5-brane web for \texorpdfstring{local $\mathbb{P}^2\cup \mathbb{F}_6 +``1\mathbf{Sym}"$}{P2 U F6 + 1Sym}}\label{sec:app-P^2F_6+1Sym}

\begin{figure}
	\includegraphics[width=14cm]{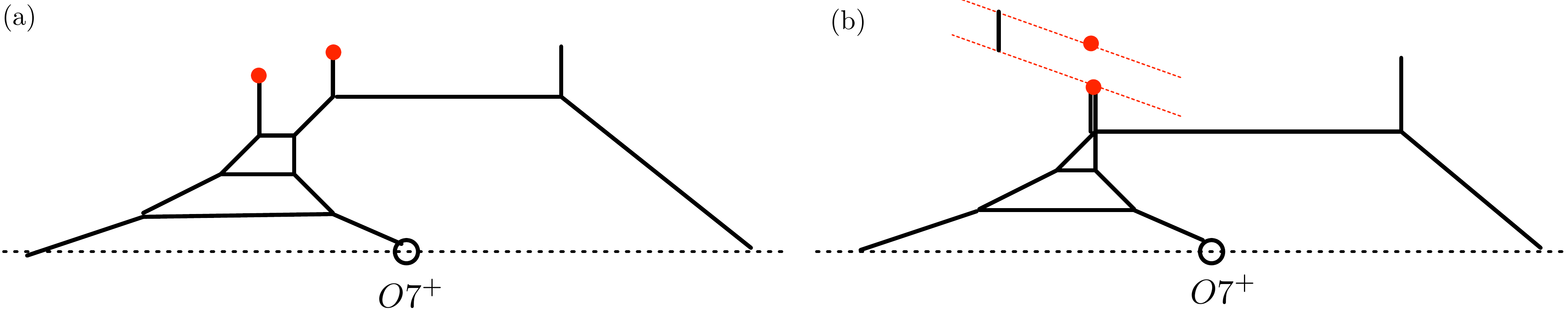}
	\centering	
	\caption{(a) A 5-brane web for $SU(4)_0+1\mathbf{Sym}$. (b) A non-perturbative Higgsing by putting two $NS5$-branes together and perform a Higgsing such that the $NS5$-brane ending two $(0,1)$ 7-branes is taken away along the $x^{7,8,9}$-directions.}
	\label{fig:SU4_0+1Sym}
\end{figure}

\begin{figure}
	\includegraphics[width=8cm]{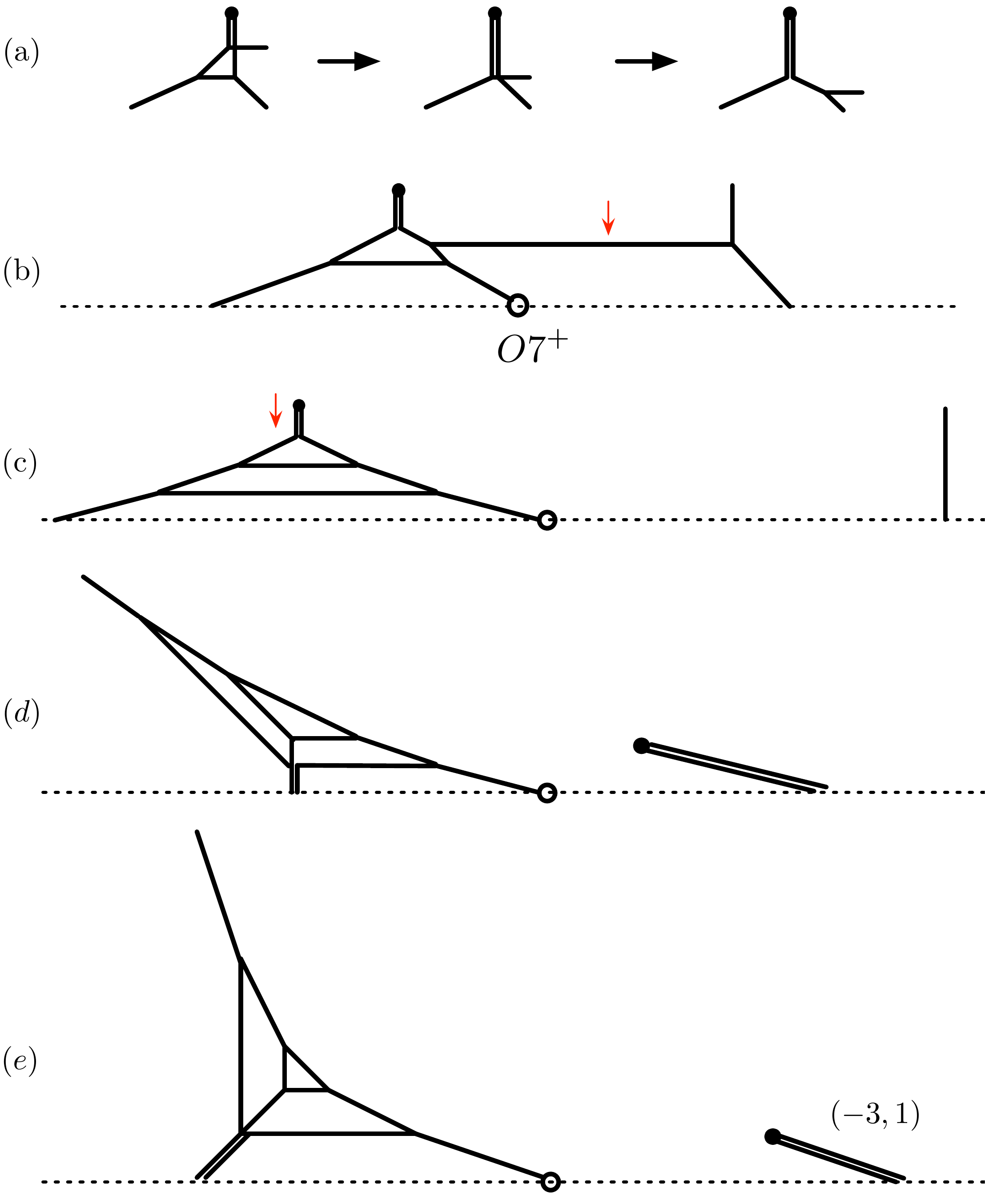}
	\centering	
	\caption{Deformations of the 5-brane web given in Figure \ref{fig:SU4_0+1Sym}(b) leading to a 5-brane web for $\mathbb{P}^2\cup \mathbb{F}_6 +``1\mathbf{Sym}"$.}
	\label{fig:P2-F6+1Sym}
\end{figure}

A 5-brane web for $\mathbb{P}^2\cup \mathbb{F}_6 +``1\mathbf{Sym}"$ is depicted in Figure~\ref{fig:P2-F6+1Sym}(e). It is worthy of noting that this 5-brane web can be obtained from a non-perturbative Higgsing of $SU(4)_0+1\mathbf{Sym}$ in Figure~\ref{fig:SU4_0+1Sym}.
By taking a non-perturbative Higgsing\footnote{A simple non-perturbative Higgsing procedure is from $SU(3)_0+5\mathbf{F}$ to $SU(2)+5\mathbf{F}$ where two parallel $NS5$-branes are bound together in such a way that the resulting 5-brane web preserves the S-rule \cite{Hayashi:2013qwa, Kim:2014nqa}.} shown in Figure~\ref{fig:SU4_0+1Sym}(b), we get a new rank-2 non-Lagrangian theory, $\mathbb{P}^2\cup \mathbb{F}_6 +``1\mathbf{Sym}"$. To see local geometry apart from the frozen singularity, consider a series of deformations of the 5-brane web given in Figure~\ref{fig:P2-F6+1Sym}. 
Lowering the $D5$-brane on the right side of an $O7^+$-plane leads to a transition like Figure \ref{fig:P2-F6+1Sym}(a) discussed in \cite{Hayashi:2017btw} giving rise to a 5-brane web in Figure \ref{fig:P2-F6+1Sym}(b). A further lowering of the $D5$-brane allocates the $D5$-brane from the right to the bottom left as depicted in Figure \ref{fig:P2-F6+1Sym}(c). The resulting 5-brane configuration then looks like locally $\mathbb{P}^2\cup \mathbb{F}_6$. Taking into account $1\mathbf{Sym}$ inherited from $SU(4)_0+1\mathbf{Sym}$, we call this $ \mathbb{P}^2\cup \mathbb{F}_6 +``1\mathbf{Sym}"$.

\begin{figure}[t]
\includegraphics[width=15cm]{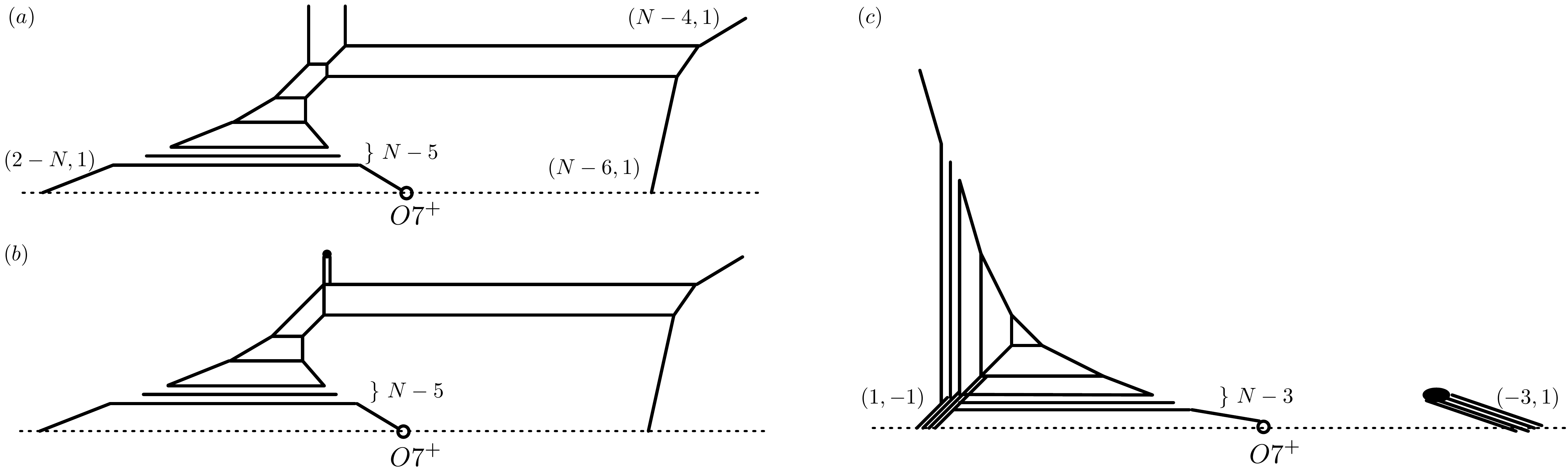}
\centering	
\caption{(a) A 5-brane web for $SU(N)_\frac{4-N}{2}+1\mathbf{Sym}$, where there are $N$ D5-branes in total. (b) A non-perturbative Higgsing. (c) 5-brane web for $\mathbb{P}^2\cup \mathbb{F}_6 \cup \mathbb{F}_8 \cup \cdots \cup \mathbb{F}_{2N}+``1\mathbf{Sym}"$ which is rank of $N-2$.}
\label{fig:P2F682N}
\end{figure}

We note that it is straightforward to implement this kind of non-perturbative Higgsing to $SU(N)_\frac{4-N}{2}+1\mathbf{Sym}$ to yield rank $N-1$ non-Lagrangian theory with a frozen singularity, $\mathbb{P}^2\cup \mathbb{F}_6 \cup \mathbb{F}_8 \cup \cdots \cup \mathbb{F}_{2N}+``1\mathbf{Sym}"$, as shown in Figure~\ref{fig:P2F682N}.
\begin{figure}[t]
\includegraphics[width=15cm]{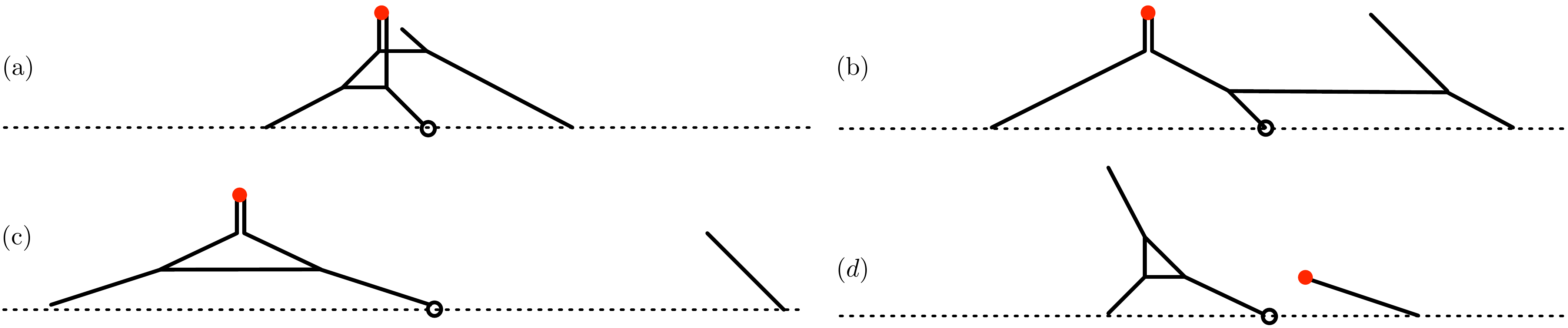}
\centering	
\caption{Various phases of 5-brane web for local $\mathbb{P}^2 +``1\mathbf{Adj}"$.}
\label{fig:localP2+1Adj-2}
\end{figure}
In particular, when $N=2$, it provides yet another way of obtaining a 5-brane web for the local $\mathbb{P}^2+``1\mathbf{Adj}"$ theory giving rise to various phases for the theory as depicted in Figure~\ref{fig:localP2+1Adj-2}(a)-(d). Note that a 5-brane configuration in Figure~\ref{fig:localP2+1Adj-2}(a) looks as if all W-bosons are legitimate so that the resulting theory is a Lagrangian theory, but it appears that some of the W-bosons would be annihilated so that the corresponding system is that of non-Lagrangian theory. \medskip \bigskip

\begin{figure}[t]
	\includegraphics[width=9cm]{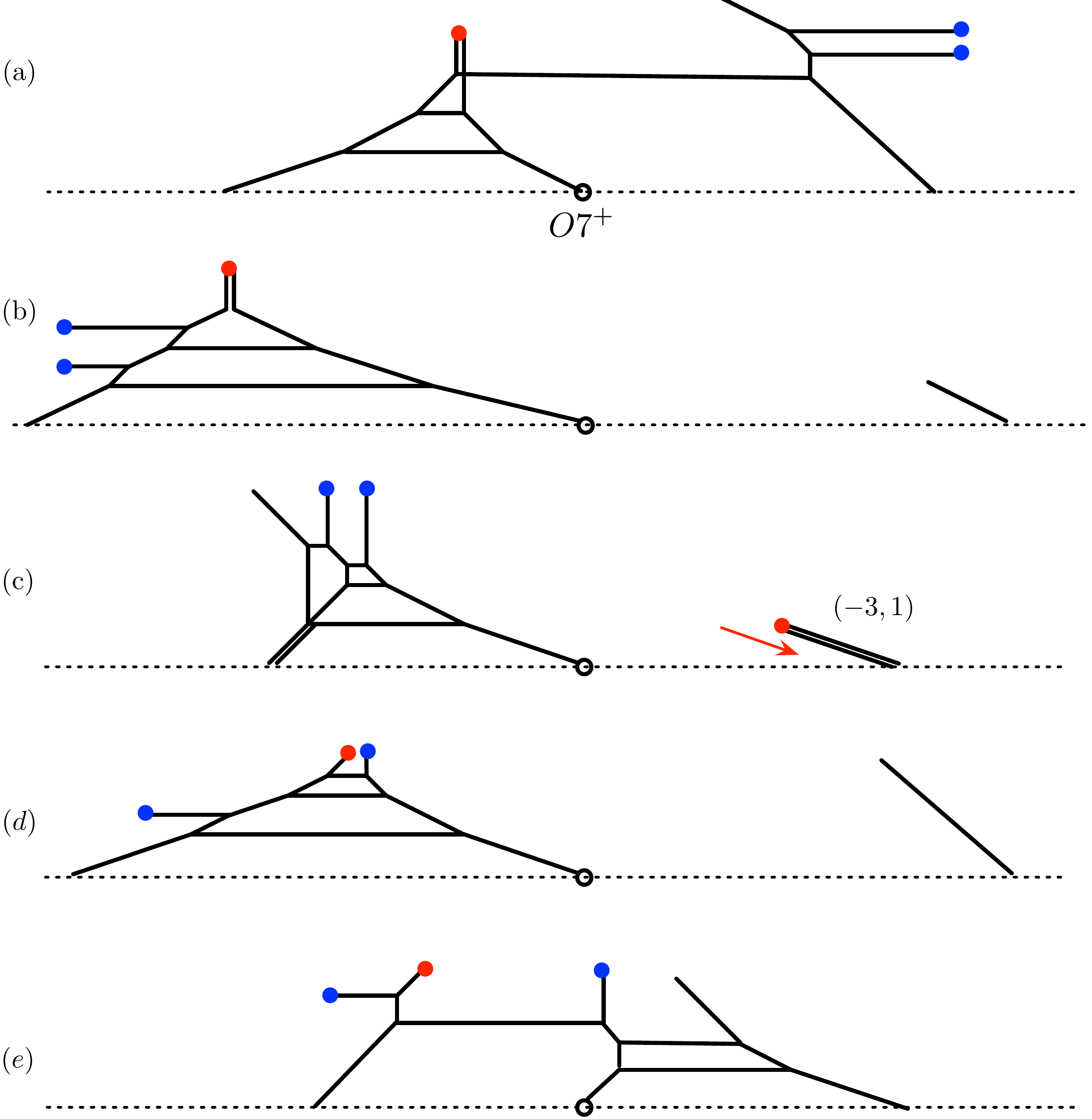}
	\centering	
	\caption{From 5-brane web for local $\mathbb{P}^2 \cup \mathbb{F}_6+``1\mathbf{Sym}+2\mathbf{F}"$ to that for $SU(3)_0 +1\mathbf{Sym}+1\mathbf{F}$. (a) A 5-brane web for local $\mathbb{P}^2 \cup \mathbb{F}_6+``1\mathbf{Sym}+2\mathbf{F}"$. (b) Locating D5-brane on the right-hand side of $O7^+$ to the left. (c) Pull down the $(0,1)$ 7-brane (in red) in figure (b) toward the cut of $O7^+$ and perform an $SL(2,\mathbb{Z})$ transformation. (d) With some mass deformations, Hanany-Witten move with the $(-3, 1)$ 7-brane in figure (c) which gives rise to an 5-brane web for $SU(3)_0 +1\mathbf{Sym}+1\mathbf{F}$. (d) A little deformation leading to the same web for $SU(3)_0 +1\mathbf{Sym}+1\mathbf{F}$ given in Figure~\ref{fig:SU3_0+1Sym+1F}.}
	\label{fig:localP2+1Adj+2F}
\end{figure}

\noindent\underline{Equivalence between $\mathbb{P}^2\cup \mathbb{F}_6 +``1\mathbf{Sym}+2\mathbf{F}"$ and $SU(3)_{0}+1\mathbf{Sym}+1\mathbf{F}$} \\
We can show that $\mathbb{P}^2\cup \mathbb{F}_6 +``1\mathbf{Sym}+2\mathbf{F}"$ and $SU(3)_{0}+1\mathbf{Sym}+1\mathbf{F}$ are equivalent by transforming a 5-brane web for $\mathbb{P}^2\cup \mathbb{F}_6 +``1\mathbf{Sym}+2\mathbf{F}"$ into that of $SU(3)_{0}+1\mathbf{Sym}+1\mathbf{F}$ that is given in Figure \ref{fig:SU3_0+1Sym+1F}. Consider a 5-brane web for $\mathbb{P}^2\cup \mathbb{F}_6 +``1\mathbf{Sym}+2\mathbf{F}"$ which is to add two flavor $D7$-branes to $\mathbb{P}^2\cup \mathbb{F}_6 +``1\mathbf{Sym}"$ in Figure~\ref{fig:SU4_0+1Sym}(b). This resulting 5-brane web is depicted in Figure~\ref{fig:localP2+1Adj+2F}(a). The deformations given in Figure~\ref{fig:localP2+1Adj+2F}(b) to \ref{fig:localP2+1Adj+2F}(d) lead to a 5-brane web for $SU(3)_0 +1\mathbf{Adj}+1\mathbf{F}$ in Figure~\ref{fig:localP2+1Adj+2F}(e) which is the same as the one in Figure~\ref{fig:SU3_0+1Sym+1F}. This clearly suggests that there is a new RG flow from $SU(3)_0+1\mathbf{Sym}+1\mathbf{F}$ to local $\mathbb{P}^2 \cup \mathbb{F}_6+``1\mathbf{Sym}+1\mathbf{F}"$.

The above equivalence relation is readily generalized to the following $SU(N)_0$ KK theory,
\begin{align}
&SU(N)_0 +1\mathbf{Sym}+(N-2)\mathbf{F} \cr
&~~~~\text{is equivalent to }	\cr
&\text{local }\mathbb{P}^2 \cup \mathbb{F}_6 \cup \mathbb{F}_8 \cup \cdots \cup \mathbb{F}_{2N-2} \cup \mathbb{F}_{2N} +``1\mathbf{Sym}+(N-1)\mathbf{F}".
\end{align}
We note that when $N=2$, the equivalence reads
\begin{align}
	SU(2)_\pi+ 1 \mathbf{Adj}\quad \Longleftrightarrow \quad
	\text{local }\mathbb{P}^2+ 1 \mathbf{Adj} + 1\mathbf{F}.
\end{align}